\RequirePackage{fix-cm}
\documentclass[3p,times,preprint,authoryear]{elsarticle}          
\usepackage{graphicx}
\usepackage{enumitem}
\usepackage{amssymb}
\usepackage{amsmath}
\usepackage{txfonts}
\usepackage{natbib}
\usepackage{wrapfig}
\usepackage{aas_macros}
\usepackage[a4paper=true,
 breaklinks=true,%
 colorlinks=true,%
 pdfauthor={Amati et al.},%
 pdftitle={The THESEUS space mission concept: science case, design and expected performances}%
 ]{hyperref}

\journal{Advances in Space Research}

\begin{document}

\begin{frontmatter}

\title{The THESEUS space mission concept: science case, design and expected performances}

\author{L. Amati; P.~O'Brien; D.~G\"otz; E.~Bozzo; C.~Tenzer; F.~Frontera; G.~Ghirlanda; C.~Labanti; J.~P.~Osborne; G.~Stratta; N.~Tanvir
% ...
}

\address{\bf Full author list and affiliations are provided at the end of the paper.}

\begin{abstract}
THESEUS is a space mission concept aimed at exploiting Gamma-Ray Bursts for investigating 
the early Universe and at providing a substantial advancement of multi-messenger and time-domain astrophysics. 
These goals will be achieved through a unique combination of instruments allowing GRB and X-ray transient detection 
over a broad field of view (more than 1sr) with 0.5-1 arcmin localization, an energy band extending from several MeV down to 0.3~keV 
and high sensitivity to transient sources in the soft X-ray domain, as well as on-board prompt (few minutes) follow-up with a 0.7~m class IR 
telescope with both imaging and spectroscopic capabilities. THESEUS will be perfectly suited for addressing the main open issues 
in cosmology such as, e.g., star formation rate and metallicity evolution of the inter-stellar and intra-galactic medium up to redshift 
$\sim$10, signatures of Pop III stars, sources and physics of re-ionization, and the faint end of the galaxy luminosity function. 
In addition, it will provide unprecedented capability to monitor the X-ray variable sky, thus detecting, localizing, and identifying the electromagnetic 
counterparts to sources of gravitational radiation, which may be routinely detected in the late '20s / early '30s by next generation facilities 
like aLIGO/ aVirgo, eLISA, KAGRA, and Einstein Telescope. THESEUS will also provide powerful synergies with the next generation of 
multi-wavelength observatories (e.g., LSST, ELT, SKA, CTA, ATHENA). 
\end{abstract}

\begin{keyword}
Gamma-ray: bursts; Cosmology: observations, dark ages, re-ionization, first stars
\end{keyword}

\end{frontmatter}

\parindent=0.5 cm

\section{Introduction}
\label{intro}

\begin{figure}
\centering
\includegraphics[scale=0.9]{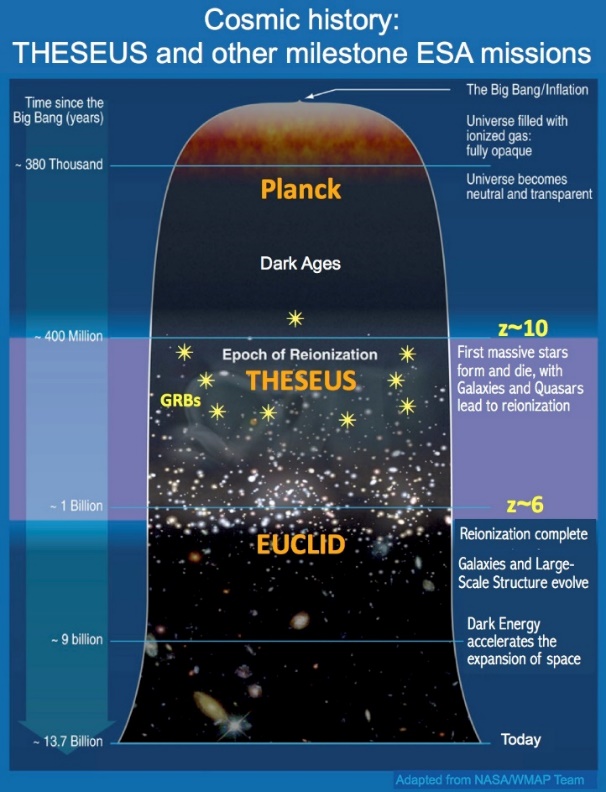}
\caption{Gamma-Ray Bursts in the cosmological context and the role of THESEUS (adapted from a picture by the NASA/WMAP Science Team).}
\label{fig:1}   
\end{figure}

\begin{figure*}[t!]
\centering
\includegraphics[scale=0.4]{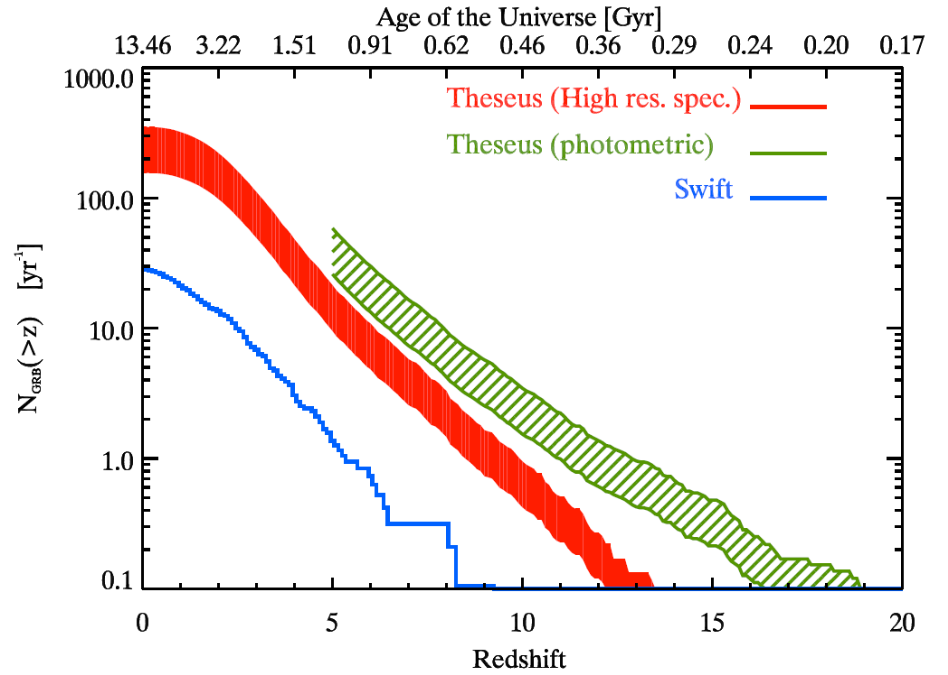}
\caption{ 
The yearly cumulative distribution of GRBs with redshift determination as a function of the 
redshift for Swift and THESEUS (see details in Sect.~\ref{science1}). 
We note that these predictions are conservative in so far as they reproduce the current GRB rate as 
a function of redshift. However, with our sensitivity, we can detect a GRB of $E_{\rm iso}\sim10^{53}$~erg 
(corresponding to the median of the GRB radiated energy distribution) up to $z$=12. Indeed, our poor knowledge 
of the GRB rate-SFR connection does not preclude the existence of a sizable number of GRBs at such high redshifts, in 
keeping with recent models of Pop III stars.}
\label{fig:4}   
\end{figure*}

The Transient High Energy Sky and Early Universe Surveyor (THESEUS) is a space mission concept 
developed by a large international collaboration in response to the calls for M-class missions by the European Space Agency (ESA).  
THESEUS is designed to vastly increase the discovery space of high energy transient phenomena over the entirety of cosmic history (Fig.~\ref{fig:1} and Fig.~\ref{fig:4}).  
Its driving science goals aim at finding answers to multiple fundamental questions of modern cosmology and astrophysics, exploiting 
the mission unique capability to: 
a) explore the physical conditions of the Early Universe (the cosmic dawn and re-ionization era) by unveiling the Gamma-Ray Burst 
(GRB) population in the first billion years; b) perform an unprecedented deep monitoring of the soft X-ray transient Universe, 
thus providing a fundamental synergy with the next-generation of gravitational wave and neutrino detectors (multi-messenger astrophysics), 
as well as the large electromagnetic (EM) facilities of the next decade. 

The most critical THESEUS targets, i.e. GRBs, are unique and powerful tools for cosmology, especially because 
of their huge luminosities, mostly emitted in X- and gamma-rays, 
their redshift (z) distribution (extending at least to z$\sim$10), and their association with the explosive death of massive stars. 
In particular, GRBs represent a unique tool to study the early Universe up to the re-ionization era. To date, there is no consensus 
on the dominant sources of re-ionization, and GRB progenitors and their hosts are very good representatives of the massive stars 
and star-forming galaxies that may have been responsible for this cosmic phase change. A statistical sample of high-z GRBs (between 30 and 80 at z$>$6, see Fig.~\ref{fig:4}) 
can provide fundamental information such as: measuring independently the cosmic star formation rate, even beyond the limits of 
current and future galaxy surveys, the number density and properties of low-mass galaxies, the neutral hydrogen fraction, 
and the escape fraction of UV photons from high-z galaxies. Even JWST and ELTs surveys, in the 2020s, will not be able to probe 
the faint end of the galaxy luminosity function (LF) at high redshifts (z$>$6-8). 
The first generation of metal-free stars (the so-called Pop III stars) and
the second generation of massive, metal-poor stars (the so-called Pop II
stars) can result in powerful GRBs \citep[see, e.g.,][]{Meszaros2010} that
thus offer a direct route to identify such elusive objects (even JWST will not be able to detect them directly) 
and study the galaxies in which they are hosted. 
Even indirectly, the  role of Pop III stars in enriching the first galaxies with metals can be studied by looking at the 
absorption features of Pop II GRBs blowing out in a medium enriched by the first Pop III supernovae \citep{Ma2017}. 
Moreover, both Pop III and massive Pop II stars may have contributed to
the reionization of the Universe due to their intensive ionizing radiation
\citep{Yoon2012, Szecsi2015}, thus detecting their final
explosion will help us better understand our cosmic history. 

\begin{figure*}[t!]
\centering
\includegraphics[scale=1.1]{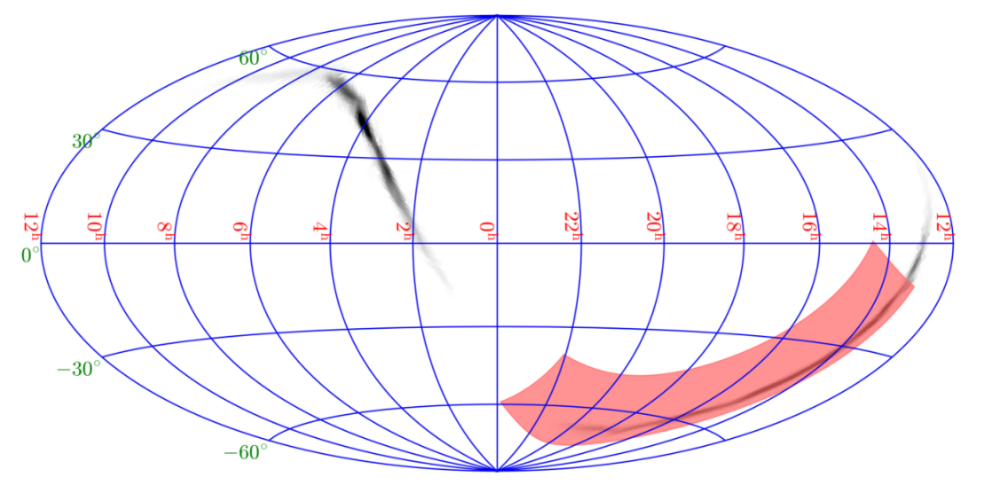}
\caption{By the end of the '20s, if Einstein Telescope will be a single detector, almost no directional information will be available for GW sources,
and a GRB-localising satellite will be essential to discover EM counterparts. If multiple third-generation detectors will be available,
the typical localization uncertainties will be comparable to the aLIGO ones. The plot shows the SXI field of view
($\sim$110$\times$30~deg$^2$) superimposed on the probability skymap of GW151226 observed by aLIGO.}
\label{fig:2}
\end{figure*}

\begin{figure*}[t!]
\centering
\includegraphics[scale=0.46]{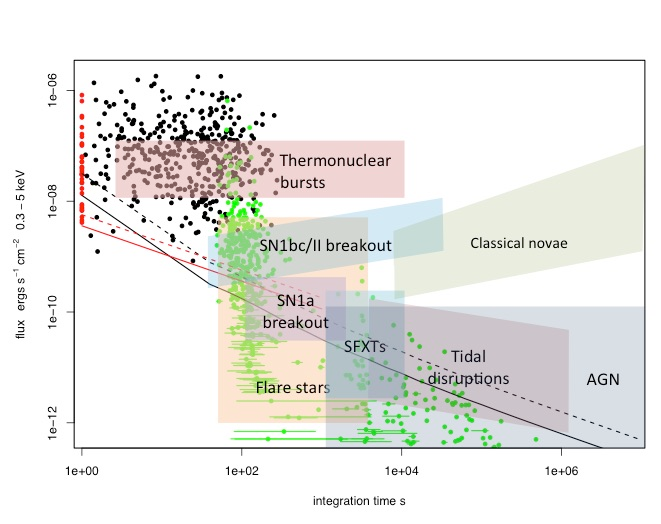}
\caption{Sensitivity of the SXI (black curves) and XGIS (red) vs. integration time. 
The solid curves assume a source column density 
of 5$\times$10$^{20}$~cm$^{-2}$ (i.e., well out of the Galactic plane and very little intrinsic absorption). The dotted curves assume a source column 
density of 10$^{22}$~cm$^{-2}$ (significant intrinsic absorption). The black dots are the peak fluxes for Swift BAT GRBs plotted against T90/2. 
The flux in the soft band 0.3-10~keV was estimated using the T90 BAT spectral fit including the absorption from the XRT spectral fit. 
The red dots are those GRBs for which T90/2 is less than 1~s. The green dots are the initial fluxes and times since trigger at 
the start of the Swift XRT GRB light-curves. The horizontal lines indicate the duration of the first time bin in the XRT light-curve. 
The various shaded regions illustrate variability and flux regions for different types of transients and variable sources.}
\label{fig:24}   
\end{figure*}

Besides high-redshift GRBs, THESEUS will serendipitously detect and localize during regular observations a large number of X-ray 
transients and variable sources (Fig.~\ref{fig:2} and Fig.~\ref{fig:24}), collecting also prompt follow-up data in the IR (see Tables~\ref{tab:1} and \ref{tab:2}). 
These observations will provide a wealth of unique science opportunities, by revealing the violent Universe as it occurs in real-time, 
exploiting an all-sky X-ray monitoring of extraordinary grasp and sensitivity carried out at high cadence. 
THESEUS will be able to locate and identify the electromagnetic counterparts to sources of gravitational radiation and neutrinos, 
which will be routinely detected in the late '20s / early '30s by next generation facilities like aLIGO/ aVirgo, eLISA, ET, 
and Km3NET. In addition, the provision of a high cadence soft X-ray monitoring capability in the 2020s together with a 0.7~m IRT in orbit will enable a 
strong synergy with transient phenomena observed with the large facilities that will be operating in the EM domain 
(e.g., ELT, SKA, CTA, JWST, ATHENA). The large number of GRBs found in the survey will permit unprecedented insights into the physics and 
progenitors of these events and their connection with peculiar core-collapse SNe. THESEUS will also substantially increase the detection rate 
and characterization of sub-energetic GRBs and X-Ray Flashes. Monitoring observations of bright and faint X-ray objects will be routinely carried out, and  
a dramatic increase in the rate of discovery of high-energy transient 
sources over the whole sky is certainly expected, including supernova shock break-outs, black hole tidal disruption events, 
and magnetar flares. 
 
The primary scientific goals of the mission thus address the Early Universe ESA Cosmic Vision theme ``How did the Universe originate and 
what is it made of?'' and, specifically, the sub-themes: 4.1) Early Universe, 4.2) The Universe taking shape, and 4.3) The evolving violent Universe. 
They also have a relevant impact on the 3.2 themes: ``The Gravitational Wave Universe'' and ``The Hot and Energetic Universe''. In addition, 
THESEUS will also automatically enable excellent observatory science opportunities, thus allowing a strong community involvement. 
It is worth remarking that THESEUS will have survey capabilities for high-energy transient phenomena complementary to the Large Synoptic Survey Telescope 
(LSST) in the optical. Their joint availability at the end of the next decade would enable a remarkable scientific synergy between them. 

\begin{figure*}
\centering
\includegraphics[scale=1.7]{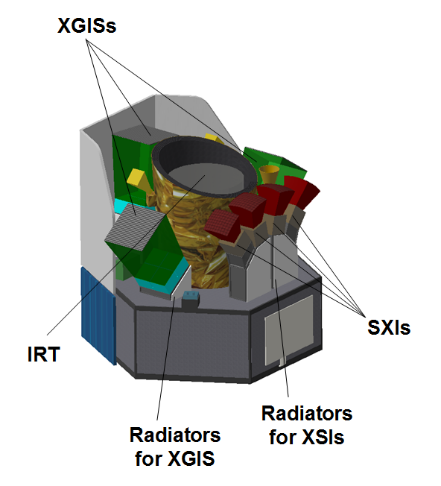}
\caption{THESEUS Satellite Baseline Configuration and Instrument suite accommodation.}
\label{fig:3}
\end{figure*}

The THESEUS scientific goals related to the full exploration of the early Universe requires 
requires the detection of many tens of GRBs from the first billion years (about 30-80), around ten times the number currently 
known at a redshift z$>$6 (Fig.~\ref{fig:4}). 
This is well beyond the capabilities of current and near future GRB detectors (Swift/BAT, the most 
sensitive one, has detected only very few GRBs above z$=$6 in 10~yrs).  As supported by intensive simulations 
performed by us and other works in the literature \citep[see, e.g.,][]{Ghirlanda2015}, the required substantial increase of high-z GRBs implies both 
an increase of $\sim$1 order of magnitude in sensitivity and an extension of the detector pass-band down to the softer  
X-rays (0.5-1~keV). Such capabilities must be provided over a broad field of view ($\sim$1~sr) 
with a source location accuracy $<$2~arcmin, in order to  allow efficient counterpart detection, on-board spectroscopy 
and redshift measurement, as well as optical and IR follow-up observations. 

The required THESEUS performance can be best obtained by including in the payload a monitor based on the lobster-eye telescope technology, 
capable of focusing soft X-rays in the 0.3-6~keV energy band over a large FOV. Such instrumentation has been under development 
for several years at the University of Leicester, has a high TRL level (e.g., BepiColombo, SVOM) and can perform all-sky monitoring 
in the soft X-rays with an unprecedented combination of FOV, source location accuracy ($<$1-2~arcmin) and sensitivity.  
An on-board infrared telescope of the 0.5-1~m class is also needed, together with 
fast slewing capability of the spacecraft (e.g., several degrees per minute), in order to provide prompt identification of 
the GRB optical/IR counterpart, refinement of the position down to $\sim$arcsec precision (thus enabling follow-up with the largest ground and space observatories), 
on-board redshift determination and spectroscopy of the counterpart and of the host galaxy. This capability will add considerably to 
the science value of ATHENA by providing a sample of known high-redshift bursts with which it will be able to sample the intervening WHIM.

The telescope may also be used for 
multiple observatory and survey science goals. Finally, the inclusion in the payload of a broad field of view hard X-ray 
detection system covering the same monitoring FOV as the lobster-eye telescopes and extending the energy band from few 
keV up to several MeV will increase significantly the capabilities of the mission.  As the lobster-eye telescopes can be 
triggered by many classes of transient phenomena (e.g., flare stars, X-ray bursts, etc), the hard X-ray detection system 
provides an efficient means to identify real GRBs and detect other transient sources (e.g., soft Gamma-repeaters). The joint data 
from the three instruments will characterize transients in terms of luminosity, spectra and timing properties over a 
broad energy band, thus getting fundamental insights into their physics. In summary, the foreseen payload of THESEUS 
includes the following instrumentation: 
\begin{itemize}
\item Soft X-ray Imager (SXI, 0.3-6~keV): a set of 4 lobster-eye telescopes units, covering a total FOV of $\sim$1~sr with source location accuracy $<$1-2~arcmin; 
\item InfraRed Telescope (IRT, 0.7-1.8~$\mu$m): a 0.7~m class IR telescope with 10$\times$10~arcmin FOV, for fast response, with both imaging and 
spectroscopy capabilities;
\item X-Gamma ray Imaging Spectrometer (XGIS, 2~keV-20~MeV): a set of coded-mask cameras using monolithic X-gamma ray detectors based on bars of Silicon 
diodes coupled with CsI crystal scintillator, granting a $\sim$1.5~sr FOV, a source location accuracy of $\sim$5~arcmin in 2-30 keV and an unprecedentedly 
broad energy band. 
\end{itemize}
 
In Fig.~\ref{fig:3} we show a sketch of the THESEUS baseline configuration and instrument suite accommodation as 
proposed for ESA/M5. The mission profile also includes: an on-board data handling units (DHUs) system capable of detecting, identifying and localizing 
likely transients in the SXI and XGIS FOV; the capability of promptly (within a few tens of seconds at most) transmitting to ground 
the trigger time and position of GRBs (and other transients of interest); a spacecraft slewing capability of $\sim$5-10~deg/min. 
The baseline launcher/orbit configuration is a launch with Vega-C to a low inclination low Earth orbit (LEO, $\sim$600~km, $<$5~deg), which has the unique 
advantages of granting a low and stable background level in the high-energy instruments, allowing the exploitation of the Earth's magnetic field for 
spacecraft fast slewing and facilitating the prompt transmission of transient triggers and positions to the ground. 

This article provides an extensive review of the THESEUS science case, scientific requirements, proposed instruments, mission profile, 
expected performances, as proposed in response to the ESA Call for M5 mission within the Cosmic Vision Programme. 
The current THESEUS collaboration involves more than 100 scientists from all over the world. The payload consortium is composed of several European 
countries, with a main role of Italy, UK, France, Germany, and Switzerland. Significant contributions are planned by Spain, Belgium, Czech Republic, 
and Poland. Additional contributions have been proposed by Ireland, Hungary, Slovenia, and Albania. International 
contributions are being discussed in several extra-European countries, including USA, Brazil, and China (already involved in the mission study). 
OHB Italia and GP Advanced Projects (Italy) kindly contributed to the study and definition of the technical aspects of 
the THESEUS mission concept.

\section{Science case}
\label{science}

The THESEUS specific science objectives can be summarized as follows: 

\begin{enumerate}[label=\Alph*.]

\item Explore the physical conditions of the Early Universe (cosmic dawn and reionization era) by unveiling a complete census 
of the GRB population in the first billion years. This will be achieved by: 
\begin{itemize}
\item Performing studies of the global star formation history of the Universe up to z$\sim$10 and possibly beyond with an unprecedented 
sensitivity for the detection of GRBs at high redshifts;
\item Detecting and studying the primordial (Pop III) and subsequent (Pop II)
star populations: when did the earliest stars form and how did they
influence their environments? 
\item Investigating the re-ionization epoch by determining the physical properties of the interstellar medium (ISM) and the 
intergalactic medium (IGM) up to $z\sim$8-10: how did re-ionization proceed as a function of environment, and was radiation from 
massive stars its primary driver? How did cosmic chemical evolution proceed as a function of time and environment?
\item Investigating the properties of the early galaxies and determining their star formation properties in the re-ionization era.
\end{itemize}

\item Perform an unprecedented deep monitoring of the X-ray transient Universe in order to: 
\begin{itemize}
\item Locate and identify the electromagnetic counterparts to sources of gravitational radiation and neutrinos, which may be routinely 
detected in the late '20s / early '30s by next generation facilities like aLIGO/ aVirgo, eLISA, ET, or Km3NET;
\item Provide real-time triggers and accurate ($\sim$1~arcmin within a few seconds; $\sim$1~arcsec within a few minutes) locations of (long/short) 
GRBs and high-energy transients for follow-up with next-generation optical-NIR (ELT, JWST if still operating), radio (SKA), X-rays (ATHENA), and TeV (CTA) telescopes;  
\item Provide a fundamental step forward in the comprehension of the physics of various classes of  Galactic and extra-Galactic transients, e.g.: 
tidal disruption events (TDE), magnetars/SGRs, SN shock break-outs, Soft X-ray Transients, thermonuclear bursts from accreting neutron stars, Novae, dwarf 
novae, stellar flares, AGNs and Blazars;
\item Fill the present gap in the discovery space of new classes of high-energy transient events, thus providing unexpected phenomena and discoveries.
\end{itemize}

\end{enumerate}

Below, we detail the THESEUS science case and describe the rich
observatory science that will come out as a result of meeting the primary mission requirements. We
show that the unique capabilities of this mission will enlarge several
research fields, and will be beneficial for multiple scientific
communities.

\subsection{Exploring the Early Universe with Gamma-Ray Bursts}
\label{science1}

A major goal of contemporary astrophysics and cosmology is to achieve a broad understanding of the formation 
of the first collapsed structures (Pop III and early Pop II stars, black holes and galaxies) 
during the first billion years in the life of the universe. This is intimately connected to the 
reionization of the IGM and build-up of global metallicity. The latter is 
very poorly constrained, and even in the JWST era will rely on crude emission line diagnostics for 
only the brightest galaxies. Regarding reionization, measurements of the Thomson scattering optical 
depth to the microwave background by the Planck satellite now suggest it substantially occurred in the 
redshift range z$\sim$7.8-8.8 \citep[see, e.g.,][]{Planck2016}, whereas the observations of the 
Gunn-Peterson trough in the spectra of distant quasars and galaxies indicate it was largely finished 
by z$\sim$6.5 \citep[see, e.g.,][]{Caruana2012, Caruana2014, Schenker2014}. Statistical measurements of the fluctuations in the redshifted 
21~cm line of neutral hydrogen by experiments such as LOFAR and SKA are expected to soon provide further 
constraints on the time history \citep[see, e.g.,][]{Patil2014}. The central question, however, remains whether 
it was predominantly radiation from massive stars that both brought about and sustained this phase change, 
or whether more exotic mechanisms must be sought. 
\begin{figure}[t!]
\centering
\includegraphics[scale=2.5]{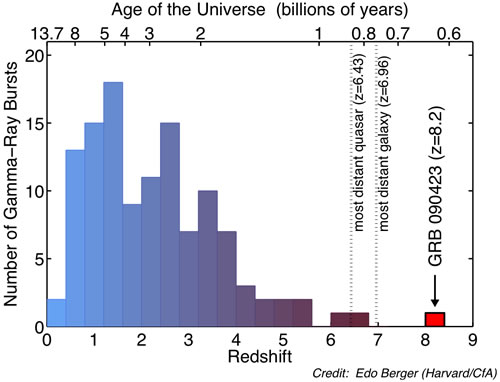}
\caption{Redshift distribution of GRBs (credits Berger, Harvard/CfA). 
}
\label{fig:4b}   
\end{figure}

Solving this problem largely splits into two subsidiary issues: how much massive star formation was occurring as a 
function of redshift? And, on average, what proportion of the ionizing radiation produced by these massive stars manages 
to escape from the immediate environs of their host galaxies? An answer to the former question can be extrapolated based on observed 
candidate $z>$7 galaxies found in Hubble Space Telescope (HST) deep fields, but there are two very significant uncertainties. The first is the completeness 
and cleanliness of the photometric redshift samples at $z>$7, while the second is the poorly constrained form and faint-end 
behaviour of the galaxy LF (at stellar masses $\lesssim$10$^8$~$M_{\odot}$), especially since galaxies below the HST detection limit 
very likely dominate the star formation budget.  Even though some constraints on fainter galaxies can be obtained 
through observations of lensing clusters \citep[see, e.g.,][]{Atek2015, Vanzella2017, Vanzella2017b}, which will be improved further by JWST, 
simulations suggest star formation was still likely occurring in considerably fainter systems \citep{Liu2016}. 
The second question concerns the Lyman continuum escape fraction and it is even more difficult to be answered, since this parameter 
cannot be determined directly at high redshifts and lower redshift studies have generally found rather low values of the escape fraction 
$f_{\rm esc}$ of only a few percent \citep{Grazian2017}. This is likely insufficient to drive reionization unless $f_{\rm esc}$ 
increases significantly at early times and/or for smaller galaxies \citep[see, e.g.,][]{Robertson2013}. 
Recently, an analysis of a sample of GRB afterglow spectra in
the range $1.6<z<6.7$ found a 98\% upper limit to the average
ionizing escape fraction of 1.5\% (Tanvir et al. submitted).  

These questions may be answered by the help of
GRBs and their host galaxies. With the help of THESEUS, we can combine
information on the cold/warm ISM gas of faint star forming galaxies
(through the afterglow spectroscopy) with that on the emission properties
of the continuum and ionised gas of the galaxy (through follow-up
observations). Therefore, we will be able to probe star formation
efficiency, escape fraction, metal enrichment and galaxy evolution during, and even
preceding, the epoch of reionization. This unique contribution will be one 
of the most important that THESEUS will provide.

\begin{figure*}[t!]
\centering
\includegraphics[scale=0.38]{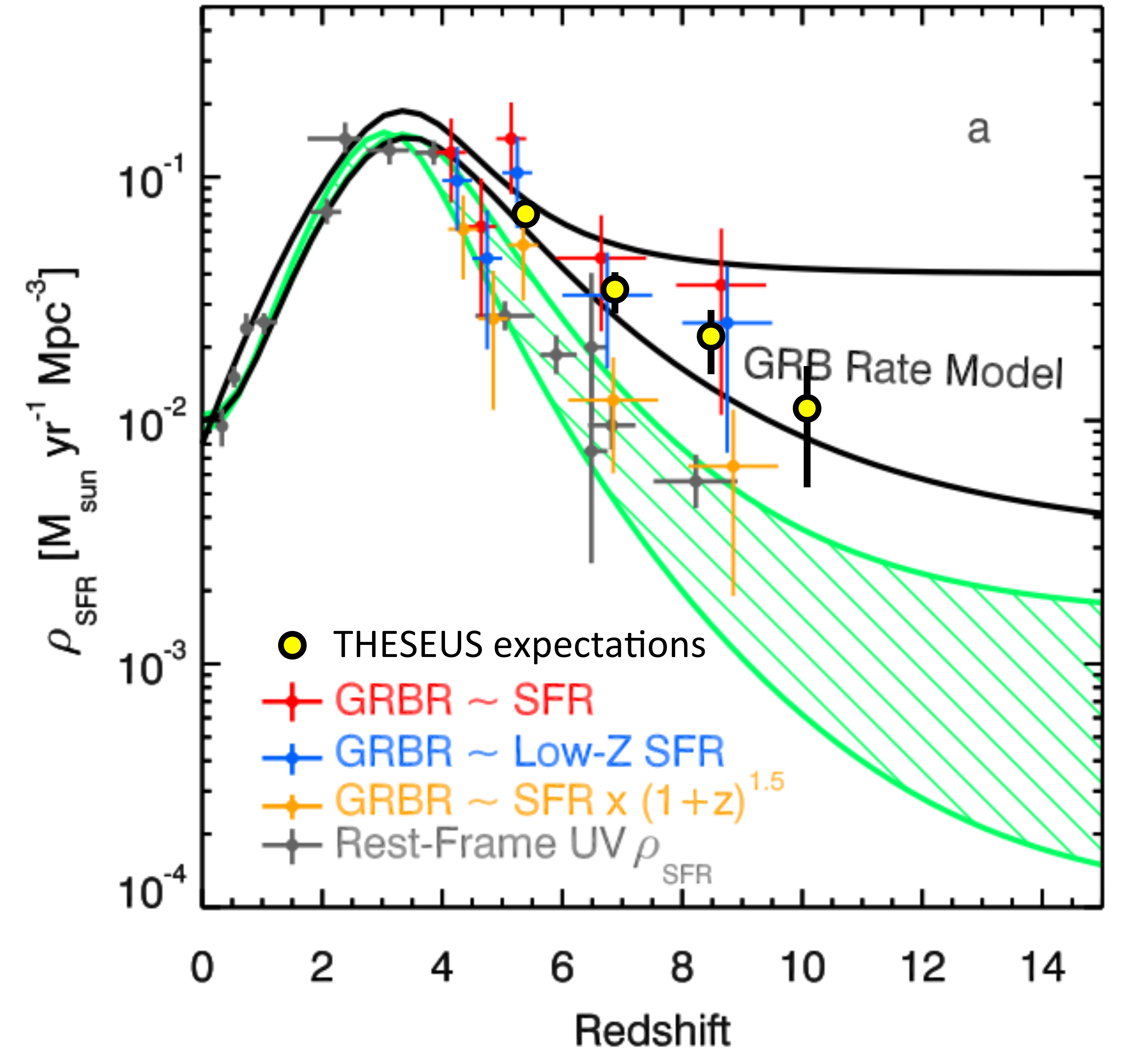}
\includegraphics[scale=0.11]{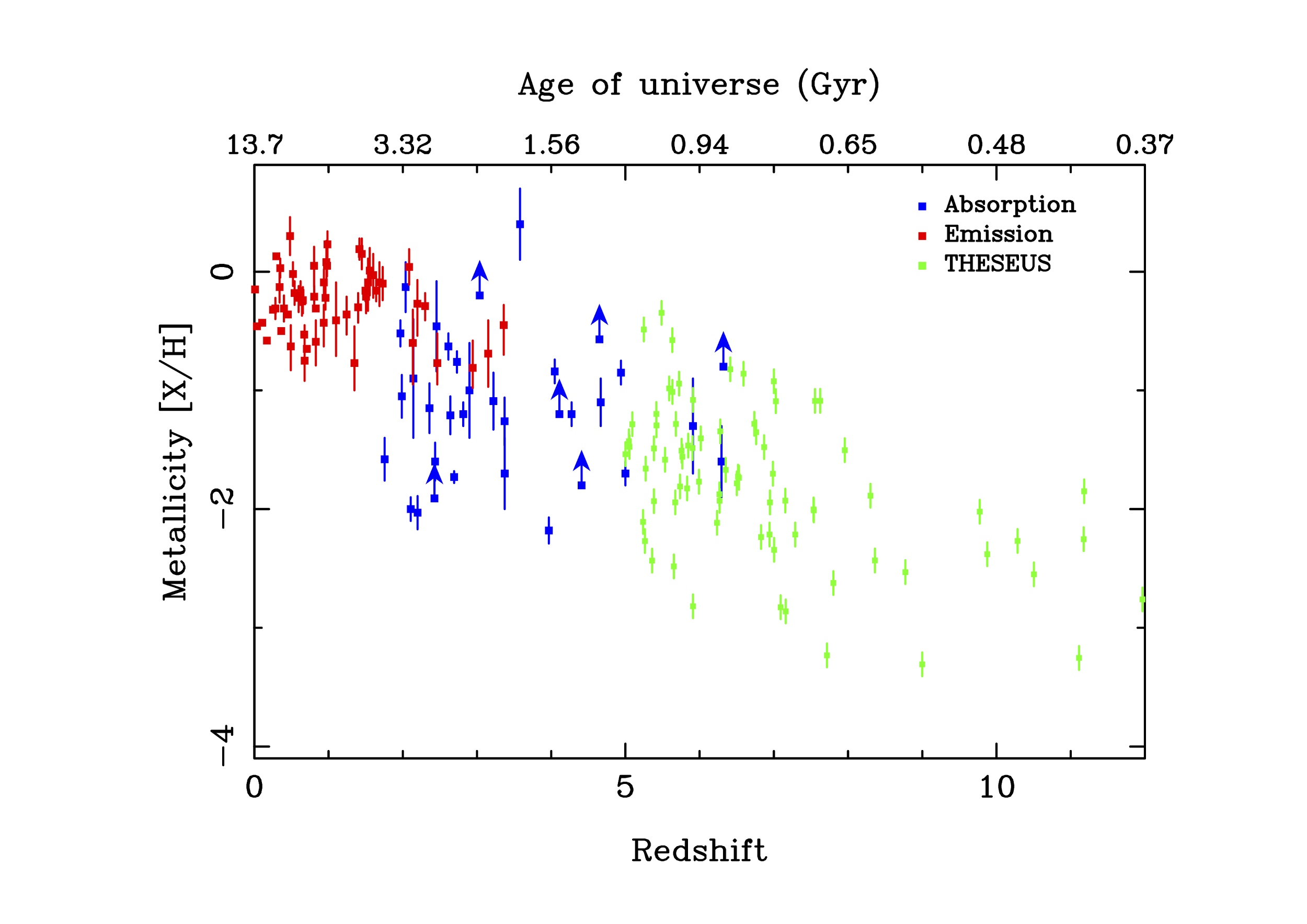}
\caption{{\it Left}: Star Formation Rate density as a function of redshift adapted from \citet{Robertson2012}. Low-$z$ data (circles) are from
\citet{Hopkins2006}, while diamonds are from \citet{Kistler2013}. The open squares show the result of
integrating the Lyman break galaxy UV luminosity functions down to the lowest measured magnitude in optical, 
while the solid squares represent the results obtained by stopping the integration at an optical magnitude of -10. The 
Salpeter initial mass function (IMF) is assumed in all cases. The plot also show with dotted lines  
the critical $\dot{\rho}^{*}$ from \citet{Madau1999} for a ratio between the clumping factor and the escape fraction of 40, 30, and 
20 (top to bottom). {\it Right}: Absorption-line based metallicities [M/H] as a function of redshift of Damped Lyman-$\alpha$ absorbers, 
for GRB-DLAs (blue symbols) and emission line metallicities for GRB hosts \citep[red symbols, adapted from][]{Sparre2014}. 
GRBs are essential to probe the evolution of ISM metallicities in the first billion years of cosmic history.}
\label{fig:5}   
\end{figure*}

\paragraph{Global star formation from GRB rate as a function of redshift \\}
Long-duration GRBs are produced by massive stars, and so track star formation, and in particular 
the populations of UV-bright stars responsible for the bulk of ionizing radiation production. Although 
there is evidence at low redshift that GRBs are disfavoured in high metallicity environments, since high redshift  
star formation is predominantly at low metallicity \citep[see, e.g.,][]{Salvaterra2013, Perley2016, Vergani2017}  
it is likely that the GRB rate to massive star formation rate ratio is approximately constant beyond $z\simeq$3. Thus simply 
establishing the GRB $N(z)$, and accounting for the instrumental selection function, provides a direct tracer of the 
global star formation rate density as a function of cosmic time (Fig.~\ref{fig:4b}). 
How selection bias is taken into account is relevant for a correct evaluation of the star formation rate \citep{Petrosian2015, Dainotti2015}.
Analyses of this sort have consistently pointed to 
a higher SFR density at redshifts $z>6$ than traditionally inferred from UV rest-frame galaxy studies (see Fig.~\ref{fig:5} left), 
which rely on counting star-forming galaxies and attempting to account for galaxies below the detection threshold. 
Although this discrepancy has been alleviated by the growing realisation of the extremely steep faint-end slope of 
the galaxy LF at $z>6$, it still appears that this steep slope must continue to very faint magnitudes, $M_{\rm AB}\simeq-11$, in order to 
provide consistency with GRB counts and indeed to achieve reionization (something that can only be quantified via a full 
census of the GRB population).

\paragraph{The high-z galaxy luminosity function \\}
The shape and normalisation of the high-redshift galaxy luminosity function is a key issue 
for understanding reionization since, to the depth achieved in the Hubble Ultra-deep 
Field (HUDF), it appears that the faint-end of the LF at $z>6$ approaches a power-law of slope $\alpha=2$. Thus the value 
of the total luminosity integral depends sensitively on the choice of the low-luminosity cut-off (and indeed the assumption 
of continued power-law form for the LF).  By conducting deep searches for the hosts of GRBs at high-z we can directly 
estimate the ratio of star formation occurring in detectable and undetectable galaxies, with the sole assumption that the 
GRB-rate is proportional to the star formation rate (see Fig.~\ref{fig:5} left).  Although currently limited by small-number statistics, 
the early application of this technique has confirmed that the majority of the star formation at $z\gtrsim6$ occurred in galaxies 
below the effective detection limit of HST \citep{Tanvir2012, McGuire2016}. Since the exact position and 
redshift of the galaxy is known via the GRB afterglow, follow-up observations are more efficient than equivalent 
deep field searches for Lyman-break galaxies.

\paragraph{The build-up of metals, molecules and dust \\}
Bright GRB afterglows with their intrinsic power-law spectra provide ideal back-lights for measuring not only the hydrogen column, 
but also obtaining exquisite abundances and gas kinematics probing to the hearts of  their host galaxies 
\citep[see Fig.~\ref{fig:5} right and, e.g.,][]{Hartoog2015}.  Further, the imprint of the local dust law, and in some 
cases observation of H2 molecular absorption, provides further detailed evidence of the state of the 
host ISM \citep[see, e.g.,][]{Friis2015}. They can thus be used to monitor cosmic metal enrichment and chemical 
evolution to early times, and search for evidence of the nucleosynthetic products of even earlier generations of stars. 
In the late 2020s, taking advantage of the availability of 30~m class ground-based telescopes (and definitely ATHENA), 
superb abundance determinations will be possible through simultaneous measurement of metal absorption lines and 
modelling the red-wing of Ly-alpha to determine host HI column density, potentially even many days after the 
burst (see Fig.~\ref{fig:7}). We emphasize that using the THESEUS on-board NIR spectroscopy capabilities 
(see Sect.~\ref{instruments} and Fig.~\ref{fig:37}) will provide the redshifts and luminosity measurements that are 
essential to optimise the time-critical follow-up observations using the highly expensive next-generation 
facilities, allowing us to select the highest priority targets and use the most appropriate telescope and 
instrument.

\paragraph{Topology of reionization \\}
It is expected that reionization proceeds in a patchy way, with ionized bubbles being created first around the highest 
density peaks where the first galaxies occur, expanding and ultimately filling the whole IGM. The topology of the 
growing network of ionized regions reflects the character of the early structure formation and the ionizing 
radiation field. With high-S/N afterglow spectroscopy, the Ly-alpha red damping wing can be decomposed into 
contributions due to the host galaxy and the IGM.  The latter provides the hydrogen neutral fraction and so 
measures the progress of reionization local to the burst.  With samples of several tens of GRBs  at $z\gtrsim7-8$, 
we can begin to statistically investigate the average and variance of the reionization process as a function 
of redshift \citep[see, e.g.,][]{McQuinn2008}.

\paragraph{Population III stars \\}

The first stars in the Universe are supposed to be very massive and
completely metal-free. However, no direct detection of such objects,
called Pop III stars, has been made so far. Thus, our theoretical understanding
of the structure and evolution, as well as the nature of the final explosion,
of the first stars still relies on many assumptions. Stellar evolution
predicts that long GRBs may be expected from fast rotating,
single Pop III stars with an initial mass of 10-80~$M_{\odot}$ \citep{Yoon2012}. 
Above this mass limit, a Pop III star may explode due to
pair-instability as a Super-luminous Supernova, or not explode at all,
falling back directly to a black hole \citep{Heger2003}. When a Pop III star explodes it changes 
the chemical composition of its environment, from which the
second stars are formed. These ideas must be examined. Also, having a full
census of GRBs with Pop III progenitors has the potential to provide an unprecedented 
enhancement of our knowledge of stellar evolution.

The same is true for the subsequent generations of stars in the early
Universe: low-metallicity massive stars, or Pop II stars, have been
simulated using our best theoretical approach, but observational evidences
are needed to confirm the theory \citep{Szecsi2015}. In particular, one
of the most promising evolutionary channels leading to fast rotating
helium stars (the so-called ``chemically homogeneous evolution'') still
needs to be tested observationally. These fast rotating helium stars at
low metallicity \citep[or TWUIN stars;][]{Szecsi2015} are predicted to
be progenitors of long GRBs if single \citep{Yoon2006} or progenitors of
short GRBs if in a close binary system \citep{Marchant2017, Szecsi2017}. 
The detection of GRBs with Pop III or Pop II progenitors provides important insights on 
stellar astrophysics.

Both Pop III and massive Pop II stars are expected to emit a large number
of ionizing photons and thus to contribute to the ionization of their
surroundings. Despite being potentially crucial in understanding cosmic
reionization, the predictions of the models are still weighted with
uncertainties due to scarce observational constraints on metal-free and
metal-poor massive stars. We also expect that interpreting the data output
of THESEUS will bring together scientists working on the so-far mostly
unrelated fields of reionization history and stellar evolution.

In addition to possible evidence for population III chemical enrichment, it has been argued that Pop III 
stars may also produce collapsar-like jetted explosions, which are likely to be of longer duration than 
typical long-GRBs, and may be detected through surveys with longer dwell times \citep{Meszaros2010}. 
To date, no direct evidence of this connection has been observationally established. The multi-wavelength 
properties of GRBs with a Pop-III progenitor are only predicted on the expected large masses and zero metallicity 
of these stars. Even the detection of a single GRB from a Pop III progenitor would put fundamental 
constraints on the unknown properties of the first stars.

\begin{figure}[t!]
\centering
\includegraphics[scale=0.42]{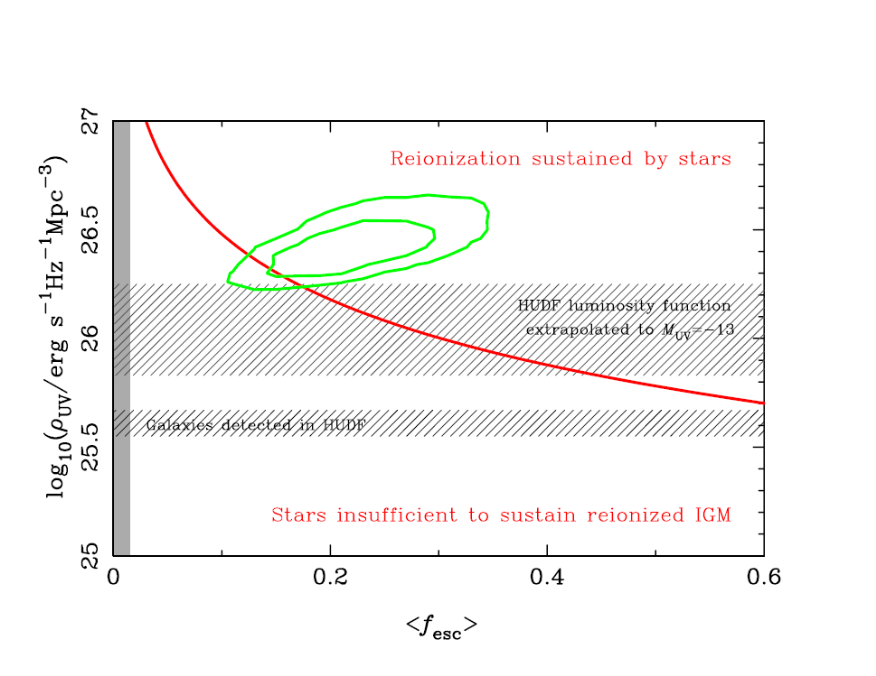}
\caption{The UV luminosity density from stars at $z\simeq8$ and average escape fraction $\langle f_{\rm esc}\rangle$ are insufficient to 
sustain reionization unless the galaxy luminosity function steepens to magnitudes fainter than $M_{\rm UV}=-13$  
(grey hatched region), and/or $\langle f_{\rm esc}\rangle$ is much higher than that typically found at $z\simeq3$ (grey shaded region). 
Even in the late 2020s, $\langle f_{\rm esc}\rangle$ at these redshifts will be largely unconstrained by direct observations. 
The green contours show the 1 and 2-$\sigma$ expectations for a sample of 25 GRBs at $z\simeq7-9$  for which deep spectroscopy 
provides the host neutral column and deep imaging constrains the fraction of star formation occurring 
in hosts below the JWST limit \citep{Robertson2013}. The input parameters were $log_{10}(\rho_{\rm UV})=26.44$ 
and $\langle f_{\rm esc}\rangle=0.23$, close to the (red) borderline for maintaining reionization by stars.}
\label{fig:6}   
\end{figure}

\begin{figure}
\centering
\includegraphics[scale=0.32]{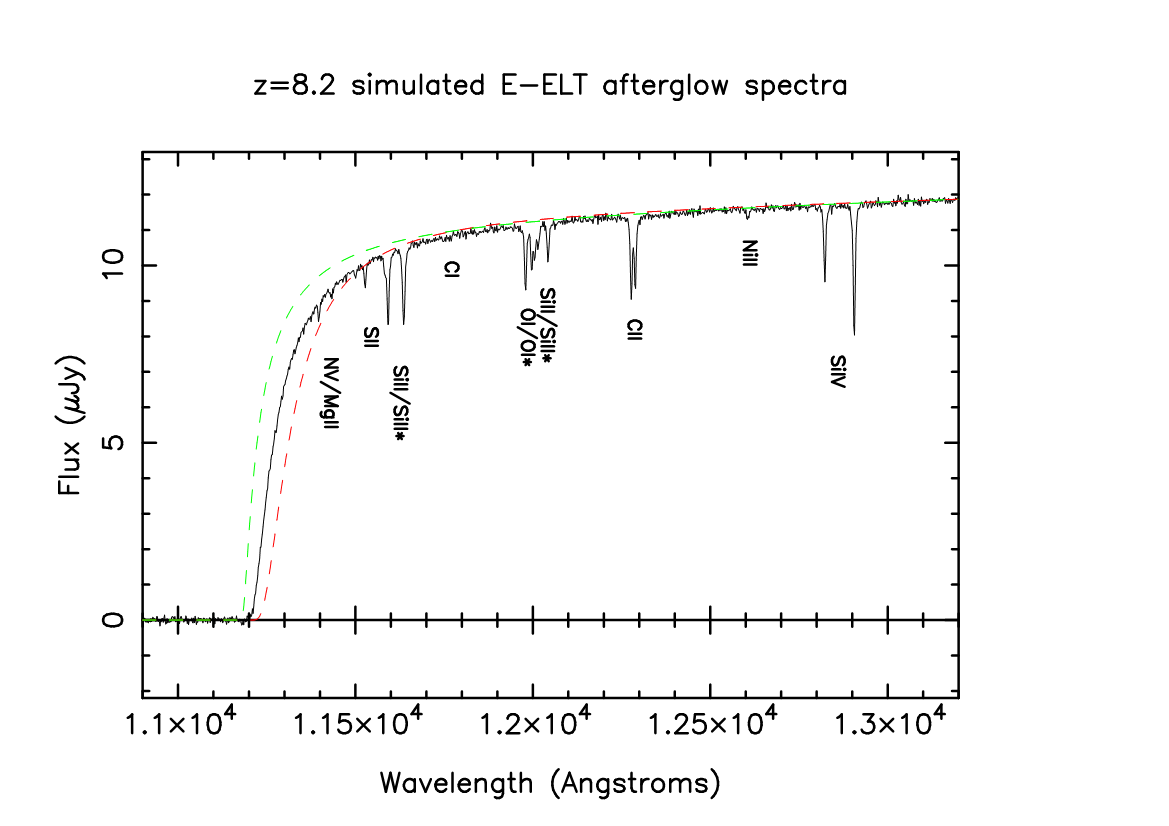}
\caption{Simulated ELT spectrum of a GRB at z=8.2 as discovered by THESEUS. 
The S/N provides exquisite abundance determinations from metal absorption lines, while fitting the Ly-a 
damping wing simultaneously fixes the IGM neutral fraction and the host HI column density, as illustrated 
by the two extreme models: a pure 100\% neutral IGM (green) and best-fit host absorption with a fully 
ionized IGM (red).}
\label{fig:7}   
\end{figure}

\paragraph{The role of THESEUS \\}
Our detailed simulations indicate that THESEUS will detect between 30 and 80 GRBs at $z>6$ over a three year mission, 
with between 10 and 25 of these at $z>8$ (and several at $z>10$). 
The on-board follow-up capability means  
that redshifts are estimated for the large majority of these, and powerful next generation ground- and space-based 
telescopes available in this era will lead to extremely deep host searches and high-S/N afterglow spectroscopy 
for many of them (e.g., using JWST, if still operating, ELT, WFIRST, EUCLID, ATHENA, etc.).  To illustrate the potential 
of such a sample, we simulate in Fig.~\ref{fig:6} the precision in constraining the product of the UV luminosity density 
and average escape fraction, $\rho_{\rm UV}f_{\rm esc}$, that would be obtained with 25 GRBs at $7<z<9$ having high-S/N afterglow 
spectroscopy and (3~hr) JWST depth host searches (for definiteness the $\rho_{\rm UV}$ axis corresponds to $z=8$).  
This will provide a much clearer answer to the question of whether stars were the dominant contributors 
to reionization. In addition, this sample will allow us to map abundance patterns across the whole range 
of the star forming galaxies in the early universe, providing multiple windows on the nature of the first 
generations of stars.

\paragraph{The great value of THESEUS to ATHENA \\}
As an example of the great relevance of THESEUS in the context of the next generation large facilities 
(e.g., SKA, CTA, ELT, ATHENA), we highlight here the THESEUS synergy with ATHENA. Two of the primary 
science goals for ATHENA are: (1) Locate the missing baryons in the Universe by probing the Warm Hot 
Intergalactic Medium (the WHIM); this requires about 10 bright GRBs per year. (2) Probe the first 
generation of stars by finding high redshift GRB; this requires about 5 high-redshift GRBs per year. 
THESEUS will enable ATHENA to achieve these goals by greatly increasing the rate of GRBs found per 
year with good localisations and redshifts. The X-ray band of the THESEUS SXI (see Sect.~\ref{SXI}) will find a greater 
proportion of high-redshift GRBs than previous missions. Hence THESEUS will: (1) localise bright 
GRBs with sufficient accuracy using the SXI to enable a rapid repointing of the ATHENA X-IFU for 
X-ray spectroscopy of the WHIM, and (2) find high-redshift GRBs using the SXI and XGIS and 
accurately localising them using the IRT for redshift determination on-board and on the ground 
to provide reliable high-redshift targets for ATHENA. Many of the other transients found by 
THESEUS, such as tidal disruption events and flaring binaries, will also be high-value 
targets for ATHENA. 

\subsection{Gravitational wave sources and multi-messenger astrophysics}
\label{gws}

The launch of THESEUS will coincide with a golden era of multi-messenger astronomy. With the first detection of gravitational 
waves (GWs) by Advanced detectors \citep{Abbott2016a, Abbott2016b}, a new window on the Universe has been opened. 
By the end of the present decade, the sky will be routinely monitored by a network of second generation GW detectors, an ensemble 
of Michelson-type interferometers, composed of the two Advanced LIGO (aLIGO) detectors in the USA and by Advanced Virgo (aVirgo) 
in Italy, plus IndIGO in India and KAGRA in Japan joining later within a few years. By the first years of 2030, more sensitive third generation 
GW detectors, such as the Einstein Telescope (ET) and Cosmic Explorer, are planned to be operational and to provide an 
increase of roughly one order of magnitude in sensitivity. 

\begin{figure}[t!]
\centering
\includegraphics[scale=0.33]{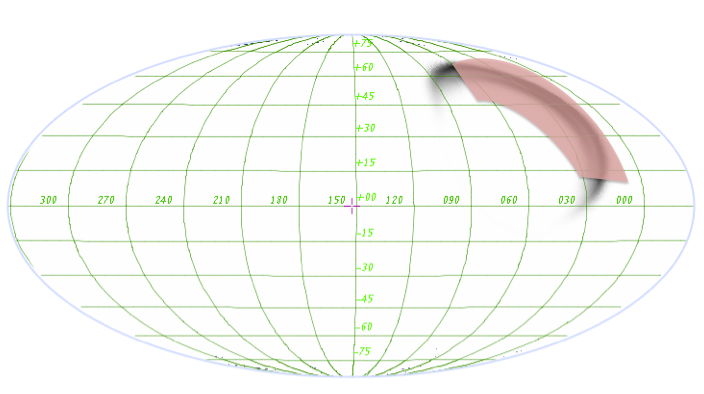}
\caption{The plot shows the SXI field of view ($\sim$110$\times$30~deg$^2$) superimposed on a simulated probability skymap of a 
merger NS-NS system observed by aLIGO and advanced Virgo \citep{Singer2016}. The SXI coverage contains nearly 100\% of the total probability.}
\label{fig:8}   
\end{figure}

Several of the most powerful transient sources of GWs predicted by general relativity, e.g. binary neutron star (NS-NS) or NS-black hole (BH) 
mergers \citep[with likely detection rate of $\sim$50 per year by the LIGO-Virgo network at design sensitivity;][]{Abadie2010}, are expected to 
produce bright electromagnetic (EM) signals across the entire EM spectrum and in particular in the X-ray and gamma-ray energy bands, as 
well as neutrinos. These expectations were met on the 17th of August 2017, when a GW signal consistent with a binary neutron star merger 
system \citep{Abbott2017a} was found shortly preceding the short gamma-ray burst GRB170817A \citep{Abbott2017b}.
GW detectors have relatively poor sky localization capabilities mainly based on triangulation methods. 
With two detectors $\sim$100-1000 square degrees are achieved; this sky localization accuracy may also apply for the third-generation GW detectors by 2030+. By 2020+ the second-generation network will count three 
up to five  GW detectors and sky localization will improve to $\sim$1-10 sq. degrees or even less \citep{Klimenko2011}. Such 
localization uncertainties will be completely covered by the THESEUS/SXI FoV (see Fig.~\ref{fig:8}).
In parallel to these advancements, IceCube and KM3nNeT and the advent of 10 km$^3$ detectors (e.g. IceCube-Gen2, \cite{Aartsen2014}) will likely revolutionize 
neutrino astrophysics. Neutrino detectors can localize to an accuracy of better than a few sq. degrees but within a smaller volume of 
the Universe \citep[see, e.g.,][and references therein]{Santander2016}. 
In order to maximize the science return of the multi-messenger investigation it is essential to have an in-orbit trigger and search 
facility that can either detect an EM signal simultaneous with a GW/neutrino event or rapidly observe with good sensitivity the large 
error boxes provided by the GW and neutrino facilities following a trigger. These combined requirements are uniquely fulfilled by 
THESEUS, which is able to trigger using XGIS or SXI and observe a very large fraction of the GW/neutrino error boxes within an orbit 
due to the large grasp of the SXI instrument (if compared to current generation X-ray facilities such as Swift, 
THESEUS/SXI has a grasp of $\sim$150 times that of Swift/XRT). 
The recent discovery of the electromagnetic counterpart of GW170187 in the form of a short GRB and an optical/NIR 
``kilonova'' emission \citep{Li_Paczynski1998,Abbott2017c,Smartt2017,Tanvir2017,Pian2017,Coulter2017} definitively open great 
perspectives for the role that missions as THESEUS would play in the multi-messenger astrophysics. Indeed, for events similar to 
GW170817, THESEUS could detect both the GRB and the kilonova emission, as well as the soft X-ray emission theoretically predicted 
for NS-NS mergers, and localize the source down to arcsecond precision (see more details in \citealt{Stratta2017}. 
For events triggered on-board with XGIS or SXI, GW searches can also be 
carried on the resultant known sky locations with lower GW detector signal-to-noise thresholds and hence an increased search distance. 
In 2020s synergies with the future facilities like JWST, WFIRST, Einstein Probe, ATHENA, ELT, TMT, 
GMT, SKA, CTA, zPTF, and LSST telescopes would leverage significant added value, extending the multi-messenger observations across 
the whole EM spectrum. The detection of EM counterparts of GW (or possibly neutrino) signals will enable a multitude of science 
programmes \citep[see, e.g.,][]{Bloom2009, Phinney2009} by allowing for parameter constraints that the GW/neutrino observations alone 
cannot fully provide. For example, finding a GW/neutrino source EM counterpart in X-rays with THESEUS/SXI will allow to localize 
the source with an accuracy good enough for optical follow-up and hence to possibly measure its redshift and luminosity. 
On the other hand, not finding an EM counterpart will constrain merger types (such as BH-BH mergers), magnetic field strengths, and astrophysical 
conditions at the time of the merger. 

\subsubsection{Gravitational wave transient sources}

\paragraph{NS-NS / NS-BH mergers: Collimated EM emission from Short GRBs \\} 
Compact binary coalescences (CBCs) are among the most promising sources of GWs that will be detected in the next decade. 
Indeed, these systems are expected to radiate GWs within the most sensitive frequency range of ground based GW detectors 
(1-2000~Hz), with large GW energy output, of the order of 10$^{-2}$~$M_{\odot}c^2$, and gravitational waveforms well predicted by General Relativity 
\citep[see, e.g.,][for a review]{Baiotti2017}. The expected rate of NS-NS systems inferred from binary pulsar observations and population 
synthesis modeling, is taken to lie between 10 and 10000 Gpc$^{-3}$ yr$^{-1}$. With the first detection of a NS-NS merger, GW170817, 
the expected rate was constrained to 320-4740 Gpc$^{-3}$ yr$^{-1}$ \citep{Abbott2017a}. 
To date, no NS-BH systems have been observed, but the rate can still be predicted through population 
synthesis, constrained by the observations of NS-NS, to be $\sim$1-1000 Gpc$^{-3}$ yr$^{-1}$ 
\citep[see, e.g.,][and references therein]{Abadie2010}. 
Before the GW170817/GRB170817 event, mounting indirect evidence associates short GRB progenitors to CBC systems with at least one neutron star \citep[see, e.g.,][]{Berger2014} and provides possible hints that the merger of magnetized NS binaries might lead to the formation of a jet with an opening angle up to of $\sim$30-40~deg \citep[so far the observed values range from 5-20~deg.;][]{Rezzolla2011, Troja2016}. 
The association of the short GRB 170817 with the GW event from the merger of a NS-NS system represent the first direct evidence 
of the progenitor nature of short GRBs \citep{Abbott2017b}. 
Further GW observations combined with multi-wavelength follow-up campaigns in the next years will likely 
confirm this scenario by simultaneous detections of CBC-GWs temporally and spatially consistent with short GRBs \citep{Stratta2017}. 

\begin{figure}[t!]
\centering
\includegraphics[scale=0.35]{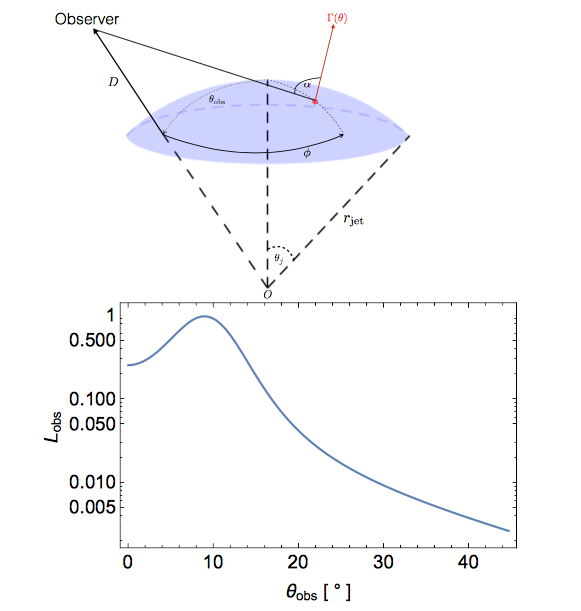}
\caption{{\it Top panel}: possible geometry of the conical jet, in  spherical coordinates with the origin at O. 
The observer is located at ($D$, $\Theta_{\rm obs}$, 0). The jet is moving with Lorentz factor $\Gamma(\Theta)$. 
{\it Bottom panel}: Luminosity fraction 
(normalized to peak emission) as a function of observer angle, $L_{\rm obs}(\Theta_{\rm obs})$, from \citet{Kathirgamaraju2017}.  
The jet luminosity at $\sim$40~deg is a factor of 300 fainter than that of the jet core. Nevertheless, such a misaligned 
jet can be detected by a $\gamma$-ray instrument if it takes place within the Advanced LIGO detectability volume.}
\label{fig:9}   
\end{figure}

Second-generation GW network can detect NS-NS and NS-BH systems up to 0.4-0.9 Gpc \citep{Abadie2010}.  
Since the estimated rate density of detected short GRBs is 1-10 Gpc$^{-3}$  yr$^{-1}$ \citep[see][and references therein]{Wanderman2015, Ghirlanda2016} 
the expected annual rate of on-axis (i.e. face-on) short GRBs with GW counterpart before the third generation detectors (see below) is rather low, of the order of a few per year or less.  However,  in a more realistic jet scenario, e.g. a structured jet scenario, prompt GRB photon flux $F$ can vary with observer viewing angle $\Theta_{obs}$. 
According to \cite{Pescalli2016}, $F \sim F_{core} (\Theta_{obs}/\Theta_{jet} )^{-s}$ with $s>2$,    
where $F_{core}$ is the typical on-axis flux of a short GRB (see Fig. \ref{fig:9}). 
Assuming $F_{core}\sim$10 ph cm$^{-2}$ s$^{-1}$ \citep[e.g.][]{Narayana2016}, since short GRBs have a median redshift of $z\sim0.5$ \citep{Berger2014}, at a distance of $\sim400$  Mpc $F_{core}$ would be $\sim10^3$  ph cm$^{-2}$ s$^{-1}$, thus even assuming a steep dependence of $F$ from $\Theta_{view}$ (e.g. $s=6$) a nearby short GRB can be detected off-axis with THESEUS/XGIS up to  $5\Theta_{jet}$ (we consider the 1 sec photon flux sensitivity of XGIS of 0.2 ph cm$^{-2}$ s$^{-1}$ , see Fig. 36). 
Table \ref{tab:1} shows the expected rate of THESESUS/XGIS short GRB detections with a GW counterpart from merging NS-NS systems (i.e. within the GW detector horizon). The quoted numbers are obtained by correcting the realistic estimate of NS-NS merger rate from the jet collimation factor by assuming a jet half-opening angle range of [10-40] deg, and by taking into account the possibility to observe off-axis short GRBs up to $5\Theta_{jet}$ (see above).  

By the time of the launch of THESEUS, gravitational radiation from such systems will be likely detectable by third-generation detectors such as the Einstein Telescope (ET) up to redshifts z$\sim$2 or larger \citep[see, e.g.,][]{Sathyaprakash2012, Punturo2010}, implying that almost all short GRBs that THESEUS will detect, 
i.e., $\sim$20 short GRBs per year, will have a detectable GW emission. Indeed, it is likely that at the typical distances at which ET detects GW events, the only EM counterparts that could feasibly be detected are SGRBs and their afterglows, making the role of THESEUS crucial for 
multi-messenger astronomy by that time. 
Almost all short GRBs are accompanied by an X-ray afterglow that SXI will detect and monitor just after the burst emission. 
Once localized with SXI, about 40\% of detected short GRBs are expected to have a detectable optical/IR counterpart. 
The IRT could point at the SXI-localised afterglow within a few minutes from the trigger. If bright enough, spectroscopic 
observations could be performed on-board, thus providing redshift estimates and information on chemical composition of 
circumburst medium. In addition, precise sky coordinates will be disseminated to ground based telescopes to perform spectroscopic observations. 
Distance measurements of a large sample of short GRBs combined with the absolute source luminosity distance provided 
by the CBC-GW signals can deliver precise measurements of the Hubble constant \citep{Schutz1986}, helping to break the degeneracies 
in determining other cosmological parameters via CMB, SNIa and BAO surveys \citep[see, e.g.,][]{Dalal2006}. Even with the 
second-generation GW detector network, the modest THESEUS EM + GW triggered coincidence number of 3-4 predicted within the 
nominal mission life (not accounting for possible ``off-axis'' prompt GRB detection), can rises to 10 or more when including 
SXI follow-up observations of GW network error boxes. With 10 GW+EM events, the Hubble constant could be constrained to 2-3\%, 
thus providing a precise independent measure of this fundamental parameter \citep{Dalal2006, Nissanke2010}. 
In addition, each individual joint GW+EM observation would provide an enormous science return from THESEUS. For example, 
the determination of the GW polarization ratio would constrain the binary orbit inclination and hence, when combined with an 
EM signal, the jet geometry and source energetics. Likewise, a better understanding of the NS equation of state can 
follow from combined GW and EM signals \citep[see, e.g.,][]{Bauswein2012, Takami2014, Lasky2014, Ciolfi2015a, Ciolfi2015b, Messenger2015, Rezzolla2016}. 

\begin{table*}
\caption{Number of NS-NS (BNS) mergers expected to be detected in the next years by second- (2020+) and 
third- (2030+) generation GW detectors and the expected number of electromagnetic counterparts as short GRBs 
(collimated) and X-ray isotropic emitting counterparts \citep[see, e.g.,][]{Ciolfi2015a, Rezzolla2015} 
with THESEUS SXI and XGIS. BNS horizon indicates the GW detector sensitivity \citep[see, e.g.,][]{Abadie2010}. 
The rate estimates of simultaneous GW+GRB detections assume that all BNS can produce a short GRB and take into account 
a combination of collimation angle range, XGIS FoV as a function of energy, and possible prompt off-axis detection. 
X-ray counterpart rate estimates assume that at least 1/3 of BNSs produce a long-lived NS remnant 
\citep[but see][]{Gao2016, Piro2017}.}
\label{tab:1}  
\begin{tabular}{llllll}
\hline\noalign{\smallskip}
\multicolumn{3}{c}{GW observations} &  \multicolumn{3}{c}{THESEUS XGIS/SXI joint GW+EM observations} \\
\noalign{\smallskip}\hline\noalign{\smallskip}
Epoch & GW detector & BNS horizon & BNS rate & XGIS/sGRB rate & SXI/X-ray isotropic \\
& & & (yr$^{-1}$) & (yr$^{-1}$) & counterpart rate  (yr$^{-1}$) \\
\noalign{\smallskip}\hline\noalign{\smallskip}
2020+ & Second-generation (advanced LIGO, & $\sim$400~Mpc & $\sim$40 & $\sim$5-15 & $\sim$1-3 (simultaneous) \\ 
& Advanced Virgo, India-LIGO, KAGRA) & & & & $\sim$6-18 (+follow-up) \\
2030+ & Second + Third-generation & $\sim$15-20~Gpc & $>$10000 & $\sim$15-25 & $\gtrsim$100 \\
& (e.g. ET, Cosmic Explorer) & & & &  \\
\noalign{\smallskip}\hline
\end{tabular}
\end{table*}

\paragraph{NS-NS / NS-BH mergers: Optical/NIR and soft X-ray isotropic emissions \\} 
Nearly isotropic EM emission is expected from NS-NS / NS-BH mergers at minutes-days time scale after the merger. 
GW emission depends only weakly on the inclination angle of the inspiral orbit, and GW detectors will mostly trigger 
on off-axis mergers (i.e. for binary systems with a non zero inclination angle).  Both serendipitous discoveries within 
the large THESEUS/SXI FoV and re-pointing of THESEUS in response to a GW trigger will allow studies of off-axis X-ray emission. 
One expected EM component is the late afterglow from the laterally spreading jet as soon as it decelerates \citep[``orphan afterglows'';][] 
{vanEerten2010}. Peak brightness is expected at 1-10~days after the trigger, with peak X-ray fluxes equal or 
below $\sim$10$^{-12}$-10$^{-13}$~erg~cm$^{-2}$~s$^{-1}$ at $\sim$200~Mpc \citep{Kanner2012}. Therefore, 
despite their low collimation, off-axis afterglows will be detected only for the most nearby CBC systems.
Another nearly-isotropic emitting component is expected if a massive millisecond magnetar is formed from two coalescing NSs. 
In this case, X-ray signals can be powered by the magnetar spin-down emission reprocessed by the matter surrounding the merger 
site (isotropically ejected during and after merger), with luminosities in the range 10$^{43}$-10$^{48}$~erg~s$^{-1}$ and time 
scales of minutes to days \citep{Metzger2014, Siegel2016a, Siegel2016b}. Alternatively, X-ray emission may 
come from direct dissipation of magnetar winds \citep[see, e.g.,][]{Zhang2013, Rezzolla2015}. Numerical simulations 
suggest that such emission is collimated but with large half-opening angles (30-40~deg, beaming factor of $\sim$0.2). 
As an additional channel in X-rays, the magnetar may accelerate the isotropically expelled matter through wind 
pressure to relativistic speeds generating a shock with ISM (``confined winds''). Synchrotron radiation produced 
in the shock is emitted nearly isotropically, with an enhanced intensity near the equator. A beaming factor 
of $\sim$0.8 is expected in this case \citep[see, e.g.,][]{Gao2013}. Overall, typical time scales for these X-ray signals 
are comparable to magnetar spin down time scales of $\sim$10$^3$-10$^5$~s, and the predicted luminosities span a wide range 
that goes from~10$^{41}$~erg~s$^{-1}$  to ~10$^{48}$~erg~s$^{-1}$. With THESEUS/SXI in combination with the second-generation 
detector network, almost all X-ray counterparts of GW events from NS-NS merging systems will be easily detected 
simultaneously with the GW trigger and/or with rapid follow-up of the GW-individuates sky region. These X-ray 
emission counterparts could be possibly detected up to large distances with the third-generation of GW 
detectors (depending on the largely uncertain intrinsic luminosity of such X-ray component, see Table~\ref{tab:1}), 
providing a unique contribution to classify X-ray emission from NS-NS systems and probes the cosmic evolution.
\begin{figure*}[t!]
\centering
\includegraphics[scale=0.75]{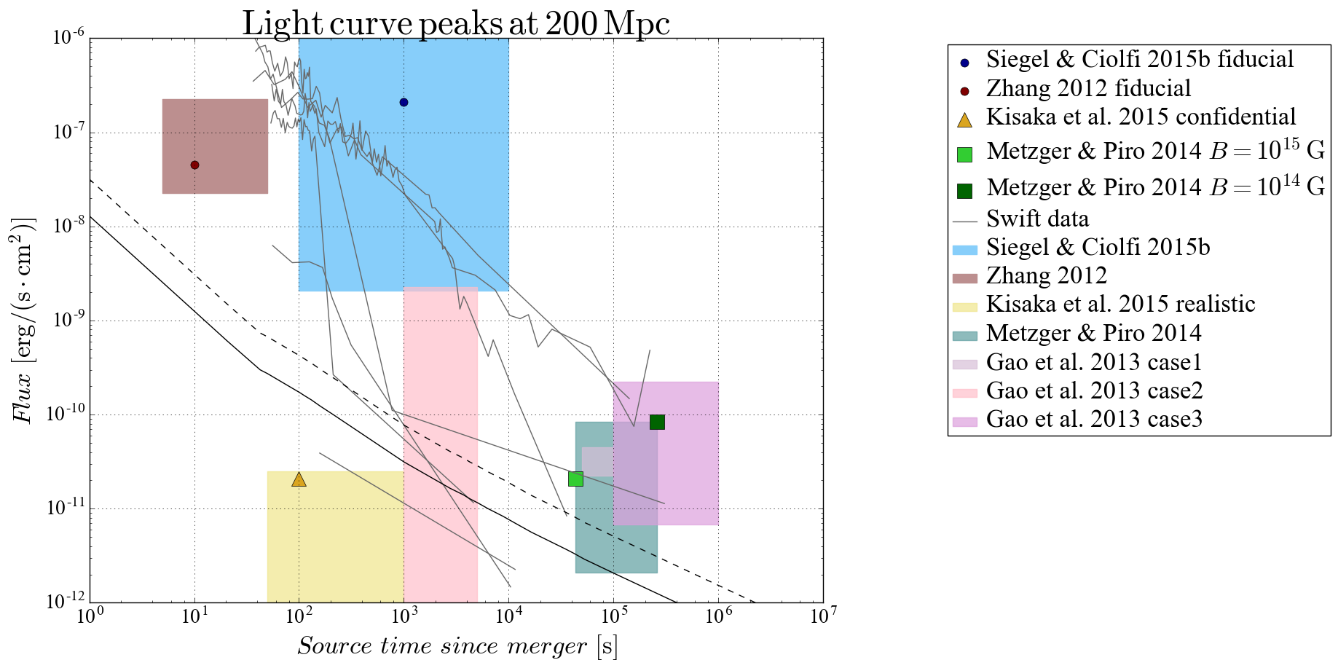}
\caption{Expected X-ray fluxes at peak luminosity from different modelling of X-ray emission from NS-NS merger systems, 
that are among the most probable GW sources that will be detected in the following years by the second- 
and third- generation GW detectors. Grey solid lines show a typical GRB X-ray afterglow observed with Swift/XRT. 
Dots show the expected flux using fiducial parameters for each model. The black and dashed line show the 
THESEUS/SXI sensitivity as a function of the exposure time (credit: S. Vinciguerra; see also Fig.~\ref{fig:25})}
\label{fig:10}   
\end{figure*}

In the optical band, the expected EM component is represented by the so-called ``macronova'' (often named ``kilonova'') emission \citep[e.g.,][]{Li_Paczynski1998}. 
During NS-NS or NS-BH mergers, a certain amount of the tidally disrupted neutron star mass is expected to become unbound and ejected into space. This matter 
has the unique conditions of high neutron density and temperature to initiate r-process nucleosynthesis of very heavy elements. 
Days after merger, the radioactive decay of such elements heats up the ejected material producing a thermal transient signal peaking 
in the optical/near-infrared (NIR) band and with luminosities of $\sim$10$^{40}$~erg~s$^{-1}$ \citep[see, e.g.,][]{Fernandez2016, Baiotti2017}. 
Interaction of the ejected matter with surrounding ISM may also produce synchrotron radiation at late times ($\sim$weeks-months) 
peaking at radio wavelengths. Another fraction of the tidally disrupted neutron star matter remains bound to the system forming 
an accreting disc. Disc wind outflows are expected to produce a more blue optical emission peaking at earlier epochs (e.g. 1-2 
days after the merger), with luminosity of the order of a few times $10^{40-41} $~erg~s$^{-1}$, depending also on the nature 
of the central remnant \citep{Kasen2015}.  

\begin{figure}[t!]
\centering
\includegraphics[scale=0.34]{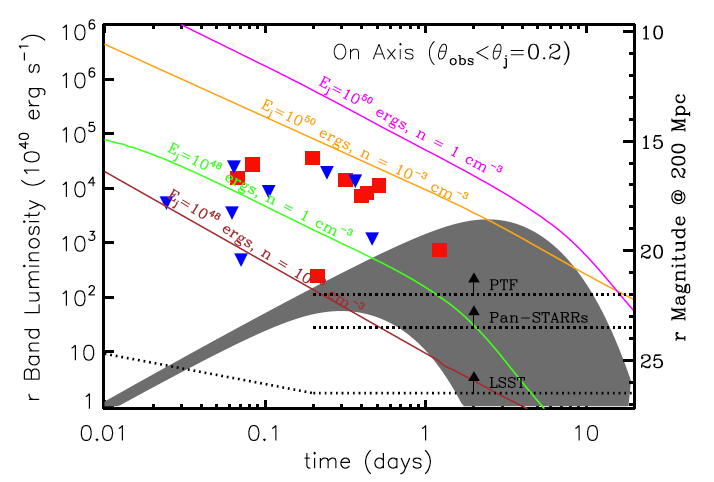}
\caption{The dark grey region shows the expected macronova r-band apparent magnitude for a source at 200~Mpc as a function of 
time from the burst onset. Solid curves show the expected GRB afterglow emission assuming different energetics and ISM densities. 
Red squares and blue triangles represent the afterglow detection (squares) and upper limits (triangles) for a sample of short GRBs 
\citep{Metzger2012}.}
\label{fig:11}   
\end{figure}

Macronovae are promising electromagnetic counterparts of binary mergers because (i) the emission is nearly isotropic and therefore the 
number of observable mergers is not limited by beaming; (ii) the week-long emission period allows for sufficient time needed 
by follow-up observations. The detectability of macronovae is currently limited by the lack of sufficiently sensitive survey 
instruments in the optical/NIR band that can provide coverage over tens of square degrees, the typical area within which GW 
events will be localized by the Advanced LIGO-Virgo network. Until the detection of GW170817, observational evidence of the macronova 
emission was obtained only in few cases during the follow-up campaigns of the optical afterglows of short GRBs 
\citep{Tanvir2013, Jin2015}. The successful electromagnetic follow-up campaign of GW170187 enabled to obtain for the first 
time a very accurate monitoring of the associated macronova, both photometric and spectroscopic \citep[e.g.][]{Smartt2017,Tanvir2017,Pian2017,Coulter2017}, 
confirming almost all the main theoretically predicted features expected from these events \citep[e.g.][]{Li_Paczynski1998,Abbott2017c,Kasen2017}. 
Figure~\ref{fig:11} shows the expected macronova apparent magnitudes for a source at 200~Mpc as well as 
the expected intrinsic luminosity \citep{Metzger2012}. Once the macronova component is identified, source location 
can be accurately recovered, allowing for the identification of the host galaxy and the search for counterparts 
in other electromagnetic bands (e.g. radio).

During the next decade (2020-2030) we expect strong synergies between the second-generation GW detector network and several 
telescopes operating at different wavelengths such as: 1) the space-based telescopes James Webb Space Telescope (JWST), 
ATHENA and WFIRST; 2) the ground-based telescope with large FOV like zPTF and LSST, which will be able to select the GW 
candidates in order to follow-up them afterwards; 3) the large telescopes, such as the 30-m class  telescopes  GMT, TMT 
and ELT, which will all follow-up the optical/NIR counterparts like macronovae; 4) the Square Kilometer Array (SKA) in 
the radio, which is well suited to detect the late-time ($\sim$weeks) signals produced by the interaction of the 
ejected matter with the interstellar medium. THESEUS/IRT will be perfectly integrated in this search, due to its photometric 
and spectroscopic capabilities and its spectral range coverage. Light curves and spectra will be acquired, thus giving the 
opportunity to disentangle expected  different components associated with macronova events \citep[e.g. disk wind + dynamical 
ejecta contributions;][]{Kasen2015}. 

\paragraph{Core collapse of massive stars: Long GRBs, Low Luminosity GRBs and Supernovae \\} 
The collapse of massive stars are expected to emit GWs since a certain degree of asymmetry in the explosion is inevitably present. 
Estimates of the GW amplitudes are still poorly constrained and the expected output in energy has enormous uncertainties, 
ranging from 10$^{-8}$ to 10$^{-2}$~$M_{\rm o}$c$^2$. If the efficiency of the GW emission is effectively very low, then only the third-generation 
GW detectors such as ET, which will be operative at the same time as THESEUS, will be able to reveal the GW emission from 
these sources and thus obtain crucial insights on the innermost mass distribution, inaccessible via EM observations. 
The collapsar scenario invoked for long GRBs requires a rapidly rotating stellar core \citep[see, e.g.,][]{Woosley1993, Paczynski1998}, 
so that the disk is centrifugally supported and able to supply the jet. This rapid rotation may lead to non-axisymmetric 
instabilities, such as the fragmentation of the collapsing core or the development of clumps in the accretion disk \citep{Giacomazzo2011}. GW/EM 
synergy plays a crucial role in the investigation of the nature of GRB progenitors and physical mechanisms. Indeed, gamma-ray 
emission and the multi-wavelength afterglow are thought to be produced at large distances from the central engine 
(i.e. $>$10$^{13}$~cm). By contrast, GWs will be produced in the immediate vicinity of the central engine, 
offering a direct probe of its physics.
Inspiral-like GW signals are predicted with an amplitude that might be observable by ET to luminosity distances of order 1~Gpc 
\citep[see, e.g.,][]{Davies2002}. With a rate of observed long GRB of $\sim$0.5~Gpc$^{-3}$~yr$^{-1}$, THESEUS could provide a simultaneous EM monitoring 
of $\sim$1 long GRBs every two years. Off-axis X-ray afterglow detections (``orphan afterglows'') can potentially increase the simultaneous 
GW+EM detection rate by a factor that strongly depends on the jet opening angle and the observer viewing angle. 
THESEUS may also observe the appearance of a NIR orphan afterglow a few days after the reception of a GW signal due to a collapsing massive star.
In addition, the possible large number of low luminosity GRBs (LLGRBs) in the nearby Universe, expected to be up to 1000 
times more numerous than long GRBs, will provide clear signatures in the GW detectors because of their much smaller distances 
with respect to long GRBs. 

\paragraph{Soft Gamma Repeaters \\} 
Fractures of the solid-crust on the surface of highly magnetized neutrons stars and dramatic magnetic-field readjustments represent 
the most widely accepted explanation to interpret X-ray sources such as giant flares and soft gamma repeaters \citep[SGRs; see, e.g.,][]{Thompson1995}. 
NS crust fractures have also been suggested to excite non radial oscillation modes that may produce detectable GWs 
\citep[see, e.g.,][]{Corsi2011, Ciolfi2011}. The most recent estimates for the energy reservoir available in a giant flare are between 
10$^{45}$~erg (about the same as the total EM emission) and 10$^{47}$~erg. The efficiency of conversion of this energy to GWs has been estimated 
in a number of recent numerical simulations and has been found to be likely too small to be within the sensitivity range of present 
GW detectors \citep{Ciolfi2012, Lasky2012}. However, at the typical frequencies of f-mode oscillations in NSs 
frequencies, ET will be sensitive to GW emissions as low as 10$^{42}$-10$^{44}$~erg at 0.8~kpc, or about 0.01\% to 1\% of the energy 
content in the EM emission in a giant flare. In the region of  20-100~Hz, ET will be able to probe emissions as low as 10$^{39}$~erg, 
i.e. as little as 10$^{-7}$ of the total energy budget \citep[see, e.g.,][and reference therein]{Chassande-Mottin2010}.

\paragraph{Neutrino sources \\}                                          
Several high-energy sources that THESEUS will monitor are also thought to be strong neutrino emitters, in particular SNe and GRBs. 
The shocks formed in the GRB ultra-relativistic jets are expected to accelerate protons to ultra-relativistic energies and that, 
after interacting with high energy photons, produce charged pions decaying as high energy neutrinos \citep[$>$10$^5$~GeV; see, e.g.,][]{Waxman1997}. 
Pulses of low energy neutrinos ($<$10~MeV) are expected to be released during core-collapse supernovae (CCSNe) with an energy 
release up to 10$^{53}$~erg. Indeed, low energy neutrinos have been detected from SN1987A at 50~kpc distance. Still significant 
uncertainties are affecting supernova models. GW and neutrino emission provide important information from the innermost 
regions as the degree of asymmetry in the matter distribution, as well the rotation rate and the strength of the magnetic 
fields, that can be used as priors in numerical simulations \citep[see, e.g.,][and reference therein]{Chassande-Mottin2010}.
Because of the neutrino very small cross-sections and low fluxes, neutrino detectors necessarily require huge amounts of 
water or liquid scintillator. Future Megaton detectors that are expected to work during the 3rd generation GW detectors, 
will reach distances up to 8~Mpc, guaranteeing simultaneous GW/neutrino and EM detection of $\sim$1~SN per year. 
Very promising for such multi-messenger studies are the LLGRBs, given their expected larger rate than for standard 
long GRBs (up to 1000 times more numerous) and their proximity. For long GRBs, joint THESEUS and GW/neutrino 
observations would further constrain progenitor models, clarifying the fraction of energy channeled via dynamical 
instabilities \citep{Fryer2002} and the relative neutrino/EM energy budgets. Neutrino observations would also 
constrain the composition of the GRB jet and the relation of GRBs to high-energy cosmic rays \citep{Abbasi2012}.

\begin{figure*}[t!]
\centering
\includegraphics[scale=0.21]{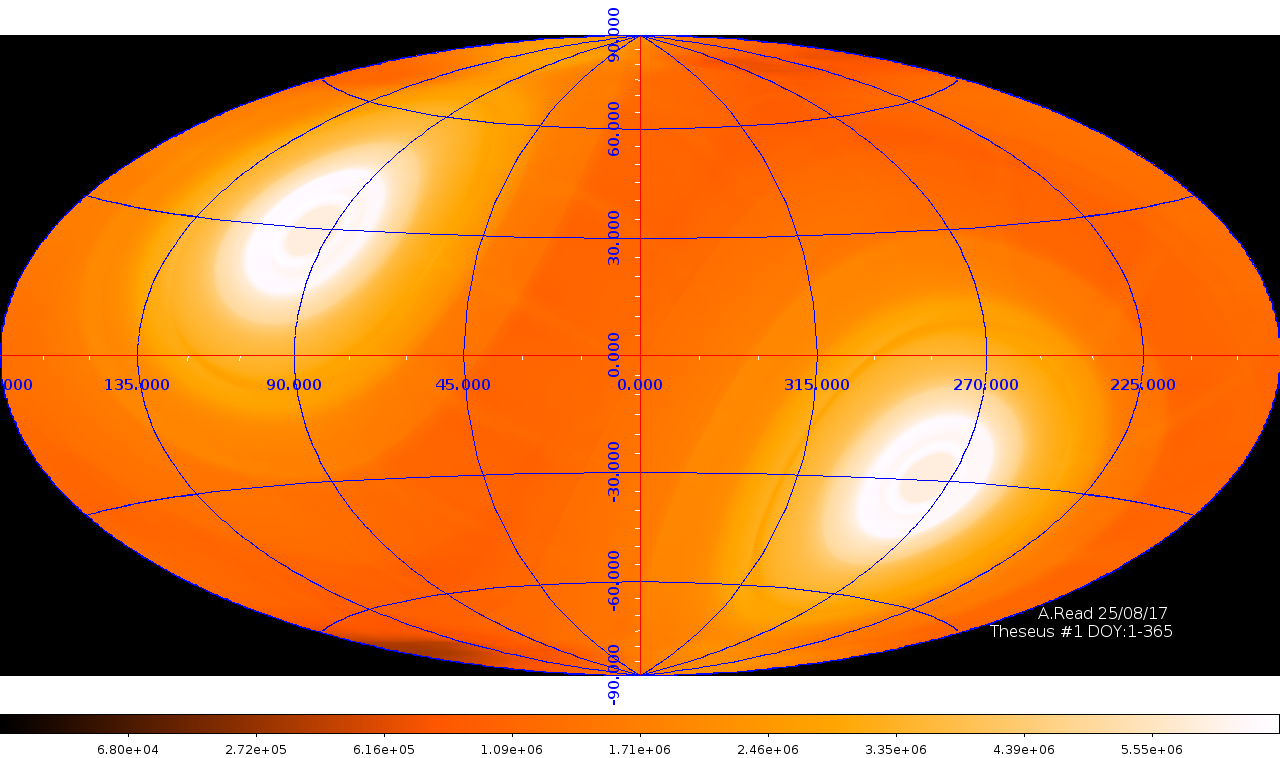}
\caption{Possible 1-year THESEUS/SXI sky exposure map (Galactic coordinates) based on the observing 
constraints and strategy described in next sections.}
\label{fig:12}   
\end{figure*}

\begin{figure*}[t!]
\centering
\includegraphics[scale=0.17]{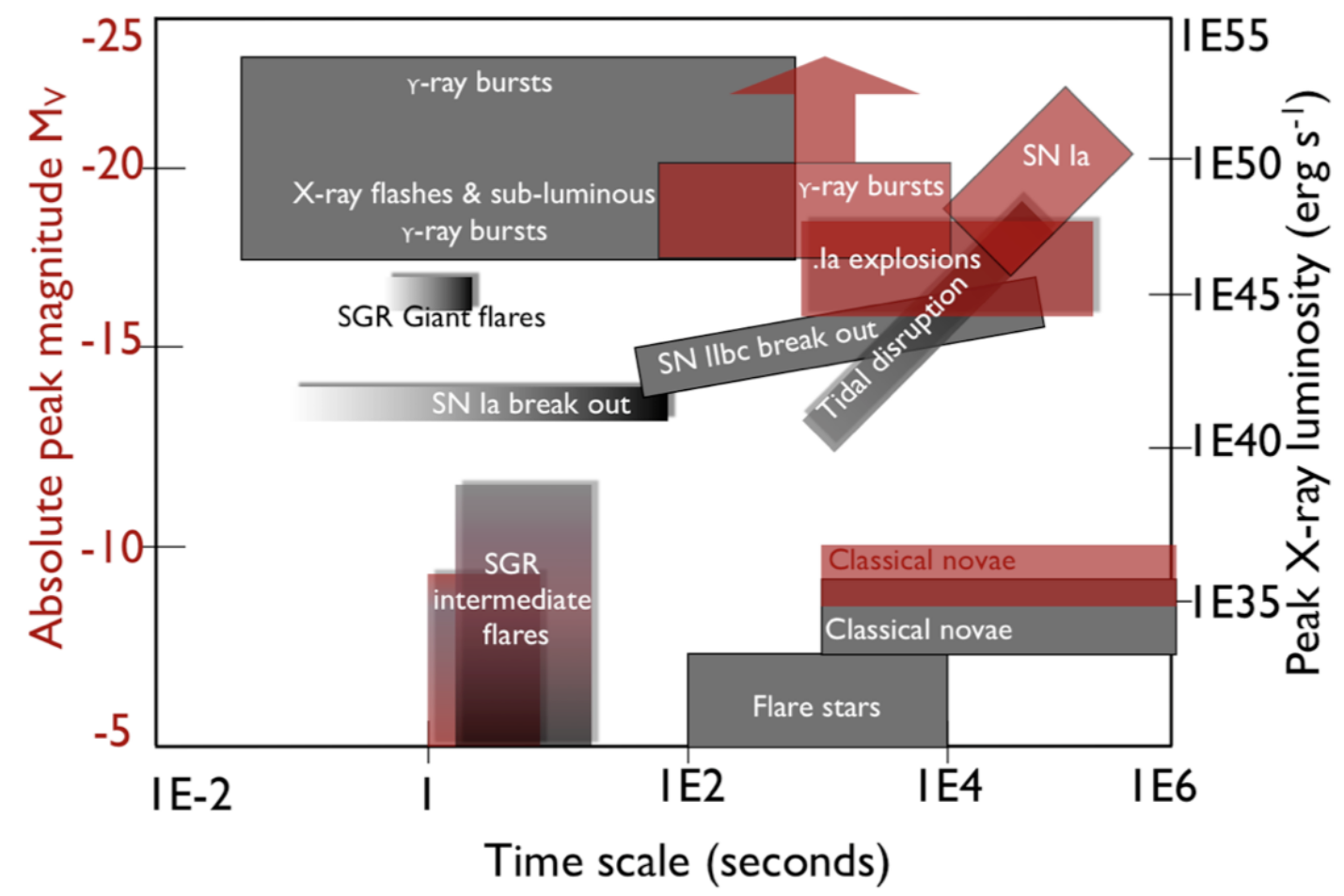}
\includegraphics[scale=0.25]{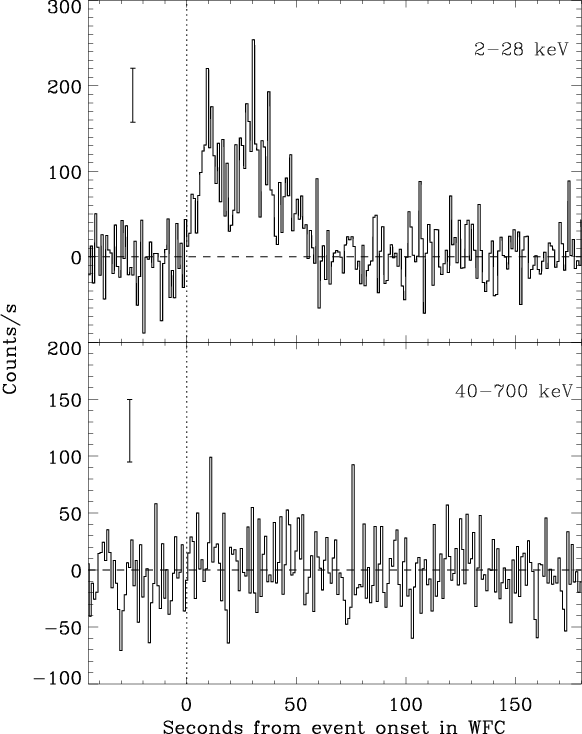}
\caption{{\it Left}: Time scales and luminosities of presently known soft X-ray transients \citep{Jonker2013}. {\it Right}: typical light curve of an X-Ray Flash, 
i.e. a GRB showing emission only in the X-ray energy range \citep[BeppoSAX XRF020427;][]{Amati2004}.}
\label{fig:13}   
\end{figure*}

\subsection{Exploring the Time-domain Universe}
The SXI and XGIS will detect a large number of both transient (Fig.~\ref{fig:13}) and steady X-ray sources serendipitously during regular observations 
(see Table~\ref{tab:2}). These data will provide a wealth of science opportunities.  Here we emphasise two primary objectives:
\begin{itemize}
\item reveal the violent Universe as it occurs in real-time, through an all-sky X-ray survey of extraordinary grasp and sensitivity carried out at high cadence. 
\item discover new high-energy transient sources over the whole sky, including supernova shock break-outs, black hole tidal disruption events, magnetar flares, 
and monitor known X-ray sources, including GRBs, with low latency observations.
\end{itemize}
These objectives relate directly to the Cosmic Vision 2015-2025 questions under (3.2), providing the crucial electromagnetic counterparts to GW sources such 
as compact star binary in-spirals and perhaps core-collapse SN, under (3.3), allowing the study of matter under extreme conditions around black holes 
and in neutron stars, and under (4.3), examining the evolving violent Universe through the study of quiescent and active massive black holes 
at the centers of galaxies. By finding huge numbers of GRBs the survey will also permit unprecedented insights in the physics and progenitors 
of GRBs and their connection with peculiar core-collapse SNe, and substantially increase the detection rate and characterization of sub-energetic 
GRBs and X-Ray Flashes. The provision of a high cadence soft X-ray survey in the 2020s together with a 0.7~m IRT in orbit 
will enable a strong synergy with transient phenomena observed with the Large Synoptic Survey Telescope (LSST).

\paragraph{Supernovae \\}
Supernovae mark the death of massive stars and are the prime process by which heavy elements are distributed through the Universe 
driving evolution in the stellar population. 
THESEUS will significantly advance our understanding of the SN explosion mechanism, 
detecting SNe at the very moment of emergence, gathering comprehensive, prompt data and alerting follow-up communities to new events. 
The birth of a new SN is revealed by a burst of high-energy emission as the shock breaks out of the star (giving access to measurements of the 
progenitor star radius). This has been spectacularly 
captured just once, in a serendipitous Swift XRT observation of SN2008D \citep{Soderberg2008}: SNe are usually found only days to 
weeks after the explosion, as radioactive heating powers optical brightening.  Theoretical calculations of shock breakout show that 
bright X-ray bursts of $L_{\rm X}=10^{43}-10^{46}~erg~s^{-1}$ are expected for both Wolf-Rayet stars and red supergiants (RSG) lasting 10-1000~s 
\citep{Nakar2010}. These progenitors are likely responsible for Type Ibc and most Type II SN respectively, which occur at rates 
of 2.6$\times$10$^{-5}$ and 4.5$\times$10$^{-5}$~Mpc$^{-3}$~yr$^{-1}$ \citep{Li2011}. 

\begin{table}
\centering
\caption{THESEUS detection rates for different astrophysical transients and variables.}
\label{tab:2}  
\begin{tabular}{ll}
\hline\noalign{\smallskip}
Transient type &  SXI rate \\
\noalign{\smallskip}\hline\noalign{\smallskip}
Magnetars & 40~day$^{-1}$ \\
SN Ia shock breakout & 4~yr$^{-1}$ \\
TDE & 50~yr$^{-1}$ \\
AGN+Blazars & 350~yr$^{-1}$ \\
Thermonuclear bursts & 35~day$^{-1}$ \\
Novae & 250~yr$^{-1}$ \\
Dwarf novae & 30~day$^{-1}$ \\
SFXTs & 1000~yr$^{-1}$ \\
Stellar flares and super flares & 400~yr$^{-1}$ \\
\noalign{\smallskip}\hline
\end{tabular}
\end{table}
SXI can detect these landmark events out to galaxies beyond 50~Mpc. These 
bursts first emit hard spectra when the shock is relativistic \citep{Nakar2012}, also allowing detection by the XGIS. Shock 
breakouts from blue supergiants are expected to release comparable energy to RSGs, emitting within the XGIS band. 

With THESEUS, we aim at finding the first X-ray bursts from thermonuclear Type Ia SN shock breakouts. The detection of even a 
single Ia X-ray event is a powerful discriminator between progenitor models, and constrains the explosion physics with implications 
for using SN Ia as standard candles and to constrain dark energy \citep{Sullivan2011}. Galactic Ia shock breakouts should be 
detectable by XGIS as very short high energy pulses \citep{Nakar2010}, while those in binary with a red giant star may produce 
bright X-ray bursts lasting minutes to hours via interaction with the companion wind \citep{Kasen2010}.

Relativistic SN shock breakout can also contribute a significant fraction of the early X-ray flux seen in low-luminosity GRBs 
\citep[see the cases of GRB~060218, and GRB~100316D;][]{Campana2006, Starling2011}, and THESEUS can make a significant contribution 
to this study, bridging the temporal gap between high and low energy X-rays of current instrumentation. Fundamental 
parameters and processes, such as the radius of breakout and the driver of the light curve time scale, as well as the 
nature of the progenitor stars themselves, remain unknown. THESEUS will discover all types of SN shock breakouts opening 
up this critical and unexplored time domain in SN evolution.

\paragraph{Tidal Disruption Events \\}
Tidal disruption events (TDEs) offer a unique probe of the ubiquity of super-massive black holes (SMBHs) in galaxies, 
accretion on timescales open to direct study, and the nature and dynamics of galactic nuclei. A star is tidally disrupted 
when the tidal forces from the SMBH exceed the self-gravity of the star \citep{Rees1988}. At this point half of the star is 
unbound, while the other half falls back to form an accretion disc at a characteristic rate of t$^{-5/3}$. Such events are 
expected to be visible in UV and soft X-rays \citep[see, e.g.,][]{Komossa2004, Gezari2012}. The discovery by Swift of 
two highly luminous outbursts from galactic nuclei implies that at times a fraction of this energy is deposited in a 
new relativistic jet outflow \citep{Levan2011, Bloom2011, Burrows2011}, offering a new route to 
their identification and an opportunity to study newly-born jets. THESEUS is ideal for both the discovery and 
characterization of TDEs, opening new windows on numerous astrophysical questions. 

TDEs offer a unique probe of SMBH presence across the Universe by revealing formerly dormant SMBH in galaxies across the 
mass scale up to $\sim$10$^8$~$M_{\odot}$. The key to extracting the most from these events is to obtain identifications early, while the 
first material is falling back. This in turn allows multi-wavelength observations at the peak of the light from which one 
can infer the mass of the black hole and the nature of the in-falling star. This can be compared to the properties of the 
galaxy and the $M_{\rm BH}$-$\sigma$ relation. 
TDEs provide unique laboratories for studies of physics under extreme conditions. They allow us to study accretion in 
active galaxies from onset to a return to dormancy, on timescales of only a few years. Through this time we can study 
the behaviour of accretion at a variety of rates, and the processes of disc and jet formation. Jets in the relativistic 
TDEs are promising locations for the acceleration of ultra-high-energy cosmic-rays. Finally, the rate and nature of the TDEs provides insight into 
the central dynamics of galaxies. An exciting possibility is using TDEs as tracers of BH-mergers. The dynamical impact 
of the merger on surrounding stars increases the TDE rate by several orders of magnitude to $>0.1$~yr$^{-1}$, raising the 
possibility of observing multiple events from a single galaxy undergoing a BH merger. For the classical TDEs, the 
SXI effective horizon in a single orbit is $\sim$200~Mpc ($z\sim$0.05). The rate of relativistic TDEs is uncertain, but both 
SXI and XGIS can see them to $z\sim$1 ($L_{\rm peak}=10^{48}~erg~s^{-1}$). For moderately conservative assumptions (beaming to 5\% of 
the sky in 10\% of TDEs) it is quite possible that the relativistic TDE rate will exceed the classical one. 

\paragraph{AGN and Blazars \\}
Active galaxies are the most powerful continuous sources of energy in the Universe and are powered by supermassive BHs 
in galactic centres. They also help control the growth of the stellar population in galaxies, so understanding their 
accretion activity is crucial. The accretion activity of the central BH in active galaxies varies over a large range 
in amplitude and timescale, due to variations in the emission efficiency or accretion rate/efficiency in AGN 
\citep[see, e.g.,][]{Mushotzky1993}, or to changes in relativistic jet properties in Blazars \citep[see, e.g.,][]{Marscher2010}. 
SXI provides the capability to monitor the X-ray flux of hundreds of AGN with $\sim$10\% accuracy on daily timescales, and hundreds 
more on longer timescales \citep{Ueda2005}. The survey strategy will permit an unbiased look at the long-term variability 
of an unprecedentedly large AGN/Blazar sample at depths never reached before. SXI monitoring will enable regular 
multi-wavelength campaigns to occur in order to probe the emission mechanisms and the geometry of the central 
active regions.

The observation of correlated X-ray/radio AGN variability with THESEUS and SKA, can be strongly diagnostic, 
allowing searches for mass-related lags and radio-frequency- related lags, opening up the jet physics. It is 
generally agreed that the radio to UV/X-ray emission from blazars is synchrotron emission from a relativistic jet oriented 
towards the observer, with the higher energy emission arising from Compton scattering of seed photons by particles in the 
jet, although many details of the jet structure are still unclear. The expected sensitivity of future radio observations 
is such that radio-X-ray studies of radio quiet Seyfert galaxies will then be possible. 
In particular, THESEUS will allow the detection of X-ray flaring activity in Seyferts. Connecting flare states and X-ray/radio 
variability in radio quiet AGN, would have implications for AGN evolution/feedback models.

THESEUS will monitor the bright AGN population and trigger follow-up observations by both THESEUS itself and other multi-wavelength 
facilities to measure the relation between different energy bands and the lag between bands which constrain the emission process. 
For example, Synchrotron Self-Compton emission predicts a TeV lag roughly equal to the light travel time across the emission region, 
whereas for external seed photons, assuming the source of variability is in the relativistic electrons of the jet, simultaneity is 
expected. Measurement of the complete shape of the synchrotron spectral component from radio to hard X-ray constrains the 
particle acceleration process in the jet and the balance of acceleration and radiative cooling; if energy is input via 
hadrons rather than via electrons, radiative losses will be less, so the spectrum will be harder at higher energies 
\citep[see, e.g.,][]{Tramacere2007}. In the THESEUS era, ten times greater TeV sensitivity than now will be provided by the 
Cerenkov Telescope Array, allowing the routine study of hundreds of blazars. VHE gamma ray emission is characterized 
by large outbursts during which detailed measurements can be made. Deep monitoring observations can also be made 
with XGIS whose hard X-ray spectral response is ideal for defining the synchrotron shape.

\paragraph{Accreting Binaries \\}
Thermonuclear X-ray bursts are produced by runaway nuclear burning on NS surfaces in our galaxy, often reaching the Eddington 
luminosity limit \citep{Strohmayer2006}. THESEUS probes a regime of deep nuclear carbon burning, and its thermal 
effects on neutron star crusts and cores, rarely observed before now \citep{Brown2004, Cumming2004}. These infrequent 
hours-long ``super-bursts'' are efficiently detected in the THESEUS sky survey due to its high exposure. It also captures 
seconds-long hydrogen- and helium-burning thermonuclear bursts more frequently. A large sample of burst properties will 
test models that predict nuclear burning dependence on mass accretion rate \citep{Heger2007}, and add to previous 
burst samples collected by BeppoSAX and RXTE \citep{Keek2010}. 

Classical and Recurrent Novae are produced by runaway thermonuclear burning on the surfaces of a white dwarf in a close binary system. 
SXI will, for the first time, detect the initial thermonuclear runaway burning phase, which lasts only a few hundred seconds. 
Previous X-ray studies have only been able to probe the shock-heated wind ahead of the ejecta and the steady burning phase 
that emerges later \citep[see, e.g.,][]{Osborne2011, Schwarz2011, osborne15}.
X-ray spectra and temporal profiles provide constraints on nuclear reaction processes, the white dwarf mass, and the 
underlying convective mixing \citep[see, e.g.,][]{Starrfield2009}.  
THESEUS will be able to monitor the brightnesses of the Super-Soft Sources \citep{Greiner1996}, candidate progenitors for Type Ia SNe, to give a 
significantly improved view of their accretion behaviour. Both Classical and Recurrent Novae are also sources of hard X-ray emission 
\citep{Sokoloski2006, Mukai2008} variable with time which appears later than the SSS phase. The onset of this emission
component  probes the existence of shocked material within the ejected shell. In addition the recent and new discoveries of Novae 
(both classical and recurrent) as source of particle acceleration by Fermi-LAT, and thus a class of gamma-ray emitters, challenges 
our knowledge of these transient objects \citep{Cheung2016}. The wide energy coverage of the SXI and XGIS instruments will allow us to 
track the temporal evolution of the different emission components along the outburst. All this will provide a significant improvement of 
our still poor knowledge of the tight link between accretion processes  and explosion mechanisms.

Accretion-driven outbursts and state changes are also seen in white dwarf, NS and BH binaries. Dwarf Novae outbursts occurring in white dwarf 
accreting systems (cataclysmic Variables) will be routinely observed in both soft X-rays and optical/nIR ranges. This unique opportunity will allow 
to solve the still open question on dwarf nova diversity in optical and X-ray behaviours \citep{Fertig2011}.
Furthermore, cataclysmic variables were thought  to be unable to launch jets. The recent radio discovery  of two high accretion rate systems, 
including dwarf novae \citep{Koerding2008, Koerding2011}, makes crucial the synergy with SKA to understand 
jet-launching processes irrespective of the compact object nature (WD, NS or BH). 
Truly magnetic systems are the X-ray brightest among CVs. However the state changes are not abrupt but on a long-term scale (months-yrs). 
THESEUS will provide excellent coverage of the outbursts of classical neutron star and black 
hole X-ray binaries, for example Supergiant Fast X-ray Transient variability reaches up to a factor of 10$^6$ \citep{rom2015}, with peak 
luminosities up to 10$^{38}$~erg~s$^{-1}$; the hour-timescale flares from these OB plus (presumed) NS systems have frequently triggered 
the Swift-BAT. THESEUS can constrain the temperature and optical depth of the accretion column \citep{Farinelli2012, Bozzo2016}, 
and the origin of the bright flares possibly due to wind accretion onto a magnetar \citep{Bozzo2008}. THESEUS will detect and provide 
localization for several black-hole transients, monitoring daily their X-ray spectral evolution throughout their full outburst. 
Pointed observations with IRT will provide strictly simultaneous IR photometry (and often spectroscopy) with very good statistics, 
allowing an unprecedented study of the disk-jet connection across all accretion regimes.

\paragraph{Magnetars \\}
Magnetars, young NS with external magnetic fields of 10$^{13}$-10$^{15}$~G, are among the most powerful and spectacular high-energy transients in the sky. 
They are characterized by the emission of highly super-Eddington short bursts of emission (Soft Gamma-ray Repeaters) and more rare Giant Flares 
(luminosity up to $10^{47}$~erg~s$^{-1}$).  Magnetars differ from other more common classes of neutron stars because all their emission (both persistent 
and bursting) is powered by the gradual and/or impulsive dissipation of magnetic energy, rather than by rotational energy or accretion 
\citep{Thompson1995}. There is evidence that the field in the interior of magnetars can exceed 10$^{16}$~G, probably as a result of 
their rapid rotation at birth (1-2~ms). About twenty sources believed to be magnetars are currently known in our Galaxy and in the 
Magellanic Clouds, but since most of them are transients with long quiescent periods, the total population waiting to be discovered 
is certainly much larger \citep{Mereghetti2008}.

Magnetars can produce Giant Flares thought to be due to star crustal fractures. Their EM emission consists of an initial, short 
($<$0.5~s) spike of hard X-rays followed by a tail of softer X-rays lasting minutes and modulated by the NS rotation (P$\sim$2-12~s). These 
extremely bright initial spikes can be detected with the XGIS to considerable distance. Based on the rate of the few Giant Flares observed 
to date, the XGIS's energy range will be better suited for the detection of such events than current coded-mask detectors due to its 
lower energy threshold.

The persistent X-ray emission from known magnetars is too faint and strongly absorbed to be detectable by SXI when these sources are 
in a quiescent state; however, the increased level of X-ray emission associated with flares and periods of increased bursting activity 
will be easily detectable. Thus THESEUS triggers will enable detailed observations of these events. These ``intermediate flares'' are 
more frequent than the Giant Flares. The wide field of view of SXI and the frequent sky coverage will, for the first time, allow 
detection of a large number of flares and obtain a reliable estimate of the frequency of such events. The count rate expected in 
SXI for a typical intermediate flare will allow detailed time-resolved study of flare properties.
\begin{figure*}[t!]
\centering
\includegraphics[scale=0.55]{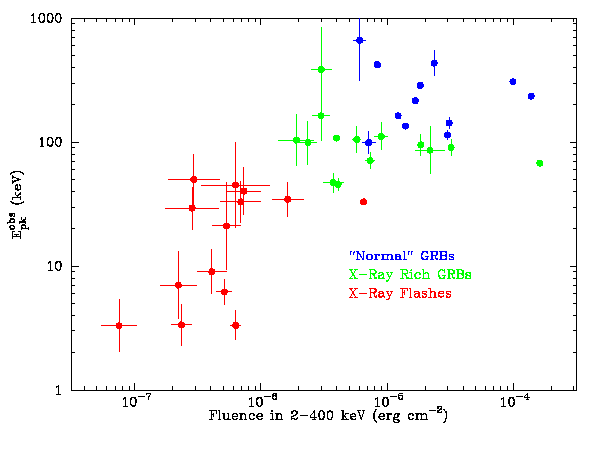}
\includegraphics[scale=0.45]{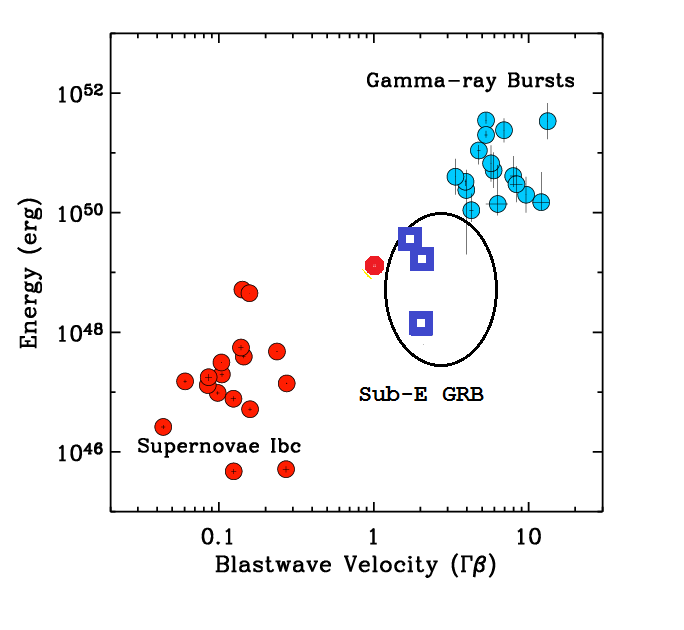}
\caption{{\it Left}: distributions in the spectral peak energy ($E_{\rm p}$) - Fluence plane of soft X-Ray Rich (XRR) GRBs and X-Ray 
Flashes (XRFs) compared to that of normal GRBs \citep{Sakamoto2005}. {\it Right}: Blast wave velocity and energy for 
massive star explosions \citep[adapted from][]{Soderberg2010}. Soft/weak GRBs may constitute the bulk of the GRB 
population and the link with other explosive events associated  to the death of massive stars.}
\label{fig:15}   
\end{figure*}

\paragraph{Stellar Coronae \\}
Stellar Flares of all sizes are important probes of coronal structures and their energetics, and thus, the underlying stellar magnetic 
dynamo \citep[see, e.g.,][]{Reale2007, Aschwanden2009}. THESEUS will quantify the extreme ``super-flares'' of nearby magnetically active 
stars $>$10$^4$ times more energetic than the Sun's largest flares. X-ray flares have long been detected and cataloged 
\citep[see, e.g.,][]{Pye1983, Favata2002, Guedel2004}, but rare ``super-flares'' can briefly ($\sim$1000~s) exceed the star's quiescent 
bolometric luminosity, and only a handful are known \citep[see, e.g., RS CVn binary II Peg and dMe star EV Lac;][]{Osten2007, Osten2010}. As an X-ray 
mission, THESEUS will directly address the high-energy ionising radiations of primary importance for solar-terrestrial interactions. 
THESEUS will provide estimates of coronal loop structure sizes and energy release in flares. Occurrence rates and luminosities 
constrain stellar models, and also affect the conditions for life in the habitable zone. Although much progress has been and continues 
to be made from studies of our own Sun, many questions remain regarding fundamental flare processes \citep[see, e.g.,][]{Guedel2004, Benz2010}; 
THESEUS will allow the theoretical models for energy release and propagation to be tested 
within a greatly expanded phase space. SXI will be the primary instrument for the stellar-flare survey, and its photon-energy 
band-pass is well-tuned to stellar-coronal emissions (which have typical characteristic temperatures $\sim$0.1-10~keV).

\subsection{GRB physics, progenitors and cosmology}
\label{grb}

\paragraph{X-ray Flashes, sub-energetic GRBs and GRB-SN connection \\}
It is well established that long GRBs are linked to SN originating from the core collapse of a stripped-envelope massive star 
(SN type Ibc). However, for GRBs the bulk of the explosion energy goes into relativistic ejecta, while the vast majority of 
SN Ibc show sub-relativistic ejecta with an energy release several orders of magnitude lower than GRBs. Some SN Ibc must 
therefore harbour  an additional key ingredient, i.e. a central engine - likely a nascent black hole - that drives the 
explosion and launches a relativistic outflow producing the GRB. Why this happens is not clear. The missing link between 
the two classes of explosions may be found amongst the X-Ray Flashes (XRFs, see Fig.~\ref{fig:15}), commonly considered 
as a softer sub-class of GRBs (but see also \cite{Ciolfi2016}), and LL-GRBs, likely related to a population of only slightly 
relativistic SN which have recently been found in the radio. 
Current X-ray facilities are quite insensitive to such events, which are typically characterized 
by soft spectra (low $E_{\rm peak}$). These events will make up $\sim$1/3 of the THESEUS GRBs, populating the existing gap between GRBs 
and ordinary SN (Fig.~\ref{fig:13}). Because of their low luminosity, we expect that most of these events will be at $z<$0.5 and that, 
in terms of rate density, they constitute the bulk of GRB population. Present surveys are in fact biased towards harder, 
more luminous events \citep[less than 5\% of long Swift bursts are at $z<$0.5;][]{Jakobsson2012}. The detection of intrinsically 
faint, X-ray soft GRBs can only be done at low X-ray energies, when there are enough photons for their detection. 
An important goal of THESEUS is to understand the paths of stellar evolution leading to the production of GRBs and of 
SNe, as described in below.

\begin{figure*}[t!]
\centering
\includegraphics[scale=0.35]{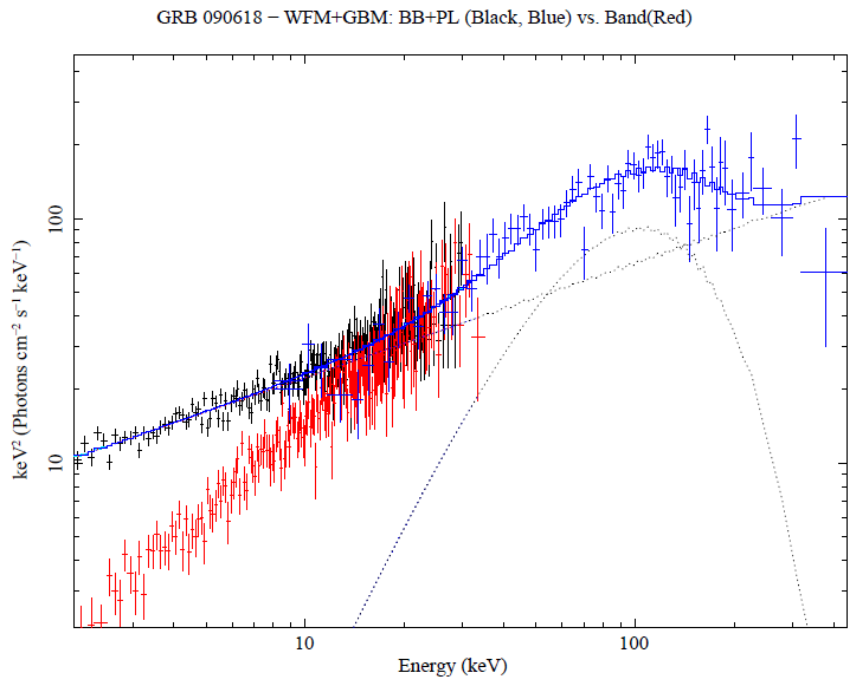}
\includegraphics[scale=0.33]{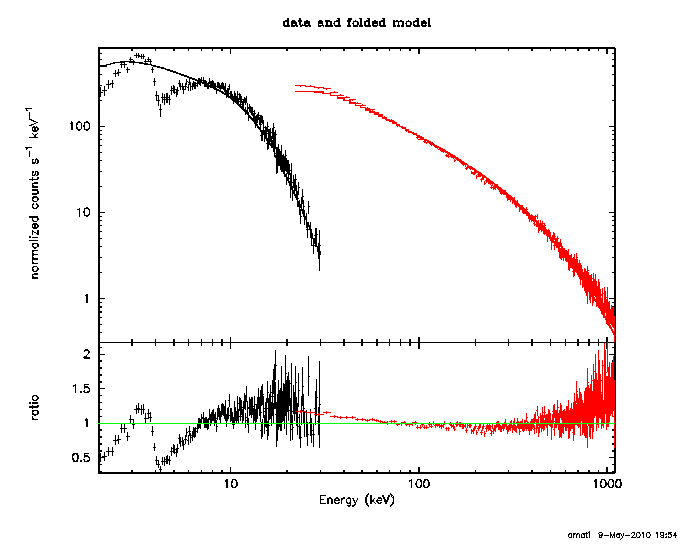}
\caption{By satisfying the requirements for the main science (Fig.~\ref{fig:20}), in particular the extension of its energy band 
down to 2~keV with large area and 300~eV energy resolution, the THESEUS/XGIS will show unique capabilities for 
discriminating among different GRB emission models (left) and detecting absorption features expected at energies 
$<$10~keV (circum-burst environments and X-ray redshift determination). {\it Left}: Simulated XGIS low-energy (SDD) spectra 
of the first 50~s of GRB~090618 \citep{Izzo2012} obtained by assuming either the Band function (black) or the power-law 
plus black-body model (red) which equally fit the Fermi/GBM measured spectrum, which is also shown (blue). The black-body 
plus power-law model components best-fitting the Fermi/GBM spectrum are also shown (black dashed lines). {\it Right}: Simulation 
of the transient absorption feature in the X-ray energy band detected by BeppoSAX/WFC in the first 8~s of GRB~990705 
\citep{Amati2000} as would be measured by the XGIS.}
\label{fig:16}   
\end{figure*}

\paragraph{GRB physics and circum-burst environment \\}
Current GRB facilities do not permit prompt X-ray observations in the spectral band where most of the photons are emitted. 
Around 60\% of THESEUS GRBs will be simultaneously detected by SXI and XGIS during the prompt emission, allowing for the first 
time measures of the energy spectrum from 0.3~keV to 10~MeV, in a domain little explored, but crucial to discriminate 
among different emission models and determine the effects of intrinsic absorption in the GRB environment (Fig.~\ref{fig:16}). 
Such observations are needed to validate models based on synchrotron emission and to measure the impact of the GRB on its surrounding. 

It is worth noting that a number of recent studies have found evidence of a sub-dominant thermal emission in the prompt emission spectrum 
\citep{ryde2010, guiriec2011, br2015}. The thermal emission is also reported until very late emission phase of the longest ultra-long 
GRB~130925A and it has been shown to have a very similar spectral evolution as a ordinary long GRB~090618 \citep{br2015}. The former was 
observed with the fine resolution data of \emph{Swift}/XRT, \emph{Nustar} and \emph{Chandra}, while for the later XRT data was available 
at the late prompt emission phase. As such studies rely on the focusing observations, they are rare and miss the evolution in the initial 
phase of emission. With THESEUS such studies will be routinely done and a broadband, fine resolution spectrum will be obtained from 
very early emission phase. Firstly, it will be an enormous step forward to study the evolution of the individual spectral components from the early 
phase, secondly, this will also increase the sample size and help in finding the spectral diversity or a unification across the GRB catalog. 

\paragraph{GRB physics in EM-GW\\}
GRBs and, more generally, core-collapse supernovae may be multi-messenger sources of  electromagnetic (EM) and gravitational waves (GW) 
by their potential association with neutron  stars and black holes. These alternatives may leave their imprint in high-frequency modulations  
of associated high energy emission, in particular, by misalignment (alignment) of magnetic  moments with the spin of a magnetar (black hole). 

\begin{figure}[t!]
\centering
\includegraphics[scale=0.2]{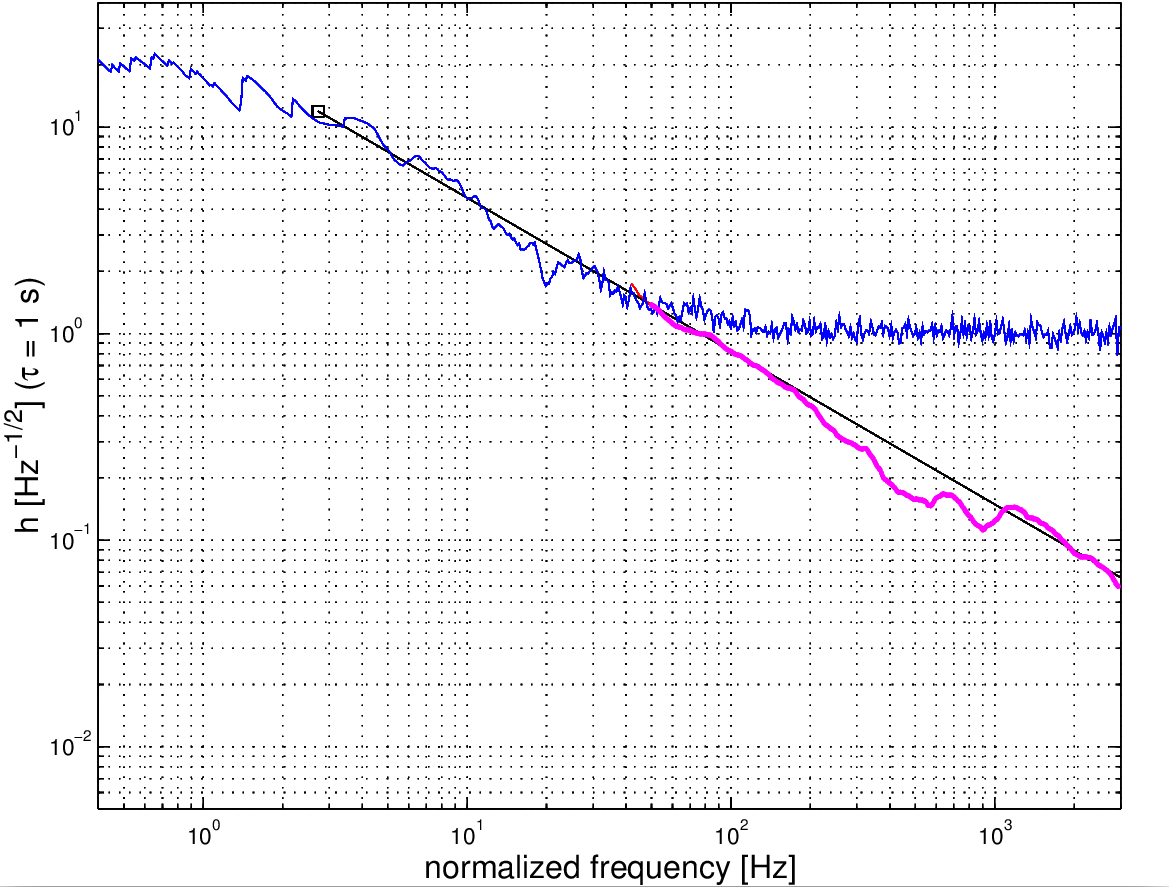}
\caption{Broadband spectrum obtained as an average of 42 relatively bright bursts from the BeppoSAX catalog through butterfly filtering 
(purple curve) using a bank of 8.64~million chirp templates. The broadband Kolmogorov spectrum extends standard Fourier-based 
spectra (blue curve) up to a few kHz in the co-moving frame. The absence of 
any high frequency bump favors inner engines harboring black holes rather than magnetars \citep[from][]{vanPutten2014}.}
\label{fig:16bis}   
\end{figure}

High resolution light curves of relatively bright GRBs are hereby expected to show non-smooth (smooth) broadband Kolmogorov spectra. Smooth 
broadband Kolmogorov spectra have  been found up to the co-moving frequency of a few kHz of bright LGRBs sampled at 2~kHz from the BeppoSAX 
catalog (see Fig.~\ref{fig:16bis}). 

THESEUS exceeds BeppoSAX in collecting area and time resolution, allowing a more detailed analysis of individual events and/or events at 
higher redshifts. Smooth broadband spectra from THESEUS would further evidence the black hole inner engines, absent any high  frequency 
modulations in spectra of gamma-ray light curves.

Results such as these provide potentially powerful priors to our searches for GW  emission accompanying LGRBs. LGRBs from rotating black 
holes, for instance, will  produce their highest GW frequencies from non-axisymmetric mass motion at the Inner Most Stable Circular Orbit 
(ISCO) powered by the ample reservoir in angular momentum of a stellar mass black hole. Efficient extraction of broadband 
spectra from gamma-ray light curves is made possible by  {\em Graphics Processor Units} (GPUs), allowing deep searches by using banks 
of millions of chirp templates developed to extract chirp-based spectrograms from noisy time-series of GRBs  (BeppoSAX, THESEUS) 
and LIGO-Virgo or KAGRA alike.

\paragraph{Complete samples \\}
In order to study properties of GRBs, their X-ray and optical/NIR afterglows, and their hosts is important to handle a sample 
that is not biased towards classes (i.e. a complete sample), because of limitation during the observations. 
Until now, several GRB complete samples have being created \citep[see, e.g.,][]{Perley2016, Greiner2001, 
Salvaterra2012}. In addition to these, other tools to overcome the problems of unbiased distributions with robust 
and sophisticated statistical techniques have already been successfully applied to GRB prompt and afterglow emission \citep{Dainotti2013a, Dainotti2015b}. 
The capability of THESEUS to detect most afterglows in the IR, excluding the few highly extinguished or at extreme redshift 
($<$10\%), will allow us for the first time to build a complete sample of GRB afterglows observed in X-rays and IR. The fact that 
THESEUS will not be limited by weather conditions and visibility constraints, but only from pointing limitation and foreground 
Galactic extinction, is a strong advantage in respect  to ground based facilities dedicated to GRB follow-up (e.g. RATIR, GROND, REM).

\paragraph{Multi-wavelength prompt emission \\}
\begin{figure}
\centering
\includegraphics[scale=0.5]{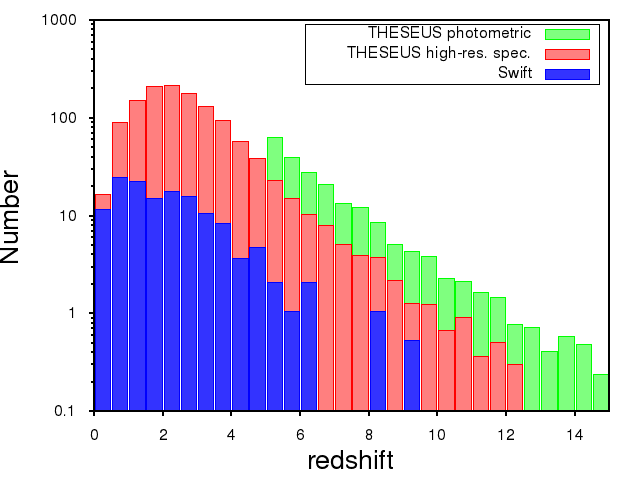}
\caption{The redshift distribution of THESEUS GRBs during a 5~yr mission lifetime compared to the actual distribution of Swift GRBs 
(blue) during the same period. GRBs with a photometric redshift are in green, and those with a spectroscopic redshift are in red.}
\label{fig:17}   
\end{figure}
Right now, thanks to Swift/UVOT and ground based optical/NIR telescopes dedicated to the rapid follow-up of GRB afterglows, 
it is possible to simultaneously follow-up the prompt emission from optical to X-ray to Gamma-rays. 
These observations allow us not 
only to better constrain the spectral energy distribution from optical to Gamma-rays and their lightcurves, testing the standard 
model, but also to test the nature of the central engine. Indeed, the observer may see simultaneously photons that have been emitted 
in different times and regions of the flow, and also with different physical origin, e.g., synchrotron or synchrotron self-Compton 
emission. 
However, there are only a few tens of bursts in 12~years of Swift activity that could be long 
and bright enough to be detected in optical \citep{Levan2014}. Indeed, up to now only five events have been studied in such detail 
\citep[see, e.g.,][]{Bloom2009, Rossi2011, Stratta2013, Troja2017}, and while they probe that standard fireball 
model can explain the observations, in some cases the optical and high-energy emission seems unrelated, or require a more complex 
modeling of the jet structure. Moreover, these observations have been performed in the optical, which is more affected by foreground 
and host line of sight dust extinction.
With THESEUS/IRT capability of starting to obtain the first images within the first 5~min from the trigger, it will be possible to detect optical 
prompt emission for the longest GRBs, roughly 10 to 20~GRBs per year. This will dramatically increase the number of events to study, 
and allowing us to statistically explore the parameter space of several models, shedding light on the structure of the jet during its 
first phases.

\paragraph{X-ray early and late afterglow \\} 
Opposed to what was initially believed, X-ray afterglows do not have a simple power-law decay. On the contrary, the vast majority 
shows a canonical behavior \citep{Nousek2006}: an initial rapid decay during the first few hundred seconds, 
a slow fading plateau phase that can persist even longer than 10$^4$~s, and a final power-law decay, with a possible achromatic 
break. While it is clear that the initial rapid decay corresponds to the tail of the prompt emission, the plateau phase is difficult 
to explain within the collapsar scenario, because it requires a long activity of the central engine. Alternatively, the necessary energy may 
come from the spin-down activity of a magnetar formed during the collapse \citep[see, e.g.,][]{Zhang2001, DallOsso2011, Rowlinson2014, Rea2015}. To complicate 
this view, X-ray flares (not visible in gamma-rays), probably due to late emission, indicate that the central engine is still active.
The relativistic outflow of a GRB is likely collimated in a jet \citep{Sari1999}. Since the jet slows down, at some point the 
relativistic open-angle becomes larger than the relativistic jet-opening angle. Thus the observer starts to see a deficit of photons 
compared to the case of isotropic emission, leading to an achromatic break in the light curve. These breaks, together with the 
knowledge of the redshift and the measure of the isotropic energy, allow us to estimate the jet-opening angle and the collimated 
energy (and used as cosmological indicator). Achromatic jet-breaks have been observed in optical/IR, however they often do not 
coincide with breaks observed in X-rays, indicating that our knowledge of the afterglows emission is still limited, and questioning 
if the jet-angle and collimated energy that have been estimated are right. Even if from the theoretical side many explanations 
(energy injections, double jet, structured jets) have been proposed to interpret the afterglow emission from IR/optical to X-rays, 
more multi-frequency observations are necessary \citep[see, e.g.,][]{Willingale2017}. Compared to today, the larger number of THESEUS 
GRBs and the more sensitive spectra observed with XGIS will allow us to better understand the nature of the afterglow and of the 
central engine of GRBs.
The study of the optical/NIR and X-ray afterglows unveils the properties of the environment. It is well known that the circumburst 
density profile influences the shape of the GRB light curves and spectra \citep[see, e.g.,][]{Racusin2009} distinguishing by ISM and wind 
environments \citep[see, e.g.,][]{Schulze2011}. Moreover, dust and gas in the line of sight dim the optical/NIR and X-ray afterglows, 
respectively \citep[see, e.g.,][]{Greiner2011}. Their systematic study will unveil the properties of the environment where GRBs explode. 
Unfortunately, up to today this has been limited by the different time coverage of X-ray and optical/NIR  observations and 
sensitivity to the late afterglows. THESEUS will likely solve this problem thanks to the simultaneous observations of 
optical/NIR and X-ray afterglows. 

\paragraph{Optical/IR afterglow detection with IRT \\}
As shown by \citet{Kann2017}, within the first hour all known optical afterglows 
have R$<$22. A classical optical afterglow has a spectral slope $\beta$$\sim$1, which translates in a color R-H$\sim$1~mag (AB photometric system), 
thus within the first hour all known afterglows have H$<$21. IRT will observe optical afterglows longer than 30~min within 1~hour from 
the trigger, reaching $H_{\rm AB}$$\sim$20.6. The optical/NIR imager GROND, reaching 1~mag fainter limits only, has been able to 
detect $\sim$90\% of all GRBs detected by Swift within 4~hours from the trigger \citep{Greiner2011}. Note that the host extinction will 
mostly have a negligible effect, with only a few cases ($\sim$10\%) with $A_{\rm V}$$>$0.5, which will noticeably dim ($>$1~mag) the observed NIR afterglow 
when the redshift is z$>$4, and still obtaining a detection rate of $\sim$90\%. However, at these redshift dusty environments are less common, because 
dust did not have the time to accumulate in the star forming regions. 
Notably, the higher rate of THESEUS GRBs will allow us to better understand the shape of dust extinction curve at high redshift which 
is now unexplored, the presence of 2175A absorption in GRB SEDs in high redshift environments and to test different models for dust 
grains.
\begin{figure}
\centering
\includegraphics[scale=0.25]{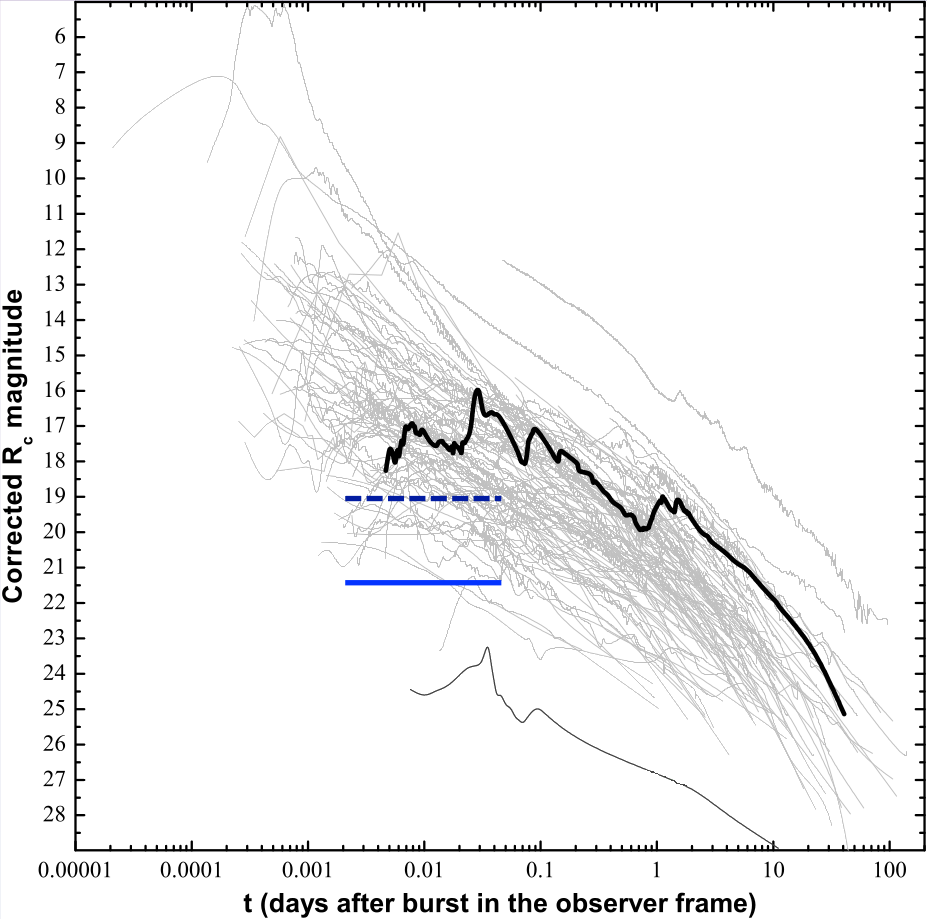}
\caption{Observed R band lightcurves of long GRBs. Highlighted is the ultra long GRB~111209A and the extremely extinguished GRB~130925A (bottom curve). 
Data is corrected for Galactic extinction. Adapted from \citet{Kann2017}. We also show in the figure with blue lines the THESEUS/IRT sensitivity 
(dashed line spectroscopy, solid line imaging).}
\label{fig:18}   
\end{figure}

\paragraph{Short GRBs \\}
Short GRBs are the least understood class of GRBs. Our understanding of the nature and range of progenitors of short GRBs remains hampered by their small 
number, and THESEUS will contribute importantly to increasing the sample for study. More fundamentally, the observations of THESEUS 
will be crucial for the interpretation of candidate signals in advanced detectors of gravitational waves. As discussed above, THESEUS/XGIS 
will find $\sim$20~SGRBs~yr$^{-1}$, most of which will be localized more precisely with SXI follow-up. At least 1/3 of the short GRBs are followed 
by a period of soft X-ray emission lasting 10-100~s. This emission carries an energy comparable and often larger than the initial spike, 
and it will be easily detected by THESEUS. THESEUS will extend the search of this intriguing feature down to a factor of $\sim$10 below present 
upper limits, allowing also for the search of ``orphan extended tails'' of SGRBs if they are not beamed \citep{Bucciantini2012}.
\begin{figure*}[t!]
\centering
\includegraphics[scale=0.22]{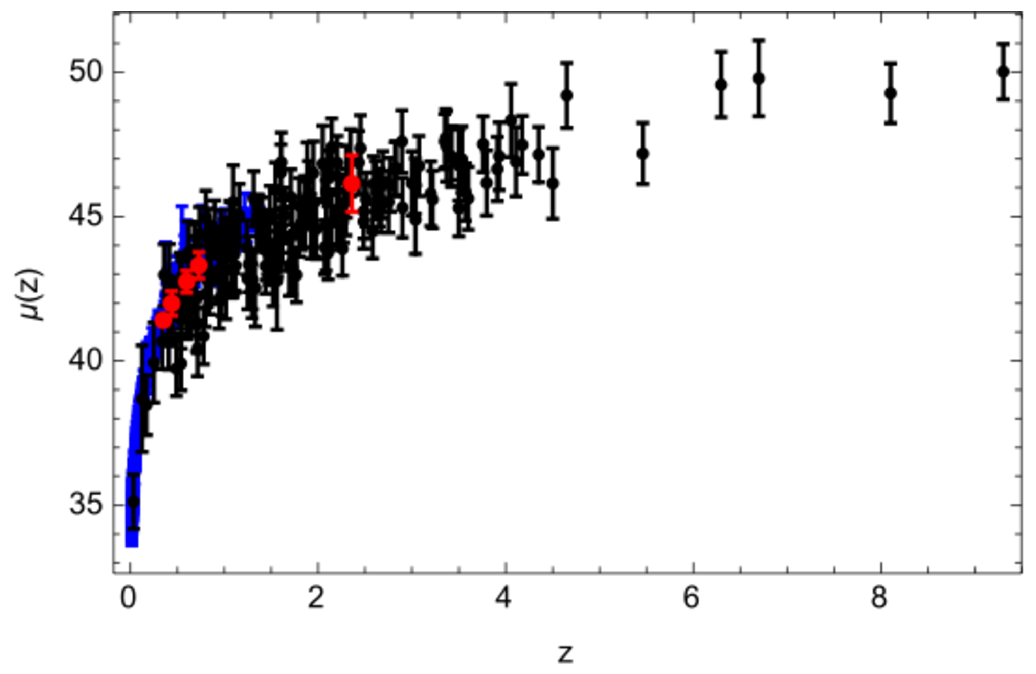}
\includegraphics[scale=0.25]{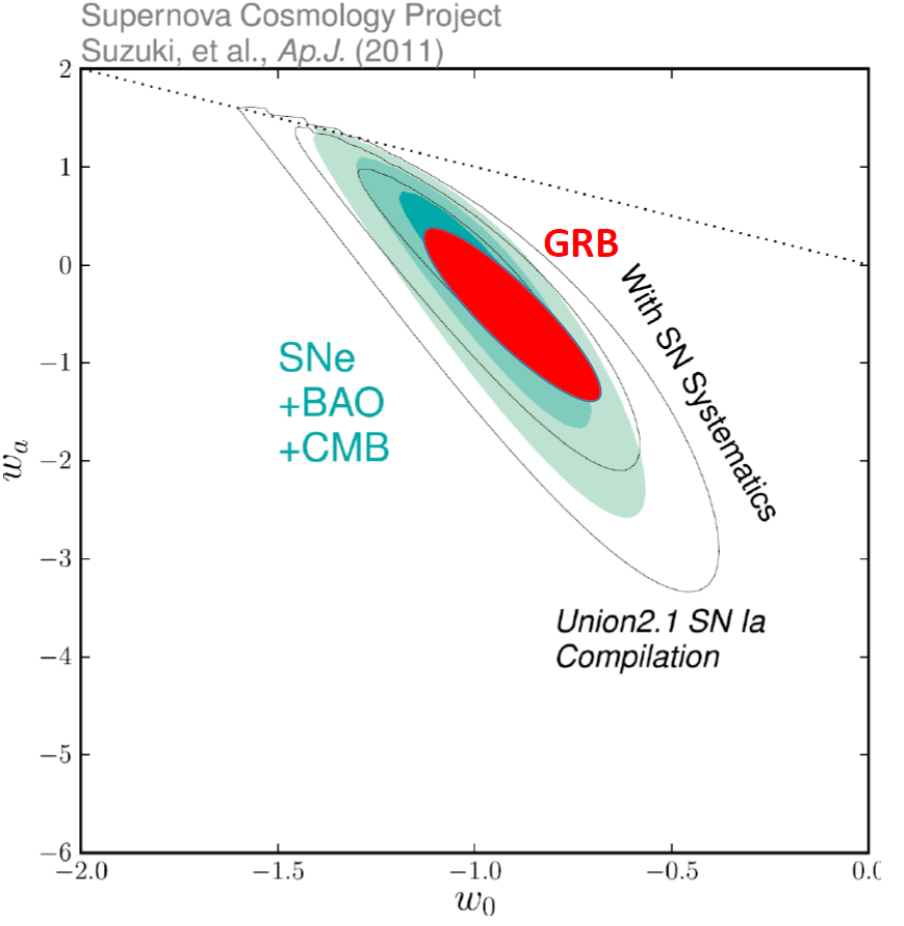}
\caption{{\it Left}: SNe-Ia + GRB Hubble diagram  obtained by exploiting the correlation between spectral peak energy ($E_{\rm p}$) and radiated 
energy or luminosity ($E_{\rm iso}$, $L_{\rm iso}$) in GRBs \citep{Demianski2017}. {\it Right}: determination of the dark energy equation of state parameters 
$w_0$ and $w_{\rm a}$ with the expected sample of GRBs form THESEUS by using the same \citep{amati13}.}
\label{fig:19}   
\end{figure*}

\paragraph{Probing expansion history of the universe and dark energy with GRBs \\}
In a few dozen seconds, GRBs emit up to 10$^{54}$~erg in terms of an equivalent isotropically radiated energy $E_{\rm iso}$, so they can be observed 
to z$\sim$10 and beyond. 
By using the spectral peak energy - radiated energy (or luminosity) correlation $E_{\rm p,i}$ - $E_{\rm iso}$ \citep{Amati2002, Yonetoku2004}, the luminosity 
at the end of the plateau emission and its rest frame duration correlation \citep{Dainotti2008} and the fundamental plane relation, an extension of the 
previous one by adding the peak luminosity in the prompt emission \citep{Dainotti2016a} are robust correlations studied and discussed for many years. 
Through these correlations, it has been demonstrated that GRBs offer a very
promising tool to probe the expansion rate history of the universe beyond the current limit of $z=2$ (Type-Ia SNe and Baryonic Acoustic Oscillations from QSO absorbers).
Additionally, the $E_{\rm p,i}$ - $E_{\rm iso}$ correlation has been shown to hold within the individual pulses of GRB prompt emission which is a 
strong argument against any instrumental selection bias, and conveniently increases the sample size for such studies \citep{br2013}. 

With the present data set of GRBs, cosmological parameters consistent with the concordance cosmology can already be 
derived \citep[see, e.g.,][]{Ghirlanda2004, Amati2008, Dainotti2013b}. Current (e.g., Swift, Fermi/GBM, Konus-WIND) and 
forthcoming GRB experiments (e.g., CALET/ GBM, SVOM, Lomonosov/ UFFO, eXTP/ WFM) will allow us to constrain $\Omega_{\rm M}$ and the dark energy 
equation of state parameters $w_0$ and $w_{\rm a}$, describing the evolution of $w$ according to $w=w_0+w_{\rm a}(1+z)$, with an accuracy comparable 
to that currently obtained with Type Ia supernovae (Fig.~\ref{fig:19}).  The order of magnitude improvement provided by THESEUS on the sample 
of GRBs, with measured redshifts and spectral parameters, will allow us to further refine the reliability of this method. This will 
offer the unique opportunity to constrain the geometry, and therefore the mass-energy content of the universe back to z$\sim$5, thus 
even extending the investigations of EUCLID and of the next generation large scale structure surveys to the entire cosmic history.
\begin{figure*}[t!]
\centering
\includegraphics[scale=0.6]{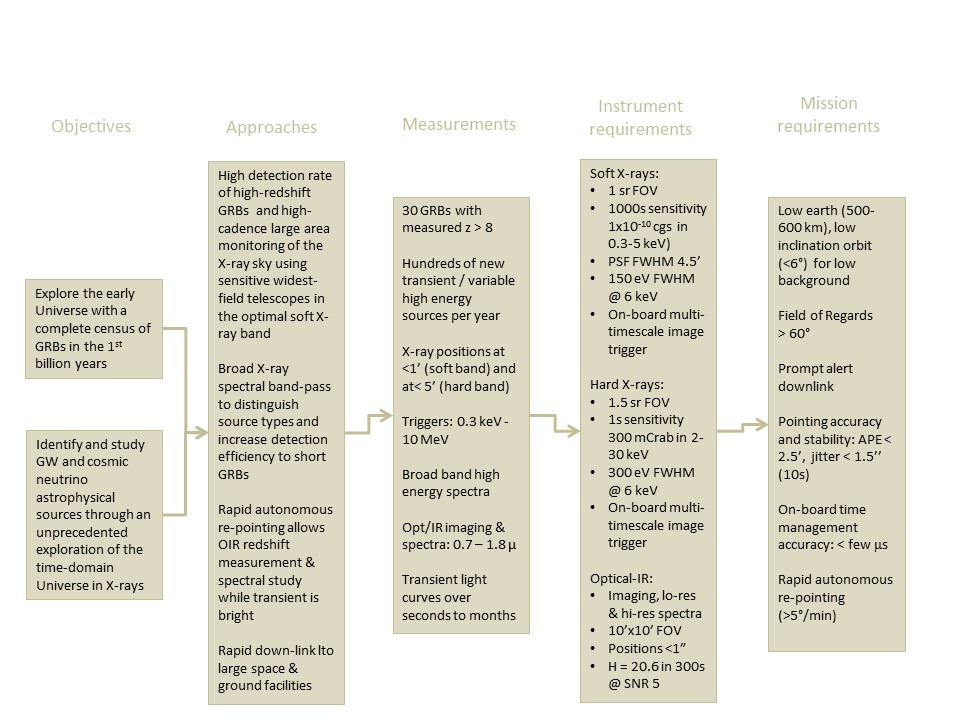}
\caption{Conversion from THESEUS science goals to instrument and spacecraft requirements.}
\label{fig:20}   
\end{figure*}

\paragraph{Synergies with JWST and E-ELT \\} 
Recently, it has been shown that GRB~111209A is linked to a SN with properties dissimilar to any known GRB-SN, with a spectrum more 
in accordance with Super-luminous supernovae \citep[SLSN; see, e.g.,][]{Greiner2015}, and like them was probably not powered by the standard 
collapsar central engine. The high rate of detection of THESEUS GRBs will significantly increase the number of long GRBs linked to SLSNe 
which can be studied with the planned JWST and E-ELT, shedding light on the GRB progenitors and their driving mechanisms. 
With the exception of the brightest hosts at low redshift (z$<$2), IRT will not allow to detect hosts in NIR. However, their contribution 
will be visible as emission and absorption lines in the afterglow spectra \citep{Hjorth2012, Savaglio2009}, and therefore we will be able to measure 
their spectroscopic redshift. Moreover, the localization of the NIR afterglow and its immediate distribution to the astronomical community, 
will permit others to follow-up the event with other ground and space based facilities, in particular the future instruments on the JWST and 30~m 
ground based telescope (e.g., ELT, GMT, TMT). We thus summarize the most relevant synergies in the following two points: 
\begin{itemize}
\item The Environment of GRBs with the future JWST and 30~m class telescopes. THESEUS will enhance the rate of GRBs discovered at high redshift. 
At present the sample of GRB host galaxies at z$>$6 is limited to 9 events. 
In the late 20's the rate of GRBs at z$>$6 is expected to be $>$5 per year. Considering visibility constraints, we expect to have $\sim$2-3~targets~yr$^{-1}$. 
JWST and future 30~m class telescope will allow us to search even for dwarf host galaxies with $H_{\rm AB}\sim31$~mag by using few hours imaging with E-ELT. 
If the host will be brighter than H$\sim$27, spectroscopy will be possible.
A prompt alert will permit to use the optical afterglow as a reference star for Adaptive Optics (AO) in 30~m class telescope and perform IFU 
studies of the environment of the GRB, with angular resolution six times better than with JWST. This will provide information on the 
metallicity/dust/gas gradients close the GRB explosion sites, shedding light on the most dusty environments \citep[see, e.g.,][]{Rossi2012}, 
on the aversion of long GRBs for high metallicities \citep[see, e.g.,][]{Schulze2016}, and in general on the properties of the environment 
(e.g., star formation, gas inflow, dust content) needed for the formation of  GRB progenitors. 
One of the driving mechanisms for the formation of the massive stars at high redshift, including progenitors of long GRBs, may be the 
inflow of pristine gas during galaxy interactions \citep[see also][]{Michalowski2015}. The high angular resolution and sensitivity achievable 
with imaging and spectroscopy instruments mounted on JWST and 30~m class telescopes will permit us to study the morphological properties and 
the UV emission of the host of high-z GRBs. These observations will allow us to test the hypothesis that galaxy interactions at high redshift 
induce the formation of very massive stars and GRB progenitors.
\item The missing host galaxies of short GRBs. The short GRB offsets normalized by host-galaxy size are 
larger than those of long GRBs, core-collapse SNe, and Type Ia SNe, with 
only 20\% located at within the galaxy radius. These results are indicative of natal kicks or an origin in globular clusters, both of which 
point to compact object binary mergers \citep{Berger2014}. About 20\% of well localized SGRBs ($\sim$7\% of the total number of SGRBs) have been 
classified as host-less \citep{Fong2013}. The non-optimal X-ray localizations do not permit us to solve this problem, because several galaxies 
lie within the X-ray error circle, as it is shown in deep optical-NIR HST images. THESEUS/IRT will allow us to localize sGRB afterglows within 
a region ($<$1~arcsec error circle) small enough to search for the host galaxies using the future JWST and 30~m class telescopes and investigate 
if these might be inside very faint galaxies or may have been ejected by natal kicks or dynamics. Distinguishing between these two scenarios 
gives constraints on the formation of NS binaries. 
\end{itemize}

\subsection{Observatory science with IRT}

Fielding an IR-specified spectrograph in space, THESEUS would provide a unique resource for understanding the evolution of large samples of obscured 
galaxies and AGN. With a rapid slewing capability, and substantial mission duration, the mission will provide a very flexible opportunity for several 
fields of astrophysics, in a way similar to what is currently done by Swift/XRT in the X-rays. For instance, it will be possible to take efficient 
images and spectra of large samples of galaxies with minute-to-many-hour-long cumulative integrations. A continuous spectral coverage with no blockages 
due to atmospheric opacity ensures that identical species~R lines can be tracked in extensive samples. The capability to cover the redshift range 
from 0.07$<$z$<$1.74 for H$\alpha$ and 0.44$<$z$<$2.29 for H$\beta$ enables Balmer decrement measurements of the extensive evolution of the AGN and galaxy 
luminosity functions at redshift $\sim$0.5-1.5, a spectral region that simply cannot be covered from the ground. These key diagnostic rest-optical 
emission lines will be observed for galaxies in this substantial range of redshifts, reaching out towards the peak of AGN and galaxy formation 
activity, over a continuous redshift range where the bulge-blackhole mass relation is being built up and established, and the main sequence of 
star formation is well-studied. With excellent image quality, THESEUS R$\sim$500 grism can also provide spatially-resolved spectral information to 
highlight AGN emission, and identify galaxy asymmetries.

The imaging sensitivity of THESEUS is about 6 magnitudes lower than for JWST in the same exposure time; nevertheless, its availability ensures that many 
important statistical samples of active and evolved galaxies, selected from a wide range of sources can be compiled and diagnosed in detail at these 
interesting redshifts. Samples can be drawn from the very large WISE- and Herschel-selected infrared samples of galaxies, from EUCLID 24-mag large-area 
near-infrared galaxy survey, augmented by near-infrared selection in surveys from UKIDSS (whose deepest field reaches approximately 1 mag deeper than 
EUCLID wide-area survey in the H band) and VISTA, and in the optical from LSST and SDSS.
Spectra for rare and unusual galaxies and AGNs selected from wide-field imaging surveys can be obtained using the wide-field of THESEUS grism, thus 
building an extensive reference sample for studying the environments of the selected galaxies and AGNs, identifying large-scale structures and 
allowing over-densities to be measured. The parallel acquisition of spectral and imaging data over substantial areas would build up a clear 
picture of the environments, including serendipitously-selected spectra, all taking advantage of the spectral resolution delivered for THESEUS's 
primary science of investigating GRB afterglows.

\section{Scientific Requirements} 

THESEUS is designed to achieve two primary scientific goals: 
\begin{enumerate}
\item Explore the physical conditions of the early Universe by providing a complete census of GRBs in the first billion years.
\item Perform an unprecedentedly deep monitoring of the X-ray transient Universe thus playing a fundamental role in the coming era 
of multi-messenger and time-domain astrophysics
\end{enumerate}
These goals are very demanding in terms of technology and require a combination of on-board capability to perform wide-field X-ray imaging, 
the ability to obtain broad band-pass X-ray spectra and to localise and characterise the high-energy transients in the optical-IR. The conversion 
from THESEUS science goals to instrument and spacecraft requirements are shown in Fig.~\ref{fig:20}. 

To meet the science requirements requires the provision of three instruments on board: a wide-field soft X-ray monitor with imaging capability (the SXI); 
a harder X-ray, non-imaging spectroscopic instrument with the same field of view as the SXI (the XGIS); and an optical/near-IR telescope with both 
imaging and spectroscopic capability (the IRT). 
The spacecraft needs to be agile (fast response to enable the IRT to detect the source) and be able to rapidly communicate triggers to the ground 
so as to enable other observatories to also follow-up the new transients. The ability to point the spacecraft into the night sky (anti-solar) 
direction for part of the orbit enhances rapid ground follow-up capability to provide additional (multi-wavelength) data. Capability for 
stable 3-axis pointing for $>$1~ksec is required to detect the longest duration transients.

\subsection{Wide field monitoring in soft X-rays with deep sensitivity} 

To greatly increase the rate of GRB detection at high redshifts and detect large numbers of other transients while simultaneously providing 
accurate localisations requires the provision of a large field-of-view soft X-ray imaging instrument (see Sect.~\ref{SXI}). We have performed simulations 
\citep{Ghirlanda2015} which show the scientific goals can be met with an instrument field of view of 1~sr, a sensitivity in 1000~s of 
10$^{-10}$~erg~cm$^{-2}$~s$^{-1}$ (0.3-6~keV) and imaging capability sufficient to provide 0.5-1~arcmin localisations (2~arcmin worst case at 
90\% c.l.; this requires a PSF FWHM 4.5~arcmin). 

\begin{figure}
\centering
\includegraphics[scale=0.3]{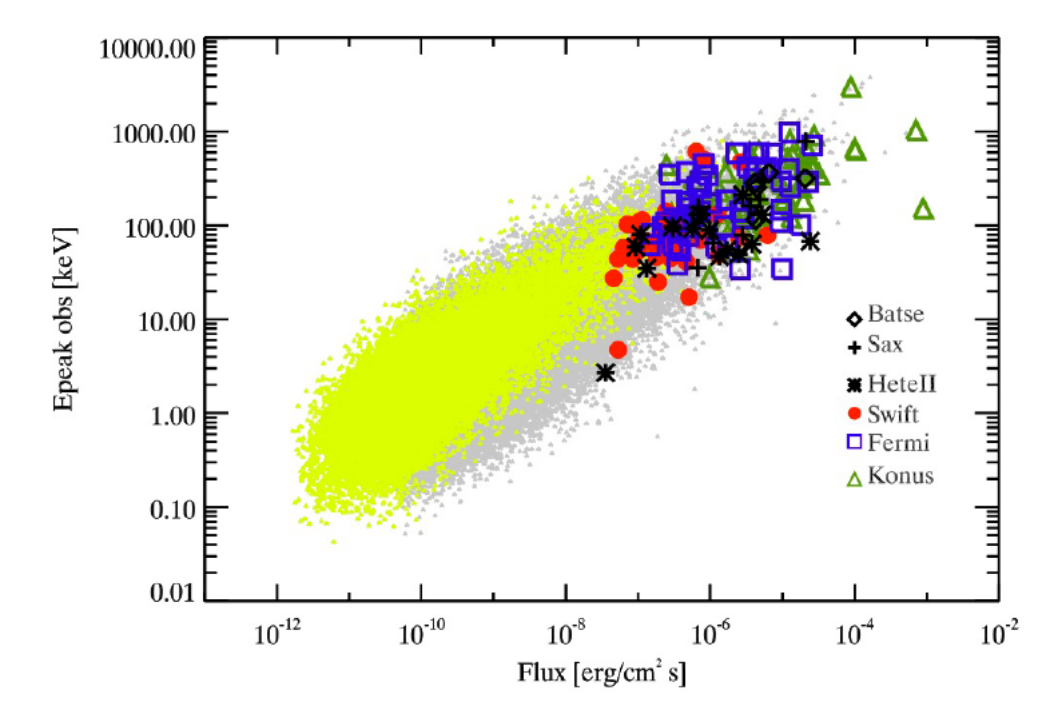}
\caption{Observer frame peak energy versus bolometric flux of GRBs with well constrained redshifts and spectra 
detected by various missions with a cloud of points from GRB population simulations \citep{Ghirlanda2015}. 
Yellow points are those at z$>$5 The use of a softer X-ray band permits the 
detection of GRBs with lower fluence and hence enhances the detection of higher redshift objects.}
\label{fig:21}   
\end{figure}
Multiple timescale software triggers are required to find the range of flux versus duration transient events. 
Using such an instrument (the THESEUS SXI) and taking into account the soft X-ray background, Fig.~\ref{fig:21} shows the expected annual 
rate of GRBs as a function of redshift. 
Also plotted is the rate of GRBs found by Swift (where the redshift distribution has been 
linearly scaled up based on those with redshift determinations - only approximately one third of Swift discovered GRBs have redshifts, 
all determined from the ground). The predicted annual rate of GRB detections by THESEUS SXI is 300-700 per year, with a very high 
($>$5-10) increased rate relative to Swift at the highest redshifts. 

As discussed below in the section on IR follow-up, imaging and 
photometric redshifts will be obtained on-board for the highest redshift GRBs and spectroscopic redshifts for the majority. For those 
GRBs detected on board but without spectroscopy triggers sent to ground, ground-based telescopes can be used to obtain spectra - giving 
priority to those with photometric indication of high redshift. THESEUS alone will obtain more spectroscopic redshifts on board in a 
year than Swift has provided in a decade. The search for high-z GRBs is part of a more general unprecedentedly deep monitoring of the 
X-ray transient Universe, whose motivation have been detailed in previous sections.
The predicted rate of detection of electromagnetic counterparts of GW signals and of other transient and variable source types 
during the survey is shown in Table~\ref{tab:1} and \ref{tab:2}. The very large detection rate of other transient types is due to the high sensitivity 
of THESEUS. This is illustrated in Fig.~\ref{fig:24}, where the source detection sensitivity of the proposed SXI and XGIS instruments 
are plotted verses integration time and overlaid are various sources types.

\subsection{Provision of broad-band X-ray spectroscopy}

\begin{figure}[t!]
\centering
\includegraphics[scale=0.33]{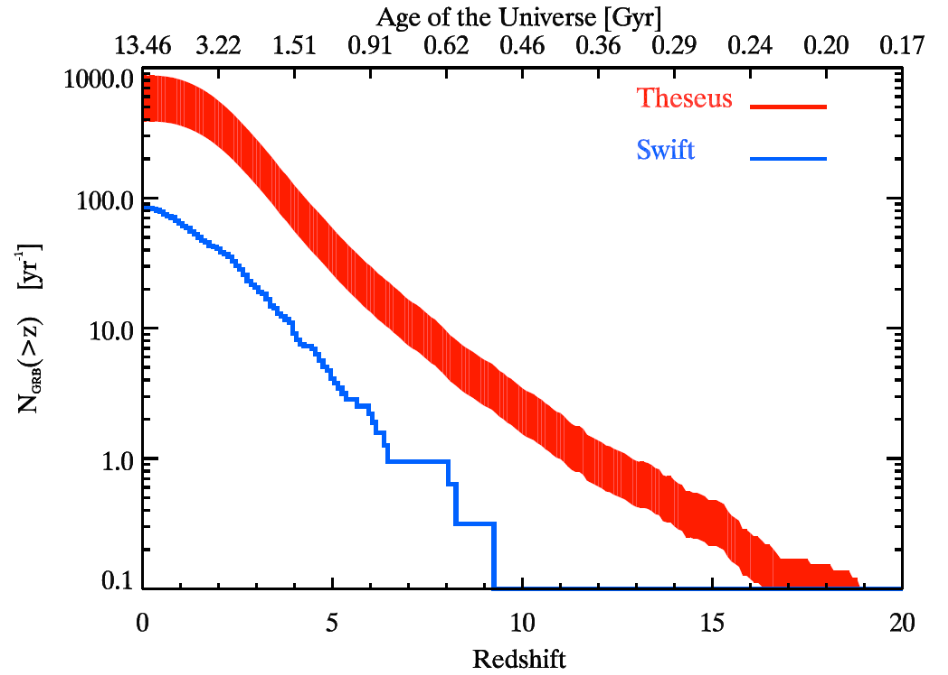}
\caption{The annual rate of GRBs predicted for THESEUS SXI (red) compared to Swift (blue). The upper scale 
shows the age of the Universe. For Swift the actual number of known redshifts is approximately one third that 
plotted and none were determined on board (the blue curve has been linearly scaled upwards to match the total 
Swift trigger rate). For THESEUS the red region uses the simulations from \citet{Ghirlanda2015} and adopts 
the expected sensitivity for the SXI and XGIS instruments.}
\label{fig:22}   
\end{figure}
The scientific objectives of THESEUS require the secure identification of sources types, in particular GRB triggers. The soft X-ray 
instrument is the primary source locator and has a high sensitivity to a wide variety of source types, as discussed above, 
as required to achieve the scientific goals.
This instrument will trigger on a large number of known sources which should not result in a spacecraft slew, but also other transient 
sources only some of which are GRBs. To reliably identify GRBs as well as spectroscopically characterise other sources and reduce the 
number of demanded slews requires the provision of a sensitive broad-band X-gamma ray instrument well matched to the sensitivity of 
the soft X-ray instrument and with source location capabilities of a few arcmin in the X-ray energy band. To meet the scientific goals 
we require an instrument with the following characteristics:
\begin{itemize}
\item extend the soft X-ray band of the imaging instrument up to the MeV band;
\item identify and localize with a few arcmin accuracy the GRBs by providing simultaneous triggers and by providing higher energy 
light curves and spectra to determine the luminosity of the GRB;
\item reduce the demanded number of spacecraft slews to observe with the IRT and act as a crucial filter to reduce soft X-ray 
trigger volume for ground/space telescope follow-up;
\item measure unbiased GRB/transient X- and gamma-ray spectra down to short time scales (ms time scales for the strongest events) 
to probe GRB physics.
\end{itemize}
\begin{figure*}[t!]
\centering
\includegraphics[scale=0.27]{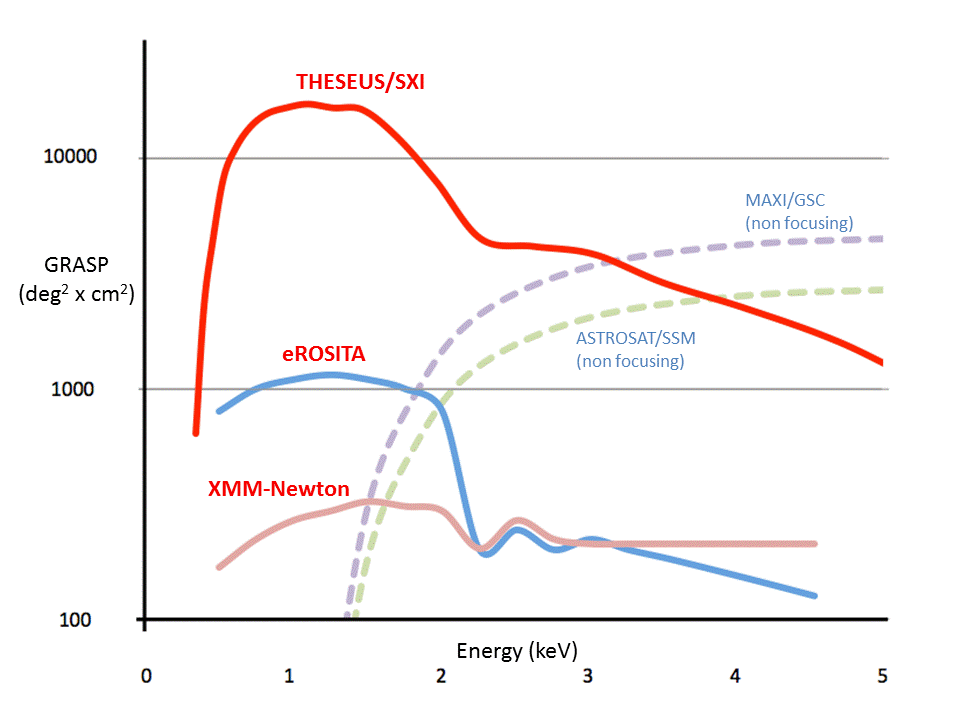}
\includegraphics[scale=0.5]{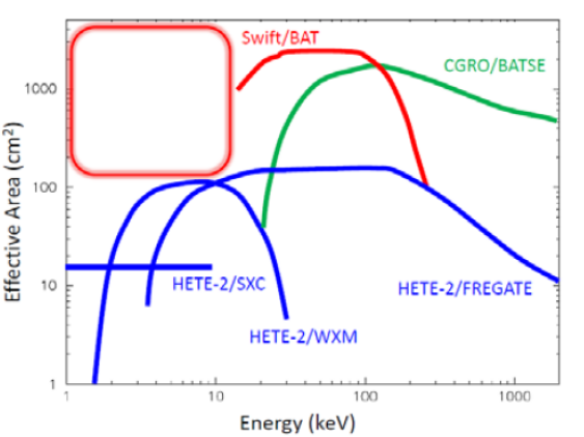}
\caption{The left-hand panel shows the GRASP (FOV$\times$Effective Area) of the THESEUS/SXI in the soft X-ray energy band compared to XMM-Newton 
and eROSITA. The GRASP of X-ray monitors on-board MAXI and ASTROSAT are also show for completeness, even though these are not focusing 
and their sensitivity for a given effective area is substantially worse than that of focusing telescopes. The leap in monitoring/survey 
of the soft X-ray sky allowed by THESEUS/SXI is outstanding. The right panel shows THESEUS/XGIS filling the parameter space in the 
top-left corner of the right-hand panel where other instruments have either too high an X-ray threshold or too low effective 
area, and will still provide 1000-1550~cm$^2$ effective area up to several MeVs \citep{yonetoku14}.}
\label{fig:23}   
\end{figure*}
The proposed XGIS instrument provides the required sensitivity and band-pass (Fig.~\ref{fig:23}). The sensitivity of the 
SXI and XGIS are well matched over the typical durations of GRBs (few to few tens of seconds). The XGIS will provide spectroscopy 
over 2-20000~keV with monitoring timescales down to ms. Despite advances during the Swift and Fermi era to identify 
and characterise GRB  phenomenon requires study of the prompt emission. Planned future missions (e.g. SVOM) do not provide the 
required combination of sensitivity and band-pass. 

\subsection{Optical-IR follow-up}

The scientific goals of THESEUS require the following on-board capabilities for an optical/near IR telescope  (IRT) to follow-up GRBs after 
a demanded spacecraft slew: 
\begin{enumerate}
\item Identify and localize the GRBs found by the SXI and XGIS to arcsecond accuracy in the visible and near IR domain (0.7-1.8~$\mu$m); 
\item Autonomously determine the photometric redshift of GRBs for z$>$4 and provide redshift upper limits for those at lower redshift;
\item Provide precise spectroscopic redshift measure for bright GRBs, together with limits on the intrinsic NH and metallicity for the majority of 
GRBs at z$>$4. 
\end{enumerate} 
The requirement number 1 is justified by the fact that the goal of the THESEUS mission is to study the Universe at $z>6$ in order to study 
the epoch of reionization. CMB experiments suggest that reionization was underway at z$\sim$9, while it appears to be completed by z$\sim$6.5. The question is 
whether massive stars could sustain a largely re-ionized Universe at z=6-9, and beyond. GRB afterglow spectra are power laws and, due to the
Ly-alpha drop-out (i.e. the Lyman alpha absorption within the GRB host galaxies and intervening IGM), a very attenuated signal is expected at wavelengths 
shorter than the Lyman alpha break, providing an unmistakable feature. Due to cosmological expansion the Ly-alpha wavelength (1216~$\AA$=0.126~$\mu$m) moves 
along the energy band, and in order to measure GRB redshifts between z=4 and z=10 the telescope detector has to be sensitive in the 0.7 to 1.8~$\mu$m  
range. GRBs with z$<$4 can be used as comparison to evaluate how massive stars evolve along the history of the Universe, and they can 
be easily followed up from ground. It is for high redshift GRBs that a NIR telescope in space really takes advantage of the absence of background 
due to the atmosphere. The field of view of the IRT telescope shall be larger than 4x4 arcmin, given that the SXI will provide error boxes which 
are at worst 2~arcmin (90\% c.l.) radius. 
For requirement number 2 the telescope will be operated in low resolution mode (R$\sim$10-20), and the Ly-alpha drop-off will be searched for. 
A fit of the sources' low resolution spectra, done on-board, should be capable of identifying high-redshift candidates. If we focus on GRBs 
at redshift larger than 6 the detector shall be optimized in terms of QE the 0.8-1.5~$\mu$m wavelength range. To obtain reliable results the 
detector QE shall be known within 10-20\%. 
\begin{figure*}[t!]
\centering
\includegraphics[scale=0.31]{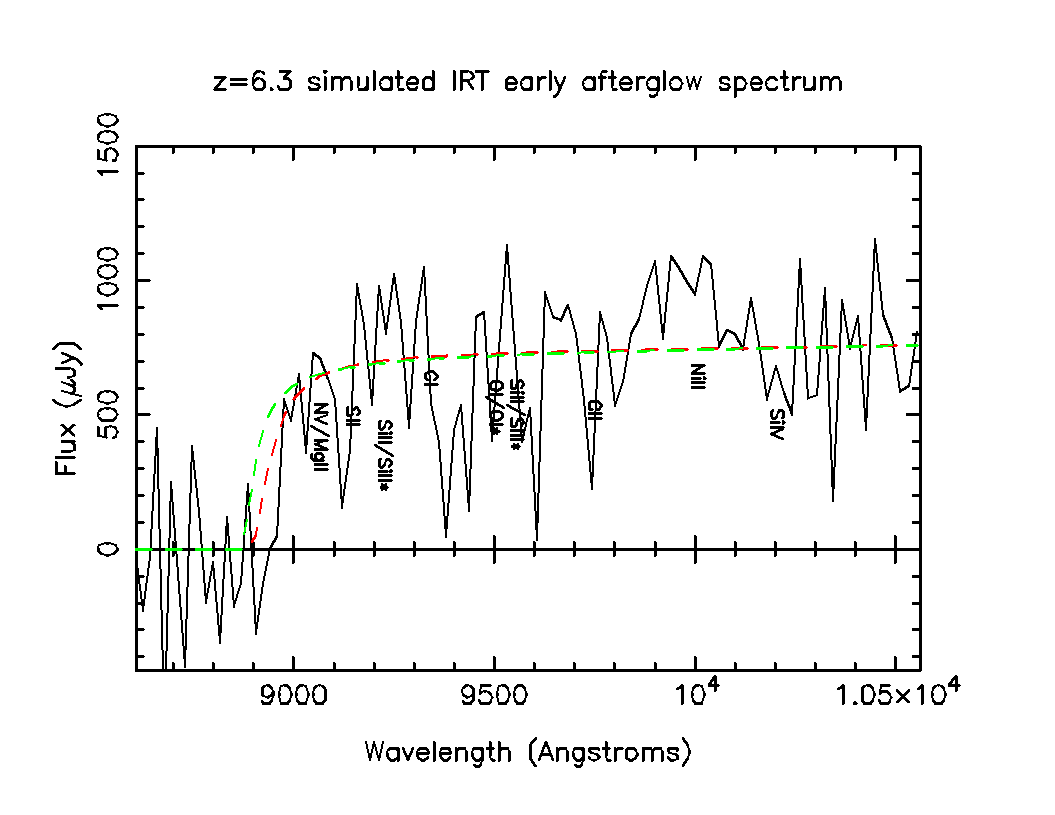}
\includegraphics[scale=0.23]{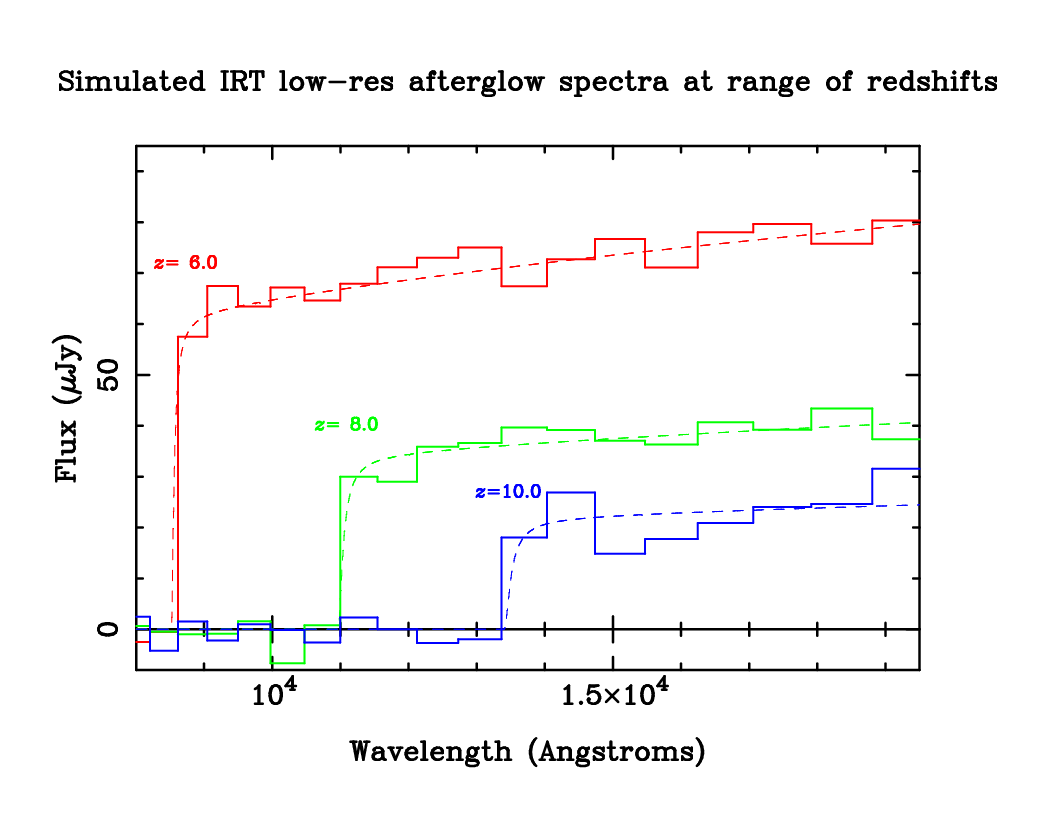}
\caption{{\it Left}: a simulated IRT high resolution (R=500) spectrum for a GRB at z=6.3 observed at 1~hr post trigger assuming 
a GRB similar to GRB~050904. The spectrum has host log($N_{\rm H}$)=21 and neutral fraction $F_{\rm x}$=0.5 (and metallicity 0.1~solar). 
The two models are: Red: log($N_{\rm H}$)=21.3, $F_{\rm x}$=0 Green: log($N_{\rm H}$)=20.3, $F_{\rm x}$=1. 
The IRT spectra provide accurate redshifts. {\it Right}: simulated IRT low resolution (R=20) spectra as a function of redshift 
for a GRB at the limiting $M_{\rm AB}$=20.8~mag at z=10 (brighter bursts can be expected at this redshift), 
and by assuming a 20~min exposure. The underlying (noise-free) 
model spectra in each case are shown as smooth, dashed lines. Even for difficult cases the low-res spectroscopy should 
provide redshifts to a few percent precision or better.  For many applications this is fine - e.g. star formation rate 
evolution.}
\label{fig:25}   
\end{figure*}
\begin{table}
\centering
\caption{THESEUS yearly detection and redshift measurements rates. The redshift measurements indicated are those 
that would be achieved directly on-board and do not include refined and/or additional measurements on-ground 
(as it is currently done for other observatories as Swift). Note that photometric redshifts are possible only 
at z$\gtrsim$5, when the Lyman ``dropout'' or ``break'' gets inside the IRT band.}
\label{tab:3}  
\begin{tabular}{lllll}
\hline\noalign{\smallskip}
THESEUS GRB/yr & All & z$>$5 & z$>$8 & z$>$10 \\
\noalign{\smallskip}\hline\noalign{\smallskip}
Detections & 387-870 & 25-60 & 4-10 & 2-4  \\
Photometric z &  & 25-60 & 4-10 & 2-4  \\
Spectroscopic z & 156-350 & 10-20 & 1-3 & 0.5-1 \\
\noalign{\smallskip}\hline
\end{tabular}
\end{table} 
The requirement number 3 deals with the high-resolution spectra (R$\sim$500). As shown in the left panel of Fig.~\ref{fig:25}, a resolution of the order 
of R$\sim$500 is good enough to identify the main absorption lines in bright GRB NIR afterglow spectra.  Such detections would (i) enable a 
more precise measure of the GRB redshift, that goes beyond the ``mere'' detection of the Ly-alpha drop off achievable with lower resolution 
spectra and (ii) help in discriminating between highly-extinguished and high-redshift (z$>$5) events. Besides the redshift measure, with the IR 
resolution spectroscopy it will be possible to derive limits on element relative abundances and metallicity (together with a measure 
of $N_{\rm H}$ for the brightest events, obtained by fitting the red wing of the Ly-alpha). 
Such information will be vital to optimize ground-based follow-up (with the large aperture 
facilities that will be available in 2028, like ELT), aimed at precise optical/IR spectroscopic studies for a detailed characterization 
of the GRB environment through the measure of chemical abundances, metallicity, and SFR. 

The sensitivity of the SXI triggering system in the 0.3-6~keV energy band probes a fluence range of 10$^{-8}$ and 10$^{-9}$ erg~cm$^{-2}$. 
Based on \citep{Ghirlanda2015} the proposed THESEUS IRT with a limiting sensitivity of 20.6~mag in the H (1.6$\pm$0.15~$\mu$m) filter 
is expected to detect all the GRB counterparts in imaging and low-resolution spectroscopy, if pointed early after the GRB trigger, 
in a 300~s exposure, and for the large majority of GRBs high-resolution spectra can be taken even 1-2 hours after the GRB (first or 
second spacecraft orbit) with the 19$^{\rm th}$ magnitude sensitivity IRT.  

In Fig.~\ref{fig:4} we show the rate of GRBs whose redshift will be spectroscopically determined by THESEUS on-board as a function 
of redshift. For the other GRBs detected by the SXI/XGIS, the IRT will provide a location and a redshift limit and thus provide a 
redshift estimate for the entire sample detected on-board. The cumulative distribution represents the rate (number of GRBs per year) 
that can be detected by THESEUS (red solid filled region). The width of the distribution accounts for the uncertainties of the 
population synthesis code adopted. For comparison, the rate of detection of GRBs by Swift is shown with a blue line. 
THESEUS out-performs Swift by about an order of magnitude at all redshifts and by more at the highest redshifts. Using the IRT 
to follow-up the SXI and XGIS will identify the highest priority high-redshift targets for the early Universe science goals.
This rate is derived from the actual population of GRBs detected by Swift and with measured redshift multiplied by a factor of 3. Indeed, 
approximately only 1/3 of the GRBs detected by Swift have their redshift measured. The upper axis shows the age of the 
Universe (5 Lobster modules with focal length 300~mm and individual field of view 0.16~sr are assumed). The detection and 
redshift-estimate annual rates expected form THESEUS are also summarized in Table~\ref{tab:3}.

\section{Scientific instruments} 
\label{instruments}

Following the scientific requirements described in the previous section, the baseline Instrument suite  configuration of THESEUS payload includes 4 
lobster-eye modules (F=300~mm), a 70~cm class IR telescope and 3 X-ray/soft gamma-ray coded-mask cameras based on Si+CsI(Tl) coupling 
technology covering  twice the  FOV of the lobster-eye modules. In summary, the scientific payload of THESEUS will be composed of:
\begin{itemize}
\item Soft X-ray Imager (SXI): a set of 4  ``Lobster-Eye'' X-ray (0.3-6~keV) telescopes covering a total FOV of $\sim$1~sr with 0.5-1~arcmin source 
location accuracy, provided by a UK led consortium;
\item InfraRed Telescope (IRT): a 70~cm class near-infrared (up to 2~$\mu$m) telescope with imaging and moderate spectral capabilities provided by 
a France led consortium (including ESA, Switzerland, and Germany);
\item X-Gamma ray Imaging Spectrometer (XGIS): a spectrometer comprising 3 detection units based on SDD+CsI(Tl) modules (2~keV-20~MeV), covering twice 
the FOV of the SXI. This instrument will be provided by an Italian led consortium (including Spain). 
\end{itemize}

All instruments are equipped with an Instrument Data Handling Unit (I-DHU), interfacing each of the three instruments with the spacecraft 
(provided by a German led consortium and Poland). The general avionic block diagram of the THESEUS PLM as well as SVM is shown in Fig.~\ref{fig:42}.

\subsection{The Soft X-Ray Imager (SXI)}
\label{SXI} 

\begin{figure*}[t!]
\centering
\includegraphics[scale=0.9]{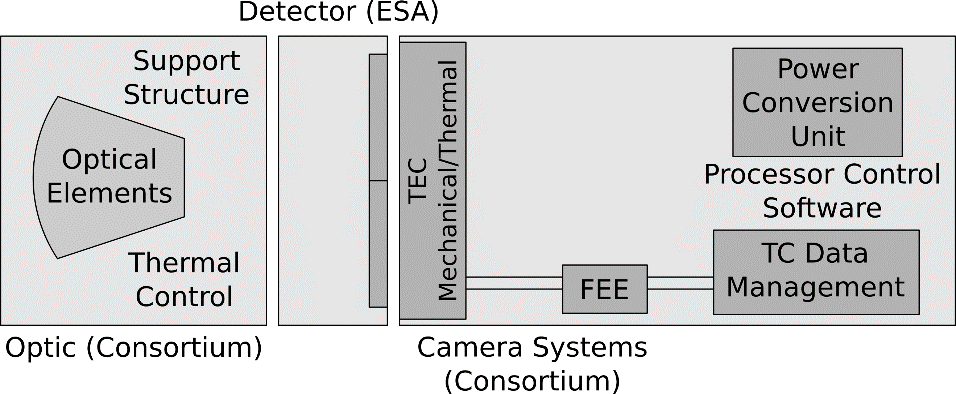}
\includegraphics[scale=0.9]{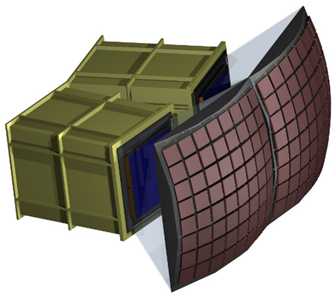}
\caption{The SXI  block diagram concept (left) and optical elements (right).}
\label{fig:26}   
\end{figure*}

The THESEUS Soft X-ray Imager (SXI) comprises 4~detector units (DUs). Each DU is a wide field lobster eye telescope using the optical 
principle first described by \citet{Angel1979} with an optical bench as shown in Fig.~\ref{fig:26}. 

\begin{figure*}[t!]
\centering
\includegraphics[scale=0.45]{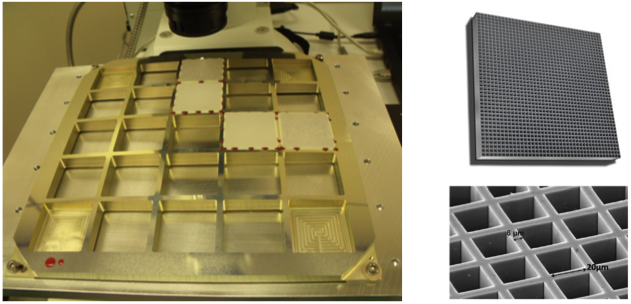}
\caption{{\it Left}: The SVOM MXT lobster eye optic aperture frame. {\it Top right}: A schematic of a single square pore MCP. {\it Bottom right}: 
A micrograph of a square pore MCP showing the pore structure. This plate has a pore size d=20~$\mu$m and a wall thickness w=6~$\mu$m.}
\label{fig:27}   
\end{figure*}

\begin{table}
\centering
\scriptsize
\caption{SXI detector unit main physical characteristics.}
\label{tab:4}  
\begin{tabular}{ll}
\hline\noalign{\smallskip}
Energy band (keV) & 0.3-6  \\
Telescope type & Lobster eye \\
Optics aperture & 320$\times$320~mm$^2$  \\
Optics configuration & 8$\times$8 square pore MCPs  \\
MCP size & 40$\times$40~mm$^2$ \\
Focal length & 300~mm \\
Focal plane shape & spherical \\
Focal plane detectors & CCD array \\
Size of each CCD & 81.2$\times$67.7~mm$^2$ \\
Pixel size & 18~$\mu$m \\
Number of pixel & 4510$\times$3758 per CCD \\
Number of CCDs & 4 \\
Field of View & $\sim$1~sr \\
Angular accuracy (best, worst) & ($<$10, 105)~arcsec\\
\noalign{\smallskip}\hline
\end{tabular}
\end{table}
The optics aperture is formed 
by an array of 8$\times$8 square pore Micro Channel Plates (MCPs). 
The MCPs are 40$\times$40~mm$^2$ and are mounted on a spherical frame with radius of curvature 600~mm 
(2 times the focal length of 300~mm). 
Table~\ref{tab:4} summarizes the SXI characteristics. The mechanical 
envelope of a SXI module has a square cross-section 320$\times$320~mm$^2$ at the optics end tapering to 200$\times$200~mm$^2$ at detector. 
The depth of the detector housing is 200~mm giving an overall module length of 500~mm. 
The left-hand side of Fig.~\ref{fig:27} shows the optics frame of the breadboard model for the SVOM MXT lobster eye telescope which comprises 
21 square MCPs mounted over a  5$\times$5 grid (the corners are unoccupied for this instrument). The front surface is spherical with radius 
of curvature 2000~mm giving a focal length of 1000~mm. The design proposed for SXI uses the same plate size and exactly the same mounting 
principle but a shorter focal length, 300~mm, so the radius of curvature of the front surface must be 600~mm. The right-hand panels of 
Fig.~\ref{fig:27} shows a schematic of a single plate and a micrograph that reveals the square pore glass structure. The focal plane of each SXI 
module is a spherical surface of radius of curvature 600~mm situated a distance 300~mm (the focal length) from the optics aperture. The 
detectors for each module comprise a 2$\times$2 array of large format detectors tilted to approximate to the spherical focal surface.

\paragraph{Calibration \\} 
The following calibrations are envisaged: 
\begin{itemize}
\item SXI optic: X-ray beam line testing to measure the focal length, the effective area and the point spread function as a function of 
off-axis angle and energy. 
\item SXI detector: Vacuum test facility to measure the gain and energy resolution as a function of energy. 
\item SXI end-to-end: X-ray beam line facility confirm the focal length and measure the instrument effective area and PSF as a function 
of photon energy and off-axis angle. 
\item SXI in orbit calibration: use cosmic sources to confirm in-flight alignment, plate scale, 
point spread function, effective area, vignetting and energy resolution. 
A regular monitoring of cosmic sources is planned to verify the status and correctness of the instrument calibration. 
\end{itemize}

\begin{figure}
\centering
\includegraphics[scale=0.35]{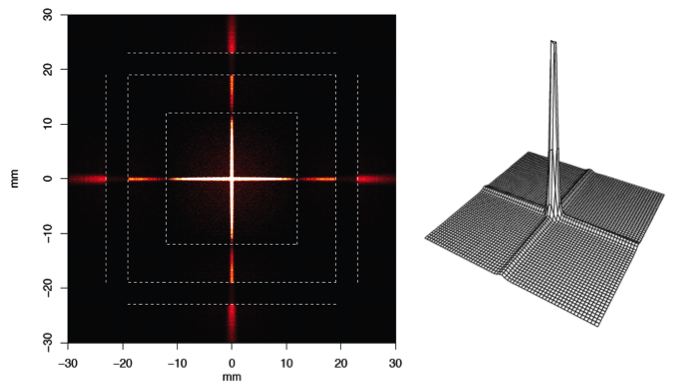}
\caption{The point spread function of the SXI.}
\label{fig:28}   
\end{figure}

\begin{figure*}[t!]
\centering
\includegraphics[scale=0.24]{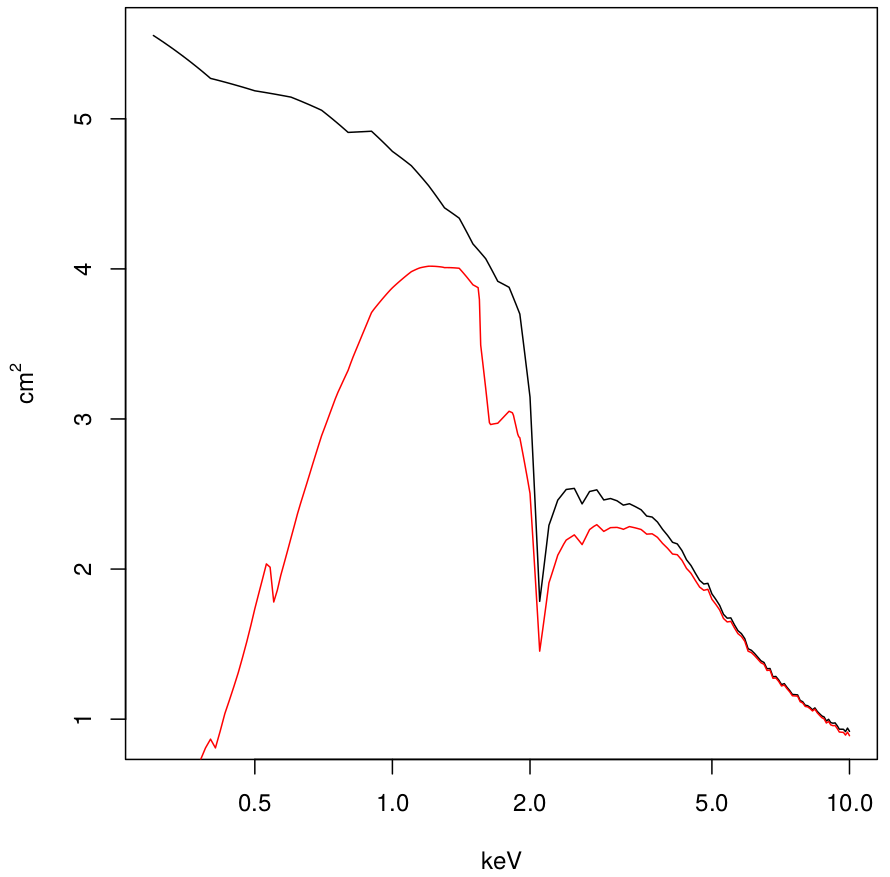}
\includegraphics[scale=1.2]{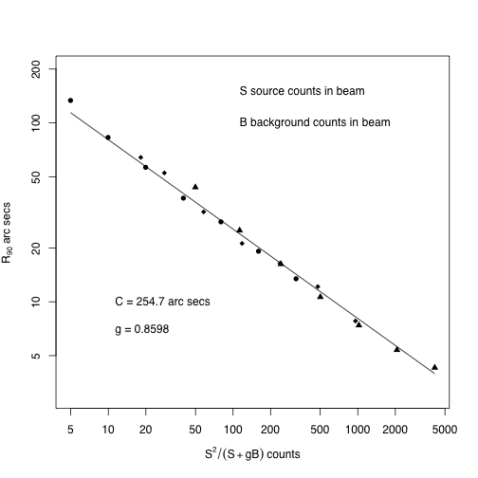}
\caption{{\it Left}: Collecting area as a function of energy (assuming a focal length of 300~mm and including the contributions 
of the central spot, the 2 cross-arms, and the straight through flux). The black line represents the optics only. 
The red curve includes the quantum efficiency of the CCD 
and the transmission of the optical blocking filter. {\it Right}: The position accuracy of the SXI as a function of source 
and background count. R90 is the error radius that contains 90\% of the derived positions.}
\label{fig:29}   
\end{figure*}

\paragraph{SXI Performance, Sensitivity and Data Rate \\} 
The imaging area of the CCDs sets the field of view of each module. A compliment of 4 SXI modules has a total field of 
view of 3200~square degrees (0.9~sr). 
The point spread function is shown in Fig.~\ref{fig:28}. The inner dotted square shows the off-axis angle at which the 
cross arms go to zero as determined by the L/d ratio of the pores. For optimum performance at 1~keV we require L/d=50. 
The outer dotted square indicates the shadowing of the cross-arms introduced by the gap between the individual MCPs in 
the aperture. The central true-focus spot is illustrated by the projection plot to the left. The FWHM is 4.5~arcmin  
and all the true-focus flux is contained by a circular beam of diameter 10~arcmin. The collecting area, 
within a 10~arcmin beam centered on the central focus, as a function of energy is shown in Fig.~\ref{fig:29}. The optics 
provides the area plotted in black. The red curve includes the quantum efficiency of the CCD and the transmission of 
the optical blocking filter comprising a 60~nm Aluminium film deposited over the front of the MCPs and 260~nm of 
Aluminium plus 500~nm of parylene on the surface of the detectors. Because the angular width of the optics MCP-array 
is 2.3~deg larger than the CCD-array the field of view is unvignetted at 1~keV and above so the collecting area 
shown in Fig.~\ref{fig:29} is constant across the field of view. 
Using the ROSAT All-sky Survey data we can estimate the count rate expected from the diffuse sky (Galactic and Cosmic) and point sources. 

The sensitivity to transient sources using this background rate and a false detection probability of 1.0$\times$10$^{-10}$ is shown as a 
function of integration time in Fig.~\ref{fig:24}. For longer integration times the source count required 
rises, e.g., to 30 counts for a 1~ks integration. We find that 94\% of the Swift BAT bursts (before 2010 September 16) 
would be detected by the SXI. The X-ray light curve of the afterglow would be detected to $>$1~ks after the trigger 
for a large fraction of the bursts.

\begin{figure*}[t!]
\centering
\includegraphics[scale=1.3]{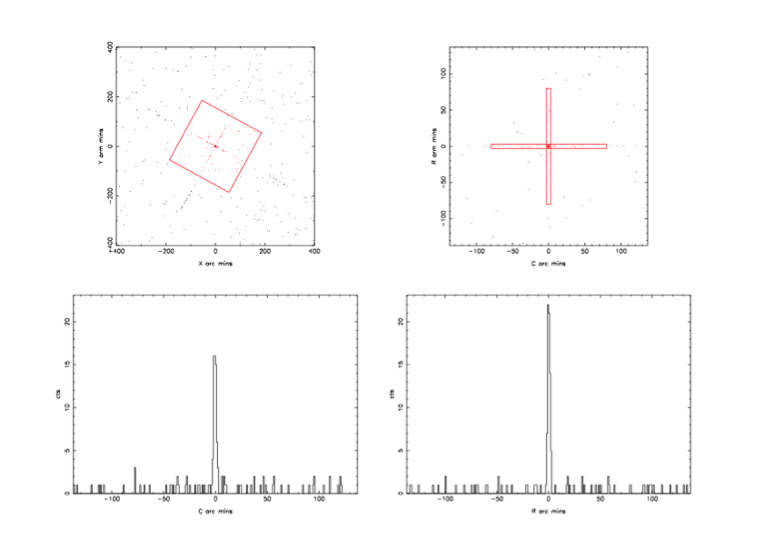}
\caption{The two stage trigger algorithm. {\it Top left-hand panel}: the detected event distribution $\Delta$T=4 seconds and a 
source count rate of 40 cts s-1 over the full PSF. The cross-arms are rotated wrt the detector axes to demonstrate how this 
can be handled. {\it Top right-hand panel}: the detected event distribution in the patch of sky aligned to the cross-arm axes 
of the PSF (shown as the red rectangle in the top left-hand panel). 
The red cross-patch indicates the area used for the second stage of the algorithm. 
{\it Bottom panels}: the histograms along columns and rows in the patch.}
\label{fig:30}   
\end{figure*}

\paragraph{Trigger Algorithm \\}
The ideal algorithm would be some form of matched filtering using the full PSF distribution but because of the extent of 
the PSF this would be far too computationally heavy. At the other extreme a simple scheme would be to search for significant 
peaks using the cell size commensurate with the central peak in the PSF. This would be faster but utilizes only $\sim$25\% of the total 
flux detected. The scheme described below is a two stage process which exploits the cross-arm geometry but avoids computationally 
expensive 2-D cross-correlation. For the 1st stage the focal plane is divided into square patches with angular side length 
$\sim4d/L=1/12$ radians aligned with the cross-arm axes. The dotted central square shown in Fig.~\ref{fig:30} indicates the size and 
orientation of such a patch. The optimum size of such patches depends on the HEW of the lobster-eye optic and the background 
count rate. The patches could correspond to detector elements or tiles in the focal plane, e.g. CCD arrays.  The peak profile 
is the line spread profile of the central spot and cross-arms of the PSF. The remainder of the histogram distribution arises 
from events in the cross-arm parallel to the histogram direction, the diffuse component of the PSF and any diffuse background 
events not associated with the source. Because we are looking for transient sources the fixed pattern of the steady sources 
in the field of view at the time would have to be subtracted from the histogram distributions. As the pointing changes this 
fixed pattern background would have to be updated. A transient source is detected if a significant peak is seen in both 
histograms.

The sensitivity of detection and accuracy of the derived position of the source within the patch depends on the bin size of 
the histograms, the HEW of the central peak of the PSF and the background. For the most sensitive detection the bin size 
should be approximately equal to the HEW but this will limit the accuracy of the position. If the bin size of the histograms 
is chosen to be significantly smaller than the width of the HEW then the histograms can be smoothed by cross-correlation with 
the expected line width profile of the peak-cross-arm combination. Using the smaller bin size the histograms can also be used 
to estimate the position centroid of the source within the patch. The significance used for this first stage should be low, 
e.g. 2.5~$\sigma$. This will provide candidate positions for the second stage.

For each of the candidate positions identified in the first stage a cross-arm patch is set up to cover just the detector 
area which is expected to contain a fraction of the full cross-arms and the central peak in the PSF. The cross-patch dimensions 
are changed depending on the integration time $\Delta$T. For short integration times the total background count will be small and 
the cross-patch size is set large to capture a large fraction of the counts from the cross-arms and central peak. 

The above considers a single value of $\Delta$T. We envisage that a series of searches would be run in parallel each using a 
different integration time so that the sensitivity limit as a function of $\Delta$T is covered. The basic scheme is illustrated 
in Fig.~\ref{fig:30}. The total source count assumed for this illustration was 40 spread over the full PSF as plotted in Fig.~\ref{fig:28}. 
The bin size used for the histograms was 1~arcmin, and the HEW of the central peak of the PSF is approximately 4~arcmins. 
We have tested the algorithm over a range of integration times and background conditions.  It achieves the sensitivity 
limit plotted in the science requirements section. When a significant transient peak is identified the position must 
be converted to sky coordinates using the current aspect solution (from Payload star trackers). Positions of all 
transients found must be cross-correlated with known source catalogues, e.g. ROSAT All-Sky Survey, Flare stars, 
Swift BAT catalogue etc. Any position which does not match a known position must be passed to the Space Craft as 
a potential trigger position.

The processing required to implement the above is as follows:
\begin{enumerate}
\item Extract frames from the CCD at $\Delta$T=2~s (this is the fastest rate set by the frame time). 
\item Apply event reconstruction algorithm to the frames to give an event list with positions in CCD pixels and a pulse height. 
\item Convert the pixel positions into a local module coordinate frame which is aligned to the cross-arms of the PSF. 
\item Accumulate counts in the 1-D histograms. 
\item Subtract the fixed source/background pattern from the histograms. 
\item Scan the histograms for significant peaks and extract candidate positions for further analysis. 
\item Set up the cross-arm mask at candidate positions to look for significant peaks. Calculate an accurate position in the local 
module coordinate frame for the peak. 
\item Convert this position into global sky coordinates (quaternion). 
\item Check positions against on-board catalogues to weed out known sources. 
\item Communicate unidentified transients to the Spacecraft. 
\end{enumerate}
Note that points 4-7 above must be repeated for different $\Delta$T values (e.g., 2, 20, 200, 2000~s).

\subsection{The X-Gamma ray Imaging Spectrometer (XGIS)}

\begin{figure*}[t!]
\centering
\includegraphics[scale=0.35]{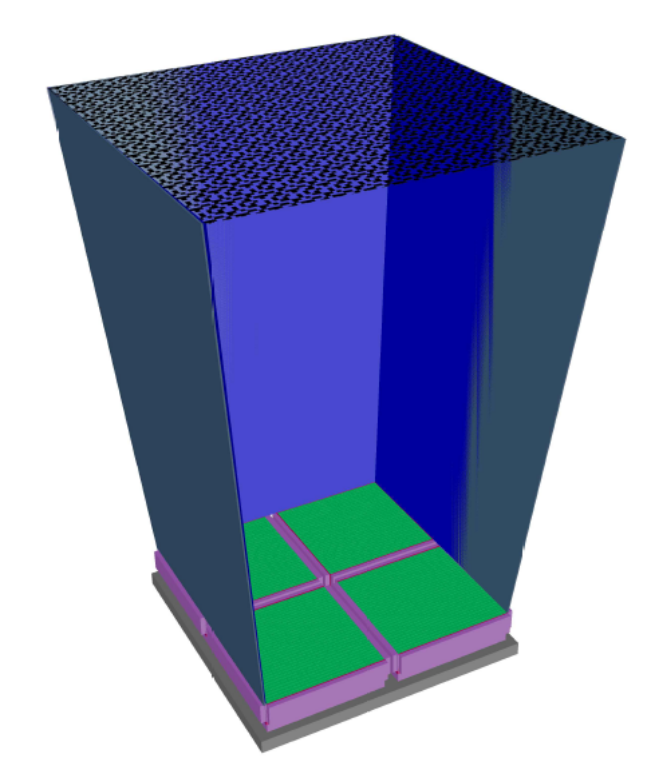}
\includegraphics[scale=0.5]{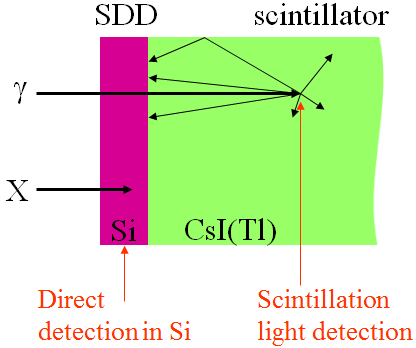}
\caption{{\it Left}: Sketch of the XGIS Unit. {\it Right}: Principle of operation of the XGIS  detection units: low-energy X-rays 
interact in Silicon, higher energy photons 
interact in the scintillator, providing an energy range covering three orders of magnitude. A pulse shape discriminator 
determines if the interaction has occurred in Si or in the crystal.}
\label{fig:31}   
\end{figure*}

The X-Gamma ray Imaging Spectrometer (XGIS) comprises 3 units (telescopes). The three units are pointed at offset directions 
in such a way that their FOV partially overlap. Each unit (Fig.~\ref{fig:31} and Table~\ref{tab:5}) has imaging capabilities in the low energy 
band (2-30~keV) thanks to the combination of an opaque mask superimposed to a position sensitive detector. A passive shield 
placed on the mechanical structure between the mask and the detector plane will determine the FOV of the XGIS unit for X-rays 
up to about 150~keV energy. Furthermore the detector plane energy range is extended up to 20~MeV without imaging capabilities. 
The main performance of an XGIS unit are reported in Table~\ref{tab:6}. 
\begin{table*}[t!]
\centering
\caption{XGIS detector unit main physical characteristics.}
\label{tab:5}  
\begin{tabular}{ll}
\hline\noalign{\smallskip}
Energy band & 2~keV-20~MeV \\
Detection plane modules & 4 \\
detector pixel/module & 32$\times$32 \\
pixel size (= mask element size) & 5$\times$5~mm \\
Low-energy detector (2-30~keV) & Silicon Drift Detector 450~$\mu$m thick \\
High energy detector ($>$30~keV) & CsI(Tl) 3~cm thick \\
Discrimination Si/CsI(Tl) detection & Pulse shape analysis \\
Dimension & 50$\times$50$\times$85~cm \\
Power & 30.0~W \\
Mass & 37.3~kg \\
\noalign{\smallskip}\hline
\end{tabular}
\end{table*}
\begin{table*}[t!]
\centering
\caption{XGIS unit characteristics vs energy range.}
\label{tab:6}  
\begin{tabular}{llll}
\hline\noalign{\smallskip}
& 2-30 keV & 30-150 keV & $>$150 keV \\
\hline\noalign{\smallskip}
Fully coded  FOV & 9$\times$9~deg$^2$ &  &  \\
Half sens. FOV & 50$\times$50~deg$^2$ & 50$\times$50~deg$^2$ (FWHM) &  \\
Total FOV & 64$\times$64~deg$^2$ & 85$\times$85~deg$^2$ (FWZR) & 2$\pi$sr \\
Ang. res & 25~arcmin & & \\
Source location accuracy & $\sim$5~arcmin (for $>$6~$\sigma$ source) & & \\
Energy res & 200~eV FWHM at 6~keV & 18\% FWHM at 60~keV & 6\% FWHM at 500~keV  \\
Timing res. & 1~$\mu$s & 1~$\mu$s & 1~$\mu$s  \\
On axis useful area & 512~cm$^2$ & 1024~cm$^2$ & 1024~cm$^2$ \\
\noalign{\smallskip}\hline
\end{tabular}
\end{table*}
The detection plane of each unit is made of 4 detector modules each one about 195$\times$195$\times$50~mm in size detecting X and 
gamma rays in the range 2~keV-20~MeV. For each  energy loss in the module, whatever  procured by EM radiation or 
ionizing particle,  the energy released, the 3 spatial coordinates and the energy deposit of the interaction and  time of occurrence 
will be recorded. 

The basic element of a module (Fig.~\ref{fig:31}) is a bar made of scintillating 
crystal 5$\times$5$\times$30~mm$^3$ in size. 
Each extreme of the bar is covered with a Photo Diode (PD) for the read-out of the 
scintillation light, while the other sides of the bar are wrapped with a light reflecting material conveying the 
scintillation light towards the PDs. The scintillator material is CsI(Tl) peaking its light emission at about 560~nm. The PD is 
realized with the technique of Silicon Drift Detectors \citep[SDD-PD;][]{Gatti1984} with an active area of 5$\times$5~mm$^2$ matching the 
scintillator cross section.  
Crystals are tightly packed in an array of 32$\times$32 elements to form the module.  The SDD and scintillator 
detect X- and gamma-rays. The operating principle (see Fig.~\ref{fig:31} right) is the following. 
The top  SDD-PD, facing  the X-/gamma-ray entrance window,  is operated both as X-ray detector for low energy X-ray photons interacting in 
Silicon and as a read-out system of the scintillation light resulting from X-/gamma-ray interactions in the scintillator. 
The bottom SDD-PD at the other extreme of the crystal bar operates only as a read-out system for the scintillators. The 
discrimination between energy losses in Si and CsI is based on the different  shape of charge pulses.
While the electron-hole pair creation from X-ray interaction in Silicon generates a fast signal (about 10~ns rise time), 
the scintillation light collection is dominated by the fluorescent states de-excitation time (0.68~$\mu$s, 64\%, and 3.34~$\mu$s, 36\%, for CsI(Tl)) 
and a few $\mu$s shaping time is needed in this case to avoid significant ballistic deficit. Pulse shape analysis (PSA) 
techniques are adopted to discriminate between signals due to energy losses in Si or CsI. The results we obtained 
for the separation of the energy losses  in the case of an $^{241}$Am source \citep{Marisaldi2004} are shown in Fig.~\ref{fig:34}. 
As can be seen from the left panel of this figure, the ratio of the two pulse heights is approximately constant for pulses of common 
shape and allows discrimination between interactions taking place in Silicon or in the scintillator. For gamma-rays interacting 
in the scintillator, combining the signals from the two PDs at the extreme of each bar allows  to determine the energy and 
the depth of the interaction inside the crystal \citep{Labanti2008}. 
\begin{figure}[t!]
\centering
\includegraphics[scale=0.3]{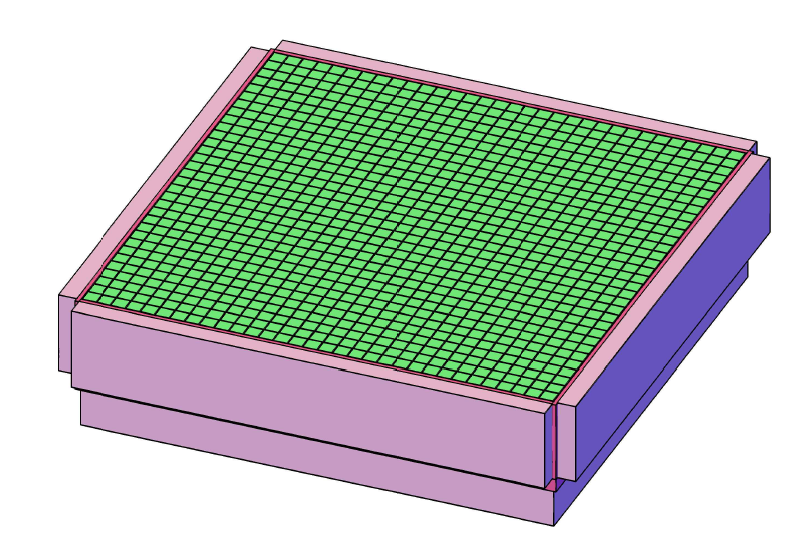}
\includegraphics[scale=0.4]{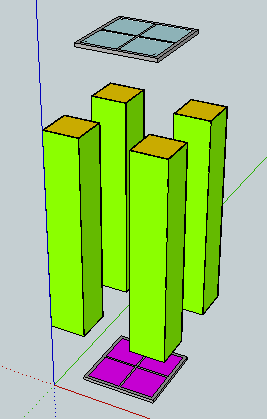}
\caption{Sketch of the XGIS module. A module (right) is made of an array of 32$\times$32 scintillator bars with Si PD read-out at 
both ends (left). Both PD and scintillators are used as active detectors. The PDs readout electronics consist of an ASIC 
pre-amp mounted near each PD's anode while the rest of the processing chain is placed at  the module sides and bottom.}
\label{fig:32}   
\end{figure}

\paragraph{XGIS building blocks \\}
In the XGIS HW, the main building blocks (see Fig.~\ref{fig:35} for one XGIS unit) are: 
\begin{enumerate}
\item the mask and the FOV delimeter; 
\item the scintillator detectors; 
\item the FEE in both its analogue part (with SDD-PD, ASIC1 and ASIC2), digital part (DFEE) and services (TLM, TLC Power supply). 
\end{enumerate}

The coded mask of each XGIS unit, placed 70~cm above the detector modules is made of stainless steel of 0.5~mm thick. 
The mask overall size is 50$\times$50~cm and will have a pattern allowing  self-support in order to guarantee the maximum transparency of the open 
elements. The mechanical  structure connecting the mask  with the detector is also made of stainless steel 0.1~mm thick and supports 4  
Tungsten slats 45~cm high with a variable thickness (0.5-0.3~mm). This structure will act as a lateral passive shield for the imager system 
(1-30~keV) and as a FOV delimiter at energies $>$150~keV. In the latter range, the resulting FOV is 50$\times$50~deg$^2$ (FWHM) and 85$\times$85~deg$^2$ (FWZR). 
By combining the three units, with an offset of $\pm$35~deg for two of them, the FOV delimiter guarantees an average XGIS effective area of 
$\sim$1400~cm$^2$ in the SXI FOV (104$\times$31~deg$^2$).
\begin{figure*}[t!]
\centering
\includegraphics[scale=0.16]{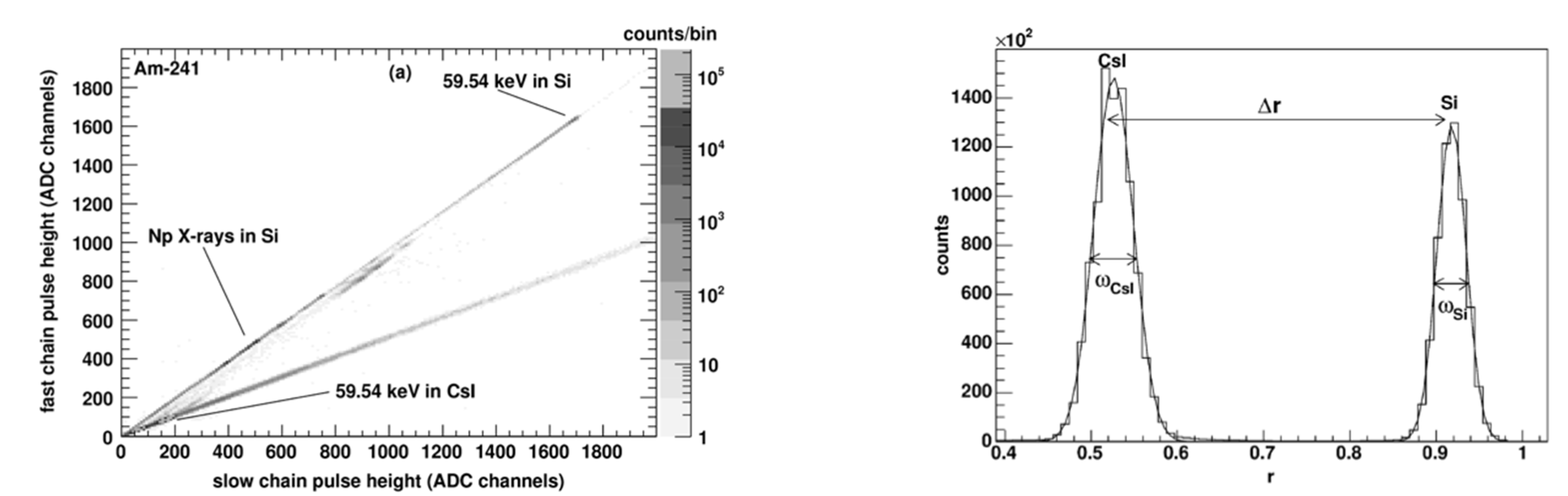}
\includegraphics[scale=0.16]{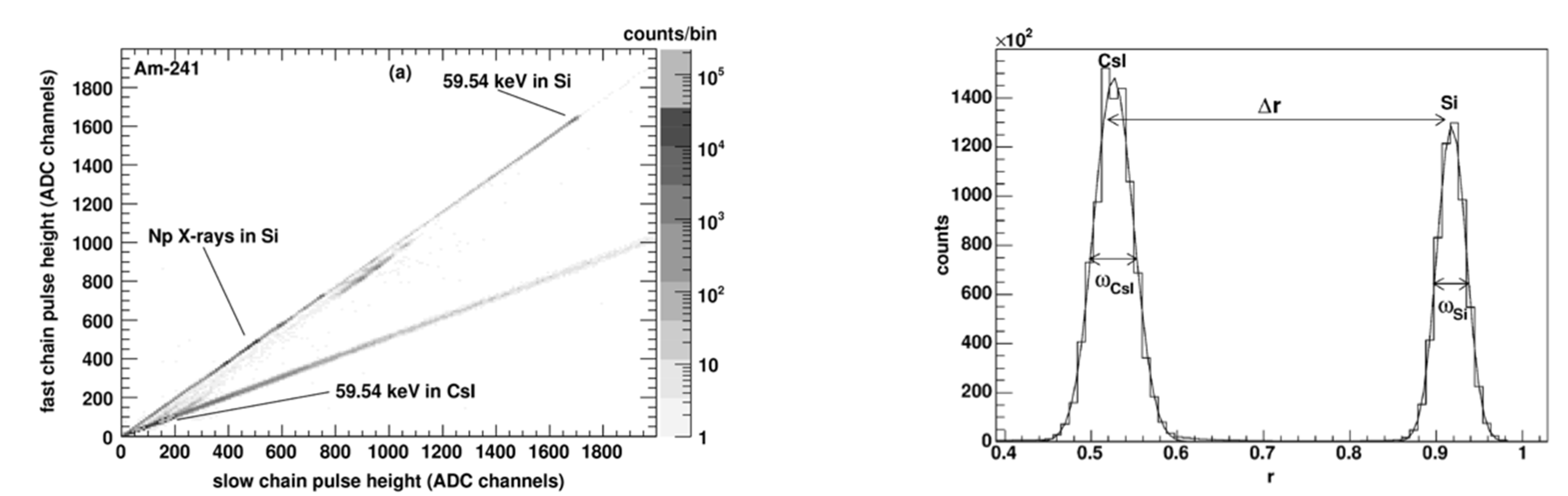}
\caption{{\it Left}: bi-dimensional spectrum of a $^{241}$Am source showing interaction in Silicon (top line) and in the scintillator (bottom line). 
{\it Right}: distribution of the ratio of the two processing chains for events in Silicon and in the scintillator. The large peak separation indicates 
the optimal discrimination performance.}
\label{fig:34}   
\end{figure*}

\paragraph{Data Handling Unit (DHU) and its functions   \\}
The whole XGIS background data rate (3 units) towards the DHU is of the order of 6000~event/sec in the 2-30~keV range and about 3700~event/sec 
above 30~keV. Each event received by DHU will be identified with a word of 64 bits (4 for module address, 10 for bar address, 
10 for signal amplitude of the fast top channel, 11 for signal amplitude of the slow top channel, 11 for signal amplitude of the slow bottom 
channel, 18 for time). DHU functions will be:
\begin{itemize}
\item discriminates between Si and CsI events. 
\item For CsI events, evaluates the interaction position inside the bar by weighting the signals of the 2~PDs (a few mm resolution expected). 
Combining this information with the address of the bar (5$\times$5~mm  in size) each module becomes a 3D position sensitive detector. 
\item Exploiting the 3D capabilities background can be minimized.
\item It continuously calculates along the orbit the event rate of each module in different energy bands (typically 2-30~keV, 30-200~keV and $>$200~keV) 
and on 5 different times scales (e.g., 1~ms, 10~ms, 100~ms, 1~s, and 10~s). 
\item In the 2-30~keV range and for each unit, it  produces images of  the FOV in a defined integration time.
\item It holds in a memory buffer  all the XGIS data, rates and images  of the last 100~s (typical) with respect to the current time.
\item Produce maps of the three unit planes with event pixel by pixel histograms in different energy bands (typically 32 with E width varying 
logarithmically) and with selectable integration times (min 1~s).
\end{itemize}
  
\paragraph{XGIS and the GRB trigger system \\}
XGIS will contribute to the THESEUS's GRB trigger system in different ways:
\begin{enumerate}
\item Qualification of the SXI triggers. The primary role of XGIS is to qualify the SXI triggers as true GRB. The basic algorithm for GRB validation 
is based on an evaluation of the significance of the count rate variation, calculated as described in the sub-section below. The procedure will be the following: 
\begin{itemize}
\item from the SXI direction given to the event, it is identified one of the three XGIS units in which the event has potentially been detected;
\item look for an excess of the rates in the modules of this unit in the bands 2-30~keV and 30-200~keV with respect to the average count-rate 
continuously calculated by DHU. 
\end{itemize} 
\item Autonomous XGIS GRB trigger based on data rate. The autonomous GRB trigger for XGIS inherits the experience acquired with the Gamma 
Ray Burst Monitor (GRBM) aboard BeppoSAX and that acquired with the Mini-calorimeter aboard AGILE \citep{Fuschino2008} and concerns all the modules 
of the 3 XGIS units. For each module the above energy intervals (2-30 and 30-200~keV) are considered for the trigger. The mean count rate of each 
module in each of these bands is continuously evaluated on different time scales (e.g., 10~ms, 100~ms, 1~s, and 10~s). A trigger condition is satisfied when, 
in one or both of these energy bands, at least a given fraction (typically $\gtrsim$3) of detection modules sees a simultaneous excess with a significance 
level of typically 5~$\sigma$ on at least one time scale with respect to the mean count rate. 
\item Autonomous XGIS GRB trigger based on images. For each XGIS unit, the 2-30~keV actual images will be confronted with reference images 
derived averaging $n$ (typically 30) previous images, and a spot emerging from the comparison at a significance level of 5~$\sigma$ typically will appear. 
If one of the above trigger condition is satisfied,  event by event data, starting from 100~s before the trigger are transmitted to ground, the duration 
of this mode lasting until the counting rate becomes consistent with background level. 
\end{enumerate}
\begin{figure*}[t!]
\centering
\includegraphics[scale=0.32]{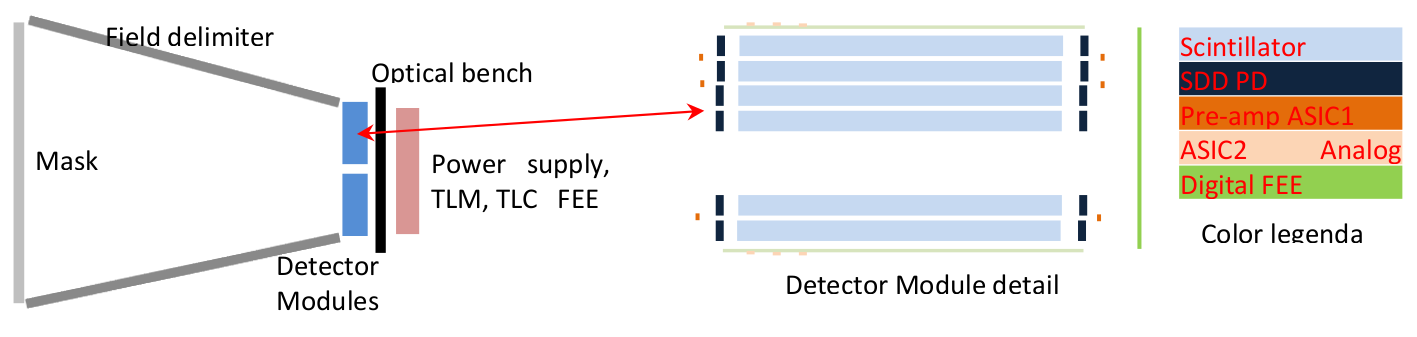}
\caption{XGIS unit main building blocks.}
\label{fig:35}   
\end{figure*}

\paragraph{Telemetry requirements  \\}
For the study  of transient or persistent sources different transmission mode will be selected starting from the photon list and the histogram maps 
of the  units. Typically the TLM load will be maintained below 2~Gbit/orbit transmitting: (i) at low energies ($<$30 keV) pixel by pixel histograms in 
various Energy channels (e.g. 32 ch)  with variable integration times (e.g. 64 sec); (ii) above the 30~keV the whole photon list.
In particular observations (e.g., crowded fields) a photon by photon transmission in the whole energy range will be selected for a total maximum 
telemetry load of 3~Gbit/orbit. In the case of a GRB trigger all the information available photon by photon is transmitted with a maximum 
telemetry load of 1~Gbit.

\begin{figure*}[t!]
\centering
\includegraphics[scale=0.32]{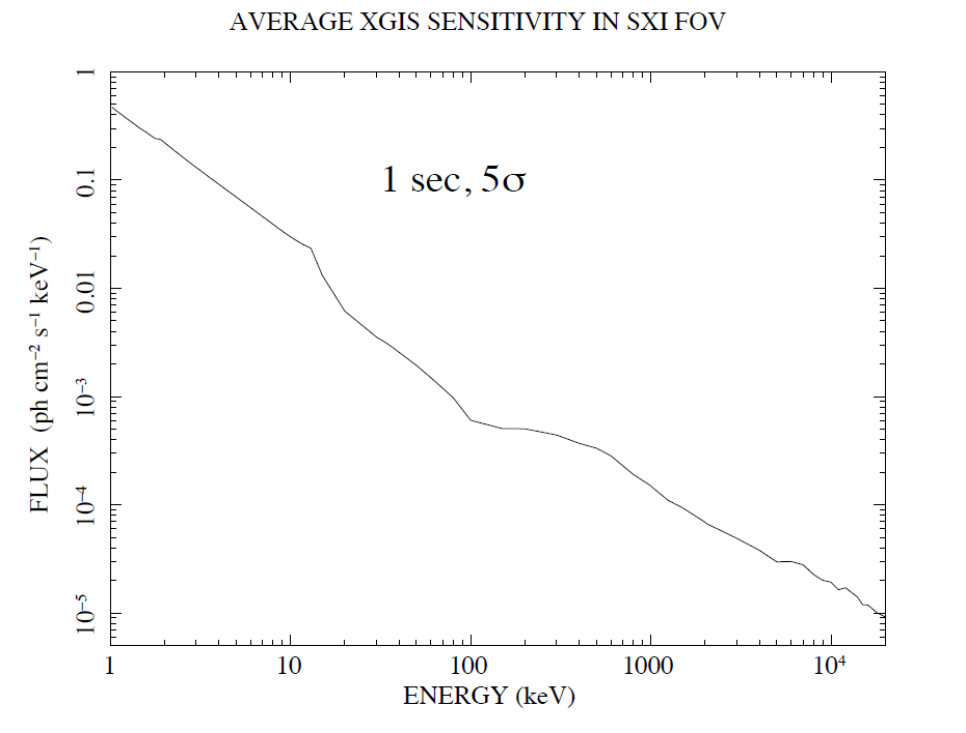}
\includegraphics[scale=0.25]{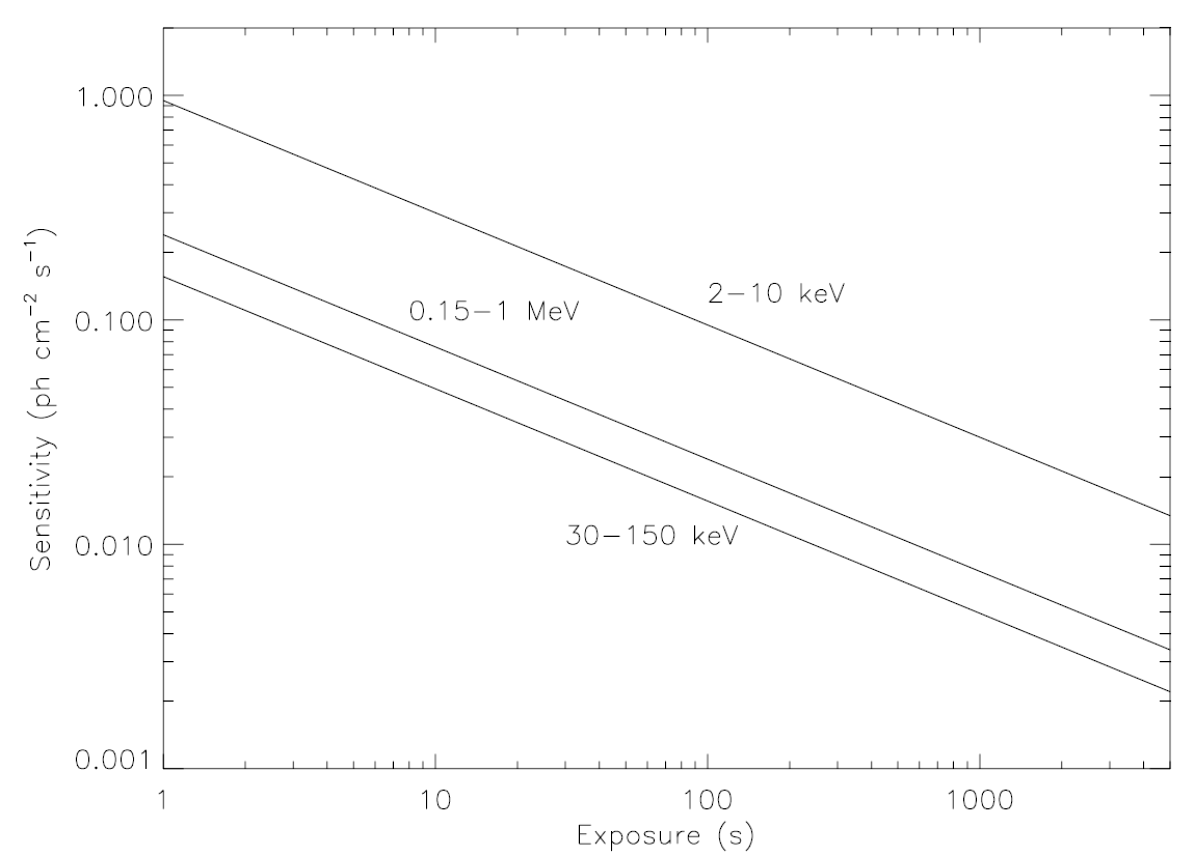}
\caption{{\it Left}: XGIS sensitivity vs. energy for an integration of 1~s. 
{\it Right}: XGIS sensitivity as a function of exposure time in different energy bands.}
\label{fig:36}   
\end{figure*}

\begin{figure*}[t!]
\centering
\includegraphics[scale=0.30]{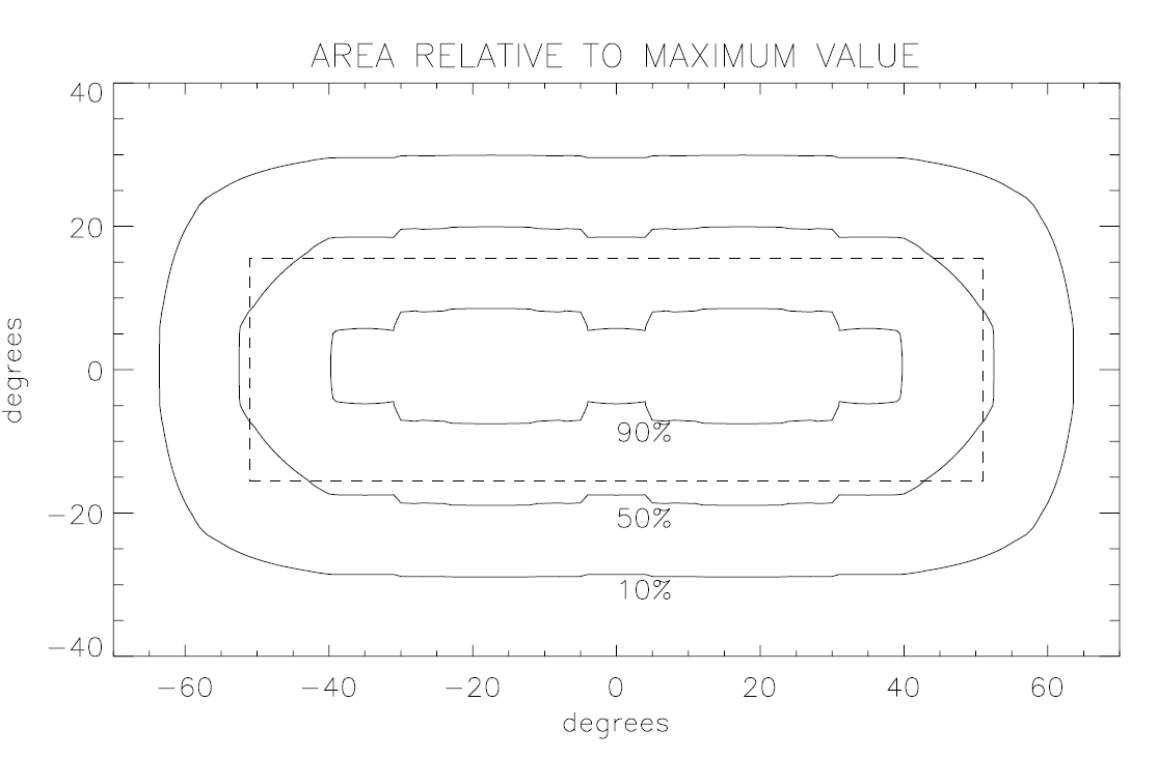}
\includegraphics[scale=1.40]{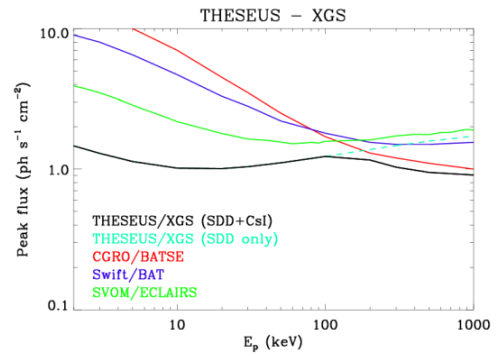}
\caption{{\it Left}: fractional variation of the effective area in the FOV of the XGIS. {\it Right}: Sensitivity of 
the XGIS to GRBs in terms of minimum detectable photon peak flux in 1s (5~$\sigma$) in the 1-1000~keV energy 
band as a function of the spectral peak energy \citep[a method proposed by][]{Band2003}. As can be seen, the combination 
of large effective area and unprecedented large energy band provides a much higher sensitivity w/r to previous 
(e.g., CGRO/BATSE), present (e.g., Swift/BAT) and next future (e.g., SVOM/ECLAIRS) in the soft energy range, 
while keeping a very good sensitivity up to the MeV range.}
\label{fig:37}   
\end{figure*}

\paragraph{XGIS  sensitivity \\}
The 5~$\sigma$ XGIS sensitivity for an integration of 1~s with energy in the SXI FOV is shown in Fig.~\ref{fig:36}, 
along with the XGIS flux sensitivity versus observation time at a significance of 5~$\sigma$ in different energy ranges. 
In Fig.~\ref{fig:37}, the FOV of the XGIS in the 2-30~keV band is compared with the SXI FOV, and the XGIS sensitivity 
vs. GRB peak energy is compared with that of other instruments. 

\subsection{The InfraRed Telescope (IRT)} 

The InfraRed Telescope (IRT) on board THESEUS is designed in order to identify, localize and study the transients and especially the afterglows 
of the GRBs detected by the Soft X-ray Imager (SXI) and the X and Gamma Imaging Spectrometer (XGIS).
The telescope (optics and tube assembly) can be made of SiC, a material that has been used in other space missions (such as Gaia, Herschel, 
Sentinel2 and SPICA study). Simulations using a 0.7~m aperture Cassegrain space borne NIR telescope (with a 0.23~m secondary mirror 
and a 10$\times$10~arcmin imaging flied of view), using a space qualified Teledyne Hawaii-2RG (2048$\times$2048~pixels) HgCdTe detector (18~$\mu$m/pixels, 
resulting in 0.3~arcsec/pix plate scale) show that for a 20.6 (H, AB) point like source and 300~s integration time one could expect 
a SNR of $\sim$5. The telescope sensitivity is limited by the platform jitter. In addition, due to the APE capability of the 
platform (2~arcmin), the high resolution spectroscopy mode cannot make use of a fine slit, and a slit-less mode over a 5$\times$5~arcmin area 
of the detector will be implemented (similarly to what is done for the WFC3 on board the Hubble Space Telescope), with the idea of making 
use of the rest of the image to locate bright sources in order to correct the frames a posteriori for the telescope jitter. The same goal 
could also be obtained by making use of the information provided by payload the high precision star trackers mounted on the IRT.
Hence the maximum limiting resolution that can be achieved by such a system for spectroscopy is limited to R$\sim$500 for a sensitivity limit 
of about 17.5 (H, AB) considering a total integration time of 1800~s. The IRT expected performances are summarized in Table~\ref{tab:7}. 
In order to achieve such performances (i.e., in conditions such that thermal background represents less than 20\% of sky background) 
the telescope needs to be cooled at 240$\pm$3~K, and this can be achieved by passive means. Concerning the IRT camera, the optics box needs 
to be cooled to 190$\pm$5~K and the IR detector itself to 95$\pm$10~K: this allows the detector dark current to be kept at an acceptable level.
The cooling of the detector at these low temperatures can hardly be achieved with a passive system in a low Earth orbit such as the one 
foreseen for THESEUS, due to the irradiation of the radiators of the infrared flux by the Earth atmosphere. A TRL 5 cooling solution for 
space applications is represented by the use of a Miniature Pulse Tube Cooler (MPTC). 

\begin{figure}[t!]
\centering
\includegraphics[scale=0.92]{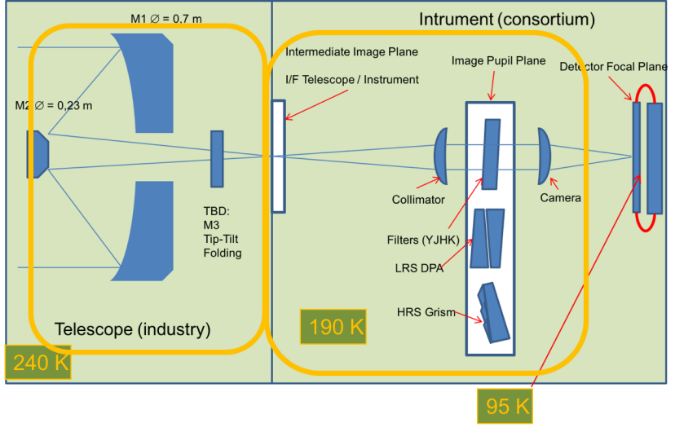}
\caption{The IRT Telescope block diagram concept.}
\label{fig:38}   
\end{figure}

In order to keep the camera design as simple as possible (e.g., avoiding to implement too many mechanisms, like tip-tilting mirrors, 
moving slits, etc.), we could implement a design with an intermediate focal plane making the interface between the telescope provided 
by ESA/industry and the IRT instrument provided by the consortium, as shown in the block diagram in Fig.~\ref{fig:38}. The focal plane 
instrument is composed of a spectral wheel and a filter wheel in which the ZYJH filters, a prism, and a volume phase holographic 
(VPH) grating will be mounted, in order to provide the expected scientific product (imaging, low and high-resolution spectra of 
GRB afterglows and other transients).

Specifications of the entire system are given in Table~\ref{tab:7}. The mechanical envelope of IRT is a cylinder with 80~cm diameter and 180~cm 
height. A sun-shield is placed on top of the telescope baffle for IRT stray-light protection.  The thermal hardware is composed of a 
pulse tube cooling the Detector and FEE electronics and a set of thermal straps extracting the heat from the electronic boxes and 
camera optics coupled to a radiator located on the spacecraft structure. The overall telescope mass is 112.6~kg and the total 
power supply is 115~W. 
\begin{figure*}[t!]
\centering
\includegraphics[scale=0.9]{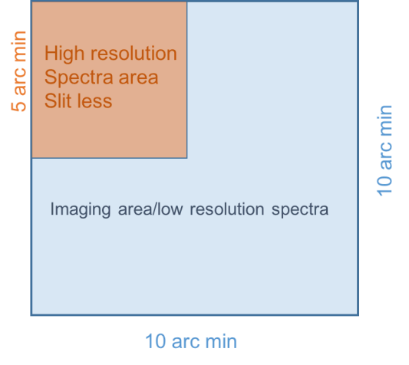}
\includegraphics[scale=1.0]{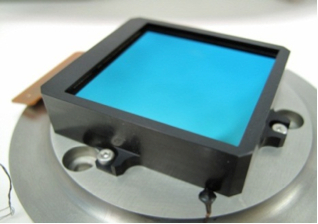}
\caption{{\it Left}: the IRT focal plane division. The blue area (10$\times$10~arcmin) is used for imaging and low resolution spectra. 
The orange area (5$\times$5~arcmin) is used for high-resolution slit-less spectra. The size of the high-resolution spectral 
area is limited by the satellite pointing capabilities. {\it Right}: Teledyne Hawaii 2RG detector at ESA Payload Technology 
Validation section during the tests for the NISP instrument on board the EUCLID mission (Credits ESA).}
\label{fig:39}   
\end{figure*}
\begin{table*}[t!]
\centering
\caption{IRT specifications.}
\label{tab:7}  
\begin{tabular}{lccc}
\hline\noalign{\smallskip}
Telescope type & \multicolumn{3}{c}{Cassegrain} \\
Primary and Secondary size & \multicolumn{3}{c}{700~mm \& 230~mm} \\
Material & \multicolumn{3}{c}{SiC (for both optics and optical tube assembly)} \\
Detector type & \multicolumn{3}{c}{Teledyne Hawaii-2RG 2048$\times$2048 pixels (18~$\mu$m each)} \\
Imaging plate scale & \multicolumn{3}{c}{0''.3/pixel} \\
Field of view & 10$\times$10~arcmin & 10$\times$10~arcmin & 5$\times$5~arcmin  \\
Resolution ($\lambda$/$\Delta\lambda$) & 2-3 (imaging) & 20 (low-res) & 500 (high-res),  goal 1000 \\
Sensitivity (AB mag) & H = 20.6 (300~s) & H = 18.5 (300~s) & H = 17.5 (1800~s) \\
Filters & \multicolumn{3}{c}{ZYJH} \\
Prism & \multicolumn{3}{c}{VPH grating} \\
Wavelength range & 0.7-1.8~$\mu$m (imaging) & 0.7-1.8~$\mu$m (low-res) & 0.7-1.8~$\mu$m (high-res, TBC) \\
Total envelope size & \multicolumn{3}{c}{800 (diameter) $\times$1800~mm} \\
Power & \multicolumn{3}{c}{115~W (50~W for thermal control)} \\
Mass & \multicolumn{3}{c}{112.6~kg}  \\
\noalign{\smallskip}\hline
\end{tabular}
\end{table*}

\paragraph{IRT Observing sequence \\}
The IRT observing sequence is as follow: 
\begin{enumerate}
\item The IRT will observe the GRB error box in imaging mode as soon as the satellite is stabilized within 1~arcsec. 
Three initial frames in the ZJH-bands will be taken (10~s each, goal 19 AB 5~$\sigma$ sensitivity limit in H) to establish 
the astrometry and determine the detected sources colours. 
\item IRT will enter the spectroscopy mode (Low Resolution Spectra, LRS) for a total integration time of 5~min  
(expected 5~$\sigma$ sensitivity limit in H 18.5 (AB)). 
\item Sources with peculiar colours and/or variability (such as GRB afterglows) should have been pinpointed while the low-res 
spectra were obtained and IRT will take a deeper (20~mag sensitivity limit (AB)) H-band image for a total of 60~s. These images 
will be then added/subtracted on board in order to identify bright variable sources with one of them possibly matching one of 
the peculiar colour ones. NIR catalogues will also be used in order to exclude known sources from the GRB candidates.
\begin{itemize}
\item In case a peculiar colour source or/and bright ($<$17.5 H (AB)) variable source is found in the imaging step, the IRT computes 
its redshift (a numerical value if 5$<$z$<$10 or an upper limit z$<$5) from the low resolution spectra obtained at point (1) and determines 
its position. Both the position and redshift estimate will be sent to ground for follow-up observations. The derived position will 
then be used in order to ask the satellite to slew to it so that the source is placed in the high resolution part of the detector 
plane (see below) where the slit-less high resolution mode spectra are acquired. Following the slew, the IRT enters the High Resolution 
Spectra (HRS) mode where it shall acquire at least three spectra of the source (for a total exposure time of 1800~s) covering the 0.7-1.8~$\mu$m range. 
Then it goes back to imaging mode (H-band) for at least another 1800~s. Note that while acquiring the spectra, continuous imaging is 
performed on the rest of the detector (Fig~\ref{fig:38}). This will allow to the on board software to correct the astrometry of the individual 
frames for satellite drift and jitter and allow a final correct reconstruction of the spectra by limiting the blurring effects.
\item In case that a faint ($>$17.5 H (AB)) variable source is found, IRT computes its redshift from the low resolution spectra, determines 
its position and sends both information to the ground (as for 3a). In this case IRT does not ask for a slew to the platform and stays in 
imaging mode for a 3600~s time interval to establish the GRB photometric light curve (covering any possible flaring) and leading the light 
curve to be known with an accuracy of $<$5\%.
\end{itemize}
\end{enumerate}

\subsection{The Instrument Data Handling Units (I-DHUs)} 

Following the concept behind the organization of the THESEUS instruments as well as the decentralized avionic scheme, each of the 
three instrument payloads will be equipped with a dedicated Instrument Data Handling Unit (I-DHU) that will serve as their TM/TC and 
power interface to the spacecraft. The aim of this scheme is to provide sufficient computing power and data storage to the individual 
instruments and thus to realize a decentralized data handling function. 

The mechanical and electrical design of the I-DHUs (described below) will be the same for all three instruments. Also the operating 
system and basic software that is running on the Processor Board inside each I-DHU will be the same. In addition, an instrument 
specific data processing software will be run on each I-DHU, implementing e.g. the above mentioned trigger algorithms and event 
detection codes.

The computational load on the I-DHU is relatively low, which allows us to use a much simpler, off-the-shelf processor with flight 
heritage on the Processor Board with the load being easily sustainable with still a large margin. The tasks of the I-DHUs can be 
separated into three main categories, namely data processing, instrument controlling and power distribution. In order to acquire/process 
the scientific data, the I-DHUs will:
\begin{itemize}
\item collect, process and store the data stream of the respective instrument;
\item implement the burst trigger algorithm on SXI and XGIS data;
\item implement the IRT burst follow up observation. 
\end{itemize}

The I-DHU consists of two main boards that are mounted inside an aluminum case. In addition, each board exists in a cold redundant 
(identical, non operating) version inside the box, that can replace the nominal board in case of a failure. Switching between the 
nominal and redundant chain is done from ground via a dedicated flight-proven circuitry. 
\begin{figure*}[t!]
\centering
\includegraphics[scale=0.2]{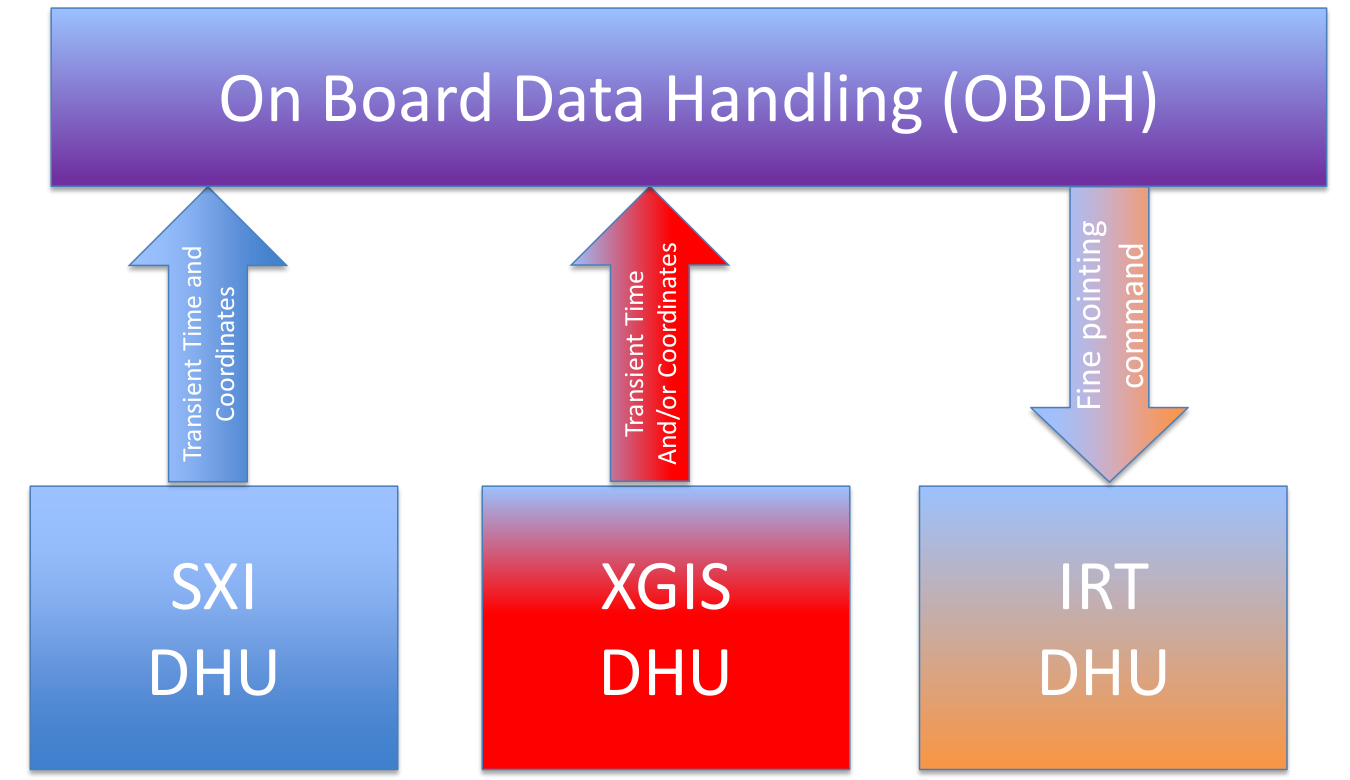}
\caption{Overview of the I-DHU operation modes and transitions.}
\label{fig:40}   
\end{figure*}

At the heart of the I-DHU design is the Processor Board. It hosts the central CPU, the mass memory, time synchronization and distribution 
circuits and the HK/health monitoring acquisition chain.  The Processor Board is connected to the spacecraft via one SpaceWire link 
through which it will receive the telecommands (TC) and send the science and HK/health data. On the other hand, the I-DHU is connected 
via another SpaceWire link to the respective instrument to relay the TCs and acquire the science and HK data. The SpaceWire interface 
communication is handled directly through the main CPU (description see below), via its existing dedicated hardware interfaces. 
The CPU is running an RTEMS (Real Time Executive Management System), an operation system (OS) on which the individual software 
tasks (detailed description below) of this I-DHU will be run in parallel. Dedicated circuitry is foreseen on the Processor Board 
for the collection, digitization and organization of HK and health data from various voltage, current and temperature sensors. 
This will reduce the HK tasks on the main CPU, leaving only the science meta data (like rate meters, counters) and dedicated 
instrument data processing to be done there. The Processor Board will be developed by the IAAT in T\"{u}bingen, Germany.
The Power Board within the I-DHU will be developed by the Centrum Badan Kosmicznych, Poland. It will generate the voltages 
for the Processor Board and distribute the power to the instrument. 

The main functions of the I-DHU on-board software are: instrument control, health monitoring and science data processing, formatting. 
The software will be designed in order to allow the instrument to have the complex functionality that it requires to allow itself to 
be updated and work around problems automatically and with input from the ground. There will be a common software that is the same 
for all I-DHUs and an extended software part with modules specific for a given instrument. An example of a software module common 
to all I-DHUs is the determination of the location of a burst or transient event. The time and location of the transient will be 
transmitted to ground using the on-board VHF system in a $<$1~kbit message. The design of the trigger software benefits from the 
heritage of the SVOM mission concept as well as past team experience on similar systems on BeppoSAX, HETE-2, AGILE as well as 
the INTEGRAL burst alert system.

Instrument control will be possible through the software via telecommands from the ground (e.g., power on \& off for individual 
units, loading parameters for processing \& on-board calibration, investigations) and autonomously on-board (e.g., mode switching, 
FDIR and diagnostic data collection). The software will implement the standard ECSS-style PUS service telecommand packets 
for housekeeping, memory maintenance, monitoring etc. and some standard telemetry packets for command acceptance, 
housekeeping, event reporting, memory management, time management, science data etc. The software will be able to 
send setup information to the instrument and receive and process the housekeeping data coming back and organize 
these data in a configurable way for a lower rate transmission to the ground. Figure~\ref{fig:40} shows the commandable 
operational modes managed by the I-DHU.

\subsection{The Trigger Broadcasting Unit (TBU)}

\begin{figure}[t!]
\centering
\includegraphics[scale=0.93]{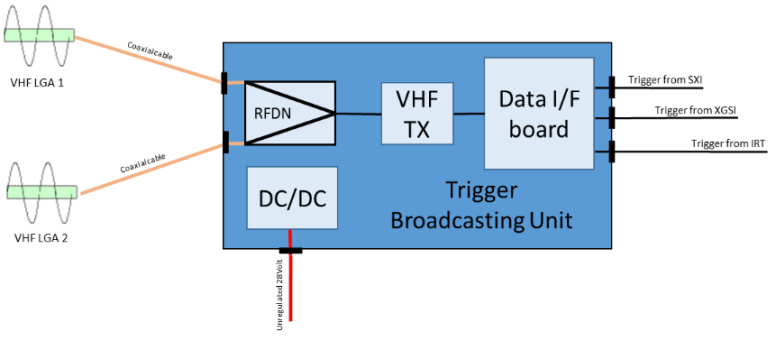}
\caption{General block diagram of the Trigger Broadcast Unit.}
\label{fig:41}   
\end{figure}

In case of trigger event  it is necessary to provide the trigger data to ground in a short time. It is expected triggered events occurred 
at the rate of one event per orbit and the data to be sent to ground is very low: the amount is $<$1~kb/trigger event. 
To support the prompt transmission of such event data to ground segment it is necessary  a link with Earth independent from 
the satellite TT\&C, which can have a link with ground station only once per orbit. The solutions evaluated for an independent and prompt 
burst position broadcast  to ground are by the means of:
\begin{itemize}
\item VHF equatorial network (SVOM and HETE-2);
\item Orbcom satellite network (implemented in Agile satellite);
\item Iridium satellite network (tested by INAF/IASF on balloon and ready to be tested  in FEES/IOD n-sat);
\item TDRSS link.
\end{itemize}
The preferred solution is the well proven  VHF broadcasting based in the utilization of the SVOM equatorial network. Link with the SVOM ground segment 
will be made of 40 ground stations located around the Earth inside a $\pm$30~deg strip. 
The satellite to SVOM network link  shall be carried out by the Trigger Broadcasting Unit (TBU, see Fig.~\ref{fig:41}) proposed in the baseline as a  
unit of the Payload Module.   The antennas are miniaturised for a better accommodation on PLM and SVM.  The VHF ground network is the same 
as the SVOM mission, extensively described in the ground segment section. For the definition of the VHF frequency range a possible choice, 
according to ITU Article~5, could be 137-137.175~MHz, reserved to space research. This band sub-system for trigger transmission 
is consequently proposed for compatibility with SVOM ground segment. 

\section{Satellite configuration and mission profile}

\begin{figure*}[t!]
\centering
\includegraphics[scale=1.4]{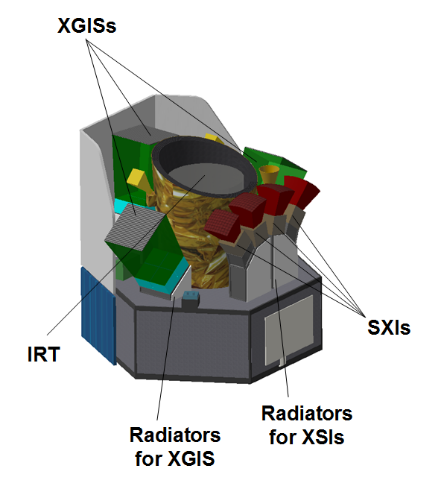}
\includegraphics[scale=1.4]{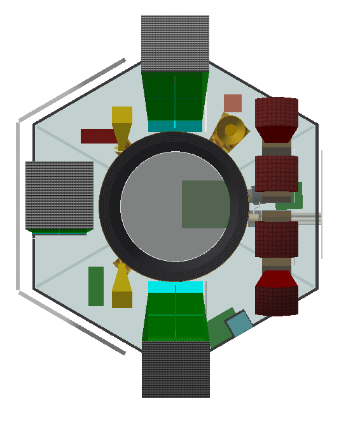}
\caption{THESEUS Satellite Baseline Configuration and Instrument suite accommodation.}
\label{fig:42}   
\end{figure*}

The satellite configuration and design take into account a modular approach.
The spacecraft platform is divided in two modules, the Payload Module (PLM) and Service Module (SVM). 
The Payload Module will mechanically support the  Instruments  DUs (SXI, XGIS and IRT) and will host internally 
the Instruments ICUs. The instruments SXI and XGIS DUs are accommodated externally on the Payload Module to which 
they are connected by means of a structural pedestals. The Instrument DUs mechanical fixing to this structure will 
be designed in order to guarantee thermo-structural decoupling from the rest of satellite.
The Service Module contains all the platform subsystems and provides the mechanical interface with the Launcher.  
Figure~\ref{fig:42} shows the spacecraft baseline configuration.  

\begin{figure}[t!]
\centering
\includegraphics[scale=1.3]{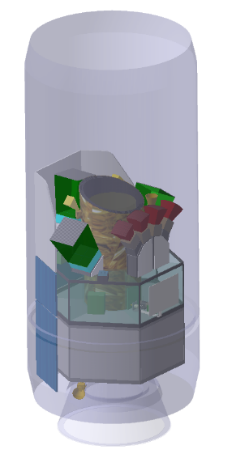}
\includegraphics[scale=1.3]{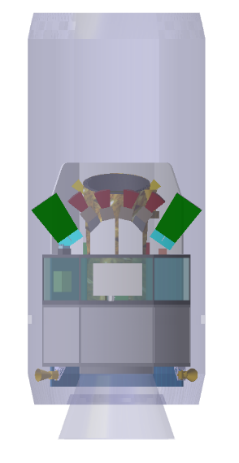}
\caption{THESEUS accommodation within VEGA C fairing.}
\label{fig:43}   
\end{figure}

The Payload Module is constructed around the IRT instrument, which is partially embedded inside the PLM structure 
and aligned with respect the S/C symmetry axis. The module top plane is the mounting base of the other instruments 
DUs  (SXI and XGIS) which are distributed around the IRT axis in order to minimize the satellite Moment of Inertia 
(MoI) but also to support efficient load transfer from the spacecraft to the launch vehicle,  respecting their 
accommodation constraints and thermal requirements. 
The IRT LoS is the reference of the overall payload. The telescope is protected from stray-light by a baffle; the 
profile of the baffle defines the position of the entrance window plane of the other instruments.
The 4 SXI DUs are nominally mounted on the opposite side of the solar panels in order to keep them in the coldest 
side of the satellite and to have the largest area of the observable sky when THESEUS lies between Sun and Earth. 
The SXI DUs positions and orientations are determined in such a way that no X-rays reflected by IRT tube or any 
other satellite structure can enter into SXI FOV. The 3 XGIS units are tilted in such a way that the FOV of the 
units partially overlap.  The resulting  overall FOV (i.e., that of the combination of the 3 units)  covers and 
center the FOV of the 4 SXI modules. 
The proposed accommodation is shown in Fig.~\ref{fig:43}. This configuration guarantees the required nominal SXI, XGIS 
and IRT Field of view combination (REQ-MIS-050).
Solar array proposed baseline consists of modular panels in fixed configuration, the panels are composed of a 
photovoltaic module  and a mechanical on which photovoltaic module is mounted.  At the level of the payload 
module top plane a Sun shield is mounted, in order to protect the instruments from solar radiation. 
The structure of the Payload module is provided of reinforcing shear panels and of an internal cylinder for the 
IRT telescope and detector accommodation. 
The internal cylinder has the function of structural support and it provides also a thermal separation  for IRT 
instrument from the rest of the module. In order to assure radiative thermal decoupling, the telescope is covered 
by multi-layer insulation.

At the level of focal plane of IRT, where the Detector is placed, a Miniature Pulse Tube Cooler (MPTC) system is 
provided, acting with a heat-pipe system and a dedicated radiator. A dedicated radiator is also provided for each 
of the other instruments (SXI and XIGS) on the pedestals. Every radiator is positioned on the support structure 
of each instrument. The Service module is currently conceived as composed of a single module  accommodating 
all the platform units (with the exception of one boresight star tracker which is integrated on the IRT telescope 
tube to minimize relative misalignment, thus improving pointing knowledge performances. In addition to the platform 
star tracker, two star trackers are foreseen in support to IRT instrument, to allow astrometric measurements 
independent from the system.

The AOCS subsystem will provide the required attitudes to support the payload observation requirements, guaranteeing:
\begin{itemize}
\item the satellite agility with: 
\begin{itemize}
\item fast slewing  ($\lesssim$60~deg/10~min) for  IRT  LoS pointing to the direction of GRBs and other transient of interest for low resolution spectra;
\item fine slewing inside the IRT full FoV to achieve the IRT LoS fine pointing necessary for IR high resolution spectra. 
\end{itemize}
\item the satellite  pointing requirements are those of REQ-MIS-060. 
\end{itemize}
The subsystem is made up by software, running on the on board computer, and by a set of sensors and actuators. ACS software is organised 
in operative status, each of them dedicated to the execution of a different mission task.  The 3-axis stabilized attitude control is based 
on a set of 4 reaction wheels used in zero momentum mode and 3 magnetic torquers mainly aimed to perform wheels desaturation.
Sensors used to reach the required attitude knowledge consist of star sensors and fine rate sensors; in addition magnetometers and a set 
of solar sensors are available to cover all the mission needs. 

\begin{table}
\centering
\caption{Summary of Instrument Suite temperatures.}
\label{tab:8}  
\begin{tabular}{lll}
\hline\noalign{\smallskip}
Instrument Element & Operative range (°C) & Cooling \\
SXI- structure/optics & (-20, +20) passive \\
SXI- detectors & -65 & active \\
XGIS-detectors & (-20, +10) & passive \\
IRT-structure & -30 & active \\
IRT-optics & -83 & active \\
\noalign{\smallskip}\hline
\end{tabular}
\end{table} 
The algorithms implemented in the ACS SW will process the sensor data in order to apply a proper control able to guarantee the required 
pointing accuracy and stability.

The THESEUS Telemetry and Telecommand subsystem includes a X-band TM/TC link between S/C and Earth for standard S/C operations in charge 
of the RF links with ground stations with uplinks for satellite telecommands (TC) and downlinks for satellite telemetries 
(TM) and the ranging function to allow range measurements. During downlink operations, 
stored data, read and formatted within the PDHU, are transmitted towards the X-Band transmission assembly, where modulation, 
up conversion to X-Band and power amplification will be executed. The X-Band antenna assembly will provide for transmission 
to the receiving ground station. 

The Thermal Control System (TCS) of the platform can be implemented following a standard approach. 
It is mainly of passive type, based on the capability of dissipating the internal generated thermal power through radiators, 
where the heat is conducted and emitted to deep space. As general thermal control philosophy, all surfaces thermo-optical 
properties are controlled by means of paints, multi layer blankets and/or materials with well-known characteristics. High 
emissivity coatings (black paint) are used for the spacecraft interior in order to uniformly distribute the generated power 
and to avoid hot spots. Lateral closing panels having the function of radiators are covered with adequate coatings 
(e.g. white paint) to optimize the heat rejection to the space. The TCS service to Instrument Suite will consider the 
temperatures summarized in Table~\ref{tab:8}.  Detailed thermal and orbital analyses will optimize the  thermal design.  
\begin{figure*}[t!]
\centering
\includegraphics[scale=1.05]{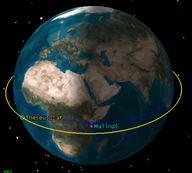}
\includegraphics[scale=1.0]{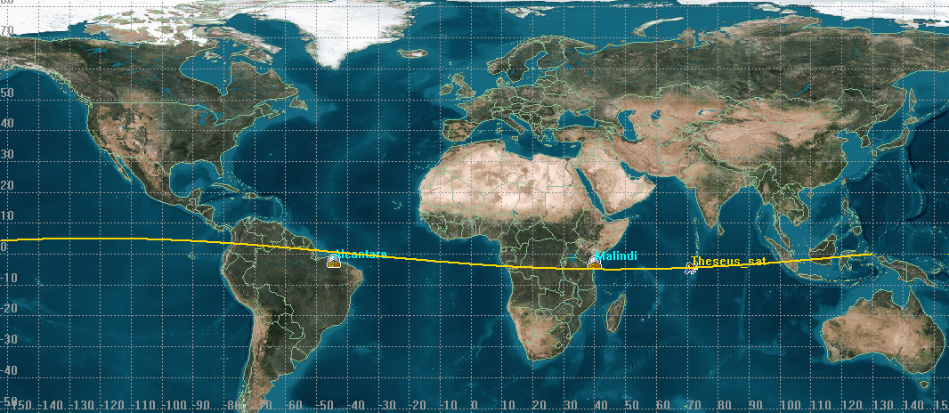}
\caption{THESEUS Orbit Configuration (Left) and THESEUS Orbit Ground Track (Right).}
\label{fig:44}   
\end{figure*}
THESEUS satellite will operate in a low equatorial orbit (altitude $<$600~km, inclination $<$5~deg).
This orbital configuration will guarantee a low and stable background level in the high-energy instruments. 
The mission has been evaluated assuming as baseline a launch with Vega-C.

The THESEUS satellite will be equipped with a suite of three instruments: SXI, XGIS and IRT. In summary the THESEUS  satellite will be capable to: 
\begin{itemize}
\item to monitor a large sky sector for detecting, identifying and localizing likely transients/burst  in the SXI and XGIS FOV; 
\item of promptly (within a few tens of seconds at most) transmitting to ground the trigger time and position of GRBs and other transients of interest;
\item of autonomous (via SXI, XGIS or IRT trigger) orientation in the sky direction of interest;
\item to perform long observation of the sky direction of interest. 
\end{itemize}
THESEUS requires a 3-axis stabilized attitude. During its orbital period THESEUS will have distinct operational modes:
\begin{itemize}
\item Survey (burst hunting) mode:  relevant to normal operation when SXI and XGIS are searching for transients. 
The accessible sky for this kind of operation will be determined by the requirement that: 
\begin{itemize}
\item THESEUS will have a Field of Regards (FoR) defining  the fraction of  sky which can be monitored of 64\%.
\item When monitoring the sky in normal operations, the number of re-pointings per orbit will be of the order of 3, resulting in observations 
with the SXI and XGIS of about the whole FoR every 3 orbits. 
\end{itemize}
\item Burst  mode: after detecting a GRB or other transient of interest, the satellite is triggered to this mode by  SXI and/or XGIS which 
transmit to satellite computer the quaternion of the area of interest. The satellite will autonomously fast repoint to place the transient 
within the field of view of the IRT according to the following steps: 
fast slewing ($\lesssim$60~deg/10~min) for  IRT  LoS pointing to the direction of GRBs and other transient of interest for low resolution spectra, 
satellite stabilization (RPE) within less than 0.5~arcsec fast data link for GRB coordinate communication to ground within a few min. 
\item Follow-up mode: the IRT shall observes inside the full FoV of 10$\times$10~arcmin  the target with the pre-scripted imaging – spectroscopy sequence.  
In case of  IRT high resolution spectra acquisition a further  satellite fine slewing (based on IRT source localization) shall  be activated to place 
the IRT LoS  inside a reduced  FoV of 5$\times$5~arcmin. XGIS and SXI are in specific follow-up data acquisition mode.  In this mode the GRB observation shall 
be performed in a time of 30~min.  After completion of the transient observation, THESEUS will return to Survey Mode to monitor the sky.  
\item IRT observatory mode: the IRT may be used as an observatory for pre-selected targets through a GO programme, driving the pointing of the satellite. 
XGIS and SXI are observing as in survey mode, with the possibility of triggering the burst mode
\end{itemize}
No specific orbit parameter change is required during the mission lifetime.  THESEUS mission can be supported as baseline by a dedicated ground 
station located in Malindi (3~deg~S, 40~deg~E).  Another ground station, located in Alcantara (2~deg~S, 40~deg~W), is supposed as possible back-up in 
case of Brazilian participation. In conformity with the selected equatorial orbit, both stations will allow highly frequent accesses to 
the satellite  by  a contact per orbit. 
Contact analysis between the satellite and the ground stations has been performed considering a minimum station elevation angle over local 
horizon of 10~deg.

\section{Scientific operations, quick-look activities and data distribution}

THESEUS being a GRB mission, it is expected that the SOC will take care of nominal routine mission observation planning with personnel 
in shift 7 days a week, 365 days per year. Starting with the list of successful proposals submitted at each yearly released announcement 
of opportunity, a long-term planning schedule is created using the standard science planning utilities. The core of the THESEUS observing 
program is, however, expected to be dominated by the follow-up of the detected GRB and other transient events. Therefore, detailed planning 
constraints (e.g., sky visibility, and mission resources such as power and telemetry) will need to be identified by the SOC on a daily 
time scale in order to shape the short term observational schedule. Payload operations exclusion windows, on-board resources envelopes 
for payload operations and final detailed checks against mission, environmental and resource constraints will have to be defined by the 
MOC on a similar timescale. The basic Mission Planning approach for all the routine science operations phases will be built on the experience 
of previous missions, such as XMM-Newton, Herschel and INTEGRAL. The rapid rescheduling imposed by the GRB nature of the mission will take 
advantage of the heritage gained through the Swift mission. 
The science observation plan will be modified for Target of Opportunities (ToOs) coming primarily from the XGIS and SXI transient event 
detections, but additional triggers can also be raised by other observatories at any wavelength. The mission will be designed to 
automatically repoint all classes of selected events detected by the SXI or the XGIS within minutes from the trigger. Longer reaction 
times are expected for external triggers. For non-GRB targets, a significant part of the traditional ToOs will be pre-proposed as part 
of the calls for observing proposals (i.e., Announcement of Opportunities, AOs). The proposal will include trigger criteria, plus 
planning information like coordinates, exposure times, instruments observational modes, etc. Also, ToO proposals can be made to the 
SOC for an observation of a target outside the AO process using the so-called Directors Discretionary Time (DDT). 
It is expected that the THESEUS Launch and Early Operations Phase (LEOP) will be concluded with the successful injection into the 
chosen orbit. During Commissioning Phase the instruments will gradually be configured to operating status, with end-to-end test of 
functionality of systems without relying on the cosmic X-ray signals. The planning cycles, up- and down-link functionality will be 
commissioned and the basic instrument modes exercised. Scientific commissioning and basic calibration verification will be completed 
with selected targets, including pointing and sensitivity stability demonstrations.  During Performance Verification Phase (PV) 
selected targets will be observed to instruments capabilities, together with ToO, burst alert and ground segment capability 
confirmation. It is anticipated that normal science operations of the first Announcement of Opportunity (AO) and the Guaranteed 
Time Observations (GTO) targets will start after 6 months when satisfactory performance has been demonstrated. 
\begin{figure}
\centering
\includegraphics[scale=0.7]{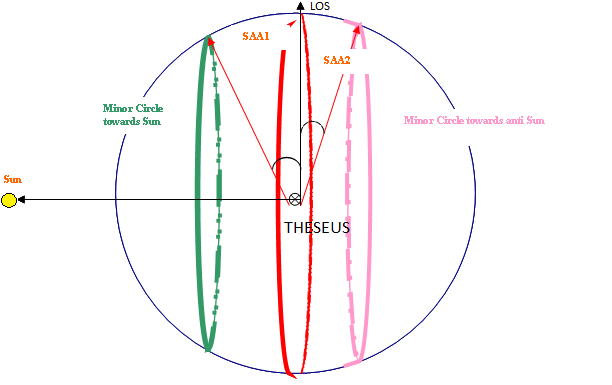}
\caption{Schematic view of the Sun (SAA1) and Earh (SAA2) avoidance angles for the expected 
orbit of THESEUS; the resulting Field of Regard (FoR) is 67\% of the full sky.}
\label{fig:45}   
\end{figure}

During normal science operations the data from the spacecraft (excluding the short alert messages broadcasted from the TBAS - THESEUS burst alert system - through 
the TBAGS - THESEUS burst alert ground segment) is received at the MOC and sent in real time to the SOC where they are automatically processed into the lower level manageable FITS event files 
(including all relevant auxiliary files). Higher level data, including a set of pre-defined scientific products to be used for quick-look activities, 
are produced by a consortium-led Science Data Center (SDC). 
Data collected during the 6 months of the PV phase will be made public only after the validation of the SDC and the instrument operation centers, 
who lead the performance and health verification of the THESEUS instrument (together with calibration activities, supported by the SOC).  
All THESEUS data will be made promptly accessible to the community on a centralized archive (THESEUS science data archive, TSDA) 
as soon as they have been processed. The Swift and Fermi experience 
demonstrated that this strategy helps in maximizing the scientific usage of the data, and is anyway mandated by the urgency of time-domain astrophysics. 
The centralized archive, to be hosted at SOC, will thus the primary repository for the science data products of all levels, and will be used as the 
central ``hub'' during all phases of the mission. It will also guarantee the long term exploitability of the mission heritage.
The alerts provided by the THESEUS on-board detection and localization system will be monitored and analyzed by a dedicated alert center, but also distributed 
to the interested community immediately after the PV phase. Relevant events discovered in the THESEUS data will be eventually communicated by the SDC to the 
community through Astronomer Telegrams or GCNs. 
\begin{figure*}[t!]
\centering
\includegraphics[scale=1.0]{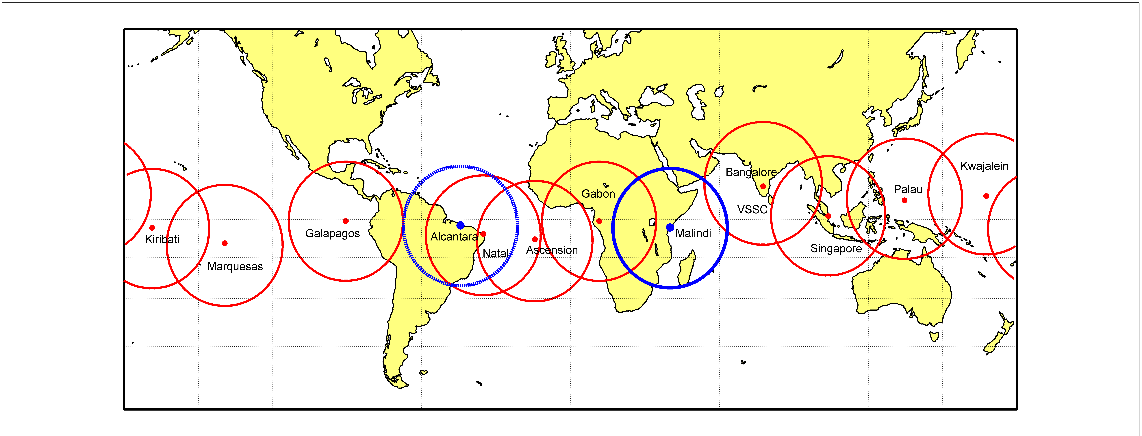}
\caption{The main THESEUS ground station Malindi is indicated, together with a possible sub-set of the SVOM 
VHF ground antennas needed to receive and broadcast the THESEUS alert messages generated on-board.}
\label{fig:46}   
\end{figure*}

\section*{Acknowledgements} 
S.E. acknowledges the financial support from contracts ASI-INAF I/009/10/0, NARO15 ASI-INAF I/037/12/0 and ASI 2015-046-R.0. 
R.H. acknowledges GA CR grant 13-33324S. 
S.V. research leading to these results has received funding from the
European Union's Seventh Framework Programme for research, technological
development and demonstration under grant agreement no 606176. 
D.S. was supported by the Czech grant 16-01116S GA \v{C}R. 
Maria Giovanna Dainotti acknowledges funding from the European Union through the Marie 
Curie Action FP7-PEOPLE-2013-IOF, under grant agreement No. 626267 (``Cosmological Candles'').

\appendix

\section*{Complete author list and affiliations}

\noindent {\bf C.~Adami}, Aix-Marseille Univ., CNRS, LAM, Laboratoire d’Astrophysique de Marseille, 13388 Marseille, France; \\  
{\bf L.~Amati}, INAF-IASF Bologna, via P. Gobetti, 101. I-40129 Bologna, Italy; \\
{\bf A.~Antonelli}, ASDC, Via del Politecnico snc - 00133 Rome, Italy; \\  
{\bf A.~Argan}, INAF-IAPS-Roma via Fosso del Cavaliere, 100, 00133, Rome, Italy; \\  
{\bf J.-L.~Atteia}, IRAP, Universit\'e de Toulouse, CNRS, UPS, CNES, Toulouse, France; \\  
{\bf P.~Attina}, GP Advanced Projects, Italy; \\               
{\bf N.~Auricchio}, INAF-IASF Bologna, via P. Gobetti, 101. I-40129 Bologna, Italy; \\  
{\bf Z.~Bagoly}, E\"otv\"os University, Budapest, Hungary; \\  
{\bf L.~Balazs}, MTA CSFK Konkoly Observatory, Konkoly-Thege M. \'ut 13-17, Budapest, 1121, Hungary; \\  
{\bf G.~Baldazzi}, INFN - Sezione di Bologna, Viale Berti Pichat 6/2, I-40127 Bologna, Italy; Department of Physics, University of Bologna, Viale Berti Pichat 6/2, I-40127 Bologna, Italy; \\  
{\bf S.~Basa}, Aix-Marseille Univ., CNRS, LAM, Laboratoire d’Astrophysique de Marseille, 13388 Marseille, France; \\  
{\bf R.~Basak}, The Oskar Klein Centre for Cosmoparticle Physics, AlbaNova, SE-106 91 Stockholm, Sweden; Department of Physics, KTH Royal Institute of Technology, AlbaNova University Center, SE-106 91 Stockholm, Sweden; \\  
{\bf P.~Bellutti}, FBK, via Sommarive, 18, 38123 Povo, Trento, Italy; \\  
{\bf M.~G.~Bernardini}, INAF - Osservatorio astronomico di Brera, Via E. Bianchi 46, Merate (LC), I-23807, Italy; \\  
{\bf G.~Bertuccio}, Politecnico di Milano, Via Anzani 42, I-22100, Como, Italy; INFN Milano, Via Celoria 16, I-20133, Milano, Italy; \\  
{\bf F.~Bianco}, Center for Cosmology and Particle Physics, New York University, 4 Washington Place, New York, NY 10003, USA; \\  
{\bf A.~Blain}, Department of Physics and Astronomy, University of Leicester, Leicester LE1 7RH, UK; \\  
{\bf S.~Boci}, Department of Physics, University of Tirana, Tirana, Albania; \\  
{\bf M.~Boer}, ARTEMIS, CNRS UMR 5270, Universit\'{e} C\^{o}te d'Azur, Observatoire de la C\^{o}te d'Azur, boulevard de l'Observatoire, CS 34229, F-06304 Nice Cedex 04, France; \\  
{\bf M.~T.~Botticella}, INAF - Capodimonte Astronomical observatory Naples, Via Moiariello 16 I-80131, Naples, Italy; \\ 
{\bf E.~Bozzo}, Department of Astronomy, University of Geneva, ch. d'\'Ecogia 16, CH-1290 Versoix, Switzerland; \\ 
{\bf O.~Boulade}, IRFU/D\'epartement d'Astrophysique,  CEA, Universit\'e  Paris-Saclay, F-91191, Gif-sur-Yvette,  France; \\  
{\bf J.~Braga}, INPE, Av. dos Astronautas 1758, 12227-010, S.J.Campos-SP, Brazil; \\  
{\bf M.~Branchesi}, Università degli Studi di Urbino Carlo Bo, via A. Saffi 2, 61029, Urbino; INFN, Sezione di Firenze, via G. Sansone 1, 50019, Sesto Fiorentino, Italy; \\  
{\bf S.~Brandt}, DTU Space - National Space Institute Elektrovej, Building 327, DK-2800 Kongens Lyngby, Denmark; \\  
{\bf M.~Briggs},  Center for Space Plasma and Aeronomic Research, University of Alabama in Huntsville, 320 Sparkman Drive, Huntsville, AL 35805, USA; \\             
{\bf E.~Brocato}, INAF - Astronomico di Teramo, Mentore Maggini s.n.c., 64100 Teramo, Italy; \\  
{\bf C.~Budtz-Jorgensen}, DTU Space - National Space Institute Elektrovej, Building 327, DK-2800 Kongens Lyngby, Denmark; \\  
{\bf A.~Bulgarelli}, INAF/IASF - Bologna, Via Gobetti 101, I-40129 Bologna, Italy; \\  
{\bf L.~Burderi}, Dipartimento di Fisica, Universit\'a degli Studi di Cagliari, SP Monserrato-Sestu km 0.7, 09042 Monserrato, Italy; \\  
{\bf C.~Butler}, INAF-IASF Bologna, via P. Gobetti, 101. I-40129 Bologna, Italy; \\  
{\bf P.~Callanan}, Department of Physics, University College Cork, Ireland; \\ 
{\bf J.~Camp}, Astrophysics Science Division, Goddard Space Flight Center, Greenbelt, Md 20771; \\  
{\bf R.~Campana}, INAF/IASF-Bologna, via Piero Gobetti 101, I-40129, Bologna, Italy; \\  
{\bf S.~Campana},  INAF - Osservatorio astronomico di Brera, Via E. Bianchi 46, Merate (LC), I-23807, Italy; \\  
{\bf E.~Campolongo}, OHB-Italia, Via Gallarate, 150 20151 Milano, ITALY; \\ 
{\bf F.~Capitanio}, INAF-IAPS-Roma via Fosso del Cavaliere, 100, 00133, Rome, Italy; \\  
{\bf S.~Capozziello}, Dipartimento di Fisica, Università di Napoli Federico II, Via Cinthia, I-80126, Napoli, Italy; \\  
{\bf J.~Caruana}, Department of Physics, University of Malta, Msida MSD 2080, Malta; Institute of Space Sciences \& Astronomy, University of Malta, Msida MSD 2080, Malta; \\  
{\bf P.~Casella}, INAF-Osservatorio Astronomico di Roma, Via Frascati 33, I-00040 Monte Porzio Catone, Italy; \\  
{\bf A.~Castro-Tirado}, IAA-CSIC, P.O. Box 03004, E-18080, Granada, Spain; \\  
{\bf B.~Cenko}, Astrophysics Science Division, NASA Goddard Space Flight Center, Mail Code 661, Greenbelt, MD 20771, USA; Joint Space-Science Institute, University of Maryland, College Park, MD 20742, USA; \\  
{\bf A.~Celotti}, SISSA, via Bonomea 265, I-34136 Trieste, Italy; INAF—Osservatorio Astronomico di Brera, via Bianchi 46, I-23807 Merate (LC), Italy; INFN - Sezione di Trieste, via Valerio 2, I-34127 Trieste, Italy; \\  
{\bf P.~Chardonnet}, LAPTh, Univ. de Savoie, CNRS, B.P. 110, Annecy-le-Vieux F-74941, France; National Research Nuclear University MEPhI, 31 Kashirskoe Sh., Moscow 115409, Russia; \\  
{\bf Y.~Chen}, Institute of High Energy Physics, Beijing 100049, China; \\  
{\bf B.~Ciardi}, Max Planck Institute for Astrophysics, Karl-Schwarzschild-Str. 1, 85741 Garching, Germany; \\  
{\bf R.~Ciolfi}, INAF, Osservatorio Astronomico di Padova, Vicolo dell' Osservatorio 5, I-35122 Padova, Italy; INFN-TIFPA, Trento Institute for Fundamental Physics and Applications, via Sommarive 14, I-38123 Trento, Italy; \\
{\bf S.~Colafrancesco}, School of Physics, University of Witwatersrand, Private Bag 3, Wits-2050, Johannesburg, South Africa; \\  
{\bf M.~Colpi}, Dipartimento di Fisica G. Occhialini, Università degli Studi di Milano Bicocca, Piazza della Scienza 3, 20126 Milano, Italy; INFN, Sezione di Milano-Bicocca, Piazza della Scienza 3, 20126 Milano, Italy; \\ 
{\bf A.~Comastri}, INAF, Osservatorio Astronomico di Bologna, Via Piero Gobetti, 93/3, 40129 Bologna, Italy; \\  
{\bf V.~Connaughton}, Universities Space Research Association, NSSTC, 320 Sparkman Drive, Huntsville, AL 35805, USA; \\             
{\bf B.~Cordier}, IRFU/D\'epartement d'Astrophysique,  CEA, Universit\'e  Paris-Saclay, F-91191, Gif-sur-Yvette,  France; \\  
{\bf C.~Contini}, OHB-Italia, Via Gallarate, 150 20151 Milano, ITALY; \\ 
{\bf S.~Covino}, INAF-Brera Astronomical Observatory, Via Bianchi 46, 23807, Merate (LC), Italy; \\  
{\bf J.-G.~Cuby}, LAM, Laboratoire d’Astrophysique de Marseille, 13388 Marseille, France; \\  
{\bf P.~D'Avanzo}, INAF - Osservatorio astronomico di Brera, Via E. Bianchi 46, Merate (LC), I-23807, Italy; \\ 
{\bf M.~Dadina}, INAF-IASF Bologna, via P. Gobetti, 101. I-40129 Bologna, Italy; \\  
{\bf M.~G.~Dainotti}, Department of Physics \& Astronomy, Stanford University, Via Pueblo Mall 382, Stanford CA, 94305-4060, USA; \\  
{\bf V.~D'Elia}, Space Science Data Center (SSDC), Agenzia Spaziale Italiana, via del Politecnico, s.n.c., I-00133, Roma, Italy; INAF-Osservatorio Astronomico di Roma, Via Frascati 33, I-00040 Monte Porzio Catone, Italy; \\   
{\bf A.~De~Luca}, INAF - Istituto di Astrofisica Spaziale e Fisica Cosmica Milano, Via E. Bassini 15, I-20133 Milano, Italy; \\  
{\bf D.~De~Martino}, INAF - Capodimonte Astronomical observatory Naples, Via Moiariello 16 I-80131, Naples, Italy; \\  
{\bf M.~De~Pasquale}, Department of Astronomy and Space Sciences, Istanbul University, Beyazit, 34119, Istanbul, Turkey; \\  
{\bf E.~Del~Monte}, INAF-IAPS-Roma via Fosso del Cavaliere, 100, 00133, Rome, Italy; \\  
{\bf M.~Della~Valle}, INAF-Osservatorio Astronomico di Capodimonte, salita Moiariello 16, 80131, Napoli, Italy; International Center for Relativistic Astrophysics, Piazzale della Repubblica 2, 65122, Pescara, Italy; \\  
{\bf A.~Drago}, INFN, Via Enrico Fermi 40, Frascati, Italy; \\ 
{\bf Y.-W.~Dong}, Institute of High Energy Physics, Beijing 100049, China; \\  
{\bf G.~Erdos}, Wigner Research Centre for Physics, Hungarian Academy of Sciences, P.O. Box 49, H-1525 Budapest, Hungary; \\             
{\bf S.~Ettori}, INAF, Osservatorio Astronomico di Bologna, Via Piero Gobetti, 93/3, 40129 Bologna, Italy; INFN, Sezione di Bologna, viale Berti Pichat 6/2, 40127 Bologna, Italy; \\  
{\bf Y.~Evangelista}, INAF-IAPS-Roma via Fosso del Cavaliere, 100, 00133, Rome, Italy; \\  
{\bf M.~Feroci}, INAF-IAPS-Roma via Fosso del Cavaliere, 100, 00133, Rome, Italy; \\  
{\bf A.~Ferrara}, Scuola Normale Superiore, Piazza dei Cavalieri 7, I-56126 Pisa, Italy; Kavli IPMU, The University of Tokyo, 5-1-5 Kashiwanoha, Kashiwa 277-8583, Japan; \\  
{\bf F.~Finelli}, INAF-IASF Bologna, via P. Gobetti, 101. I-40129 Bologna, Italy; \\  
{\bf M.~Fiorini}, INAF - Istituto di Astrofisica Spaziale e Fisica Cosmica Milano, Via E. Bassini 15, I-20133 Milano, Italy; \\  
{\bf F.~Frontera}, Department of Physics and Earth Sciences, University of Ferrara, Via Saragat 1, I-44122 Ferrara, Italy, and INAF-IASF, Via Gobetti, 101, I-40129 Bologna, Italy; \\  
{\bf F.~Fuschino}, INAF-IASF Bologna, via P. Gobetti, 101. I-40129 Bologna, Italy; \\  
{\bf J.~Fynbo}, Dark Cosmology Centre, Niels Bohr Institute, University of Copenhagen, Juliane Maries Vej 30, DK-2100 Copenhagen, Denmark; \\  
{\bf A.~Gal-Yam}, Department of Particle Physics and Astrophysics, Faculty of Physics, Weizmann Institute of Science, Rehovot 76100, Israel; \\ 
{\bf P.~Gandhi}, Department of Physics \& Astronomy, University of Southampton, Highfield, Southampton SO17\,1BJ, UK; \\  
{\bf B.~Gendre}, University of the Virgin Islands, 2 John Brewer's Bay, 00802 St Thomas, US Virgin Islands; Etelman Observatory, Bonne Resolution, St Thomas, US Virgin Islands; \\  
{\bf G.~Ghirlanda}, INAF - Osservatorio astronomico di Brera, Via E. Bianchi 46, Merate (LC), I-23807, Italy; \\  
{\bf G.~Ghisellini}, INAF - Osservatorio astronomico di Brera, Via E. Bianchi 46, Merate (LC), I-23807, Italy; \\  
{\bf P.~Giommi}, Italian Space Agency, ASI, via del Politecnico snc, 00133 Roma, Italy; \\  
{\bf A.~Gomboc}, Centre for Astrophysics and Cosmology, University of Nova Gorica, Vipavska 11c, 5270 Ajdov \v s\v cina, Slovenia; \\  
{\bf D.~G\"otz},  IRFU/D\'epartement d'Astrophysique,  CEA, Universit\'e  Paris-Saclay, F-91191, Gif-sur-Yvette,  France; \\ 
{\bf A.~Grado}, INAF - Capodimonte Astronomical observatory Naples, Via Moiariello 16 I-80131, Naples, Italy; \\ 
{\bf J.~Greiner}, Max Planck Institute for Astrophysics, Karl-Schwarzschild-Str. 1, 85741 Garching, Germany; \\  
{\bf C.~Guidorzi}, Department of Physics and Earth Sciences, University of Ferrara, Via Saragat 1, I-44122 Ferrara, Italy; \\  
{\bf S.~Guiriec}, Department of Physics, The George Washington University, 725 21st Street NW, Washington, DC 20052, USA; NASA Goddard Space Flight Center, Greenbelt, MD 20771, USA; Department of Astronomy, University of Maryland, College Park, MD 20742, USA; Center for Research and Exploration in Space Science and Technology (CRESST), Greenbelt, MD 20771, USA; \\  
{\bf M.~Hafizi}, Department of Physics, University of Tirana, Tirana, Albania; \\  
{\bf L.~Hanlon}, Space Science Group, School of Physics, University College Dublin, Belfield, Dublin 4, Ireland; \\  
{\bf J.~Harms}, Università degli Studi di Urbino Carlo Bo, I-61029 Urbino, Italy; \\  
{\bf M.~Hernanz}, Institute of Space Sciences (IEEC–CSIC), Carrer de Can Magrans s/n, E-08193 Barcelona, Spain; \\  
{\bf J.~Hjorth}, Dark Cosmology Centre, Niels Bohr Institute, University of Copenhagen, Juliane Maries Vej 30, DK-2100 Copenhagen, Denmark; \\  
{\bf A.~Hornstrup}, DTU Space - National Space Institute Elektrovej, Building 327, DK-2800 Kongens Lyngby, Denmark; \\  
{\bf R.~Hudec}, Czech Technical University, Faculty of Electrical Engineering, Prague 16627, Czech Republic; Kazan Federal University, Kazan 420008, Russian Federations; \\  
{\bf I.~Hutchinson}, Department of Physics and Astronomy, University of Leicester, Leicester LE1 7RH, UK; \\  
{\bf G.~Israel}, INAF-Osservatorio Astronomico di Roma, Via Frascati 33, I-00040 Monte Porzio Catone, Italy; \\  
{\bf L.~Izzo}, Instituto de Astrofisica de Andalucia (IAA-CSIC), Glorieta de la Astronomia s/n, 18008 Granada, Spain; \\  
{\bf P.~Jonker}, SRON, Netherlands Institute for Space Research, Sorbonnelaan 2, NL-3584~CA Utrecht, The Netherlands; Department of Astrophysics/IMAPP, Radboud University, P.O.~Box 9010, NL-6500 GL Nijmegen, The Netherlands; \\  
{\bf Y.~Kaneko}, Faculty of Engineering and Natural Sciences, Sabancı University, Orhanlı Tuzla, Istanbul 34956, Turkey; \\  
{\bf N.~Kawai}, Department of Physics, Tokyo Institute of Technology, 2-12-1 Ookayama, Meguro-ku, Tokyo 152-8551; \\ 
{\bf L.~Kiss}, Konkoly Observatory, Research Centre for Astronomy and Earth Sciences, Hungarian Academy of Sciences, Konkoly Thege Mikl\'os \'ut 15-17, H-1121 Budapest, Hungary; \\  
{\bf K.~Wiersema}, Department of Physics and Astronomy, University of Leicester, Leicester LE1 7RH, UK ; \\  
{\bf S.~Korpela}, University of Helsinki, Department of Physics, P.O.Box 48 FIN-00014 University of Helsinki, Finland; \\  
{\bf P.~Kumar}, Department of Astronomy, University of Texas at Austin, Austin, TX 78712, USA; \\  
{\bf I.~Kuvvetli}, DTU Space - National Space Institute Elektrovej, Building 327, DK-2800 Kongens Lyngby, Denmark; \\  
{\bf C.~Labanti}, INAF-IASF Bologna, via P. Gobetti, 101. I-40129 Bologna, Italy; \\  
{\bf M.~Lavagna}, Politecnico di Milano, Via La Masa 1, 20156 Milano, Italy; \\  
{\bf V.~Lebrun}, LAM, Laboratoire d’Astrophysique de Marseille, 13388 Marseille, France; \\  
{\bf E.~Le~Floch}, IRFU/D\'epartement d'Astrophysique,  CEA, Universit\'e  Paris-Saclay, F-91191, Gif-sur-Yvette,  France; \\  
{\bf T.~Li}, Department of Engineering Physics and Center for Astrophysics, Tsinghua University, Beijing, China; \\  
{\bf F.~Longo}, Department of Physics, University of Trieste, via Valerio 2, Trieste, Italy; INFN Trieste, via Valerio 2, Trieste, Italy; \\  
{\bf F.~Lu}, Institute of High Energy Physics, Beijing 100049, China; \\  
{\bf M.~Lyutikov}, Department of Physics, Purdue University, 525 Northwestern Avenue, West Lafayette, IN 47907-2036 and Department of Physics and McGill Space Institute, McGill University, 3600; University Street, Montreal, Quebec H3A 2T8, Canada; \\  
{\bf A.~MacFadyen}, Center for Cosmology and Particle Physics, New York University, New York, NY, USA; \\  
{\bf U.~Maio}, Leibniz Institut for Astrophysics, An der Sternwarte 16, 14482 Potsdam, Germany; INAF-Osservatorio Astronomico di Trieste, via G.~Tiepolo 11, 34131 Trieste, Italy; \\  
{\bf E.~Maiorano}, INAF-IASF Bologna, via Piero Gobetti 101, I-40129 Bologna, Italy; \\  
{\bf G.~Malaguti}, INAF-IASF Bologna, via Piero Gobetti 101, I-40129 Bologna, Italy; \\  
{\bf P.~Malcovati}, Department of Electrical, Computer, and Biomedical Engineering, University of Pavia, Pavia, Italy; \\  
{\bf D.~Malesani}, Dark Cosmology Centre, Niels Bohr Institute, University of Copenhagen, Juliane Maries Vej 30, DK-2100 Copenhagen, Denmark; \\  
{\bf L.~Maraschi}, INAF - Osservatorio astronomico di Brera, Via E. Bianchi 46, Merate (LC), I-23807, Italy; \\  
{\bf R.~Margutti}, Center for Interdisciplinary Exploration and Research in Astrophysics (CIERA) and Department of Physics and Astrophysics, Northwestern University, Evanston, IL 60208, USA; \\  
{\bf M.~Marisaldi}, INAF-IASF Bologna, via Piero Gobetti 101, I-40129 Bologna, Italy; \\  
{\bf A.~Martin-Carrillo}, Space Science Group, School of Physics, University College Dublin, Belfield, Dublin 4, Ireland; \\  
{\bf N.~Masetti}, INAF-IASF Bologna, via Piero Gobetti, 101, I-40129 Bologna, Italy; Departamento de Ciencias F\'{i}sicas, Universidad Andr\'es Bello, Fern\'andez Concha 700, Las Condes, Santiago, Chile; \\  
{\bf S.~McBreen}, School of Physics, University College Dublin, Belfield, Stillorgan Road, Dublin 4, Ireland; \\  
{\bf A.~Melandri}, INAF - Osservatorio Astronomico di Brera, via E. Bianchi 36, I-23807 Merate (LC), Italy; \\  
{\bf S.~Mereghetti}, INAF - IASF Milano, Via E. Bassini 15, 20133 Milano, Italy; \\  
{\bf R.~Mignani}, INAF - Istituto di Astrofisica Spaziale e Fisica Cosmica Milano, via E. Bassini 15, 20133, Milano, Italy; Janusz Gil Institute of Astronomy, University of Zielona G\'ora, Lubuska 2, 65-265, Zielona G\'ora, Poland; \\  
{\bf M.~Modjaz}, Center for Cosmology and Particle Physics, Department of Physics, New York University, 726 Broadway office 1044, New York, NY 10003, USA ; \\  
{\bf G.~Morgante}, INAF-IASF Bologna, via Piero Gobetti, 101, I-40129 Bologna, Italy; \\  
{\bf B.~Morelli}, OHB-Italia, Via Gallarate, 150 20151 Milano, ITALY; \\ 
{\bf D.~Morris}, Etelman Observatory, St. Thomas, United States Virgin Islands 00802, USA; College of Science and Math, University of Virgin Islands, St. Thomas, United States Virgin Islands 00802, USA; \\  
{\bf C.~Mundell}, Department of Physics, University of Bath, Claverton Down, Bath BA2 7AY, UK; \\  
{\bf H.~U.~Nargaard-Nielsen}, DTU Space - National Space Institute Elektrovej, Building 327, DK-2800 Kongens Lyngby, Denmark; \\  
{\bf S.~Nagataki}, Astrophysical Big Bang Laboratory (ABBL), RIKEN, Saitama 351-0198, Japan; \\ 
{\bf L.~Nicastro}, INAF - Instituto di Astrofisica Spaziale e Fisica Cosmica, Via Piero Gobetti 101, I-40129 Bologna, Italy; \\  
{\bf P.~O'Brien}, Department of Physics and Astronomy, University of Leicester, Leicester LE1 7RH, UK; \\
{\bf N.~Omodei}, W. W. Hansen Experimental Physics Laboratory, Kavli Institute for Particle Astrophysics and Cosmology, Department of Physics and SLAC National Accelerator Laboratory, Stanford University, Stanford, CA 94305, USA; \\  
{\bf M.~Orlandini}, INAF-IASF Bologna, via Piero Gobetti, 101, I-40129 Bologna, Italy; \\                        
{\bf P.~Orleanski}, Space Research Center of the Polish Academy of Sciences, Warsaw, Poland; \\  
{\bf J.~P.~Osborne}, Department of Physics and Astronomy, University of Leicester, Leicester LE1 7RH, UK; \\         
{\bf A.~Paizis}, INAF - IASF Milano, Via E. Bassini 15, 20133 Milano, Italy; \\  
{\bf E.~Palazzi}, INAF-IASF Bologna, via Piero Gobetti, 101, I-40129 Bologna, Italy; \\  
{\bf S.~Paltani}, Department of Astronomy, University of Geneva, ch. d'\'Ecogia 16, CH-1290 Versoix, Switzerland; \\  
{\bf F.~Panessa}, INAF-IAPS-Roma via Fosso del Cavaliere, 100, 00133, Rome, Italy; \\  
{\bf G.~Pareschi}, INAF - Osservatorio astronomico di Brera, Via E. Bianchi 46, Merate (LC), I-23807, Italy; \\  
{\bf P.~P\'{a}ta}, Department of Radioelectronics, Faculty of Electrical Engineering, Czech Technical University in Prague, Technick\'{a}  2, 166 27 Prague 6, Czech Republic; \\  
{\bf A.~Pe'er}, Department of Physics, University College Cork, Ireland; \\  
{\bf A.~V.~Penacchioni}, ICRA Net, Piazza della Repubblica 10, I-65122 Pescara, Italy; ASI Science Data Center, Via del Politecnico s.n.c., I-00133 Rome, Italy; Department of Physical Sciences, Earth and Environment, University of Siena, Via Roma 56, I-53100 Siena, Italy; \\  
{\bf V.~Petrosian}, W. W. Hansen Experimental Physics Laboratory, Kavli Institute for Particle Astrophysics and Cosmology, Department of Physics and SLAC National Accelerator Laboratory, Stanford University, Stanford, CA 94305, USA; \\  
{\bf E.~Pian}, Scuola Normale Superiore, Piazza dei Cavalieri 7, I-56126 Pisa, Italy; \\  
{\bf E.~Piedipalumbo}, Dipartimento di Fisica, Universit\'a degli Studi di Napoli Federico II, Compl. Univ. Monte S. Angelo, 80126 Naples, Italy; INFN, Sez. di Napoli, Compl. Univ. Monte S. Angelo, Edificio 6, via Cinthia, 80126 Napoli, Italy; \\  
{\bf T.~Piran}, Racah Institute of Physics, The Hebrew University of Jerusalem, Jerusalem 91904, Israel; \\  
{\bf P.~Piranomonte}, INAF-Osservatorio Astronomico di Roma, Via Frascati 33, I-00040 Monte Porzio Catone, Italy; \\  
{\bf L.~Piro}, INAF-IAPS-Roma via Fosso del Cavaliere, 100, 00133, Rome, Italy; \\  
{\bf A.~Rachevski}, INFN - Sezione di Trieste, via Valerio 2, I-34127 Trieste, Italy; \\  
{\bf G.~Rauw}, Universit\'e de Li\`ege, Quartier Agora, All\'ee du 6 Ao\^ut 19c, B-4000 Sart Tilman, Li\`ege, Belgium; \\  
{\bf M.~Razzano}, Department of Physics, University of Pisa and INFN-Pisa,Pisa,  I-56127; \\  
{\bf A.~Read}, Department of Physics and Astronomy, University of Leicester, Leicester LE1 7RH, UK; \\  
{\bf V.~Reglero}, Image Processing Laboratory, University of Valencia C/Catedratico Jose Beltran, 2, 46980 Paterna (Valencia), Spain; \\  
{\bf E.~Renotte}, Centre Spatial de Li\`ege, Parc Scientifique du Sart Tilman Avenue du Pr\`e-Aily, 4031 Angleur-Li\`ege, Belgium; \\  
{\bf L.~Rezzolla}, Institut f{\"u}r Theoretische Physik, Johann Wolfgang Goethe-Universit{\"a}t, Max-von-Laue-Stra{\ss}e 1, 60438 Frankfurt, Germany; Frankfurt Institute for Advanced Studies, Ruth-Moufang-Stra{\ss}e 1, 60438 Frankfurt, Germany; \\  
{\bf J.~Rhoads}, School of Earth and Space Exploration, Arizona State University, Tempe, AZ 85287, USA; Astrophysics Science Division, Goddard Space Flight Center, 8800 Greenbelt Road, Greenbelt, MD 20771, USA; \\  
{\bf T.~Rodic}, SPACE-SI, Slovenian Centre of Excellence for Space Sciences and Technologies, Ljubljana, Slovenia; \\  
{\bf P.~Romano}, INAF-Osservatorio Astronomico di Brera, via E.\ Bianchi, 46.   I-23807 Merate, Italy; \\  
{\bf P.~Rosati}, Universita degli Studi di Ferrara, Via Saragat 1, Ferrara, Italy; \\  
{\bf A.~Rossi}, INAF-IASF Bologna, via P. Gobetti, 101. I-40129 Bologna, Italy; \\  
{\bf R.~Ruffini}, ICRANet, P.zza della Repubblica 10, 65122 Pescara, Italy; ICRA and Dipartimento di Fisica, Sapienza Universit\`a di Roma, P.le Aldo Moro 5, 00185 Rome, Italy; \\  
{\bf F.~Ryde}, The Oskar Klein Centre for Cosmoparticle Physics, AlbaNova, SE-106 91 Stockholm, Sweden; Department of Physics, KTH Royal Institute of Technology, AlbaNova University Center, SE-106 91 Stockholm, Sweden; \\  
{\bf L.~Sabau-Graziati}, Division de Ciencias del Espacio (INTA), Torrejon de Ardoz, Madrid, Spain; \\  
{\bf R.~Salvaterra},  INAF - Istituto di Astrofisica Spaziale e Fisica Cosmica Milano, via E. Bassini 15, 20133, Milano, Italy ; \\  
{\bf A.~Santangelo}, Institut f\"ur Astronomie und Astrophysik, Abteilung Hochenergieastrophysik, Kepler Center for Astro and Particle Physics, Eberhard Karls Universit\"at, Sand 1, D 72076 T\"ubingen, Germany; \\  
{\bf S.~Savaglio}, Physics Dept., University of Calabria, via P. Bucci, 87036, Arcavacata di Rende, Italy; \\  
{\bf V.~Sguera}, INAF-IASF Bologna, via P. Gobetti, 101. I-40129 Bologna, Italy; \\  
{\bf P.~Schady}, Max Planck Institute for extraterrestrial Physics, Giessenbachstrasse 1, 85748 Garching, Germany; \\  
{\bf M.~Sims}, Department of Physics and Astronomy, University of Leicester, Leicester LE1 7RH, UK; \\  
{\bf W.~Skidmore}, Warren Skidmore, Thirty Meter Telescope International Observatory, 100 W. Walnut St., Suite 300, Pasadena, CA 91124, USA; \\  
{\bf L.~Song}, Key Laboratory of Particle Astrophysics, Institute of High Energy Physics, Chinese Academy of Sciences, Beijing 100049, China; \\  
{\bf J.~Soomin}, Instituto de Astrofisica de Andalucia (IAA-CSIC), Glorieta de la Astronomia s/n, 18008 Granada, Spain; \\  
{\bf E.~Stanway}, Department of Physics, University of Warwick, Gibbet Hill Road, Coventry, CV4 7AL, UK; \\  
{\bf R.~Starling}, University of Leicester, Department of Physics and Astronomy and Leicester Institute of Space and Earth Observation, University Road, Leicester LE1 7RH, UK; \\  
{\bf G.~Stratta}, Universit\'a degli Studi di Urbino Carlo Bo, I-61029 Urbino, Italy; \\                
{\bf D.~Sz\'ecsi}, Astronomical Institute of the Czech Academy of Sciences, Fri\v{c}ova 298, 25165 Ond\v{r}ejov, Czech Republic; School of Physics and Astronomy and Institute of Gravitational Wave Astronomy, University of Birmingham, Edgbaston, Birmingham B15 2TT, UK; \\  
{\bf G.~Tagliaferri}, INAF - Osservatorio astronomico di Brera, Via E. Bianchi 46, Merate (LC), I-23807, Italy; \\  
{\bf N.~Tanvir}, University of Leicester, Department of Physics and Astronomy and Leicester Institute of Space \& Earth Observation, University Road, Leicester, LE1 7RH, UK; \\  
{\bf C.~Tenzer}, Institut f\"ur Astronomie und Astrophysik, Abteilung Hochenergieastrophysik, Kepler Center for Astro and Particle Physics, Eberhard Karls Universit\"at, Sand 1, D 72076 T\"ubingen, Germany  \\  
{\bf M.~Topinka}, Dublin Institute for Advanced Studies, School of Cosmic Physics, 31 Fitzwilliam Place, Dublin 2, Ireland; \\  
{\bf E.~Troja}, Department of Astronomy, University of Maryland, College Park, Maryland 20742-4111, USA; NASA Goddard Space Flight Center, 8800 Greenbelt Rd, Greenbelt, Maryland 20771, USA; \\  
{\bf Y.~Urata}, Institute of Astronomy, National Central University, Chung-Li 32054, Taiwan \\
{\bf M.~Uslenghi}, INAF-IASF, via E. Bassini 15, 20133 Milano, Italy; \\  
{\bf A.~Vacchi}, INFN Trieste, via Valerio 2, Trieste, Italy; \\  
{\bf L.~Valenziano}, INAF/IASF Bologna, via Gobetti 101, Bologna, Italy; \\  
{\bf M.~van~Putten}, Sejong University, 98 Gunja-Dong Gwangin-gu, Seoul 143-747, Korea; \\  
{\bf E.~Vanzella}, INAF Osservatorio Astronomico di Bologna, via Ranzani 1, 40127 Bologna, Italy; \\  
{\bf S.~Vercellone}, INAF-Osservatorio Astronomico di Brera, via E.\ Bianchi, 46.   I-23807 Merate, Italy; \\  
{\bf S.~Vergani}, GEPI, Observatoire de Paris, PSL Research University, CNRS, Place Jules Janssen, 92190 Meudon, France; INAF/Osservatorio Astronomico di Brera, via Bianchi 46, 23807 Merate (LC), Italy; \\  
{\bf G.~Vianello}, SLAC National Accelerator Laboratory, Stanford University, Stanford, CA 94305, USA; \\  
{\bf S.~Vinciguerra}, Institute of Gravitational Wave Astronomy \& School of Physics and Astronomy, University of Birmingham, Birmingham, B15 2TT, United Kingdom; \\  
{\bf D.~Watson}, Dark Cosmology Centre, Niels Bohr Institute, University of Copenhagen, Juliane Maries Vej 30, DK-2100 Copenhagen, Denmark; \\  
{\bf R.~Willingale}, Department of Physics and Astronomy, University of Leicester, Leicester LE1 7RH, UK \\             
{\bf C.~Wilson-Hodge}, NASA/Marshall Space Flight Center, Huntsville, AL, USA; \\  
{\bf S.~Vojtech}, Astronomical Institute Academy of Sciences of the Czech Republic, Fricova 1, Ondrejov, CZ-25165, Czech Republic; \\ 
{\bf D.~Yonetoku}, Faculty of Mathematics and Physics, Kanazawa University, Ishikawa 920-1192, Japan; \\  
{\bf G.~Zampa}, INFN Trieste, via Valerio 2, Trieste, Italy; \\  
{\bf N.~Zampa}, INFN Trieste, via Valerio 2, Trieste, Italy; \\  
{\bf B.~Zhang}, Department of Physics and Astronomy, University of Nevada, Las Vegas, NV 89154, USA; \\
{\bf B.~B.~Zhang}, IAA-CSIC, P.O. Box 03004, E-18080, Granada, Spain; \\  
{\bf S.~Zhang}, Institute of High Energy Physics, Beijing 100049, China; \\  
{\bf S.-N.~Zhang}, Institute of High Energy Physics, Beijing 100049, China; \\  
{\bf J.~Zicha}, Department of Instrumentation and Control Engineering, Faculty of Mechanical Engineering, Czech Technical University in Prague, Technicka 4, 166 07 Praha 6, Czech Republic

\bibliographystyle{model5-names}
\bibliography{theseus}                

\begin{thebibliography}{206}
\expandafter\ifx\csname natexlab\endcsname\relax\def\natexlab#1{#1}\fi
\providecommand{\url}[1]{\texttt{#1}}
\providecommand{\href}[2]{#2}
\providecommand{\path}[1]{#1}
\providecommand{\DOIprefix}{doi:}
\providecommand{\ArXivprefix}{arXiv:}
\providecommand{\URLprefix}{URL: }
\providecommand{\Pubmedprefix}{pmid:}
\providecommand{\doi}[1]{\href{http://dx.doi.org/#1}{\path{#1}}}
\providecommand{\Pubmed}[1]{\href{pmid:#1}{\path{#1}}}
\providecommand{\bibinfo}[2]{#2}
\ifx\xfnm\relax \def\xfnm[#1]{\unskip,\space#1}\fi
\bibitem[{{Abadie} et~al.(2010){Abadie}, {Abbott}, {Abbott}, {Accadia},
  {Acernese}, {Adhikari}, {Ajith}, {Allen}, {Allen}, {Amador Ceron} \&
  et~al.}]{Abadie2010}
\bibinfo{author}{{Abadie}, J.}, \bibinfo{author}{{Abbott}, B.~P.},
  \bibinfo{author}{{Abbott}, R.}, \bibinfo{author}{{Accadia}, T.},
  \bibinfo{author}{{Acernese}, F.}, \bibinfo{author}{{Adhikari}, R.},
  \bibinfo{author}{{Ajith}, P.}, \bibinfo{author}{{Allen}, B.},
  \bibinfo{author}{{Allen}, G.}, \bibinfo{author}{{Amador Ceron}, E.}, \&
  \bibinfo{author}{et~al.} (\bibinfo{year}{2010}).
\newblock \bibinfo{title}{{All-sky search for gravitational-wave bursts in the
  first joint LIGO-GEO-Virgo run}}.
\newblock {\it \bibinfo{journal}{\prd}\/},  {\it \bibinfo{volume}{81}\/},
  \bibinfo{pages}{102001}. \DOIprefix\doi{10.1103/PhysRevD.81.102001}.
  \href{http://arxiv.org/abs/1002.1036}{\tt arXiv:1002.1036}.
\bibitem[{{Abbasi} et~al.(2012){Abbasi}, {Abdou}, {Abu-Zayyad}, {Ackermann},
  {Adams}, {Aguilar}, {Ahlers}, {Altmann}, {Andeen}, {Auffenberg} \&
  et~al.}]{Abbasi2012}
\bibinfo{author}{{Abbasi}, R.}, \bibinfo{author}{{Abdou}, Y.},
  \bibinfo{author}{{Abu-Zayyad}, T.}, \bibinfo{author}{{Ackermann}, M.},
  \bibinfo{author}{{Adams}, J.}, \bibinfo{author}{{Aguilar}, J.~A.},
  \bibinfo{author}{{Ahlers}, M.}, \bibinfo{author}{{Altmann}, D.},
  \bibinfo{author}{{Andeen}, K.}, \bibinfo{author}{{Auffenberg}, J.}, \&
  \bibinfo{author}{et~al.} (\bibinfo{year}{2012}).
\newblock \bibinfo{title}{{An absence of neutrinos associated with cosmic-ray
  acceleration in gamma-ray bursts}}.
\newblock {\it \bibinfo{journal}{\nat}\/},  {\it \bibinfo{volume}{484}\/},
  \bibinfo{pages}{351--354}. \DOIprefix\doi{10.1038/nature11068}.
  \href{http://arxiv.org/abs/1204.4219}{\tt arXiv:1204.4219}.
\bibitem[{{Abbott} et~al.(2016{\natexlab{a}}){Abbott}, {Abbott}, {Abbott},
  {Abernathy}, {Acernese}, {Ackley}, {Adams}, {Adams}, {Addesso}, {Adhikari} \&
  et~al.}]{Abbott2016b}
\bibinfo{author}{{Abbott}, B.~P.}, \bibinfo{author}{{Abbott}, R.},
  \bibinfo{author}{{Abbott}, T.~D.}, \bibinfo{author}{{Abernathy}, M.~R.},
  \bibinfo{author}{{Acernese}, F.}, \bibinfo{author}{{Ackley}, K.},
  \bibinfo{author}{{Adams}, C.}, \bibinfo{author}{{Adams}, T.},
  \bibinfo{author}{{Addesso}, P.}, \bibinfo{author}{{Adhikari}, R.~X.}, \&
  \bibinfo{author}{et~al.} (\bibinfo{year}{2016}{\natexlab{a}}).
\newblock \bibinfo{title}{{GW151226: Observation of Gravitational Waves from a
  22-Solar-Mass Binary Black Hole Coalescence}}.
\newblock {\it \bibinfo{journal}{Physical Review Letters}\/},  {\it
  \bibinfo{volume}{116}\/}, \bibinfo{pages}{241103}.
  \DOIprefix\doi{10.1103/PhysRevLett.116.241103}.
  \href{http://arxiv.org/abs/1606.04855}{\tt arXiv:1606.04855}.
\bibitem[{{Abbott} et~al.(2016{\natexlab{b}}){Abbott}, {Abbott}, {Abbott},
  {Abernathy}, {Acernese}, {Ackley}, {Adams}, {Adams}, {Addesso}, {Adhikari} \&
  et~al.}]{Abbott2016a}
\bibinfo{author}{{Abbott}, B.~P.}, \bibinfo{author}{{Abbott}, R.},
  \bibinfo{author}{{Abbott}, T.~D.}, \bibinfo{author}{{Abernathy}, M.~R.},
  \bibinfo{author}{{Acernese}, F.}, \bibinfo{author}{{Ackley}, K.},
  \bibinfo{author}{{Adams}, C.}, \bibinfo{author}{{Adams}, T.},
  \bibinfo{author}{{Addesso}, P.}, \bibinfo{author}{{Adhikari}, R.~X.}, \&
  \bibinfo{author}{et~al.} (\bibinfo{year}{2016}{\natexlab{b}}).
\newblock \bibinfo{title}{{Properties of the Binary Black Hole Merger
  GW150914}}.
\newblock {\it \bibinfo{journal}{Physical Review Letters}\/},  {\it
  \bibinfo{volume}{116}\/}, \bibinfo{pages}{241102}.
  \DOIprefix\doi{10.1103/PhysRevLett.116.241102}.
  \href{http://arxiv.org/abs/1602.03840}{\tt arXiv:1602.03840}.
\bibitem[{{Abbott} et~al.(2017{\natexlab{a}}){Abbott}, {Abbott}, {Abbott},
  {Acernese}, {Ackley}, {Adams}, {Adams}, {Addesso}, {Adhikari}, {Adya} \&
  et~al.}]{Abbott2017b}
\bibinfo{author}{{Abbott}, B.~P.}, \bibinfo{author}{{Abbott}, R.},
  \bibinfo{author}{{Abbott}, T.~D.}, \bibinfo{author}{{Acernese}, F.},
  \bibinfo{author}{{Ackley}, K.}, \bibinfo{author}{{Adams}, C.},
  \bibinfo{author}{{Adams}, T.}, \bibinfo{author}{{Addesso}, P.},
  \bibinfo{author}{{Adhikari}, R.~X.}, \bibinfo{author}{{Adya}, V.~B.}, \&
  \bibinfo{author}{et~al.} (\bibinfo{year}{2017}{\natexlab{a}}).
\newblock \bibinfo{title}{{Gravitational Waves and Gamma-Rays from a Binary
  Neutron Star Merger: GW170817 and GRB 170817A}}.
\newblock {\it \bibinfo{journal}{\apjl}\/},  {\it \bibinfo{volume}{848}\/},
  \bibinfo{pages}{L13}. \DOIprefix\doi{10.3847/2041-8213/aa920c}.
  \href{http://arxiv.org/abs/1710.05834}{\tt arXiv:1710.05834}.
\bibitem[{{Abbott} et~al.(2017{\natexlab{b}}){Abbott}, {Abbott}, {Abbott},
  {Acernese}, {Ackley}, {Adams}, {Adams}, {Addesso}, {Adhikari}, {Adya} \&
  et~al.}]{Abbott2017a}
\bibinfo{author}{{Abbott}, B.~P.}, \bibinfo{author}{{Abbott}, R.},
  \bibinfo{author}{{Abbott}, T.~D.}, \bibinfo{author}{{Acernese}, F.},
  \bibinfo{author}{{Ackley}, K.}, \bibinfo{author}{{Adams}, C.},
  \bibinfo{author}{{Adams}, T.}, \bibinfo{author}{{Addesso}, P.},
  \bibinfo{author}{{Adhikari}, R.~X.}, \bibinfo{author}{{Adya}, V.~B.}, \&
  \bibinfo{author}{et~al.} (\bibinfo{year}{2017}{\natexlab{b}}).
\newblock \bibinfo{title}{{GW170817: Observation of Gravitational Waves from a
  Binary Neutron Star Inspiral}}.
\newblock {\it \bibinfo{journal}{Physical Review Letters}\/},  {\it
  \bibinfo{volume}{119}\/}, \bibinfo{pages}{161101}.
  \DOIprefix\doi{10.1103/PhysRevLett.119.161101}.
  \href{http://arxiv.org/abs/1710.05832}{\tt arXiv:1710.05832}.
\bibitem[{{Abbott} et~al.(2017{\natexlab{c}}){Abbott}, {Abbott}, {Abbott},
  {Acernese}, {Ackley}, {Adams}, {Adams}, {Addesso}, {Adhikari}, {Adya} \&
  et~al.}]{Abbott2017c}
\bibinfo{author}{{Abbott}, B.~P.}, \bibinfo{author}{{Abbott}, R.},
  \bibinfo{author}{{Abbott}, T.~D.}, \bibinfo{author}{{Acernese}, F.},
  \bibinfo{author}{{Ackley}, K.}, \bibinfo{author}{{Adams}, C.},
  \bibinfo{author}{{Adams}, T.}, \bibinfo{author}{{Addesso}, P.},
  \bibinfo{author}{{Adhikari}, R.~X.}, \bibinfo{author}{{Adya}, V.~B.}, \&
  \bibinfo{author}{et~al.} (\bibinfo{year}{2017}{\natexlab{c}}).
\newblock \bibinfo{title}{{Multi-messenger Observations of a Binary Neutron
  Star Merger}}.
\newblock {\it \bibinfo{journal}{\apjl}\/},  {\it \bibinfo{volume}{848}\/},
  \bibinfo{pages}{L12}. \DOIprefix\doi{10.3847/2041-8213/aa91c9}.
  \href{http://arxiv.org/abs/1710.05833}{\tt arXiv:1710.05833}.
\bibitem[{{Amati} \& {Della Valle}(2013)}]{amati13}
\bibinfo{author}{{Amati}, L.}, \& \bibinfo{author}{{Della Valle}, M.}
  (\bibinfo{year}{2013}).
\newblock \bibinfo{title}{{Measuring Cosmological Parameters with Gamma Ray
  Bursts}}.
\newblock {\it \bibinfo{journal}{International Journal of Modern Physics D}\/},
   {\it \bibinfo{volume}{22}\/}, \bibinfo{pages}{1330028}.
  \DOIprefix\doi{10.1142/S0218271813300280}.
  \href{http://arxiv.org/abs/1310.3141}{\tt arXiv:1310.3141}.
\bibitem[{{Amati} et~al.(2004){Amati}, {Frontera}, {in't Zand}, {Capalbi},
  {Landi}, {Soffitta}, {Vetere}, {Antonelli}, {Costa}, {Del Sordo}, {Feroci},
  {Guidorzi}, {Heise}, {Masetti}, {Montanari}, {Nicastro}, {Palazzi} \&
  {Piro}}]{Amati2004}
\bibinfo{author}{{Amati}, L.}, \bibinfo{author}{{Frontera}, F.},
  \bibinfo{author}{{in't Zand}, J.~J.~M.}, \bibinfo{author}{{Capalbi}, M.},
  \bibinfo{author}{{Landi}, R.}, \bibinfo{author}{{Soffitta}, P.},
  \bibinfo{author}{{Vetere}, L.}, \bibinfo{author}{{Antonelli}, L.~A.},
  \bibinfo{author}{{Costa}, E.}, \bibinfo{author}{{Del Sordo}, S.},
  \bibinfo{author}{{Feroci}, M.}, \bibinfo{author}{{Guidorzi}, C.},
  \bibinfo{author}{{Heise}, J.}, \bibinfo{author}{{Masetti}, N.},
  \bibinfo{author}{{Montanari}, E.}, \bibinfo{author}{{Nicastro}, L.},
  \bibinfo{author}{{Palazzi}, E.}, \& \bibinfo{author}{{Piro}, L.}
  (\bibinfo{year}{2004}).
\newblock \bibinfo{title}{{Prompt and afterglow X-ray emission from the X-Ray
  Flash of 2002 April 27}}.
\newblock {\it \bibinfo{journal}{\aap}\/},  {\it \bibinfo{volume}{426}\/},
  \bibinfo{pages}{415--423}. \DOIprefix\doi{10.1051/0004-6361:20047146}.
  \href{http://arxiv.org/abs/astro-ph/0407166}{\tt arXiv:astro-ph/0407166}.
\bibitem[{{Amati} et~al.(2002){Amati}, {Frontera}, {Tavani}, {in't Zand},
  {Antonelli}, {Costa}, {Feroci}, {Guidorzi}, {Heise}, {Masetti}, {Montanari},
  {Nicastro}, {Palazzi}, {Pian}, {Piro} \& {Soffitta}}]{Amati2002}
\bibinfo{author}{{Amati}, L.}, \bibinfo{author}{{Frontera}, F.},
  \bibinfo{author}{{Tavani}, M.}, \bibinfo{author}{{in't Zand}, J.~J.~M.},
  \bibinfo{author}{{Antonelli}, A.}, \bibinfo{author}{{Costa}, E.},
  \bibinfo{author}{{Feroci}, M.}, \bibinfo{author}{{Guidorzi}, C.},
  \bibinfo{author}{{Heise}, J.}, \bibinfo{author}{{Masetti}, N.},
  \bibinfo{author}{{Montanari}, E.}, \bibinfo{author}{{Nicastro}, L.},
  \bibinfo{author}{{Palazzi}, E.}, \bibinfo{author}{{Pian}, E.},
  \bibinfo{author}{{Piro}, L.}, \& \bibinfo{author}{{Soffitta}, P.}
  (\bibinfo{year}{2002}).
\newblock \bibinfo{title}{{Intrinsic spectra and energetics of BeppoSAX
  Gamma-Ray Bursts with known redshifts}}.
\newblock {\it \bibinfo{journal}{\aap}\/},  {\it \bibinfo{volume}{390}\/},
  \bibinfo{pages}{81--89}. \DOIprefix\doi{10.1051/0004-6361:20020722}.
  \href{http://arxiv.org/abs/astro-ph/0205230}{\tt arXiv:astro-ph/0205230}.
\bibitem[{{Amati} et~al.(2000){Amati}, {Frontera}, {Vietri}, {in't Zand},
  {Soffitta}, {Costa}, {Del Sordo}, {Pian}, {Piro}, {Antonelli}, {Fiume},
  {Feroci}, {Gandolfi}, {Guidorzi}, {Heise}, {Kuulkers}, {Masetti},
  {Montanari}, {Nicastro}, {Orlandini} \& {Palazzi}}]{Amati2000}
\bibinfo{author}{{Amati}, L.}, \bibinfo{author}{{Frontera}, F.},
  \bibinfo{author}{{Vietri}, M.}, \bibinfo{author}{{in't Zand}, J.~J.~M.},
  \bibinfo{author}{{Soffitta}, P.}, \bibinfo{author}{{Costa}, E.},
  \bibinfo{author}{{Del Sordo}, S.}, \bibinfo{author}{{Pian}, E.},
  \bibinfo{author}{{Piro}, L.}, \bibinfo{author}{{Antonelli}, L.~A.},
  \bibinfo{author}{{Fiume}, D.~D.}, \bibinfo{author}{{Feroci}, M.},
  \bibinfo{author}{{Gandolfi}, G.}, \bibinfo{author}{{Guidorzi}, C.},
  \bibinfo{author}{{Heise}, J.}, \bibinfo{author}{{Kuulkers}, E.},
  \bibinfo{author}{{Masetti}, N.}, \bibinfo{author}{{Montanari}, E.},
  \bibinfo{author}{{Nicastro}, L.}, \bibinfo{author}{{Orlandini}, M.}, \&
  \bibinfo{author}{{Palazzi}, E.} (\bibinfo{year}{2000}).
\newblock \bibinfo{title}{{Discovery of a Transient Absorption Edge in the
  X-ray Spectrum of GRB 990705}}.
\newblock {\it \bibinfo{journal}{Science}\/},  {\it \bibinfo{volume}{290}\/},
  \bibinfo{pages}{953--955}. \DOIprefix\doi{10.1126/science.290.5493.953}.
  \href{http://arxiv.org/abs/astro-ph/0012318}{\tt arXiv:astro-ph/0012318}.
\bibitem[{{Amati} et~al.(2008){Amati}, {Guidorzi}, {Frontera}, {Della Valle},
  {Finelli}, {Landi} \& {Montanari}}]{Amati2008}
\bibinfo{author}{{Amati}, L.}, \bibinfo{author}{{Guidorzi}, C.},
  \bibinfo{author}{{Frontera}, F.}, \bibinfo{author}{{Della Valle}, M.},
  \bibinfo{author}{{Finelli}, F.}, \bibinfo{author}{{Landi}, R.}, \&
  \bibinfo{author}{{Montanari}, E.} (\bibinfo{year}{2008}).
\newblock \bibinfo{title}{{Measuring the cosmological parameters with the
  E$_{p,i}$-E$_{iso}$ correlation of gamma-ray bursts}}.
\newblock {\it \bibinfo{journal}{\mnras}\/},  {\it \bibinfo{volume}{391}\/},
  \bibinfo{pages}{577--584}. \DOIprefix\doi{10.1111/j.1365-2966.2008.13943.x}.
  \href{http://arxiv.org/abs/0805.0377}{\tt arXiv:0805.0377}.
\bibitem[{{Angel}(1979)}]{Angel1979}
\bibinfo{author}{{Angel}, J.~R.~P.} (\bibinfo{year}{1979}).
\newblock \bibinfo{title}{{Lobster eyes as X-ray telescopes}}.
\newblock {\it \bibinfo{journal}{\apj}\/},  {\it \bibinfo{volume}{233}\/},
  \bibinfo{pages}{364--373}. \DOIprefix\doi{10.1086/157397}.
\bibitem[{{Aschwanden} \& {Tsiklauri}(2009)}]{Aschwanden2009}
\bibinfo{author}{{Aschwanden}, M.~J.}, \& \bibinfo{author}{{Tsiklauri}, D.}
  (\bibinfo{year}{2009}).
\newblock \bibinfo{title}{{The Hydrodynamic Evolution of Impulsively Heated
  Coronal Loops: Explicit Analytical Approximations}}.
\newblock {\it \bibinfo{journal}{\apjs}\/},  {\it \bibinfo{volume}{185}\/},
  \bibinfo{pages}{171--185}. \DOIprefix\doi{10.1088/0067-0049/185/1/171}.
\bibitem[{{Atek} et~al.(2015){Atek}, {Richard}, {Jauzac}, {Kneib}, {Natarajan},
  {Limousin}, {Schaerer}, {Jullo}, {Ebeling}, {Egami} \& {Clement}}]{Atek2015}
\bibinfo{author}{{Atek}, H.}, \bibinfo{author}{{Richard}, J.},
  \bibinfo{author}{{Jauzac}, M.}, \bibinfo{author}{{Kneib}, J.-P.},
  \bibinfo{author}{{Natarajan}, P.}, \bibinfo{author}{{Limousin}, M.},
  \bibinfo{author}{{Schaerer}, D.}, \bibinfo{author}{{Jullo}, E.},
  \bibinfo{author}{{Ebeling}, H.}, \bibinfo{author}{{Egami}, E.}, \&
  \bibinfo{author}{{Clement}, B.} (\bibinfo{year}{2015}).
\newblock \bibinfo{title}{{Are Ultra-faint Galaxies at z = 6-8 Responsible for
  Cosmic Reionization? Combined Constraints from the Hubble Frontier Fields
  Clusters and Parallels}}.
\newblock {\it \bibinfo{journal}{\apj}\/},  {\it \bibinfo{volume}{814}\/},
  \bibinfo{pages}{69}. \href{http://arxiv.org/abs/1509.06764}{\tt
  arXiv:1509.06764}.
\bibitem[{{Baiotti} \& {Rezzolla}(2017)}]{Baiotti2017}
\bibinfo{author}{{Baiotti}, L.}, \& \bibinfo{author}{{Rezzolla}, L.}
  (\bibinfo{year}{2017}).
\newblock \bibinfo{title}{{Binary neutron star mergers: a review of Einstein
  richest laboratory}}.
\newblock {\it \bibinfo{journal}{Reports on Progress in Physics}\/},  {\it
  \bibinfo{volume}{80}\/}, \bibinfo{pages}{096901}.
  \DOIprefix\doi{10.1088/1361-6633/aa67bb}.
  \href{http://arxiv.org/abs/1607.03540}{\tt arXiv:1607.03540}.
\bibitem[{{Band}(2003)}]{Band2003}
\bibinfo{author}{{Band}, D.~L.} (\bibinfo{year}{2003}).
\newblock \bibinfo{title}{{Comparison of the Gamma-Ray Burst Sensitivity of
  Different Detectors}}.
\newblock {\it \bibinfo{journal}{\apj}\/},  {\it \bibinfo{volume}{588}\/},
  \bibinfo{pages}{945--951}. \DOIprefix\doi{10.1086/374242}.
  \href{http://arxiv.org/abs/astro-ph/0212452}{\tt arXiv:astro-ph/0212452}.
\bibitem[{{Basak} \& {Rao}(2013)}]{br2013}
\bibinfo{author}{{Basak}, R.}, \& \bibinfo{author}{{Rao}, A.~R.}
  (\bibinfo{year}{2013}).
\newblock \bibinfo{title}{{Pulse-wise Amati correlation in Fermi gamma-ray
  bursts}}.
\newblock {\it \bibinfo{journal}{\mnras}\/},  {\it \bibinfo{volume}{436}\/},
  \bibinfo{pages}{3082--3088}. \DOIprefix\doi{10.1093/mnras/stt1790}.
  \href{http://arxiv.org/abs/1309.5233}{\tt arXiv:1309.5233}.
\bibitem[{{Basak} \& {Rao}(2015)}]{br2015}
\bibinfo{author}{{Basak}, R.}, \& \bibinfo{author}{{Rao}, A.~R.}
  (\bibinfo{year}{2015}).
\newblock \bibinfo{title}{{Thermal Emissions Spanning the Prompt and the
  Afterglow Phases of the Ultra-long GRB 130925A}}.
\newblock {\it \bibinfo{journal}{\apj}\/},  {\it \bibinfo{volume}{807}\/},
  \bibinfo{pages}{34}. \DOIprefix\doi{10.1088/0004-637X/807/1/34}.
  \href{http://arxiv.org/abs/1505.03296}{\tt arXiv:1505.03296}.
\bibitem[{{Bauswein} \& {Janka}(2012)}]{Bauswein2012}
\bibinfo{author}{{Bauswein}, A.}, \& \bibinfo{author}{{Janka}, H.-T.}
  (\bibinfo{year}{2012}).
\newblock \bibinfo{title}{{Measuring Neutron-Star Properties via Gravitational
  Waves from Neutron-Star Mergers}}.
\newblock {\it \bibinfo{journal}{Physical Review Letters}\/},  {\it
  \bibinfo{volume}{108}\/}, \bibinfo{pages}{011101}.
  \DOIprefix\doi{10.1103/PhysRevLett.108.011101}.
  \href{http://arxiv.org/abs/1106.1616}{\tt arXiv:1106.1616}.
\bibitem[{{Benz} \& {G{\"u}del}(2010)}]{Benz2010}
\bibinfo{author}{{Benz}, A.~O.}, \& \bibinfo{author}{{G{\"u}del}, M.}
  (\bibinfo{year}{2010}).
\newblock \bibinfo{title}{{Physical Processes in Magnetically Driven Flares on
  the Sun, Stars, and Young Stellar Objects}}.
\newblock {\it \bibinfo{journal}{\araa}\/},  {\it \bibinfo{volume}{48}\/},
  \bibinfo{pages}{241--287}.
  \DOIprefix\doi{10.1146/annurev-astro-082708-101757}.
\bibitem[{{Berger}(2014)}]{Berger2014}
\bibinfo{author}{{Berger}, E.} (\bibinfo{year}{2014}).
\newblock \bibinfo{title}{{Short-Duration Gamma-Ray Bursts}}.
\newblock {\it \bibinfo{journal}{\araa}\/},  {\it \bibinfo{volume}{52}\/},
  \bibinfo{pages}{43--105}.
  \DOIprefix\doi{10.1146/annurev-astro-081913-035926}.
  \href{http://arxiv.org/abs/1311.2603}{\tt arXiv:1311.2603}.
\bibitem[{{Bloom} et~al.(2011){Bloom}, {Giannios}, {Metzger}, {Cenko},
  {Perley}, {Butler}, {Tanvir}, {Levan}, {O'Brien}, {Strubbe}, {De Colle},
  {Ramirez-Ruiz}, {Lee}, {Nayakshin}, {Quataert}, {King}, {Cucchiara},
  {Guillochon}, {Bower}, {Fruchter}, {Morgan} \& {van der Horst}}]{Bloom2011}
\bibinfo{author}{{Bloom}, J.~S.}, \bibinfo{author}{{Giannios}, D.},
  \bibinfo{author}{{Metzger}, B.~D.}, \bibinfo{author}{{Cenko}, S.~B.},
  \bibinfo{author}{{Perley}, D.~A.}, \bibinfo{author}{{Butler}, N.~R.},
  \bibinfo{author}{{Tanvir}, N.~R.}, \bibinfo{author}{{Levan}, A.~J.},
  \bibinfo{author}{{O'Brien}, P.~T.}, \bibinfo{author}{{Strubbe}, L.~E.},
  \bibinfo{author}{{De Colle}, F.}, \bibinfo{author}{{Ramirez-Ruiz}, E.},
  \bibinfo{author}{{Lee}, W.~H.}, \bibinfo{author}{{Nayakshin}, S.},
  \bibinfo{author}{{Quataert}, E.}, \bibinfo{author}{{King}, A.~R.},
  \bibinfo{author}{{Cucchiara}, A.}, \bibinfo{author}{{Guillochon}, J.},
  \bibinfo{author}{{Bower}, G.~C.}, \bibinfo{author}{{Fruchter}, A.~S.},
  \bibinfo{author}{{Morgan}, A.~N.}, \& \bibinfo{author}{{van der Horst},
  A.~J.} (\bibinfo{year}{2011}).
\newblock \bibinfo{title}{{A Possible Relativistic Jetted Outburst from a
  Massive Black Hole Fed by a Tidally Disrupted Star}}.
\newblock {\it \bibinfo{journal}{Science}\/},  {\it \bibinfo{volume}{333}\/},
  \bibinfo{pages}{203}. \DOIprefix\doi{10.1126/science.1207150}.
  \href{http://arxiv.org/abs/1104.3257}{\tt arXiv:1104.3257}.
\bibitem[{{Bloom} et~al.(2009){Bloom}, {Holz}, {Hughes}, {Menou}, {Adams},
  {Anderson}, {Becker}, {Bower}, {Brandt}, {Cobb}, {Cook}, {Corsi}, {Covino},
  {Fox}, {Fruchter}, {Fryer}, {Grindlay}, {Hartmann}, {Haiman}, {Kocsis},
  {Jones}, {Loeb}, {Marka}, {Metzger}, {Nakar}, {Nissanke}, {Perley}, {Piran},
  {Poznanski}, {Prince}, {Schnittman}, {Soderberg}, {Strauss}, {Shawhan},
  {Shoemaker}, {Sievers}, {Stubbs}, {Tagliaferri}, {Ubertini} \&
  {Wozniak}}]{Bloom2009}
\bibinfo{author}{{Bloom}, J.~S.}, \bibinfo{author}{{Holz}, D.~E.},
  \bibinfo{author}{{Hughes}, S.~A.}, \bibinfo{author}{{Menou}, K.},
  \bibinfo{author}{{Adams}, A.}, \bibinfo{author}{{Anderson}, S.~F.},
  \bibinfo{author}{{Becker}, A.}, \bibinfo{author}{{Bower}, G.~C.},
  \bibinfo{author}{{Brandt}, N.}, \bibinfo{author}{{Cobb}, B.},
  \bibinfo{author}{{Cook}, K.}, \bibinfo{author}{{Corsi}, A.},
  \bibinfo{author}{{Covino}, S.}, \bibinfo{author}{{Fox}, D.},
  \bibinfo{author}{{Fruchter}, A.}, \bibinfo{author}{{Fryer}, C.},
  \bibinfo{author}{{Grindlay}, J.}, \bibinfo{author}{{Hartmann}, D.},
  \bibinfo{author}{{Haiman}, Z.}, \bibinfo{author}{{Kocsis}, B.},
  \bibinfo{author}{{Jones}, L.}, \bibinfo{author}{{Loeb}, A.},
  \bibinfo{author}{{Marka}, S.}, \bibinfo{author}{{Metzger}, B.},
  \bibinfo{author}{{Nakar}, E.}, \bibinfo{author}{{Nissanke}, S.},
  \bibinfo{author}{{Perley}, D.~A.}, \bibinfo{author}{{Piran}, T.},
  \bibinfo{author}{{Poznanski}, D.}, \bibinfo{author}{{Prince}, T.},
  \bibinfo{author}{{Schnittman}, J.}, \bibinfo{author}{{Soderberg}, A.},
  \bibinfo{author}{{Strauss}, M.}, \bibinfo{author}{{Shawhan}, P.~S.},
  \bibinfo{author}{{Shoemaker}, D.~H.}, \bibinfo{author}{{Sievers}, J.},
  \bibinfo{author}{{Stubbs}, C.}, \bibinfo{author}{{Tagliaferri}, G.},
  \bibinfo{author}{{Ubertini}, P.}, \& \bibinfo{author}{{Wozniak}, P.}
  (\bibinfo{year}{2009}).
\newblock \bibinfo{title}{{Astro2010 Decadal Survey Whitepaper: Coordinated
  Science in the Gravitational and Electromagnetic Skies}}.
\newblock {\it \bibinfo{journal}{ArXiv e-prints}\/}, .
  \href{http://arxiv.org/abs/0902.1527}{\tt arXiv:0902.1527}.
\bibitem[{{Bozzo} et~al.(2016){Bozzo}, {Bhalerao}, {Pradhan}, {Tomsick},
  {Romano}, {Ferrigno}, {Chaty}, {Oskinova}, {Manousakis}, {Walter}, {Falanga},
  {Campana}, {Stella}, {Ramolla} \& {Chini}}]{Bozzo2016}
\bibinfo{author}{{Bozzo}, E.}, \bibinfo{author}{{Bhalerao}, V.},
  \bibinfo{author}{{Pradhan}, P.}, \bibinfo{author}{{Tomsick}, J.},
  \bibinfo{author}{{Romano}, P.}, \bibinfo{author}{{Ferrigno}, C.},
  \bibinfo{author}{{Chaty}, S.}, \bibinfo{author}{{Oskinova}, L.},
  \bibinfo{author}{{Manousakis}, A.}, \bibinfo{author}{{Walter}, R.},
  \bibinfo{author}{{Falanga}, M.}, \bibinfo{author}{{Campana}, S.},
  \bibinfo{author}{{Stella}, L.}, \bibinfo{author}{{Ramolla}, M.}, \&
  \bibinfo{author}{{Chini}, R.} (\bibinfo{year}{2016}).
\newblock \bibinfo{title}{{Multi-wavelength observations of IGR J17544-2619
  from quiescence to outburst}}.
\newblock {\it \bibinfo{journal}{\aap}\/},  {\it \bibinfo{volume}{596}\/},
  \bibinfo{pages}{A16}. \DOIprefix\doi{10.1051/0004-6361/201629311}.
  \href{http://arxiv.org/abs/1610.02648}{\tt arXiv:1610.02648}.
\bibitem[{{Bozzo} et~al.(2008){Bozzo}, {Falanga} \& {Stella}}]{Bozzo2008}
\bibinfo{author}{{Bozzo}, E.}, \bibinfo{author}{{Falanga}, M.}, \&
  \bibinfo{author}{{Stella}, L.} (\bibinfo{year}{2008}).
\newblock \bibinfo{title}{{Are There Magnetars in High-Mass X-Ray Binaries? The
  Case of Supergiant Fast X-Ray Transients}}.
\newblock {\it \bibinfo{journal}{\apj}\/},  {\it \bibinfo{volume}{683}\/},
  \bibinfo{pages}{1031--1044}. \DOIprefix\doi{10.1086/589990}.
  \href{http://arxiv.org/abs/0805.1849}{\tt arXiv:0805.1849}.
\bibitem[{{Brown}(2004)}]{Brown2004}
\bibinfo{author}{{Brown}, E.~F.} (\bibinfo{year}{2004}).
\newblock \bibinfo{title}{{Superburst Ignition and Implications for Neutron
  Star Interiors}}.
\newblock {\it \bibinfo{journal}{\apjl}\/},  {\it \bibinfo{volume}{614}\/},
  \bibinfo{pages}{L57--L60}. \DOIprefix\doi{10.1086/425562}.
  \href{http://arxiv.org/abs/astro-ph/0409032}{\tt arXiv:astro-ph/0409032}.
\bibitem[{{Bucciantini} et~al.(2012){Bucciantini}, {Metzger}, {Thompson} \&
  {Quataert}}]{Bucciantini2012}
\bibinfo{author}{{Bucciantini}, N.}, \bibinfo{author}{{Metzger}, B.~D.},
  \bibinfo{author}{{Thompson}, T.~A.}, \& \bibinfo{author}{{Quataert}, E.}
  (\bibinfo{year}{2012}).
\newblock \bibinfo{title}{{Short gamma-ray bursts with extended emission from
  magnetar birth: jet formation and collimation}}.
\newblock {\it \bibinfo{journal}{\mnras}\/},  {\it \bibinfo{volume}{419}\/},
  \bibinfo{pages}{1537--1545}.
  \DOIprefix\doi{10.1111/j.1365-2966.2011.19810.x}.
  \href{http://arxiv.org/abs/1106.4668}{\tt arXiv:1106.4668}.
\bibitem[{{Burrows} et~al.(2011){Burrows}, {Kennea}, {Ghisellini}, {Mangano},
  {Zhang}, {Page}, {Eracleous}, {Romano}, {Sakamoto}, {Falcone}, {Osborne},
  {Campana}, {Beardmore}, {Breeveld}, {Chester}, {Corbet}, {Covino},
  {Cummings}, {D'Avanzo}, {D'Elia}, {Esposito}, {Evans}, {Fugazza}, {Gelbord},
  {Hiroi}, {Holland}, {Huang}, {Im}, {Israel}, {Jeon}, {Jeon}, {Jun}, {Kawai},
  {Kim}, {Krimm}, {Marshall}, {P.~M{\'e}sz{\'a}ros}, {Negoro}, {Omodei},
  {Park}, {Perkins}, {Sugizaki}, {Sung}, {Tagliaferri}, {Troja}, {Ueda},
  {Urata}, {Usui}, {Antonelli}, {Barthelmy}, {Cusumano}, {Giommi}, {Melandri},
  {Perri}, {Racusin}, {Sbarufatti}, {Siegel} \& {Gehrels}}]{Burrows2011}
\bibinfo{author}{{Burrows}, D.~N.}, \bibinfo{author}{{Kennea}, J.~A.},
  \bibinfo{author}{{Ghisellini}, G.}, \bibinfo{author}{{Mangano}, V.},
  \bibinfo{author}{{Zhang}, B.}, \bibinfo{author}{{Page}, K.~L.},
  \bibinfo{author}{{Eracleous}, M.}, \bibinfo{author}{{Romano}, P.},
  \bibinfo{author}{{Sakamoto}, T.}, \bibinfo{author}{{Falcone}, A.~D.},
  \bibinfo{author}{{Osborne}, J.~P.}, \bibinfo{author}{{Campana}, S.},
  \bibinfo{author}{{Beardmore}, A.~P.}, \bibinfo{author}{{Breeveld}, A.~A.},
  \bibinfo{author}{{Chester}, M.~M.}, \bibinfo{author}{{Corbet}, R.},
  \bibinfo{author}{{Covino}, S.}, \bibinfo{author}{{Cummings}, J.~R.},
  \bibinfo{author}{{D'Avanzo}, P.}, \bibinfo{author}{{D'Elia}, V.},
  \bibinfo{author}{{Esposito}, P.}, \bibinfo{author}{{Evans}, P.~A.},
  \bibinfo{author}{{Fugazza}, D.}, \bibinfo{author}{{Gelbord}, J.~M.},
  \bibinfo{author}{{Hiroi}, K.}, \bibinfo{author}{{Holland}, S.~T.},
  \bibinfo{author}{{Huang}, K.~Y.}, \bibinfo{author}{{Im}, M.},
  \bibinfo{author}{{Israel}, G.}, \bibinfo{author}{{Jeon}, Y.},
  \bibinfo{author}{{Jeon}, Y.-B.}, \bibinfo{author}{{Jun}, H.~D.},
  \bibinfo{author}{{Kawai}, N.}, \bibinfo{author}{{Kim}, J.~H.},
  \bibinfo{author}{{Krimm}, H.~A.}, \bibinfo{author}{{Marshall}, F.~E.},
  \bibinfo{author}{{P.~M{\'e}sz{\'a}ros}}, \bibinfo{author}{{Negoro}, H.},
  \bibinfo{author}{{Omodei}, N.}, \bibinfo{author}{{Park}, W.-K.},
  \bibinfo{author}{{Perkins}, J.~S.}, \bibinfo{author}{{Sugizaki}, M.},
  \bibinfo{author}{{Sung}, H.-I.}, \bibinfo{author}{{Tagliaferri}, G.},
  \bibinfo{author}{{Troja}, E.}, \bibinfo{author}{{Ueda}, Y.},
  \bibinfo{author}{{Urata}, Y.}, \bibinfo{author}{{Usui}, R.},
  \bibinfo{author}{{Antonelli}, L.~A.}, \bibinfo{author}{{Barthelmy}, S.~D.},
  \bibinfo{author}{{Cusumano}, G.}, \bibinfo{author}{{Giommi}, P.},
  \bibinfo{author}{{Melandri}, A.}, \bibinfo{author}{{Perri}, M.},
  \bibinfo{author}{{Racusin}, J.~L.}, \bibinfo{author}{{Sbarufatti}, B.},
  \bibinfo{author}{{Siegel}, M.~H.}, \& \bibinfo{author}{{Gehrels}, N.}
  (\bibinfo{year}{2011}).
\newblock \bibinfo{title}{{Relativistic jet activity from the tidal disruption
  of a star by a massive black hole}}.
\newblock {\it \bibinfo{journal}{\nat}\/},  {\it \bibinfo{volume}{476}\/},
  \bibinfo{pages}{421--424}. \DOIprefix\doi{10.1038/nature10374}.
  \href{http://arxiv.org/abs/1104.4787}{\tt arXiv:1104.4787}.
\bibitem[{{Campana} et~al.(2006){Campana}, {Mangano}, {Blustin}, {Brown},
  {Burrows}, {Chincarini}, {Cummings}, {Cusumano}, {Della Valle}, {Malesani},
  {M{\'e}sz{\'a}ros}, {Nousek}, {Page}, {Sakamoto}, {Waxman}, {Zhang}, {Dai},
  {Gehrels}, {Immler}, {Marshall}, {Mason}, {Moretti}, {O'Brien}, {Osborne},
  {Page}, {Romano}, {Roming}, {Tagliaferri}, {Cominsky}, {Giommi}, {Godet},
  {Kennea}, {Krimm}, {Angelini}, {Barthelmy}, {Boyd}, {Palmer}, {Wells} \&
  {White}}]{Campana2006}
\bibinfo{author}{{Campana}, S.}, \bibinfo{author}{{Mangano}, V.},
  \bibinfo{author}{{Blustin}, A.~J.}, \bibinfo{author}{{Brown}, P.},
  \bibinfo{author}{{Burrows}, D.~N.}, \bibinfo{author}{{Chincarini}, G.},
  \bibinfo{author}{{Cummings}, J.~R.}, \bibinfo{author}{{Cusumano}, G.},
  \bibinfo{author}{{Della Valle}, M.}, \bibinfo{author}{{Malesani}, D.},
  \bibinfo{author}{{M{\'e}sz{\'a}ros}, P.}, \bibinfo{author}{{Nousek}, J.~A.},
  \bibinfo{author}{{Page}, M.}, \bibinfo{author}{{Sakamoto}, T.},
  \bibinfo{author}{{Waxman}, E.}, \bibinfo{author}{{Zhang}, B.},
  \bibinfo{author}{{Dai}, Z.~G.}, \bibinfo{author}{{Gehrels}, N.},
  \bibinfo{author}{{Immler}, S.}, \bibinfo{author}{{Marshall}, F.~E.},
  \bibinfo{author}{{Mason}, K.~O.}, \bibinfo{author}{{Moretti}, A.},
  \bibinfo{author}{{O'Brien}, P.~T.}, \bibinfo{author}{{Osborne}, J.~P.},
  \bibinfo{author}{{Page}, K.~L.}, \bibinfo{author}{{Romano}, P.},
  \bibinfo{author}{{Roming}, P.~W.~A.}, \bibinfo{author}{{Tagliaferri}, G.},
  \bibinfo{author}{{Cominsky}, L.~R.}, \bibinfo{author}{{Giommi}, P.},
  \bibinfo{author}{{Godet}, O.}, \bibinfo{author}{{Kennea}, J.~A.},
  \bibinfo{author}{{Krimm}, H.}, \bibinfo{author}{{Angelini}, L.},
  \bibinfo{author}{{Barthelmy}, S.~D.}, \bibinfo{author}{{Boyd}, P.~T.},
  \bibinfo{author}{{Palmer}, D.~M.}, \bibinfo{author}{{Wells}, A.~A.}, \&
  \bibinfo{author}{{White}, N.~E.} (\bibinfo{year}{2006}).
\newblock \bibinfo{title}{{The association of GRB 060218 with a supernova and
  the evolution of the shock wave}}.
\newblock {\it \bibinfo{journal}{\nat}\/},  {\it \bibinfo{volume}{442}\/},
  \bibinfo{pages}{1008--1010}. \DOIprefix\doi{10.1038/nature04892}.
  \href{http://arxiv.org/abs/astro-ph/0603279}{\tt arXiv:astro-ph/0603279}.
\bibitem[{{Caruana} et~al.(2012){Caruana}, {Bunker}, {Wilkins}, {Stanway},
  {Lacy}, {Jarvis}, {Lorenzoni} \& {Hickey}}]{Caruana2012}
\bibinfo{author}{{Caruana}, J.}, \bibinfo{author}{{Bunker}, A.~J.},
  \bibinfo{author}{{Wilkins}, S.~M.}, \bibinfo{author}{{Stanway}, E.~R.},
  \bibinfo{author}{{Lacy}, M.}, \bibinfo{author}{{Jarvis}, M.~J.},
  \bibinfo{author}{{Lorenzoni}, S.}, \& \bibinfo{author}{{Hickey}, S.}
  (\bibinfo{year}{2012}).
\newblock \bibinfo{title}{{No evidence for Lyman-alpha emission in spectroscopy
  of z 7 candidate galaxies}}.
\newblock {\it \bibinfo{journal}{\mnras}\/},  {\it \bibinfo{volume}{427}\/},
  \bibinfo{pages}{3055--3070}.
  \DOIprefix\doi{10.1111/j.1365-2966.2012.21996.x}.
  \href{http://arxiv.org/abs/1208.5987}{\tt arXiv:1208.5987}.
\bibitem[{{Caruana} et~al.(2014){Caruana}, {Bunker}, {Wilkins}, {Stanway},
  {Lorenzoni}, {Jarvis} \& {Ebert}}]{Caruana2014}
\bibinfo{author}{{Caruana}, J.}, \bibinfo{author}{{Bunker}, A.~J.},
  \bibinfo{author}{{Wilkins}, S.~M.}, \bibinfo{author}{{Stanway}, E.~R.},
  \bibinfo{author}{{Lorenzoni}, S.}, \bibinfo{author}{{Jarvis}, M.~J.}, \&
  \bibinfo{author}{{Ebert}, H.} (\bibinfo{year}{2014}).
\newblock \bibinfo{title}{{Spectroscopy of z 7 candidate galaxies: using
  Lyman-alpha to constrain the neutral fraction of hydrogen in the
  high-redshift universe}}.
\newblock {\it \bibinfo{journal}{\mnras}\/},  {\it \bibinfo{volume}{443}\/},
  \bibinfo{pages}{2831--2842}. \DOIprefix\doi{10.1093/mnras/stu1341}.
  \href{http://arxiv.org/abs/1311.0057}{\tt arXiv:1311.0057}.
\bibitem[{{Chassande-Mottin} et~al.(2010){Chassande-Mottin}, {LIGO Scientific
  Collaboration} \& {Virgo Collaboration}}]{Chassande-Mottin2010}
\bibinfo{author}{{Chassande-Mottin}, E.}, \bibinfo{author}{{LIGO Scientific
  Collaboration}}, \& \bibinfo{author}{{Virgo Collaboration}}
  (\bibinfo{year}{2010}).
\newblock \bibinfo{title}{{Joint searches for gravitational waves and
  high-energy neutrinos}}.
\newblock In {\it \bibinfo{booktitle}{Journal of Physics Conference Series}\/}
  (p. \bibinfo{pages}{012002}).
\newblock volume \bibinfo{volume}{243} of {\it \bibinfo{series}{Journal of
  Physics Conference Series}\/}.
\newblock \DOIprefix\doi{10.1088/1742-6596/243/1/012002}.
\bibitem[{{Cheung} et~al.(2016){Cheung}, {Jean}, {Shore}, {Stawarz}, {Corbet},
  {Kn{\"o}dlseder}, {Starrfield}, {Wood}, {Desiante}, {Longo}, {Pivato} \&
  {Wood}}]{Cheung2016}
\bibinfo{author}{{Cheung}, C.~C.}, \bibinfo{author}{{Jean}, P.},
  \bibinfo{author}{{Shore}, S.~N.}, \bibinfo{author}{{Stawarz}, {\L}.},
  \bibinfo{author}{{Corbet}, R.~H.~D.}, \bibinfo{author}{{Kn{\"o}dlseder}, J.},
  \bibinfo{author}{{Starrfield}, S.}, \bibinfo{author}{{Wood}, D.~L.},
  \bibinfo{author}{{Desiante}, R.}, \bibinfo{author}{{Longo}, F.},
  \bibinfo{author}{{Pivato}, G.}, \& \bibinfo{author}{{Wood}, K.~S.}
  (\bibinfo{year}{2016}).
\newblock \bibinfo{title}{{Fermi-LAT Gamma-Ray Detections of Classical Novae
  V1369 Centauri 2013 and V5668 Sagittarii 2015}}.
\newblock {\it \bibinfo{journal}{\apj}\/},  {\it \bibinfo{volume}{826}\/},
  \bibinfo{pages}{142}. \DOIprefix\doi{10.3847/0004-637X/826/2/142}.
  \href{http://arxiv.org/abs/1605.04216}{\tt arXiv:1605.04216}.
\bibitem[{{Ciolfi}(2016)}]{Ciolfi2016}
\bibinfo{author}{{Ciolfi}, R.} (\bibinfo{year}{2016}).
\newblock \bibinfo{title}{{X-ray Flashes Powered by the Spindown of Long-lived
  Neutron Stars}}.
\newblock {\it \bibinfo{journal}{\apj}\/},  {\it \bibinfo{volume}{829}\/},
  \bibinfo{pages}{72}. \DOIprefix\doi{10.3847/0004-637X/829/2/72}.
  \href{http://arxiv.org/abs/1606.01743}{\tt arXiv:1606.01743}.
\bibitem[{{Ciolfi} et~al.(2011){Ciolfi}, {Lander}, {Manca} \&
  {Rezzolla}}]{Ciolfi2011}
\bibinfo{author}{{Ciolfi}, R.}, \bibinfo{author}{{Lander}, S.~K.},
  \bibinfo{author}{{Manca}, G.~M.}, \& \bibinfo{author}{{Rezzolla}, L.}
  (\bibinfo{year}{2011}).
\newblock \bibinfo{title}{{Instability-driven Evolution of Poloidal Magnetic
  Fields in Relativistic Stars}}.
\newblock {\it \bibinfo{journal}{\apjl}\/},  {\it \bibinfo{volume}{736}\/},
  \bibinfo{pages}{L6}. \DOIprefix\doi{10.1088/2041-8205/736/1/L6}.
  \href{http://arxiv.org/abs/1105.3971}{\tt arXiv:1105.3971}.
\bibitem[{{Ciolfi} \& {Rezzolla}(2012)}]{Ciolfi2012}
\bibinfo{author}{{Ciolfi}, R.}, \& \bibinfo{author}{{Rezzolla}, L.}
  (\bibinfo{year}{2012}).
\newblock \bibinfo{title}{{Poloidal-field Instability in Magnetized
  Relativistic Stars}}.
\newblock {\it \bibinfo{journal}{\apj}\/},  {\it \bibinfo{volume}{760}\/},
  \bibinfo{pages}{1}. \DOIprefix\doi{10.1088/0004-637X/760/1/1}.
  \href{http://arxiv.org/abs/1206.6604}{\tt arXiv:1206.6604}.
\bibitem[{{Ciolfi} \& {Siegel}(2015{\natexlab{a}})}]{Ciolfi2015b}
\bibinfo{author}{{Ciolfi}, R.}, \& \bibinfo{author}{{Siegel}, D.~M.}
  (\bibinfo{year}{2015}{\natexlab{a}}).
\newblock \bibinfo{title}{{Short gamma-ray bursts from binary neutron star
  mergers: the time-reversal scenario}}.
\newblock {\it \bibinfo{journal}{ArXiv e-prints}\/}, .
  \href{http://arxiv.org/abs/1505.01420}{\tt arXiv:1505.01420}.
\bibitem[{{Ciolfi} \& {Siegel}(2015{\natexlab{b}})}]{Ciolfi2015a}
\bibinfo{author}{{Ciolfi}, R.}, \& \bibinfo{author}{{Siegel}, D.~M.}
  (\bibinfo{year}{2015}{\natexlab{b}}).
\newblock \bibinfo{title}{{Short Gamma-Ray Bursts in the ``Time-reversal''
  Scenario}}.
\newblock {\it \bibinfo{journal}{\apjl}\/},  {\it \bibinfo{volume}{798}\/},
  \bibinfo{pages}{L36}. \DOIprefix\doi{10.1088/2041-8205/798/2/L36}.
  \href{http://arxiv.org/abs/1411.2015}{\tt arXiv:1411.2015}.
\bibitem[{{Corsi} \& {Owen}(2011)}]{Corsi2011}
\bibinfo{author}{{Corsi}, A.}, \& \bibinfo{author}{{Owen}, B.~J.}
  (\bibinfo{year}{2011}).
\newblock \bibinfo{title}{{Maximum gravitational-wave energy emissible in
  magnetar flares}}.
\newblock {\it \bibinfo{journal}{\prd}\/},  {\it \bibinfo{volume}{83}\/},
  \bibinfo{pages}{104014}. \DOIprefix\doi{10.1103/PhysRevD.83.104014}.
  \href{http://arxiv.org/abs/1102.3421}{\tt arXiv:1102.3421}.
\bibitem[{{Coulter} et~al.(2017){Coulter}, {Foley}, {Kilpatrick}, {Drout},
  {Piro}, {Shappee}, {Siebert}, {Simon}, {Ulloa}, {Kasen}, {Madore},
  {Murguia-Berthier}, {Pan}, {Prochaska}, {Ramirez-Ruiz}, {Rest} \&
  {Rojas-Bravo}}]{Coulter2017}
\bibinfo{author}{{Coulter}, D.~A.}, \bibinfo{author}{{Foley}, R.~J.},
  \bibinfo{author}{{Kilpatrick}, C.~D.}, \bibinfo{author}{{Drout}, M.~R.},
  \bibinfo{author}{{Piro}, A.~L.}, \bibinfo{author}{{Shappee}, B.~J.},
  \bibinfo{author}{{Siebert}, M.~R.}, \bibinfo{author}{{Simon}, J.~D.},
  \bibinfo{author}{{Ulloa}, N.}, \bibinfo{author}{{Kasen}, D.},
  \bibinfo{author}{{Madore}, B.~F.}, \bibinfo{author}{{Murguia-Berthier}, A.},
  \bibinfo{author}{{Pan}, Y.-C.}, \bibinfo{author}{{Prochaska}, J.~X.},
  \bibinfo{author}{{Ramirez-Ruiz}, E.}, \bibinfo{author}{{Rest}, A.}, \&
  \bibinfo{author}{{Rojas-Bravo}, C.} (\bibinfo{year}{2017}).
\newblock \bibinfo{title}{{Swope Supernova Survey 2017a (SSS17a), the Optical
  Counterpart to a Gravitational Wave Source}}.
\newblock {\it \bibinfo{journal}{ArXiv e-prints}\/}, .
  \href{http://arxiv.org/abs/1710.05452}{\tt arXiv:1710.05452}.
\bibitem[{{Cumming} \& {Macbeth}(2004)}]{Cumming2004}
\bibinfo{author}{{Cumming}, A.}, \& \bibinfo{author}{{Macbeth}, J.}
  (\bibinfo{year}{2004}).
\newblock \bibinfo{title}{{The Thermal Evolution following a Superburst on an
  Accreting Neutron Star}}.
\newblock {\it \bibinfo{journal}{\apjl}\/},  {\it \bibinfo{volume}{603}\/},
  \bibinfo{pages}{L37--L40}. \DOIprefix\doi{10.1086/382873}.
  \href{http://arxiv.org/abs/astro-ph/0401317}{\tt arXiv:astro-ph/0401317}.
\bibitem[{{Dainotti} et~al.(2015{\natexlab{a}}){Dainotti}, {Petrosian},
  {Willingale}, {O'Brien}, {Ostrowski} \& {Nagataki}}]{Dainotti2015b}
\bibinfo{author}{{Dainotti}, M.}, \bibinfo{author}{{Petrosian}, V.},
  \bibinfo{author}{{Willingale}, R.}, \bibinfo{author}{{O'Brien}, P.},
  \bibinfo{author}{{Ostrowski}, M.}, \& \bibinfo{author}{{Nagataki}, S.}
  (\bibinfo{year}{2015}{\natexlab{a}}).
\newblock \bibinfo{title}{{Luminosity-time and luminosity-luminosity
  correlations for GRB prompt and afterglow plateau emissions}}.
\newblock {\it \bibinfo{journal}{\mnras}\/},  {\it \bibinfo{volume}{451}\/},
  \bibinfo{pages}{3898--3908}. \DOIprefix\doi{10.1093/mnras/stv1229}.
  \href{http://arxiv.org/abs/1506.00702}{\tt arXiv:1506.00702}.
\bibitem[{{Dainotti} et~al.(2008){Dainotti}, {Cardone} \&
  {Capozziello}}]{Dainotti2008}
\bibinfo{author}{{Dainotti}, M.~G.}, \bibinfo{author}{{Cardone}, V.~F.}, \&
  \bibinfo{author}{{Capozziello}, S.} (\bibinfo{year}{2008}).
\newblock \bibinfo{title}{{A time-luminosity correlation for {$\gamma$}-ray
  bursts in the X-rays}}.
\newblock {\it \bibinfo{journal}{\mnras}\/},  {\it \bibinfo{volume}{391}\/},
  \bibinfo{pages}{L79--L83}. \DOIprefix\doi{10.1111/j.1745-3933.2008.00560.x}.
  \href{http://arxiv.org/abs/0809.1389}{\tt arXiv:0809.1389}.
\bibitem[{{Dainotti} et~al.(2013{\natexlab{a}}){Dainotti}, {Cardone},
  {Piedipalumbo} \& {Capozziello}}]{Dainotti2013b}
\bibinfo{author}{{Dainotti}, M.~G.}, \bibinfo{author}{{Cardone}, V.~F.},
  \bibinfo{author}{{Piedipalumbo}, E.}, \& \bibinfo{author}{{Capozziello}, S.}
  (\bibinfo{year}{2013}{\natexlab{a}}).
\newblock \bibinfo{title}{{Slope evolution of GRB correlations and cosmology}}.
\newblock {\it \bibinfo{journal}{\mnras}\/},  {\it \bibinfo{volume}{436}\/},
  \bibinfo{pages}{82--88}. \DOIprefix\doi{10.1093/mnras/stt1516}.
  \href{http://arxiv.org/abs/1308.1918}{\tt arXiv:1308.1918}.
\bibitem[{{Dainotti} et~al.(2015{\natexlab{b}}){Dainotti}, {Del Vecchio},
  {Shigehiro} \& {Capozziello}}]{Dainotti2015}
\bibinfo{author}{{Dainotti}, M.~G.}, \bibinfo{author}{{Del Vecchio}, R.},
  \bibinfo{author}{{Shigehiro}, N.}, \& \bibinfo{author}{{Capozziello}, S.}
  (\bibinfo{year}{2015}{\natexlab{b}}).
\newblock \bibinfo{title}{{Selection Effects in Gamma-Ray Burst Correlations:
  Consequences on the Ratio between Gamma-Ray Burst and Star Formation Rates}}.
\newblock {\it \bibinfo{journal}{\apj}\/},  {\it \bibinfo{volume}{800}\/},
  \bibinfo{pages}{31}. \DOIprefix\doi{10.1088/0004-637X/800/1/31}.
  \href{http://arxiv.org/abs/1412.3969}{\tt arXiv:1412.3969}.
\bibitem[{{Dainotti} et~al.(2013{\natexlab{b}}){Dainotti}, {Petrosian},
  {Singal} \& {Ostrowski}}]{Dainotti2013a}
\bibinfo{author}{{Dainotti}, M.~G.}, \bibinfo{author}{{Petrosian}, V.},
  \bibinfo{author}{{Singal}, J.}, \& \bibinfo{author}{{Ostrowski}, M.}
  (\bibinfo{year}{2013}{\natexlab{b}}).
\newblock \bibinfo{title}{{Determination of the Intrinsic Luminosity Time
  Correlation in the X-Ray Afterglows of Gamma-Ray Bursts}}.
\newblock {\it \bibinfo{journal}{\apj}\/},  {\it \bibinfo{volume}{774}\/},
  \bibinfo{pages}{157}. \DOIprefix\doi{10.1088/0004-637X/774/2/157}.
  \href{http://arxiv.org/abs/1307.7297}{\tt arXiv:1307.7297}.
\bibitem[{{Dainotti} et~al.(2016){Dainotti}, {Postnikov}, {Hernandez} \&
  {Ostrowski}}]{Dainotti2016a}
\bibinfo{author}{{Dainotti}, M.~G.}, \bibinfo{author}{{Postnikov}, S.},
  \bibinfo{author}{{Hernandez}, X.}, \& \bibinfo{author}{{Ostrowski}, M.}
  (\bibinfo{year}{2016}).
\newblock \bibinfo{title}{{A Fundamental Plane for Long Gamma-Ray Bursts with
  X-Ray Plateaus}}.
\newblock {\it \bibinfo{journal}{\apjl}\/},  {\it \bibinfo{volume}{825}\/},
  \bibinfo{pages}{L20}. \DOIprefix\doi{10.3847/2041-8205/825/2/L20}.
  \href{http://arxiv.org/abs/1604.06840}{\tt arXiv:1604.06840}.
\bibitem[{{Dalal} et~al.(2006){Dalal}, {Holz}, {Hughes} \& {Jain}}]{Dalal2006}
\bibinfo{author}{{Dalal}, N.}, \bibinfo{author}{{Holz}, D.~E.},
  \bibinfo{author}{{Hughes}, S.~A.}, \& \bibinfo{author}{{Jain}, B.}
  (\bibinfo{year}{2006}).
\newblock \bibinfo{title}{{Short GRB and binary black hole standard sirens as a
  probe of dark energy}}.
\newblock {\it \bibinfo{journal}{\prd}\/},  {\it \bibinfo{volume}{74}\/},
  \bibinfo{pages}{063006}. \DOIprefix\doi{10.1103/PhysRevD.74.063006}.
  \href{http://arxiv.org/abs/astro-ph/0601275}{\tt arXiv:astro-ph/0601275}.
\bibitem[{{Dall'Osso} et~al.(2011){Dall'Osso}, {Stratta}, {Guetta}, {Covino},
  {De Cesare} \& {Stella}}]{DallOsso2011}
\bibinfo{author}{{Dall'Osso}, S.}, \bibinfo{author}{{Stratta}, G.},
  \bibinfo{author}{{Guetta}, D.}, \bibinfo{author}{{Covino}, S.},
  \bibinfo{author}{{De Cesare}, G.}, \& \bibinfo{author}{{Stella}, L.}
  (\bibinfo{year}{2011}).
\newblock \bibinfo{title}{{Gamma-ray bursts afterglows with energy injection
  from a spinning down neutron star}}.
\newblock {\it \bibinfo{journal}{\aap}\/},  {\it \bibinfo{volume}{526}\/},
  \bibinfo{pages}{A121}. \DOIprefix\doi{10.1051/0004-6361/201014168}.
  \href{http://arxiv.org/abs/1004.2788}{\tt arXiv:1004.2788}.
\bibitem[{{Davies} et~al.(2002){Davies}, {King}, {Rosswog} \&
  {Wynn}}]{Davies2002}
\bibinfo{author}{{Davies}, M.~B.}, \bibinfo{author}{{King}, A.},
  \bibinfo{author}{{Rosswog}, S.}, \& \bibinfo{author}{{Wynn}, G.}
  (\bibinfo{year}{2002}).
\newblock \bibinfo{title}{{Gamma-Ray Bursts, Supernova Kicks, and Gravitational
  Radiation}}.
\newblock {\it \bibinfo{journal}{\apjl}\/},  {\it \bibinfo{volume}{579}\/},
  \bibinfo{pages}{L63--L66}. \DOIprefix\doi{10.1086/345288}.
  \href{http://arxiv.org/abs/astro-ph/0204358}{\tt arXiv:astro-ph/0204358}.
\bibitem[{{Demianski} et~al.(2017){Demianski}, {Piedipalumbo}, {Sawant} \&
  {Amati}}]{Demianski2017}
\bibinfo{author}{{Demianski}, M.}, \bibinfo{author}{{Piedipalumbo}, E.},
  \bibinfo{author}{{Sawant}, D.}, \& \bibinfo{author}{{Amati}, L.}
  (\bibinfo{year}{2017}).
\newblock \bibinfo{title}{{Cosmology with gamma-ray bursts. I. The Hubble
  diagram through the calibrated E$_{p,i}$-E$_{iso}$ correlation}}.
\newblock {\it \bibinfo{journal}{\aap}\/},  {\it \bibinfo{volume}{598}\/},
  \bibinfo{pages}{A112}. \DOIprefix\doi{10.1051/0004-6361/201628909}.
  \href{http://arxiv.org/abs/1610.00854}{\tt arXiv:1610.00854}.
\bibitem[{{Farinelli} et~al.(2012){Farinelli}, {Romano}, {Mangano},
  {Ceccobello}, {Ducci}, {Vercellone}, {Esposito}, {Kennea} \&
  {Burrows}}]{Farinelli2012}
\bibinfo{author}{{Farinelli}, R.}, \bibinfo{author}{{Romano}, P.},
  \bibinfo{author}{{Mangano}, V.}, \bibinfo{author}{{Ceccobello}, C.},
  \bibinfo{author}{{Ducci}, L.}, \bibinfo{author}{{Vercellone}, S.},
  \bibinfo{author}{{Esposito}, P.}, \bibinfo{author}{{Kennea}, J.~A.}, \&
  \bibinfo{author}{{Burrows}, D.~N.} (\bibinfo{year}{2012}).
\newblock \bibinfo{title}{{Swift observations of two supergiant fast X-ray
  transient prototypes in outburst}}.
\newblock {\it \bibinfo{journal}{\mnras}\/},  {\it \bibinfo{volume}{424}\/},
  \bibinfo{pages}{2854--2863}.
  \DOIprefix\doi{10.1111/j.1365-2966.2012.21422.x}.
  \href{http://arxiv.org/abs/1205.7059}{\tt arXiv:1205.7059}.
\bibitem[{{Favata}(2002)}]{Favata2002}
\bibinfo{author}{{Favata}, F.} (\bibinfo{year}{2002}).
\newblock \bibinfo{title}{{Large stellar flares: a review of recent novel
  results}}.
\newblock In \bibinfo{editor}{F.~{Favata}}, \& \bibinfo{editor}{J.~J. {Drake}}
  (Eds.), {\it \bibinfo{booktitle}{Stellar Coronae in the Chandra and
  XMM-NEWTON Era}\/} (p. \bibinfo{pages}{115}).
\newblock volume \bibinfo{volume}{277} of {\it \bibinfo{series}{Astronomical
  Society of the Pacific Conference Series}\/}.
\bibitem[{{Fern{\'a}ndez} et~al.(2016){Fern{\'a}ndez}, {Gim}, {van Gorkom},
  {Yun}, {Momjian}, {Popping}, {Chomiuk}, {Hess}, {Hunt}, {Kreckel}, {Lucero},
  {Maddox}, {Oosterloo}, {Pisano}, {Verheijen}, {Hales}, {Chung}, {Dodson},
  {Golap}, {Gross}, {Henning}, {Hibbard}, {Jaff{\'e}}, {Donovan Meyer},
  {Meyer}, {Sanchez-Barrantes}, {Schiminovich}, {Wicenec}, {Wilcots},
  {Bershady}, {Scoville}, {Strader}, {Tremou}, {Salinas} \&
  {Ch{\'a}vez}}]{Fernandez2016}
\bibinfo{author}{{Fern{\'a}ndez}, X.}, \bibinfo{author}{{Gim}, H.~B.},
  \bibinfo{author}{{van Gorkom}, J.~H.}, \bibinfo{author}{{Yun}, M.~S.},
  \bibinfo{author}{{Momjian}, E.}, \bibinfo{author}{{Popping}, A.},
  \bibinfo{author}{{Chomiuk}, L.}, \bibinfo{author}{{Hess}, K.~M.},
  \bibinfo{author}{{Hunt}, L.}, \bibinfo{author}{{Kreckel}, K.},
  \bibinfo{author}{{Lucero}, D.}, \bibinfo{author}{{Maddox}, N.},
  \bibinfo{author}{{Oosterloo}, T.}, \bibinfo{author}{{Pisano}, D.~J.},
  \bibinfo{author}{{Verheijen}, M.~A.~W.}, \bibinfo{author}{{Hales}, C.~A.},
  \bibinfo{author}{{Chung}, A.}, \bibinfo{author}{{Dodson}, R.},
  \bibinfo{author}{{Golap}, K.}, \bibinfo{author}{{Gross}, J.},
  \bibinfo{author}{{Henning}, P.}, \bibinfo{author}{{Hibbard}, J.},
  \bibinfo{author}{{Jaff{\'e}}, Y.~L.}, \bibinfo{author}{{Donovan Meyer}, J.},
  \bibinfo{author}{{Meyer}, M.}, \bibinfo{author}{{Sanchez-Barrantes}, M.},
  \bibinfo{author}{{Schiminovich}, D.}, \bibinfo{author}{{Wicenec}, A.},
  \bibinfo{author}{{Wilcots}, E.}, \bibinfo{author}{{Bershady}, M.},
  \bibinfo{author}{{Scoville}, N.}, \bibinfo{author}{{Strader}, J.},
  \bibinfo{author}{{Tremou}, E.}, \bibinfo{author}{{Salinas}, R.}, \&
  \bibinfo{author}{{Ch{\'a}vez}, R.} (\bibinfo{year}{2016}).
\newblock \bibinfo{title}{{Highest Redshift Image of Neutral Hydrogen in
  Emission: A CHILES Detection of a Starbursting Galaxy at z = 0.376}}.
\newblock {\it \bibinfo{journal}{\apjl}\/},  {\it \bibinfo{volume}{824}\/},
  \bibinfo{pages}{L1}. \DOIprefix\doi{10.3847/2041-8205/824/1/L1}.
  \href{http://arxiv.org/abs/1606.00013}{\tt arXiv:1606.00013}.
\bibitem[{{Fertig} et~al.(2011){Fertig}, {Mukai}, {Nelson} \&
  {Cannizzo}}]{Fertig2011}
\bibinfo{author}{{Fertig}, D.}, \bibinfo{author}{{Mukai}, K.},
  \bibinfo{author}{{Nelson}, T.}, \& \bibinfo{author}{{Cannizzo}, J.~K.}
  (\bibinfo{year}{2011}).
\newblock \bibinfo{title}{{The Fall and the Rise of X-Rays from Dwarf Novae in
  Outburst: RXTE Observations of VW Hydri and WW Ceti}}.
\newblock {\it \bibinfo{journal}{\pasp}\/},  {\it \bibinfo{volume}{123}\/},
  \bibinfo{pages}{1054}. \DOIprefix\doi{10.1086/661949}.
  \href{http://arxiv.org/abs/1107.3142}{\tt arXiv:1107.3142}.
\bibitem[{{Fong} et~al.(2013){Fong}, {Berger}, {Chornock}, {Margutti}, {Levan},
  {Tanvir}, {Tunnicliffe}, {Czekala}, {Fox}, {Perley}, {Cenko}, {Zauderer},
  {Laskar}, {Persson}, {Monson}, {Kelson}, {Birk}, {Murphy}, {Servillat} \&
  {Anglada}}]{Fong2013}
\bibinfo{author}{{Fong}, W.}, \bibinfo{author}{{Berger}, E.},
  \bibinfo{author}{{Chornock}, R.}, \bibinfo{author}{{Margutti}, R.},
  \bibinfo{author}{{Levan}, A.~J.}, \bibinfo{author}{{Tanvir}, N.~R.},
  \bibinfo{author}{{Tunnicliffe}, R.~L.}, \bibinfo{author}{{Czekala}, I.},
  \bibinfo{author}{{Fox}, D.~B.}, \bibinfo{author}{{Perley}, D.~A.},
  \bibinfo{author}{{Cenko}, S.~B.}, \bibinfo{author}{{Zauderer}, B.~A.},
  \bibinfo{author}{{Laskar}, T.}, \bibinfo{author}{{Persson}, S.~E.},
  \bibinfo{author}{{Monson}, A.~J.}, \bibinfo{author}{{Kelson}, D.~D.},
  \bibinfo{author}{{Birk}, C.}, \bibinfo{author}{{Murphy}, D.},
  \bibinfo{author}{{Servillat}, M.}, \& \bibinfo{author}{{Anglada}, G.}
  (\bibinfo{year}{2013}).
\newblock \bibinfo{title}{{Demographics of the Galaxies Hosting Short-duration
  Gamma-Ray Bursts}}.
\newblock {\it \bibinfo{journal}{\apj}\/},  {\it \bibinfo{volume}{769}\/},
  \bibinfo{pages}{56}. \DOIprefix\doi{10.1088/0004-637X/769/1/56}.
  \href{http://arxiv.org/abs/1302.3221}{\tt arXiv:1302.3221}.
\bibitem[{{Friis} et~al.(2015){Friis}, {De Cia}, {Kr{\"u}hler}, {Fynbo},
  {Ledoux}, {Vreeswijk}, {Watson}, {Malesani}, {Gorosabel}, {Starling},
  {Jakobsson}, {Varela}, {Wiersema}, {Drachmann}, {Trotter}, {Th{\"o}ne}, {de
  Ugarte Postigo}, {D'Elia}, {Elliott}, {Maturi}, {Goldoni}, {FriisGreiner},
  {Haislip}, {Kaper}, {Knust}, {LaCluyze}, {Milvang-Jensen}, {Reichart},
  {Schulze}, {Sudilovsky}, {Tanvir} \& {Vergani}}]{Friis2015}
\bibinfo{author}{{Friis}, M.}, \bibinfo{author}{{De Cia}, A.},
  \bibinfo{author}{{Kr{\"u}hler}, T.}, \bibinfo{author}{{Fynbo}, J.~P.~U.},
  \bibinfo{author}{{Ledoux}, C.}, \bibinfo{author}{{Vreeswijk}, P.~M.},
  \bibinfo{author}{{Watson}, D.~J.}, \bibinfo{author}{{Malesani}, D.},
  \bibinfo{author}{{Gorosabel}, J.}, \bibinfo{author}{{Starling}, R.~L.~C.},
  \bibinfo{author}{{Jakobsson}, P.}, \bibinfo{author}{{Varela}, K.},
  \bibinfo{author}{{Wiersema}, K.}, \bibinfo{author}{{Drachmann}, A.~P.},
  \bibinfo{author}{{Trotter}, A.}, \bibinfo{author}{{Th{\"o}ne}, C.~C.},
  \bibinfo{author}{{de Ugarte Postigo}, A.}, \bibinfo{author}{{D'Elia}, V.},
  \bibinfo{author}{{Elliott}, J.}, \bibinfo{author}{{Maturi}, M.},
  \bibinfo{author}{{Goldoni}, P.}, \bibinfo{author}{{FriisGreiner}, J.},
  \bibinfo{author}{{Haislip}, J.}, \bibinfo{author}{{Kaper}, L.},
  \bibinfo{author}{{Knust}, F.}, \bibinfo{author}{{LaCluyze}, A.},
  \bibinfo{author}{{Milvang-Jensen}, B.}, \bibinfo{author}{{Reichart}, D.},
  \bibinfo{author}{{Schulze}, S.}, \bibinfo{author}{{Sudilovsky}, V.},
  \bibinfo{author}{{Tanvir}, N.}, \& \bibinfo{author}{{Vergani}, S.~D.}
  (\bibinfo{year}{2015}).
\newblock \bibinfo{title}{{The warm, the excited, and the molecular gas: GRB
  121024A shining through its star-forming galaxy}}.
\newblock {\it \bibinfo{journal}{\mnras}\/},  {\it \bibinfo{volume}{451}\/},
  \bibinfo{pages}{167--183}. \DOIprefix\doi{10.1093/mnras/stv960}.
  \href{http://arxiv.org/abs/1409.6315}{\tt arXiv:1409.6315}.
\bibitem[{{Fryer} et~al.(2002){Fryer}, {Holz} \& {Hughes}}]{Fryer2002}
\bibinfo{author}{{Fryer}, C.~L.}, \bibinfo{author}{{Holz}, D.~E.}, \&
  \bibinfo{author}{{Hughes}, S.~A.} (\bibinfo{year}{2002}).
\newblock \bibinfo{title}{{Gravitational Wave Emission from Core Collapse of
  Massive Stars}}.
\newblock {\it \bibinfo{journal}{\apj}\/},  {\it \bibinfo{volume}{565}\/},
  \bibinfo{pages}{430--446}. \DOIprefix\doi{10.1086/324034}.
  \href{http://arxiv.org/abs/astro-ph/0106113}{\tt arXiv:astro-ph/0106113}.
\bibitem[{{Fuschino} et~al.(2008){Fuschino}, {Labanti}, {Galli}, {Marisaldi},
  {Bulgarelli}, {Gianotti}, {Trifoglio}, {Argan}, {Del Monte}, {Donnarumma},
  {Feroci}, {Lazzarotto}, {Pacciani}, {Tavani} \& {Trois}}]{Fuschino2008}
\bibinfo{author}{{Fuschino}, F.}, \bibinfo{author}{{Labanti}, C.},
  \bibinfo{author}{{Galli}, M.}, \bibinfo{author}{{Marisaldi}, M.},
  \bibinfo{author}{{Bulgarelli}, A.}, \bibinfo{author}{{Gianotti}, F.},
  \bibinfo{author}{{Trifoglio}, M.}, \bibinfo{author}{{Argan}, A.},
  \bibinfo{author}{{Del Monte}, E.}, \bibinfo{author}{{Donnarumma}, I.},
  \bibinfo{author}{{Feroci}, M.}, \bibinfo{author}{{Lazzarotto}, F.},
  \bibinfo{author}{{Pacciani}, L.}, \bibinfo{author}{{Tavani}, M.}, \&
  \bibinfo{author}{{Trois}, A.} (\bibinfo{year}{2008}).
\newblock \bibinfo{title}{{Search of GRB with AGILE Minicalorimeter}}.
\newblock {\it \bibinfo{journal}{Nuclear Instruments and Methods in Physics
  Research A}\/},  {\it \bibinfo{volume}{588}\/}, \bibinfo{pages}{17--21}.
  \DOIprefix\doi{10.1016/j.nima.2008.01.004}.
\bibitem[{{Gao} et~al.(2013){Gao}, {Ding}, {Wu}, {Zhang} \& {Dai}}]{Gao2013}
\bibinfo{author}{{Gao}, H.}, \bibinfo{author}{{Ding}, X.},
  \bibinfo{author}{{Wu}, X.-F.}, \bibinfo{author}{{Zhang}, B.}, \&
  \bibinfo{author}{{Dai}, Z.-G.} (\bibinfo{year}{2013}).
\newblock \bibinfo{title}{{Bright Broadband Afterglows of Gravitational Wave
  Bursts from Mergers of Binary Neutron Stars}}.
\newblock {\it \bibinfo{journal}{\apj}\/},  {\it \bibinfo{volume}{771}\/},
  \bibinfo{pages}{86}. \DOIprefix\doi{10.1088/0004-637X/771/2/86}.
  \href{http://arxiv.org/abs/1301.0439}{\tt arXiv:1301.0439}.
\bibitem[{{Gao} et~al.(2016){Gao}, {Zhang} \& {L{\"u}}}]{Gao2016}
\bibinfo{author}{{Gao}, H.}, \bibinfo{author}{{Zhang}, B.}, \&
  \bibinfo{author}{{L{\"u}}, H.-J.} (\bibinfo{year}{2016}).
\newblock \bibinfo{title}{{Constraints on binary neutron star merger product
  from short GRB observations}}.
\newblock {\it \bibinfo{journal}{\prd}\/},  {\it \bibinfo{volume}{93}\/},
  \bibinfo{pages}{044065}. \DOIprefix\doi{10.1103/PhysRevD.93.044065}.
  \href{http://arxiv.org/abs/1511.00753}{\tt arXiv:1511.00753}.
\bibitem[{{Gatti} \& {Rehak}(1984)}]{Gatti1984}
\bibinfo{author}{{Gatti}, E.}, \& \bibinfo{author}{{Rehak}, P.}
  (\bibinfo{year}{1984}).
\newblock \bibinfo{title}{{Semiconductor drift chamber — An application of a
  novel charge transport scheme}}.
\newblock {\it \bibinfo{journal}{Nuclear Instruments and Methods in Physics
  Research}\/},  {\it \bibinfo{volume}{225}\/}, \bibinfo{pages}{608--614}.
  \DOIprefix\doi{10.1016/0167-5087(84)90113-3}.
\bibitem[{{Gezari} et~al.(2012){Gezari}, {Chornock}, {Rest}, {Huber},
  {Forster}, {Berger}, {Challis}, {Neill}, {Martin}, {Heckman}, {Lawrence},
  {Norman}, {Narayan}, {Foley}, {Marion}, {Scolnic}, {Chomiuk}, {Soderberg},
  {Smith}, {Kirshner}, {Riess}, {Smartt}, {Stubbs}, {Tonry}, {Wood-Vasey},
  {Burgett}, {Chambers}, {Grav}, {Heasley}, {Kaiser}, {Kudritzki}, {Magnier},
  {Morgan} \& {Price}}]{Gezari2012}
\bibinfo{author}{{Gezari}, S.}, \bibinfo{author}{{Chornock}, R.},
  \bibinfo{author}{{Rest}, A.}, \bibinfo{author}{{Huber}, M.~E.},
  \bibinfo{author}{{Forster}, K.}, \bibinfo{author}{{Berger}, E.},
  \bibinfo{author}{{Challis}, P.~J.}, \bibinfo{author}{{Neill}, J.~D.},
  \bibinfo{author}{{Martin}, D.~C.}, \bibinfo{author}{{Heckman}, T.},
  \bibinfo{author}{{Lawrence}, A.}, \bibinfo{author}{{Norman}, C.},
  \bibinfo{author}{{Narayan}, G.}, \bibinfo{author}{{Foley}, R.~J.},
  \bibinfo{author}{{Marion}, G.~H.}, \bibinfo{author}{{Scolnic}, D.},
  \bibinfo{author}{{Chomiuk}, L.}, \bibinfo{author}{{Soderberg}, A.},
  \bibinfo{author}{{Smith}, K.}, \bibinfo{author}{{Kirshner}, R.~P.},
  \bibinfo{author}{{Riess}, A.~G.}, \bibinfo{author}{{Smartt}, S.~J.},
  \bibinfo{author}{{Stubbs}, C.~W.}, \bibinfo{author}{{Tonry}, J.~L.},
  \bibinfo{author}{{Wood-Vasey}, W.~M.}, \bibinfo{author}{{Burgett}, W.~S.},
  \bibinfo{author}{{Chambers}, K.~C.}, \bibinfo{author}{{Grav}, T.},
  \bibinfo{author}{{Heasley}, J.~N.}, \bibinfo{author}{{Kaiser}, N.},
  \bibinfo{author}{{Kudritzki}, R.-P.}, \bibinfo{author}{{Magnier}, E.~A.},
  \bibinfo{author}{{Morgan}, J.~S.}, \& \bibinfo{author}{{Price}, P.~A.}
  (\bibinfo{year}{2012}).
\newblock \bibinfo{title}{{An ultraviolet-optical flare from the tidal
  disruption of a helium-rich stellar core}}.
\newblock {\it \bibinfo{journal}{\nat}\/},  {\it \bibinfo{volume}{485}\/},
  \bibinfo{pages}{217--220}. \DOIprefix\doi{10.1038/nature10990}.
  \href{http://arxiv.org/abs/1205.0252}{\tt arXiv:1205.0252}.
\bibitem[{{Ghirlanda} et~al.(2004){Ghirlanda}, {Ghisellini}, {Lazzati} \&
  {Firmani}}]{Ghirlanda2004}
\bibinfo{author}{{Ghirlanda}, G.}, \bibinfo{author}{{Ghisellini}, G.},
  \bibinfo{author}{{Lazzati}, D.}, \& \bibinfo{author}{{Firmani}, C.}
  (\bibinfo{year}{2004}).
\newblock \bibinfo{title}{{Gamma-Ray Bursts: New Rulers to Measure the
  Universe}}.
\newblock {\it \bibinfo{journal}{\apjl}\/},  {\it \bibinfo{volume}{613}\/},
  \bibinfo{pages}{L13--L16}. \DOIprefix\doi{10.1086/424915}.
  \href{http://arxiv.org/abs/astro-ph/0408350}{\tt arXiv:astro-ph/0408350}.
\bibitem[{{Ghirlanda} et~al.(2016){Ghirlanda}, {Salafia}, {Pescalli},
  {Ghisellini}, {Salvaterra}, {Chassande-Mottin}, {Colpi}, {Nappo}, {D'Avanzo},
  {Melandri}, {Bernardini}, {Branchesi}, {Campana}, {Ciolfi}, {Covino},
  {G{\"o}tz}, {Vergani}, {Zennaro} \& {Tagliaferri}}]{Ghirlanda2016}
\bibinfo{author}{{Ghirlanda}, G.}, \bibinfo{author}{{Salafia}, O.~S.},
  \bibinfo{author}{{Pescalli}, A.}, \bibinfo{author}{{Ghisellini}, G.},
  \bibinfo{author}{{Salvaterra}, R.}, \bibinfo{author}{{Chassande-Mottin}, E.},
  \bibinfo{author}{{Colpi}, M.}, \bibinfo{author}{{Nappo}, F.},
  \bibinfo{author}{{D'Avanzo}, P.}, \bibinfo{author}{{Melandri}, A.},
  \bibinfo{author}{{Bernardini}, M.~G.}, \bibinfo{author}{{Branchesi}, M.},
  \bibinfo{author}{{Campana}, S.}, \bibinfo{author}{{Ciolfi}, R.},
  \bibinfo{author}{{Covino}, S.}, \bibinfo{author}{{G{\"o}tz}, D.},
  \bibinfo{author}{{Vergani}, S.~D.}, \bibinfo{author}{{Zennaro}, M.}, \&
  \bibinfo{author}{{Tagliaferri}, G.} (\bibinfo{year}{2016}).
\newblock \bibinfo{title}{{Short gamma-ray bursts at the dawn of the
  gravitational wave era}}.
\newblock {\it \bibinfo{journal}{\aap}\/},  {\it \bibinfo{volume}{594}\/},
  \bibinfo{pages}{A84}. \DOIprefix\doi{10.1051/0004-6361/201628993}.
  \href{http://arxiv.org/abs/1607.07875}{\tt arXiv:1607.07875}.
\bibitem[{{Ghirlanda} et~al.(2015){Ghirlanda}, {Salvaterra}, {Ghisellini},
  {Mereghetti}, {Tagliaferri}, {Campana}, {Osborne}, {O'Brien}, {Tanvir},
  {Willingale}, {Amati}, {Basa}, {Bernardini}, {Burlon}, {Covino}, {D'Avanzo},
  {Frontera}, {G{\"o}tz}, {Melandri}, {Nava}, {Piro} \&
  {Vergani}}]{Ghirlanda2015}
\bibinfo{author}{{Ghirlanda}, G.}, \bibinfo{author}{{Salvaterra}, R.},
  \bibinfo{author}{{Ghisellini}, G.}, \bibinfo{author}{{Mereghetti}, S.},
  \bibinfo{author}{{Tagliaferri}, G.}, \bibinfo{author}{{Campana}, S.},
  \bibinfo{author}{{Osborne}, J.~P.}, \bibinfo{author}{{O'Brien}, P.},
  \bibinfo{author}{{Tanvir}, N.}, \bibinfo{author}{{Willingale}, D.},
  \bibinfo{author}{{Amati}, L.}, \bibinfo{author}{{Basa}, S.},
  \bibinfo{author}{{Bernardini}, M.~G.}, \bibinfo{author}{{Burlon}, D.},
  \bibinfo{author}{{Covino}, S.}, \bibinfo{author}{{D'Avanzo}, P.},
  \bibinfo{author}{{Frontera}, F.}, \bibinfo{author}{{G{\"o}tz}, D.},
  \bibinfo{author}{{Melandri}, A.}, \bibinfo{author}{{Nava}, L.},
  \bibinfo{author}{{Piro}, L.}, \& \bibinfo{author}{{Vergani}, S.~D.}
  (\bibinfo{year}{2015}).
\newblock \bibinfo{title}{{Accessing the population of high-redshift Gamma Ray
  Bursts}}.
\newblock {\it \bibinfo{journal}{\mnras}\/},  {\it \bibinfo{volume}{448}\/},
  \bibinfo{pages}{2514--2524}. \DOIprefix\doi{10.1093/mnras/stv183}.
  \href{http://arxiv.org/abs/1502.02676}{\tt arXiv:1502.02676}.
\bibitem[{{Giacomazzo} et~al.(2011){Giacomazzo}, {Rezzolla} \&
  {Stergioulas}}]{Giacomazzo2011}
\bibinfo{author}{{Giacomazzo}, B.}, \bibinfo{author}{{Rezzolla}, L.}, \&
  \bibinfo{author}{{Stergioulas}, N.} (\bibinfo{year}{2011}).
\newblock \bibinfo{title}{{Collapse of differentially rotating neutron stars
  and cosmic censorship}}.
\newblock {\it \bibinfo{journal}{\prd}\/},  {\it \bibinfo{volume}{84}\/},
  \bibinfo{pages}{024022}. \DOIprefix\doi{10.1103/PhysRevD.84.024022}.
  \href{http://arxiv.org/abs/1105.0122}{\tt arXiv:1105.0122}.
\bibitem[{{Grazian} et~al.(2017){Grazian}, {Giallongo}, {Paris}, {Boutsia},
  {Dickinson}, {Santini}, {Windhorst}, {Jansen}, {Cohen}, {Ashcraft},
  {Scarlata}, {Rutkowski}, {Vanzella}, {Cusano}, {Cristiani}, {Giavalisco},
  {Ferguson}, {Koekemoer}, {Grogin}, {Castellano}, {Fiore}, {Fontana},
  {Marchi}, {Pedichini}, {Pentericci}, {Amor{\'{\i}}n}, {Barro}, {Bonchi},
  {Bongiorno}, {Faber}, {Fumana}, {Galametz}, {Guaita}, {Kocevski}, {Merlin},
  {Nonino}, {O'Connell}, {Pilo}, {Ryan}, {Sani}, {Speziali}, {Testa}, {Weiner}
  \& {Yan}}]{Grazian2017}
\bibinfo{author}{{Grazian}, A.}, \bibinfo{author}{{Giallongo}, E.},
  \bibinfo{author}{{Paris}, D.}, \bibinfo{author}{{Boutsia}, K.},
  \bibinfo{author}{{Dickinson}, M.}, \bibinfo{author}{{Santini}, P.},
  \bibinfo{author}{{Windhorst}, R.~A.}, \bibinfo{author}{{Jansen}, R.~A.},
  \bibinfo{author}{{Cohen}, S.~H.}, \bibinfo{author}{{Ashcraft}, T.~A.},
  \bibinfo{author}{{Scarlata}, C.}, \bibinfo{author}{{Rutkowski}, M.~J.},
  \bibinfo{author}{{Vanzella}, E.}, \bibinfo{author}{{Cusano}, F.},
  \bibinfo{author}{{Cristiani}, S.}, \bibinfo{author}{{Giavalisco}, M.},
  \bibinfo{author}{{Ferguson}, H.~C.}, \bibinfo{author}{{Koekemoer}, A.},
  \bibinfo{author}{{Grogin}, N.~A.}, \bibinfo{author}{{Castellano}, M.},
  \bibinfo{author}{{Fiore}, F.}, \bibinfo{author}{{Fontana}, A.},
  \bibinfo{author}{{Marchi}, F.}, \bibinfo{author}{{Pedichini}, F.},
  \bibinfo{author}{{Pentericci}, L.}, \bibinfo{author}{{Amor{\'{\i}}n}, R.},
  \bibinfo{author}{{Barro}, G.}, \bibinfo{author}{{Bonchi}, A.},
  \bibinfo{author}{{Bongiorno}, A.}, \bibinfo{author}{{Faber}, S.~M.},
  \bibinfo{author}{{Fumana}, M.}, \bibinfo{author}{{Galametz}, A.},
  \bibinfo{author}{{Guaita}, L.}, \bibinfo{author}{{Kocevski}, D.~D.},
  \bibinfo{author}{{Merlin}, E.}, \bibinfo{author}{{Nonino}, M.},
  \bibinfo{author}{{O'Connell}, R.~W.}, \bibinfo{author}{{Pilo}, S.},
  \bibinfo{author}{{Ryan}, R.~E.}, \bibinfo{author}{{Sani}, E.},
  \bibinfo{author}{{Speziali}, R.}, \bibinfo{author}{{Testa}, V.},
  \bibinfo{author}{{Weiner}, B.}, \& \bibinfo{author}{{Yan}, H.}
  (\bibinfo{year}{2017}).
\newblock \bibinfo{title}{{Lyman continuum escape fraction of faint galaxies at
  z 3.3 in the CANDELS/GOODS-North, EGS, and COSMOS fields with LBC}}.
\newblock {\it \bibinfo{journal}{\aap}\/},  {\it \bibinfo{volume}{602}\/},
  \bibinfo{pages}{A18}. \DOIprefix\doi{10.1051/0004-6361/201730447}.
  \href{http://arxiv.org/abs/1703.00354}{\tt arXiv:1703.00354}.
\bibitem[{{Greiner}(1996)}]{Greiner1996}
\bibinfo{author}{{Greiner}, J.} (\bibinfo{year}{1996}).
\newblock \bibinfo{title}{{Supersoft X-ray sources. Proceedings.}}
\newblock In \bibinfo{editor}{J.~{Greiner}} (Ed.), {\it
  \bibinfo{booktitle}{Supersoft X-Ray Sources}\/}.
\newblock volume \bibinfo{volume}{472} of {\it \bibinfo{series}{Lecture Notes
  in Physics, Berlin Springer Verlag}\/}.
\bibitem[{{Greiner} et~al.(2001){Greiner}, {Cuby} \&
  {McCaughrean}}]{Greiner2001}
\bibinfo{author}{{Greiner}, J.}, \bibinfo{author}{{Cuby}, J.~G.}, \&
  \bibinfo{author}{{McCaughrean}, M.~J.} (\bibinfo{year}{2001}).
\newblock \bibinfo{title}{{An unusually massive stellar black hole in the
  Galaxy}}.
\newblock {\it \bibinfo{journal}{\nat}\/},  {\it \bibinfo{volume}{414}\/},
  \bibinfo{pages}{522--525}. \DOIprefix\doi{10.1038/35107019}.
  \href{http://arxiv.org/abs/astro-ph/0111538}{\tt arXiv:astro-ph/0111538}.
\bibitem[{{Greiner} et~al.(2011){Greiner}, {Kr{\"u}hler}, {Klose}, {Afonso},
  {Clemens}, {Filgas}, {Hartmann}, {K{\"u}pc{\"u} Yolda{\c s}}, {Nardini},
  {Olivares E.}, {Rau}, {Rossi}, {Schady} \& {Updike}}]{Greiner2011}
\bibinfo{author}{{Greiner}, J.}, \bibinfo{author}{{Kr{\"u}hler}, T.},
  \bibinfo{author}{{Klose}, S.}, \bibinfo{author}{{Afonso}, P.},
  \bibinfo{author}{{Clemens}, C.}, \bibinfo{author}{{Filgas}, R.},
  \bibinfo{author}{{Hartmann}, D.~H.}, \bibinfo{author}{{K{\"u}pc{\"u} Yolda{\c
  s}}, A.}, \bibinfo{author}{{Nardini}, M.}, \bibinfo{author}{{Olivares E.},
  F.}, \bibinfo{author}{{Rau}, A.}, \bibinfo{author}{{Rossi}, A.},
  \bibinfo{author}{{Schady}, P.}, \& \bibinfo{author}{{Updike}, A.}
  (\bibinfo{year}{2011}).
\newblock \bibinfo{title}{{The nature of ``dark'' gamma-ray bursts}}.
\newblock {\it \bibinfo{journal}{\aap}\/},  {\it \bibinfo{volume}{526}\/},
  \bibinfo{pages}{A30}. \DOIprefix\doi{10.1051/0004-6361/201015458}.
  \href{http://arxiv.org/abs/1011.0618}{\tt arXiv:1011.0618}.
\bibitem[{{Greiner} et~al.(2015){Greiner}, {Mazzali}, {Kann}, {Kr{\"u}hler},
  {Pian}, {Prentice}, {Olivares E.}, {Rossi}, {Klose}, {Taubenberger}, {Knust},
  {Afonso}, {Ashall}, {Bolmer}, {Delvaux}, {Diehl}, {Elliott}, {Filgas},
  {Fynbo}, {Graham}, {Guelbenzu}, {Kobayashi}, {Leloudas}, {Savaglio},
  {Schady}, {Schmidl}, {Schweyer}, {Sudilovsky}, {Tanga}, {Updike}, {van
  Eerten} \& {Varela}}]{Greiner2015}
\bibinfo{author}{{Greiner}, J.}, \bibinfo{author}{{Mazzali}, P.~A.},
  \bibinfo{author}{{Kann}, D.~A.}, \bibinfo{author}{{Kr{\"u}hler}, T.},
  \bibinfo{author}{{Pian}, E.}, \bibinfo{author}{{Prentice}, S.},
  \bibinfo{author}{{Olivares E.}, F.}, \bibinfo{author}{{Rossi}, A.},
  \bibinfo{author}{{Klose}, S.}, \bibinfo{author}{{Taubenberger}, S.},
  \bibinfo{author}{{Knust}, F.}, \bibinfo{author}{{Afonso}, P.~M.~J.},
  \bibinfo{author}{{Ashall}, C.}, \bibinfo{author}{{Bolmer}, J.},
  \bibinfo{author}{{Delvaux}, C.}, \bibinfo{author}{{Diehl}, R.},
  \bibinfo{author}{{Elliott}, J.}, \bibinfo{author}{{Filgas}, R.},
  \bibinfo{author}{{Fynbo}, J.~P.~U.}, \bibinfo{author}{{Graham}, J.~F.},
  \bibinfo{author}{{Guelbenzu}, A.~N.}, \bibinfo{author}{{Kobayashi}, S.},
  \bibinfo{author}{{Leloudas}, G.}, \bibinfo{author}{{Savaglio}, S.},
  \bibinfo{author}{{Schady}, P.}, \bibinfo{author}{{Schmidl}, S.},
  \bibinfo{author}{{Schweyer}, T.}, \bibinfo{author}{{Sudilovsky}, V.},
  \bibinfo{author}{{Tanga}, M.}, \bibinfo{author}{{Updike}, A.~C.},
  \bibinfo{author}{{van Eerten}, H.}, \& \bibinfo{author}{{Varela}, K.}
  (\bibinfo{year}{2015}).
\newblock \bibinfo{title}{{A very luminous magnetar-powered supernova
  associated with an ultra-long {$\gamma$}-ray burst}}.
\newblock {\it \bibinfo{journal}{\nat}\/},  {\it \bibinfo{volume}{523}\/},
  \bibinfo{pages}{189--192}. \DOIprefix\doi{10.1038/nature14579}.
  \href{http://arxiv.org/abs/1509.03279}{\tt arXiv:1509.03279}.
\bibitem[{{G{\"u}del}(2004)}]{Guedel2004}
\bibinfo{author}{{G{\"u}del}, M.} (\bibinfo{year}{2004}).
\newblock \bibinfo{title}{{X-ray astronomy of stellar coronae}}.
\newblock {\it \bibinfo{journal}{\aapr}\/},  {\it \bibinfo{volume}{12}\/},
  \bibinfo{pages}{71--237}. \DOIprefix\doi{10.1007/s00159-004-0023-2}.
  \href{http://arxiv.org/abs/astro-ph/0406661}{\tt arXiv:astro-ph/0406661}.
\bibitem[{{Guiriec} et~al.(2011){Guiriec}, {Connaughton}, {Briggs}, {Burgess},
  {Ryde}, {Daigne}, {M{\'e}sz{\'a}ros}, {Goldstein}, {McEnery}, {Omodei},
  {Bhat}, {Bissaldi}, {Camero-Arranz}, {Chaplin}, {Diehl}, {Fishman}, {Foley},
  {Gibby}, {Giles}, {Greiner}, {Gruber}, {von Kienlin}, {Kippen},
  {Kouveliotou}, {McBreen}, {Meegan}, {Paciesas}, {Preece}, {Rau}, {Tierney},
  {van der Horst} \& {Wilson-Hodge}}]{guiriec2011}
\bibinfo{author}{{Guiriec}, S.}, \bibinfo{author}{{Connaughton}, V.},
  \bibinfo{author}{{Briggs}, M.~S.}, \bibinfo{author}{{Burgess}, M.},
  \bibinfo{author}{{Ryde}, F.}, \bibinfo{author}{{Daigne}, F.},
  \bibinfo{author}{{M{\'e}sz{\'a}ros}, P.}, \bibinfo{author}{{Goldstein}, A.},
  \bibinfo{author}{{McEnery}, J.}, \bibinfo{author}{{Omodei}, N.},
  \bibinfo{author}{{Bhat}, P.~N.}, \bibinfo{author}{{Bissaldi}, E.},
  \bibinfo{author}{{Camero-Arranz}, A.}, \bibinfo{author}{{Chaplin}, V.},
  \bibinfo{author}{{Diehl}, R.}, \bibinfo{author}{{Fishman}, G.},
  \bibinfo{author}{{Foley}, S.}, \bibinfo{author}{{Gibby}, M.},
  \bibinfo{author}{{Giles}, M.~M.}, \bibinfo{author}{{Greiner}, J.},
  \bibinfo{author}{{Gruber}, D.}, \bibinfo{author}{{von Kienlin}, A.},
  \bibinfo{author}{{Kippen}, M.}, \bibinfo{author}{{Kouveliotou}, C.},
  \bibinfo{author}{{McBreen}, S.}, \bibinfo{author}{{Meegan}, C.~A.},
  \bibinfo{author}{{Paciesas}, W.}, \bibinfo{author}{{Preece}, R.},
  \bibinfo{author}{{Rau}, A.}, \bibinfo{author}{{Tierney}, D.},
  \bibinfo{author}{{van der Horst}, A.~J.}, \& \bibinfo{author}{{Wilson-Hodge},
  C.} (\bibinfo{year}{2011}).
\newblock \bibinfo{title}{{Detection of a Thermal Spectral Component in the
  Prompt Emission of GRB 100724B}}.
\newblock {\it \bibinfo{journal}{\apjl}\/},  {\it \bibinfo{volume}{727}\/},
  \bibinfo{pages}{L33}. \DOIprefix\doi{10.1088/2041-8205/727/2/L33}.
  \href{http://arxiv.org/abs/1010.4601}{\tt arXiv:1010.4601}.
\bibitem[{{Hartoog} et~al.(2015){Hartoog}, {Malesani}, {Fynbo}, {Goto},
  {Kr{\"u}hler}, {Vreeswijk}, {De Cia}, {Xu}, {M{\o}ller}, {Covino}, {D'Elia},
  {Flores}, {Goldoni}, {Hjorth}, {Jakobsson}, {Krogager}, {Kaper}, {Ledoux},
  {Levan}, {Milvang-Jensen}, {Sollerman}, {Sparre}, {Tagliaferri}, {Tanvir},
  {de Ugarte Postigo}, {Vergani}, {Wiersema}, {Datson}, {Salinas}, {Mikkelsen}
  \& {Aghanim}}]{Hartoog2015}
\bibinfo{author}{{Hartoog}, O.~E.}, \bibinfo{author}{{Malesani}, D.},
  \bibinfo{author}{{Fynbo}, J.~P.~U.}, \bibinfo{author}{{Goto}, T.},
  \bibinfo{author}{{Kr{\"u}hler}, T.}, \bibinfo{author}{{Vreeswijk}, P.~M.},
  \bibinfo{author}{{De Cia}, A.}, \bibinfo{author}{{Xu}, D.},
  \bibinfo{author}{{M{\o}ller}, P.}, \bibinfo{author}{{Covino}, S.},
  \bibinfo{author}{{D'Elia}, V.}, \bibinfo{author}{{Flores}, H.},
  \bibinfo{author}{{Goldoni}, P.}, \bibinfo{author}{{Hjorth}, J.},
  \bibinfo{author}{{Jakobsson}, P.}, \bibinfo{author}{{Krogager}, J.-K.},
  \bibinfo{author}{{Kaper}, L.}, \bibinfo{author}{{Ledoux}, C.},
  \bibinfo{author}{{Levan}, A.~J.}, \bibinfo{author}{{Milvang-Jensen}, B.},
  \bibinfo{author}{{Sollerman}, J.}, \bibinfo{author}{{Sparre}, M.},
  \bibinfo{author}{{Tagliaferri}, G.}, \bibinfo{author}{{Tanvir}, N.~R.},
  \bibinfo{author}{{de Ugarte Postigo}, A.}, \bibinfo{author}{{Vergani},
  S.~D.}, \bibinfo{author}{{Wiersema}, K.}, \bibinfo{author}{{Datson}, J.},
  \bibinfo{author}{{Salinas}, R.}, \bibinfo{author}{{Mikkelsen}, K.}, \&
  \bibinfo{author}{{Aghanim}, N.} (\bibinfo{year}{2015}).
\newblock \bibinfo{title}{{VLT/X-Shooter spectroscopy of the afterglow of the
  Swift GRB 130606A. Chemical abundances and reionisation at z \~{} 6}}.
\newblock {\it \bibinfo{journal}{\aap}\/},  {\it \bibinfo{volume}{580}\/},
  \bibinfo{pages}{A139}. \DOIprefix\doi{10.1051/0004-6361/201425001}.
  \href{http://arxiv.org/abs/1409.4804}{\tt arXiv:1409.4804}.
\bibitem[{{Heger} et~al.(2007){Heger}, {Cumming}, {Galloway} \&
  {Woosley}}]{Heger2007}
\bibinfo{author}{{Heger}, A.}, \bibinfo{author}{{Cumming}, A.},
  \bibinfo{author}{{Galloway}, D.~K.}, \& \bibinfo{author}{{Woosley}, S.~E.}
  (\bibinfo{year}{2007}).
\newblock \bibinfo{title}{{Models of Type I X-Ray Bursts from GS 1826-24: A
  Probe of rp-Process Hydrogen Burning}}.
\newblock {\it \bibinfo{journal}{\apjl}\/},  {\it \bibinfo{volume}{671}\/},
  \bibinfo{pages}{L141--L144}. \DOIprefix\doi{10.1086/525522}.
  \href{http://arxiv.org/abs/0711.1195}{\tt arXiv:0711.1195}.
\bibitem[{{Heger} et~al.(2003){Heger}, {Fryer}, {Woosley}, {Langer} \&
  {Hartmann}}]{Heger2003}
\bibinfo{author}{{Heger}, A.}, \bibinfo{author}{{Fryer}, C.~L.},
  \bibinfo{author}{{Woosley}, S.~E.}, \bibinfo{author}{{Langer}, N.}, \&
  \bibinfo{author}{{Hartmann}, D.~H.} (\bibinfo{year}{2003}).
\newblock \bibinfo{title}{{How Massive Single Stars End Their Life}}.
\newblock {\it \bibinfo{journal}{\apj}\/},  {\it \bibinfo{volume}{591}\/},
  \bibinfo{pages}{288--300}. \DOIprefix\doi{10.1086/375341}.
  \href{http://arxiv.org/abs/astro-ph/0212469}{\tt arXiv:astro-ph/0212469}.
\bibitem[{{Hjorth} et~al.(2012){Hjorth}, {Malesani}, {Jakobsson}, {Jaunsen},
  {Fynbo}, {Gorosabel}, {Kr{\"u}hler}, {Levan}, {Micha{\l}owski},
  {Milvang-Jensen}, {M{\o}ller}, {Schulze}, {Tanvir} \& {Watson}}]{Hjorth2012}
\bibinfo{author}{{Hjorth}, J.}, \bibinfo{author}{{Malesani}, D.},
  \bibinfo{author}{{Jakobsson}, P.}, \bibinfo{author}{{Jaunsen}, A.~O.},
  \bibinfo{author}{{Fynbo}, J.~P.~U.}, \bibinfo{author}{{Gorosabel}, J.},
  \bibinfo{author}{{Kr{\"u}hler}, T.}, \bibinfo{author}{{Levan}, A.~J.},
  \bibinfo{author}{{Micha{\l}owski}, M.~J.}, \bibinfo{author}{{Milvang-Jensen},
  B.}, \bibinfo{author}{{M{\o}ller}, P.}, \bibinfo{author}{{Schulze}, S.},
  \bibinfo{author}{{Tanvir}, N.~R.}, \& \bibinfo{author}{{Watson}, D.}
  (\bibinfo{year}{2012}).
\newblock \bibinfo{title}{{The Optically Unbiased Gamma-Ray Burst Host (TOUGH)
  Survey. I. Survey Design and Catalogs}}.
\newblock {\it \bibinfo{journal}{\apj}\/},  {\it \bibinfo{volume}{756}\/},
  \bibinfo{pages}{187}. \DOIprefix\doi{10.1088/0004-637X/756/2/187}.
  \href{http://arxiv.org/abs/1205.3162}{\tt arXiv:1205.3162}.
\bibitem[{{Hopkins} \& {Beacom}(2006)}]{Hopkins2006}
\bibinfo{author}{{Hopkins}, A.~M.}, \& \bibinfo{author}{{Beacom}, J.~F.}
  (\bibinfo{year}{2006}).
\newblock \bibinfo{title}{{On the Normalization of the Cosmic Star Formation
  History}}.
\newblock {\it \bibinfo{journal}{\apj}\/},  {\it \bibinfo{volume}{651}\/},
  \bibinfo{pages}{142--154}. \DOIprefix\doi{10.1086/506610}.
  \href{http://arxiv.org/abs/astro-ph/0601463}{\tt arXiv:astro-ph/0601463}.
\bibitem[{{IceCube-Gen2 Collaboration} et~al.(2014){IceCube-Gen2
  Collaboration}, {:}, {Aartsen}, {Ackermann}, {Adams}, {Aguilar}, {Ahlers},
  {Ahrens}, {Altmann}, {Anderson} \& et~al.}]{Aartsen2014}
\bibinfo{author}{{IceCube-Gen2 Collaboration}}, \bibinfo{author}{{:}},
  \bibinfo{author}{{Aartsen}, M.~G.}, \bibinfo{author}{{Ackermann}, M.},
  \bibinfo{author}{{Adams}, J.}, \bibinfo{author}{{Aguilar}, J.~A.},
  \bibinfo{author}{{Ahlers}, M.}, \bibinfo{author}{{Ahrens}, M.},
  \bibinfo{author}{{Altmann}, D.}, \bibinfo{author}{{Anderson}, T.}, \&
  \bibinfo{author}{et~al.} (\bibinfo{year}{2014}).
\newblock \bibinfo{title}{{IceCube-Gen2: A Vision for the Future of Neutrino
  Astronomy in Antarctica}}.
\newblock {\it \bibinfo{journal}{ArXiv e-prints}\/}, .
  \href{http://arxiv.org/abs/1412.5106}{\tt arXiv:1412.5106}.
\bibitem[{{Izzo} et~al.(2012){Izzo}, {Ruffini}, {Penacchioni}, {Bianco},
  {Caito}, {Chakrabarti}, {Rueda}, {Nandi} \& {Patricelli}}]{Izzo2012}
\bibinfo{author}{{Izzo}, L.}, \bibinfo{author}{{Ruffini}, R.},
  \bibinfo{author}{{Penacchioni}, A.~V.}, \bibinfo{author}{{Bianco}, C.~L.},
  \bibinfo{author}{{Caito}, L.}, \bibinfo{author}{{Chakrabarti}, S.~K.},
  \bibinfo{author}{{Rueda}, J.~A.}, \bibinfo{author}{{Nandi}, A.}, \&
  \bibinfo{author}{{Patricelli}, B.} (\bibinfo{year}{2012}).
\newblock \bibinfo{title}{{A double component in GRB 090618: a proto-black hole
  and a genuinely long gamma-ray burst}}.
\newblock {\it \bibinfo{journal}{\aap}\/},  {\it \bibinfo{volume}{543}\/},
  \bibinfo{pages}{A10}. \DOIprefix\doi{10.1051/0004-6361/201117436}.
  \href{http://arxiv.org/abs/1202.4374}{\tt arXiv:1202.4374}.
\bibitem[{{Jakobsson} et~al.(2012){Jakobsson}, {Hjorth}, {Malesani}, {Chapman},
  {Fynbo}, {Tanvir}, {Milvang-Jensen}, {Vreeswijk}, {Letawe} \&
  {Starling}}]{Jakobsson2012}
\bibinfo{author}{{Jakobsson}, P.}, \bibinfo{author}{{Hjorth}, J.},
  \bibinfo{author}{{Malesani}, D.}, \bibinfo{author}{{Chapman}, R.},
  \bibinfo{author}{{Fynbo}, J.~P.~U.}, \bibinfo{author}{{Tanvir}, N.~R.},
  \bibinfo{author}{{Milvang-Jensen}, B.}, \bibinfo{author}{{Vreeswijk}, P.~M.},
  \bibinfo{author}{{Letawe}, G.}, \& \bibinfo{author}{{Starling}, R.~L.~C.}
  (\bibinfo{year}{2012}).
\newblock \bibinfo{title}{{The Optically Unbiased GRB Host (TOUGH) Survey. III.
  Redshift Distribution}}.
\newblock {\it \bibinfo{journal}{\apj}\/},  {\it \bibinfo{volume}{752}\/},
  \bibinfo{pages}{62}. \DOIprefix\doi{10.1088/0004-637X/752/1/62}.
  \href{http://arxiv.org/abs/1205.3490}{\tt arXiv:1205.3490}.
\bibitem[{{Jin} et~al.(2015){Jin}, {Li}, {Cano}, {Covino}, {Fan} \&
  {Wei}}]{Jin2015}
\bibinfo{author}{{Jin}, Z.-P.}, \bibinfo{author}{{Li}, X.},
  \bibinfo{author}{{Cano}, Z.}, \bibinfo{author}{{Covino}, S.},
  \bibinfo{author}{{Fan}, Y.-Z.}, \& \bibinfo{author}{{Wei}, D.-M.}
  (\bibinfo{year}{2015}).
\newblock \bibinfo{title}{{The Light Curve of the Macronova Associated with the
  Long-Short Burst GRB 060614}}.
\newblock {\it \bibinfo{journal}{\apjl}\/},  {\it \bibinfo{volume}{811}\/},
  \bibinfo{pages}{L22}. \DOIprefix\doi{10.1088/2041-8205/811/2/L22}.
  \href{http://arxiv.org/abs/1507.07206}{\tt arXiv:1507.07206}.
\bibitem[{{Jonker} et~al.(2013){Jonker}, {O'Brien}, {Amati}, {Atteia},
  {Campana}, {Evans}, {Fender}, {Kouveliotou}, {Lodato}, {Osborne}, {Piro},
  {Rau}, {Tanvir} \& {Willingale}}]{Jonker2013}
\bibinfo{author}{{Jonker}, P.}, \bibinfo{author}{{O'Brien}, P.},
  \bibinfo{author}{{Amati}, L.}, \bibinfo{author}{{Atteia}, J.-L.},
  \bibinfo{author}{{Campana}, S.}, \bibinfo{author}{{Evans}, P.},
  \bibinfo{author}{{Fender}, R.}, \bibinfo{author}{{Kouveliotou}, C.},
  \bibinfo{author}{{Lodato}, G.}, \bibinfo{author}{{Osborne}, J.},
  \bibinfo{author}{{Piro}, L.}, \bibinfo{author}{{Rau}, A.},
  \bibinfo{author}{{Tanvir}, N.}, \& \bibinfo{author}{{Willingale}, R.}
  (\bibinfo{year}{2013}).
\newblock \bibinfo{title}{{The Hot and Energetic Universe: Luminous
  extragalactic transients}}.
\newblock {\it \bibinfo{journal}{ArXiv e-prints}\/}, .
  \href{http://arxiv.org/abs/1306.2336}{\tt arXiv:1306.2336}.
\bibitem[{{Kann} et~al.(2017){Kann}, {Schady}, {Olivares E.}, {Klose}, {Rossi},
  {Perley}, {Kr{\"u}hler}, {Greiner}, {Nicuesa Guelbenzu}, {Elliott}, {Knust},
  {Filgas}, {Pian}, {Mazzali}, {Fynbo}, {Leloudas}, {Afonso}, {Delvaux},
  {Graham}, {Rau}, {Schmidl}, {Schulze}, {Tanga}, {Updike} \&
  {Varela}}]{Kann2017}
\bibinfo{author}{{Kann}, D.~A.}, \bibinfo{author}{{Schady}, P.},
  \bibinfo{author}{{Olivares E.}, F.}, \bibinfo{author}{{Klose}, S.},
  \bibinfo{author}{{Rossi}, A.}, \bibinfo{author}{{Perley}, D.~A.},
  \bibinfo{author}{{Kr{\"u}hler}, T.}, \bibinfo{author}{{Greiner}, J.},
  \bibinfo{author}{{Nicuesa Guelbenzu}, A.}, \bibinfo{author}{{Elliott}, J.},
  \bibinfo{author}{{Knust}, F.}, \bibinfo{author}{{Filgas}, R.},
  \bibinfo{author}{{Pian}, E.}, \bibinfo{author}{{Mazzali}, P.},
  \bibinfo{author}{{Fynbo}, J.~P.~U.}, \bibinfo{author}{{Leloudas}, G.},
  \bibinfo{author}{{Afonso}, P.~M.~J.}, \bibinfo{author}{{Delvaux}, C.},
  \bibinfo{author}{{Graham}, J.~F.}, \bibinfo{author}{{Rau}, A.},
  \bibinfo{author}{{Schmidl}, S.}, \bibinfo{author}{{Schulze}, S.},
  \bibinfo{author}{{Tanga}, M.}, \bibinfo{author}{{Updike}, A.~C.}, \&
  \bibinfo{author}{{Varela}, K.} (\bibinfo{year}{2017}).
\newblock \bibinfo{title}{{The Optical/NIR afterglow of GRB 111209A: Complex
  yet not Unprecedented}}.
\newblock {\it \bibinfo{journal}{ArXiv e-prints}\/}, .
  \href{http://arxiv.org/abs/1706.00601}{\tt arXiv:1706.00601}.
\bibitem[{{Kanner} et~al.(2012){Kanner}, {Camp}, {Racusin}, {Gehrels} \&
  {White}}]{Kanner2012}
\bibinfo{author}{{Kanner}, J.}, \bibinfo{author}{{Camp}, J.},
  \bibinfo{author}{{Racusin}, J.}, \bibinfo{author}{{Gehrels}, N.}, \&
  \bibinfo{author}{{White}, D.} (\bibinfo{year}{2012}).
\newblock \bibinfo{title}{{Seeking Counterparts to Advanced LIGO/Virgo
  Transients with Swift}}.
\newblock {\it \bibinfo{journal}{\apj}\/},  {\it \bibinfo{volume}{759}\/},
  \bibinfo{pages}{22}. \DOIprefix\doi{10.1088/0004-637X/759/1/22}.
  \href{http://arxiv.org/abs/1209.2342}{\tt arXiv:1209.2342}.
\bibitem[{{Kasen}(2010)}]{Kasen2010}
\bibinfo{author}{{Kasen}, D.} (\bibinfo{year}{2010}).
\newblock \bibinfo{title}{{Astrophysics: The supernova has two faces}}.
\newblock {\it \bibinfo{journal}{\nat}\/},  {\it \bibinfo{volume}{466}\/},
  \bibinfo{pages}{37--38}. \DOIprefix\doi{10.1038/466037a}.
\bibitem[{{Kasen} et~al.(2015){Kasen}, {Fern{\'a}ndez} \&
  {Metzger}}]{Kasen2015}
\bibinfo{author}{{Kasen}, D.}, \bibinfo{author}{{Fern{\'a}ndez}, R.}, \&
  \bibinfo{author}{{Metzger}, B.~D.} (\bibinfo{year}{2015}).
\newblock \bibinfo{title}{{Kilonova light curves from the disc wind outflows of
  compact object mergers}}.
\newblock {\it \bibinfo{journal}{\mnras}\/},  {\it \bibinfo{volume}{450}\/},
  \bibinfo{pages}{1777--1786}. \DOIprefix\doi{10.1093/mnras/stv721}.
  \href{http://arxiv.org/abs/1411.3726}{\tt arXiv:1411.3726}.
\bibitem[{{Kasen} et~al.(2017){Kasen}, {Metzger}, {Barnes}, {Quataert} \&
  {Ramirez-Ruiz}}]{Kasen2017}
\bibinfo{author}{{Kasen}, D.}, \bibinfo{author}{{Metzger}, B.},
  \bibinfo{author}{{Barnes}, J.}, \bibinfo{author}{{Quataert}, E.}, \&
  \bibinfo{author}{{Ramirez-Ruiz}, E.} (\bibinfo{year}{2017}).
\newblock \bibinfo{title}{{Origin of the heavy elements in binary neutron-star
  mergers from a gravitational-wave event}}.
\newblock {\it \bibinfo{journal}{\nat}\/},  {\it \bibinfo{volume}{551}\/},
  \bibinfo{pages}{80--84}. \DOIprefix\doi{10.1038/nature24453}.
  \href{http://arxiv.org/abs/1710.05463}{\tt arXiv:1710.05463}.
\bibitem[{{Kathirgamaraju} et~al.(2017){Kathirgamaraju}, {Barniol Duran} \&
  {Giannios}}]{Kathirgamaraju2017}
\bibinfo{author}{{Kathirgamaraju}, A.}, \bibinfo{author}{{Barniol Duran}, R.},
  \& \bibinfo{author}{{Giannios}, D.} (\bibinfo{year}{2017}).
\newblock \bibinfo{title}{{Off-axis short GRBs from structured jets as
  counterparts to GW events}}.
\newblock {\it \bibinfo{journal}{ArXiv e-prints}\/}, .
  \href{http://arxiv.org/abs/1708.07488}{\tt arXiv:1708.07488}.
\bibitem[{{Keek} et~al.(2010){Keek}, {Galloway}, {in't Zand} \&
  {Heger}}]{Keek2010}
\bibinfo{author}{{Keek}, L.}, \bibinfo{author}{{Galloway}, D.~K.},
  \bibinfo{author}{{in't Zand}, J.~J.~M.}, \& \bibinfo{author}{{Heger}, A.}
  (\bibinfo{year}{2010}).
\newblock \bibinfo{title}{{Multi-instrument X-ray Observations of Thermonuclear
  Bursts with Short Recurrence Times}}.
\newblock {\it \bibinfo{journal}{\apj}\/},  {\it \bibinfo{volume}{718}\/},
  \bibinfo{pages}{292--305}. \DOIprefix\doi{10.1088/0004-637X/718/1/292}.
  \href{http://arxiv.org/abs/1005.3302}{\tt arXiv:1005.3302}.
\bibitem[{{Kistler} et~al.(2013){Kistler}, {Yuksel} \& {Hopkins}}]{Kistler2013}
\bibinfo{author}{{Kistler}, M.~D.}, \bibinfo{author}{{Yuksel}, H.}, \&
  \bibinfo{author}{{Hopkins}, A.~M.} (\bibinfo{year}{2013}).
\newblock \bibinfo{title}{{The Cosmic Star Formation Rate from the Faintest
  Galaxies in the Unobservable Universe}}.
\newblock {\it \bibinfo{journal}{ArXiv e-prints}\/}, .
  \href{http://arxiv.org/abs/1305.1630}{\tt arXiv:1305.1630}.
\bibitem[{{Klimenko} et~al.(2011){Klimenko}, {Vedovato}, {Drago}, {Mazzolo},
  {Mitselmakher}, {Pankow}, {Prodi}, {Re}, {Salemi} \&
  {Yakushin}}]{Klimenko2011}
\bibinfo{author}{{Klimenko}, S.}, \bibinfo{author}{{Vedovato}, G.},
  \bibinfo{author}{{Drago}, M.}, \bibinfo{author}{{Mazzolo}, G.},
  \bibinfo{author}{{Mitselmakher}, G.}, \bibinfo{author}{{Pankow}, C.},
  \bibinfo{author}{{Prodi}, G.}, \bibinfo{author}{{Re}, V.},
  \bibinfo{author}{{Salemi}, F.}, \& \bibinfo{author}{{Yakushin}, I.}
  (\bibinfo{year}{2011}).
\newblock \bibinfo{title}{{Localization of gravitational wave sources with
  networks of advanced detectors}}.
\newblock {\it \bibinfo{journal}{\prd}\/},  {\it \bibinfo{volume}{83}\/},
  \bibinfo{pages}{102001}. \DOIprefix\doi{10.1103/PhysRevD.83.102001}.
  \href{http://arxiv.org/abs/1101.5408}{\tt arXiv:1101.5408}.
\bibitem[{{Komossa} et~al.(2004){Komossa}, {Halpern}, {Schartel}, {Hasinger},
  {Santos-Lleo} \& {Predehl}}]{Komossa2004}
\bibinfo{author}{{Komossa}, S.}, \bibinfo{author}{{Halpern}, J.},
  \bibinfo{author}{{Schartel}, N.}, \bibinfo{author}{{Hasinger}, G.},
  \bibinfo{author}{{Santos-Lleo}, M.}, \& \bibinfo{author}{{Predehl}, P.}
  (\bibinfo{year}{2004}).
\newblock \bibinfo{title}{{A Huge Drop in the X-Ray Luminosity of the Nonactive
  Galaxy RX J1242.6-1119A, and the First Postflare Spectrum: Testing the Tidal
  Disruption Scenario}}.
\newblock {\it \bibinfo{journal}{\apjl}\/},  {\it \bibinfo{volume}{603}\/},
  \bibinfo{pages}{L17--L20}. \DOIprefix\doi{10.1086/382046}.
  \href{http://arxiv.org/abs/astro-ph/0402468}{\tt arXiv:astro-ph/0402468}.
\bibitem[{{K{\"o}rding} et~al.(2008){K{\"o}rding}, {Rupen}, {Knigge}, {Fender},
  {Dhawan}, {Templeton} \& {Muxlow}}]{Koerding2008}
\bibinfo{author}{{K{\"o}rding}, E.}, \bibinfo{author}{{Rupen}, M.},
  \bibinfo{author}{{Knigge}, C.}, \bibinfo{author}{{Fender}, R.},
  \bibinfo{author}{{Dhawan}, V.}, \bibinfo{author}{{Templeton}, M.}, \&
  \bibinfo{author}{{Muxlow}, T.} (\bibinfo{year}{2008}).
\newblock \bibinfo{title}{{A Transient Radio Jet in an Erupting Dwarf Nova}}.
\newblock {\it \bibinfo{journal}{Science}\/},  {\it \bibinfo{volume}{320}\/},
  \bibinfo{pages}{1318}. \DOIprefix\doi{10.1126/science.1155492}.
  \href{http://arxiv.org/abs/0806.1002}{\tt arXiv:0806.1002}.
\bibitem[{{K{\"o}rding} et~al.(2011){K{\"o}rding}, {Knigge}, {Tzioumis} \&
  {Fender}}]{Koerding2011}
\bibinfo{author}{{K{\"o}rding}, E.~G.}, \bibinfo{author}{{Knigge}, C.},
  \bibinfo{author}{{Tzioumis}, T.}, \& \bibinfo{author}{{Fender}, R.}
  (\bibinfo{year}{2011}).
\newblock \bibinfo{title}{{Detection of radio emission from a nova-like
  cataclysmic variable: evidence of jets?}}
\newblock {\it \bibinfo{journal}{\mnras}\/},  {\it \bibinfo{volume}{418}\/},
  \bibinfo{pages}{L129--L132}.
  \DOIprefix\doi{10.1111/j.1745-3933.2011.01158.x}.
\bibitem[{{Labanti} et~al.(2008){Labanti}, {Marisaldi}, {Fuschino}, {Bastia},
  {Negri}, {Perotti} \& {Soltau}}]{Labanti2008}
\bibinfo{author}{{Labanti}, C.}, \bibinfo{author}{{Marisaldi}, M.},
  \bibinfo{author}{{Fuschino}, F.}, \bibinfo{author}{{Bastia}, P.},
  \bibinfo{author}{{Negri}, B.}, \bibinfo{author}{{Perotti}, F.}, \&
  \bibinfo{author}{{Soltau}, H.} (\bibinfo{year}{2008}).
\newblock \bibinfo{title}{{Position sensitive x- and gamma-ray scintillator
  detector for new space telescopes}}.
\newblock In {\it \bibinfo{booktitle}{High Energy, Optical, and Infrared
  Detectors for Astronomy III}\/} (p. \bibinfo{pages}{702116}).
\newblock volume \bibinfo{volume}{7021} of {\it \bibinfo{series}{\procspie}\/}.
\newblock \DOIprefix\doi{10.1117/12.789383}.
\bibitem[{{Lasky} et~al.(2014){Lasky}, {Haskell}, {Ravi}, {Howell} \&
  {Coward}}]{Lasky2014}
\bibinfo{author}{{Lasky}, P.~D.}, \bibinfo{author}{{Haskell}, B.},
  \bibinfo{author}{{Ravi}, V.}, \bibinfo{author}{{Howell}, E.~J.}, \&
  \bibinfo{author}{{Coward}, D.~M.} (\bibinfo{year}{2014}).
\newblock \bibinfo{title}{{Nuclear equation of state from observations of short
  gamma-ray burst remnants}}.
\newblock {\it \bibinfo{journal}{\prd}\/},  {\it \bibinfo{volume}{89}\/},
  \bibinfo{pages}{047302}. \DOIprefix\doi{10.1103/PhysRevD.89.047302}.
  \href{http://arxiv.org/abs/1311.1352}{\tt arXiv:1311.1352}.
\bibitem[{{Lasky} et~al.(2012){Lasky}, {Zink} \& {Kokkotas}}]{Lasky2012}
\bibinfo{author}{{Lasky}, P.~D.}, \bibinfo{author}{{Zink}, B.}, \&
  \bibinfo{author}{{Kokkotas}, K.~D.} (\bibinfo{year}{2012}).
\newblock \bibinfo{title}{{Gravitational Waves and Hydromagnetic Instabilities
  in Rotating Magnetized Neutron Stars}}.
\newblock {\it \bibinfo{journal}{ArXiv e-prints}\/}, .
  \href{http://arxiv.org/abs/1203.3590}{\tt arXiv:1203.3590}.
\bibitem[{{Levan} et~al.(2011){Levan}, {Tanvir}, {Cenko}, {Perley}, {Wiersema},
  {Bloom}, {Fruchter}, {Postigo}, {O'Brien}, {Butler}, {van der Horst},
  {Leloudas}, {Morgan}, {Misra}, {Bower}, {Farihi}, {Tunnicliffe}, {Modjaz},
  {Silverman}, {Hjorth}, {Th{\"o}ne}, {Cucchiara}, {Cer{\'o}n},
  {Castro-Tirado}, {Arnold}, {Bremer}, {Brodie}, {Carroll}, {Cooper}, {Curran},
  {Cutri}, {Ehle}, {Forbes}, {Fynbo}, {Gorosabel}, {Graham}, {Hoffman},
  {Guziy}, {Jakobsson}, {Kamble}, {Kerr}, {Kasliwal}, {Kouveliotou},
  {Kocevski}, {Law}, {Nugent}, {Ofek}, {Poznanski}, {Quimby}, {Rol},
  {Romanowsky}, {S{\'a}nchez-Ram{\'{\i}}rez}, {Schulze}, {Singh}, {van
  Spaandonk}, {Starling}, {Strom}, {Tello}, {Vaduvescu}, {Wheatley}, {Wijers},
  {Winters} \& {Xu}}]{Levan2011}
\bibinfo{author}{{Levan}, A.~J.}, \bibinfo{author}{{Tanvir}, N.~R.},
  \bibinfo{author}{{Cenko}, S.~B.}, \bibinfo{author}{{Perley}, D.~A.},
  \bibinfo{author}{{Wiersema}, K.}, \bibinfo{author}{{Bloom}, J.~S.},
  \bibinfo{author}{{Fruchter}, A.~S.}, \bibinfo{author}{{Postigo}, A.~d.~U.},
  \bibinfo{author}{{O'Brien}, P.~T.}, \bibinfo{author}{{Butler}, N.},
  \bibinfo{author}{{van der Horst}, A.~J.}, \bibinfo{author}{{Leloudas}, G.},
  \bibinfo{author}{{Morgan}, A.~N.}, \bibinfo{author}{{Misra}, K.},
  \bibinfo{author}{{Bower}, G.~C.}, \bibinfo{author}{{Farihi}, J.},
  \bibinfo{author}{{Tunnicliffe}, R.~L.}, \bibinfo{author}{{Modjaz}, M.},
  \bibinfo{author}{{Silverman}, J.~M.}, \bibinfo{author}{{Hjorth}, J.},
  \bibinfo{author}{{Th{\"o}ne}, C.}, \bibinfo{author}{{Cucchiara}, A.},
  \bibinfo{author}{{Cer{\'o}n}, J.~M.~C.}, \bibinfo{author}{{Castro-Tirado},
  A.~J.}, \bibinfo{author}{{Arnold}, J.~A.}, \bibinfo{author}{{Bremer}, M.},
  \bibinfo{author}{{Brodie}, J.~P.}, \bibinfo{author}{{Carroll}, T.},
  \bibinfo{author}{{Cooper}, M.~C.}, \bibinfo{author}{{Curran}, P.~A.},
  \bibinfo{author}{{Cutri}, R.~M.}, \bibinfo{author}{{Ehle}, J.},
  \bibinfo{author}{{Forbes}, D.}, \bibinfo{author}{{Fynbo}, J.},
  \bibinfo{author}{{Gorosabel}, J.}, \bibinfo{author}{{Graham}, J.},
  \bibinfo{author}{{Hoffman}, D.~I.}, \bibinfo{author}{{Guziy}, S.},
  \bibinfo{author}{{Jakobsson}, P.}, \bibinfo{author}{{Kamble}, A.},
  \bibinfo{author}{{Kerr}, T.}, \bibinfo{author}{{Kasliwal}, M.~M.},
  \bibinfo{author}{{Kouveliotou}, C.}, \bibinfo{author}{{Kocevski}, D.},
  \bibinfo{author}{{Law}, N.~M.}, \bibinfo{author}{{Nugent}, P.~E.},
  \bibinfo{author}{{Ofek}, E.~O.}, \bibinfo{author}{{Poznanski}, D.},
  \bibinfo{author}{{Quimby}, R.~M.}, \bibinfo{author}{{Rol}, E.},
  \bibinfo{author}{{Romanowsky}, A.~J.},
  \bibinfo{author}{{S{\'a}nchez-Ram{\'{\i}}rez}, R.},
  \bibinfo{author}{{Schulze}, S.}, \bibinfo{author}{{Singh}, N.},
  \bibinfo{author}{{van Spaandonk}, L.}, \bibinfo{author}{{Starling},
  R.~L.~C.}, \bibinfo{author}{{Strom}, R.~G.}, \bibinfo{author}{{Tello},
  J.~C.}, \bibinfo{author}{{Vaduvescu}, O.}, \bibinfo{author}{{Wheatley},
  P.~J.}, \bibinfo{author}{{Wijers}, R.~A.~M.~J.}, \bibinfo{author}{{Winters},
  J.~M.}, \& \bibinfo{author}{{Xu}, D.} (\bibinfo{year}{2011}).
\newblock \bibinfo{title}{{An Extremely Luminous Panchromatic Outburst from the
  Nucleus of a Distant Galaxy}}.
\newblock {\it \bibinfo{journal}{Science}\/},  {\it \bibinfo{volume}{333}\/},
  \bibinfo{pages}{199}. \DOIprefix\doi{10.1126/science.1207143}.
  \href{http://arxiv.org/abs/1104.3356}{\tt arXiv:1104.3356}.
\bibitem[{{Levan} et~al.(2014){Levan}, {Tanvir}, {Fruchter}, {Hjorth}, {Pian},
  {Mazzali}, {Hounsell}, {Perley}, {Cano}, {Graham}, {Cenko}, {Fynbo},
  {Kouveliotou}, {Pe'er}, {Misra} \& {Wiersema}}]{Levan2014}
\bibinfo{author}{{Levan}, A.~J.}, \bibinfo{author}{{Tanvir}, N.~R.},
  \bibinfo{author}{{Fruchter}, A.~S.}, \bibinfo{author}{{Hjorth}, J.},
  \bibinfo{author}{{Pian}, E.}, \bibinfo{author}{{Mazzali}, P.},
  \bibinfo{author}{{Hounsell}, R.~A.}, \bibinfo{author}{{Perley}, D.~A.},
  \bibinfo{author}{{Cano}, Z.}, \bibinfo{author}{{Graham}, J.},
  \bibinfo{author}{{Cenko}, S.~B.}, \bibinfo{author}{{Fynbo}, J.~P.~U.},
  \bibinfo{author}{{Kouveliotou}, C.}, \bibinfo{author}{{Pe'er}, A.},
  \bibinfo{author}{{Misra}, K.}, \& \bibinfo{author}{{Wiersema}, K.}
  (\bibinfo{year}{2014}).
\newblock \bibinfo{title}{{Hubble Space Telescope Observations of the
  Afterglow, Supernova, and Host Galaxy Associated with the Extremely Bright
  GRB 130427A}}.
\newblock {\it \bibinfo{journal}{\apj}\/},  {\it \bibinfo{volume}{792}\/},
  \bibinfo{pages}{115}. \DOIprefix\doi{10.1088/0004-637X/792/2/115}.
  \href{http://arxiv.org/abs/1307.5338}{\tt arXiv:1307.5338}.
\bibitem[{{Li} \& {Paczy{\'n}ski}(1998)}]{Li_Paczynski1998}
\bibinfo{author}{{Li}, L.-X.}, \& \bibinfo{author}{{Paczy{\'n}ski}, B.}
  (\bibinfo{year}{1998}).
\newblock \bibinfo{title}{{Transient Events from Neutron Star Mergers}}.
\newblock {\it \bibinfo{journal}{\apjl}\/},  {\it \bibinfo{volume}{507}\/},
  \bibinfo{pages}{L59--L62}. \DOIprefix\doi{10.1086/311680}.
  \href{http://arxiv.org/abs/astro-ph/9807272}{\tt arXiv:astro-ph/9807272}.
\bibitem[{{Li} et~al.(2011){Li}, {Chornock}, {Leaman}, {Filippenko},
  {Poznanski}, {Wang}, {Ganeshalingam} \& {Mannucci}}]{Li2011}
\bibinfo{author}{{Li}, W.}, \bibinfo{author}{{Chornock}, R.},
  \bibinfo{author}{{Leaman}, J.}, \bibinfo{author}{{Filippenko}, A.~V.},
  \bibinfo{author}{{Poznanski}, D.}, \bibinfo{author}{{Wang}, X.},
  \bibinfo{author}{{Ganeshalingam}, M.}, \& \bibinfo{author}{{Mannucci}, F.}
  (\bibinfo{year}{2011}).
\newblock \bibinfo{title}{{Nearby supernova rates from the Lick Observatory
  Supernova Search - III. The rate-size relation, and the rates as a function
  of galaxy Hubble type and colour}}.
\newblock {\it \bibinfo{journal}{\mnras}\/},  {\it \bibinfo{volume}{412}\/},
  \bibinfo{pages}{1473--1507}.
  \DOIprefix\doi{10.1111/j.1365-2966.2011.18162.x}.
  \href{http://arxiv.org/abs/1006.4613}{\tt arXiv:1006.4613}.
\bibitem[{{Liu} et~al.(2016){Liu}, {Mutch}, {Angel}, {Duffy}, {Geil}, {Poole},
  {Mesinger} \& {Wyithe}}]{Liu2016}
\bibinfo{author}{{Liu}, C.}, \bibinfo{author}{{Mutch}, S.~J.},
  \bibinfo{author}{{Angel}, P.~W.}, \bibinfo{author}{{Duffy}, A.~R.},
  \bibinfo{author}{{Geil}, P.~M.}, \bibinfo{author}{{Poole}, G.~B.},
  \bibinfo{author}{{Mesinger}, A.}, \& \bibinfo{author}{{Wyithe}, J.~S.~B.}
  (\bibinfo{year}{2016}).
\newblock \bibinfo{title}{{Dark-ages reionization and galaxy formation
  simulation - IV. UV luminosity functions of high-redshift galaxies}}.
\newblock {\it \bibinfo{journal}{\mnras}\/},  {\it \bibinfo{volume}{462}\/},
  \bibinfo{pages}{235--249}. \DOIprefix\doi{10.1093/mnras/stw1015}.
  \href{http://arxiv.org/abs/1512.00563}{\tt arXiv:1512.00563}.
\bibitem[{{Ma} et~al.(2017){Ma}, {Maio}, {Ciardi} \& {Salvaterra}}]{Ma2017}
\bibinfo{author}{{Ma}, Q.}, \bibinfo{author}{{Maio}, U.},
  \bibinfo{author}{{Ciardi}, B.}, \& \bibinfo{author}{{Salvaterra}, R.}
  (\bibinfo{year}{2017}).
\newblock \bibinfo{title}{{Constraining the PopIII IMF with high-z GRBs}}.
\newblock {\it \bibinfo{journal}{\mnras}\/},  {\it \bibinfo{volume}{466}\/},
  \bibinfo{pages}{1140--1148}. \DOIprefix\doi{10.1093/mnras/stw3159}.
  \href{http://arxiv.org/abs/1610.03594}{\tt arXiv:1610.03594}.
\bibitem[{{Madau} et~al.(1999){Madau}, {Haardt} \& {Rees}}]{Madau1999}
\bibinfo{author}{{Madau}, P.}, \bibinfo{author}{{Haardt}, F.}, \&
  \bibinfo{author}{{Rees}, M.~J.} (\bibinfo{year}{1999}).
\newblock \bibinfo{title}{{Radiative Transfer in a Clumpy Universe. III. The
  Nature of Cosmological Ionizing Sources}}.
\newblock {\it \bibinfo{journal}{\apj}\/},  {\it \bibinfo{volume}{514}\/},
  \bibinfo{pages}{648--659}. \DOIprefix\doi{10.1086/306975}.
  \href{http://arxiv.org/abs/astro-ph/9809058}{\tt arXiv:astro-ph/9809058}.
\bibitem[{{Marchant} et~al.(2017){Marchant}, {Langer}, {Podsiadlowski},
  {Tauris}, {de Mink}, {Mandel} \& {Moriya}}]{Marchant2017}
\bibinfo{author}{{Marchant}, P.}, \bibinfo{author}{{Langer}, N.},
  \bibinfo{author}{{Podsiadlowski}, P.}, \bibinfo{author}{{Tauris}, T.~M.},
  \bibinfo{author}{{de Mink}, S.}, \bibinfo{author}{{Mandel}, I.}, \&
  \bibinfo{author}{{Moriya}, T.~J.} (\bibinfo{year}{2017}).
\newblock \bibinfo{title}{{Ultra-luminous X-ray sources and
  neutron-star-black-hole mergers from very massive close binaries at low
  metallicity}}.
\newblock {\it \bibinfo{journal}{\aap}\/},  {\it \bibinfo{volume}{604}\/},
  \bibinfo{pages}{A55}. \DOIprefix\doi{10.1051/0004-6361/201630188}.
  \href{http://arxiv.org/abs/1705.04734}{\tt arXiv:1705.04734}.
\bibitem[{{Marisaldi} et~al.(2004){Marisaldi}, {Labanti}, {Bulgarelli},
  {Celesti}, {Di Cocco}, {Gianotti}, {Mauri}, {Rossi}, {Traci} \&
  {Trifoglio}}]{Marisaldi2004}
\bibinfo{author}{{Marisaldi}, M.}, \bibinfo{author}{{Labanti}, C.},
  \bibinfo{author}{{Bulgarelli}, A.}, \bibinfo{author}{{Celesti}, E.},
  \bibinfo{author}{{Di Cocco}, G.}, \bibinfo{author}{{Gianotti}, F.},
  \bibinfo{author}{{Mauri}, A.}, \bibinfo{author}{{Rossi}, E.},
  \bibinfo{author}{{Traci}, A.}, \& \bibinfo{author}{{Trifoglio}, M.}
  (\bibinfo{year}{2004}).
\newblock \bibinfo{title}{{Calorimeter prototype based on silicon drift
  detectors coupled to scintillators for Compton telescopes application}}.
\newblock {\it \bibinfo{journal}{\nar}\/},  {\it \bibinfo{volume}{48}\/},
  \bibinfo{pages}{305--308}. \DOIprefix\doi{10.1016/j.newar.2003.11.035}.
\bibitem[{{Marscher} et~al.(2010){Marscher}, {Jorstad}, {Larionov}, {Agudo},
  {Aller}, {Aller}, {Lahteenmaki}, {Smith}, {Krichbaum} \&
  {McHardy}}]{Marscher2010}
\bibinfo{author}{{Marscher}, A.~P.}, \bibinfo{author}{{Jorstad}, S.~G.},
  \bibinfo{author}{{Larionov}, V.~M.}, \bibinfo{author}{{Agudo}, I.},
  \bibinfo{author}{{Aller}, M.~F.}, \bibinfo{author}{{Aller}, H.~D.},
  \bibinfo{author}{{Lahteenmaki}, A.}, \bibinfo{author}{{Smith}, P.~S.},
  \bibinfo{author}{{Krichbaum}, T.}, \& \bibinfo{author}{{McHardy}, I.~M.}
  (\bibinfo{year}{2010}).
\newblock \bibinfo{title}{{Comprehensive Multi-waveband Monitoring of Gamma-ray
  Blazars}}.
\newblock In {\it \bibinfo{booktitle}{AAS/High Energy Astrophysics Division
  \#11}\/} (p. \bibinfo{pages}{709}).
\newblock volume~\bibinfo{volume}{42} of {\it \bibinfo{series}{Bulletin of the
  American Astronomical Society}\/}.
\bibitem[{{McGuire} et~al.(2016){McGuire}, {Tanvir}, {Levan}, {Trenti},
  {Stanway}, {Shull}, {Wiersema}, {Perley}, {Starling}, {Bremer}, {Stocke},
  {Hjorth}, {Rhoads}, {Curtis-Lake}, {Schulze}, {Levesque}, {Robertson},
  {Fynbo}, {Ellis} \& {Fruchter}}]{McGuire2016}
\bibinfo{author}{{McGuire}, J.~T.~W.}, \bibinfo{author}{{Tanvir}, N.~R.},
  \bibinfo{author}{{Levan}, A.~J.}, \bibinfo{author}{{Trenti}, M.},
  \bibinfo{author}{{Stanway}, E.~R.}, \bibinfo{author}{{Shull}, J.~M.},
  \bibinfo{author}{{Wiersema}, K.}, \bibinfo{author}{{Perley}, D.~A.},
  \bibinfo{author}{{Starling}, R.~L.~C.}, \bibinfo{author}{{Bremer}, M.},
  \bibinfo{author}{{Stocke}, J.~T.}, \bibinfo{author}{{Hjorth}, J.},
  \bibinfo{author}{{Rhoads}, J.~E.}, \bibinfo{author}{{Curtis-Lake}, E.},
  \bibinfo{author}{{Schulze}, S.}, \bibinfo{author}{{Levesque}, E.~M.},
  \bibinfo{author}{{Robertson}, B.}, \bibinfo{author}{{Fynbo}, J.~P.~U.},
  \bibinfo{author}{{Ellis}, R.~S.}, \& \bibinfo{author}{{Fruchter}, A.~S.}
  (\bibinfo{year}{2016}).
\newblock \bibinfo{title}{{Detection of Three Gamma-ray Burst Host Galaxies at
  z$\sim$6}}.
\newblock {\it \bibinfo{journal}{\apj}\/},  {\it \bibinfo{volume}{825}\/},
  \bibinfo{pages}{135}. \DOIprefix\doi{10.3847/0004-637X/825/2/135}.
  \href{http://arxiv.org/abs/1512.07808}{\tt arXiv:1512.07808}.
\bibitem[{{McQuinn} et~al.(2008){McQuinn}, {Lidz}, {Zaldarriaga}, {Hernquist}
  \& {Dutta}}]{McQuinn2008}
\bibinfo{author}{{McQuinn}, M.}, \bibinfo{author}{{Lidz}, A.},
  \bibinfo{author}{{Zaldarriaga}, M.}, \bibinfo{author}{{Hernquist}, L.}, \&
  \bibinfo{author}{{Dutta}, S.} (\bibinfo{year}{2008}).
\newblock \bibinfo{title}{{Probing the neutral fraction of the IGM with GRBs
  during the epoch of reionization}}.
\newblock {\it \bibinfo{journal}{\mnras}\/},  {\it \bibinfo{volume}{388}\/},
  \bibinfo{pages}{1101--1110}.
  \DOIprefix\doi{10.1111/j.1365-2966.2008.13271.x}.
  \href{http://arxiv.org/abs/0710.1018}{\tt arXiv:0710.1018}.
\bibitem[{{Mereghetti}(2008)}]{Mereghetti2008}
\bibinfo{author}{{Mereghetti}, S.} (\bibinfo{year}{2008}).
\newblock \bibinfo{title}{{The strongest cosmic magnets: soft gamma-ray
  repeaters and anomalous X-ray pulsars}}.
\newblock {\it \bibinfo{journal}{\aapr}\/},  {\it \bibinfo{volume}{15}\/},
  \bibinfo{pages}{225--287}. \DOIprefix\doi{10.1007/s00159-008-0011-z}.
  \href{http://arxiv.org/abs/0804.0250}{\tt arXiv:0804.0250}.
\bibitem[{{Messenger} et~al.(2015){Messenger}, {Bulten}, {Crowder},
  {Dergachev}, {Galloway}, {Goetz}, {Jonker}, {Lasky}, {Meadors}, {Melatos},
  {Premachandra}, {Riles}, {Sammut}, {Thrane}, {Whelan} \&
  {Zhang}}]{Messenger2015}
\bibinfo{author}{{Messenger}, C.}, \bibinfo{author}{{Bulten}, H.~J.},
  \bibinfo{author}{{Crowder}, S.~G.}, \bibinfo{author}{{Dergachev}, V.},
  \bibinfo{author}{{Galloway}, D.~K.}, \bibinfo{author}{{Goetz}, E.},
  \bibinfo{author}{{Jonker}, R.~J.~G.}, \bibinfo{author}{{Lasky}, P.~D.},
  \bibinfo{author}{{Meadors}, G.~D.}, \bibinfo{author}{{Melatos}, A.},
  \bibinfo{author}{{Premachandra}, S.}, \bibinfo{author}{{Riles}, K.},
  \bibinfo{author}{{Sammut}, L.}, \bibinfo{author}{{Thrane}, E.~H.},
  \bibinfo{author}{{Whelan}, J.~T.}, \& \bibinfo{author}{{Zhang}, Y.}
  (\bibinfo{year}{2015}).
\newblock \bibinfo{title}{{Gravitational waves from Scorpius X-1: A comparison
  of search methods and prospects for detection with advanced detectors}}.
\newblock {\it \bibinfo{journal}{\prd}\/},  {\it \bibinfo{volume}{92}\/},
  \bibinfo{pages}{023006}. \DOIprefix\doi{10.1103/PhysRevD.92.023006}.
  \href{http://arxiv.org/abs/1504.05889}{\tt arXiv:1504.05889}.
\bibitem[{{M{\'e}sz{\'a}ros} \& {Rees}(2010)}]{Meszaros2010}
\bibinfo{author}{{M{\'e}sz{\'a}ros}, P.}, \& \bibinfo{author}{{Rees}, M.~J.}
  (\bibinfo{year}{2010}).
\newblock \bibinfo{title}{{Population III Gamma-ray Bursts}}.
\newblock {\it \bibinfo{journal}{\apj}\/},  {\it \bibinfo{volume}{715}\/},
  \bibinfo{pages}{967--971}. \DOIprefix\doi{10.1088/0004-637X/715/2/967}.
  \href{http://arxiv.org/abs/1004.2056}{\tt arXiv:1004.2056}.
\bibitem[{{Metzger} \& {Berger}(2012)}]{Metzger2012}
\bibinfo{author}{{Metzger}, B.~D.}, \& \bibinfo{author}{{Berger}, E.}
  (\bibinfo{year}{2012}).
\newblock \bibinfo{title}{{What is the Most Promising Electromagnetic
  Counterpart of a Neutron Star Binary Merger?}}
\newblock {\it \bibinfo{journal}{\apj}\/},  {\it \bibinfo{volume}{746}\/},
  \bibinfo{pages}{48}. \DOIprefix\doi{10.1088/0004-637X/746/1/48}.
  \href{http://arxiv.org/abs/1108.6056}{\tt arXiv:1108.6056}.
\bibitem[{{Metzger} \& {Piro}(2014)}]{Metzger2014}
\bibinfo{author}{{Metzger}, B.~D.}, \& \bibinfo{author}{{Piro}, A.~L.}
  (\bibinfo{year}{2014}).
\newblock \bibinfo{title}{{Optical and X-ray emission from stable millisecond
  magnetars formed from the merger of binary neutron stars}}.
\newblock {\it \bibinfo{journal}{\mnras}\/},  {\it \bibinfo{volume}{439}\/},
  \bibinfo{pages}{3916--3930}. \DOIprefix\doi{10.1093/mnras/stu247}.
  \href{http://arxiv.org/abs/1311.1519}{\tt arXiv:1311.1519}.
\bibitem[{{Micha{\l}owski} et~al.(2015){Micha{\l}owski}, {Gentile}, {Hjorth},
  {Krumholz}, {Tanvir}, {Kamphuis}, {Burlon}, {Baes}, {Basa}, {Berta}, {Castro
  Cer{\'o}n}, {Crosby}, {D'Elia}, {Elliott}, {Greiner}, {Hunt}, {Klose},
  {Koprowski}, {Le Floc'h}, {Malesani}, {Murphy}, {Nicuesa Guelbenzu},
  {Palazzi}, {Rasmussen}, {Rossi}, {Savaglio}, {Schady}, {Sollerman}, {de
  Ugarte Postigo}, {Watson}, {van der Werf}, {Vergani} \&
  {Xu}}]{Michalowski2015}
\bibinfo{author}{{Micha{\l}owski}, M.~J.}, \bibinfo{author}{{Gentile}, G.},
  \bibinfo{author}{{Hjorth}, J.}, \bibinfo{author}{{Krumholz}, M.~R.},
  \bibinfo{author}{{Tanvir}, N.~R.}, \bibinfo{author}{{Kamphuis}, P.},
  \bibinfo{author}{{Burlon}, D.}, \bibinfo{author}{{Baes}, M.},
  \bibinfo{author}{{Basa}, S.}, \bibinfo{author}{{Berta}, S.},
  \bibinfo{author}{{Castro Cer{\'o}n}, J.~M.}, \bibinfo{author}{{Crosby}, D.},
  \bibinfo{author}{{D'Elia}, V.}, \bibinfo{author}{{Elliott}, J.},
  \bibinfo{author}{{Greiner}, J.}, \bibinfo{author}{{Hunt}, L.~K.},
  \bibinfo{author}{{Klose}, S.}, \bibinfo{author}{{Koprowski}, M.~P.},
  \bibinfo{author}{{Le Floc'h}, E.}, \bibinfo{author}{{Malesani}, D.},
  \bibinfo{author}{{Murphy}, T.}, \bibinfo{author}{{Nicuesa Guelbenzu}, A.},
  \bibinfo{author}{{Palazzi}, E.}, \bibinfo{author}{{Rasmussen}, J.},
  \bibinfo{author}{{Rossi}, A.}, \bibinfo{author}{{Savaglio}, S.},
  \bibinfo{author}{{Schady}, P.}, \bibinfo{author}{{Sollerman}, J.},
  \bibinfo{author}{{de Ugarte Postigo}, A.}, \bibinfo{author}{{Watson}, D.},
  \bibinfo{author}{{van der Werf}, P.}, \bibinfo{author}{{Vergani}, S.~D.}, \&
  \bibinfo{author}{{Xu}, D.} (\bibinfo{year}{2015}).
\newblock \bibinfo{title}{{Massive stars formed in atomic hydrogen reservoirs:
  H I observations of gamma-ray burst host galaxies}}.
\newblock {\it \bibinfo{journal}{\aap}\/},  {\it \bibinfo{volume}{582}\/},
  \bibinfo{pages}{A78}. \DOIprefix\doi{10.1051/0004-6361/201526542}.
  \href{http://arxiv.org/abs/1508.03094}{\tt arXiv:1508.03094}.
\bibitem[{{Mukai} et~al.(2008){Mukai}, {Orio} \& {Della Valle}}]{Mukai2008}
\bibinfo{author}{{Mukai}, K.}, \bibinfo{author}{{Orio}, M.}, \&
  \bibinfo{author}{{Della Valle}, M.} (\bibinfo{year}{2008}).
\newblock \bibinfo{title}{{Novae as a Class of Transient X-Ray Sources}}.
\newblock {\it \bibinfo{journal}{\apj}\/},  {\it \bibinfo{volume}{677}\/},
  \bibinfo{pages}{1248--1252}. \DOIprefix\doi{10.1086/529362}.
\bibitem[{{Mushotzky} et~al.(1993){Mushotzky}, {Done} \&
  {Pounds}}]{Mushotzky1993}
\bibinfo{author}{{Mushotzky}, R.~F.}, \bibinfo{author}{{Done}, C.}, \&
  \bibinfo{author}{{Pounds}, K.~A.} (\bibinfo{year}{1993}).
\newblock \bibinfo{title}{{X-ray spectra and time variability of active
  galactic nuclei}}.
\newblock {\it \bibinfo{journal}{\araa}\/},  {\it \bibinfo{volume}{31}\/},
  \bibinfo{pages}{717--761}. \DOIprefix\doi{10.1146/annurev.astro.31.1.717}.
\bibitem[{{Nakar} \& {Sari}(2010)}]{Nakar2010}
\bibinfo{author}{{Nakar}, E.}, \& \bibinfo{author}{{Sari}, R.}
  (\bibinfo{year}{2010}).
\newblock \bibinfo{title}{{Early Supernovae Light Curves Following the Shock
  Breakout}}.
\newblock {\it \bibinfo{journal}{\apj}\/},  {\it \bibinfo{volume}{725}\/},
  \bibinfo{pages}{904--921}. \DOIprefix\doi{10.1088/0004-637X/725/1/904}.
  \href{http://arxiv.org/abs/1004.2496}{\tt arXiv:1004.2496}.
\bibitem[{{Nakar} \& {Sari}(2012)}]{Nakar2012}
\bibinfo{author}{{Nakar}, E.}, \& \bibinfo{author}{{Sari}, R.}
  (\bibinfo{year}{2012}).
\newblock \bibinfo{title}{{Relativistic Shock Breakouts - A Variety of
  Gamma-Ray Flares: From Low-luminosity Gamma-Ray Bursts to Type Ia
  Supernovae}}.
\newblock {\it \bibinfo{journal}{\apj}\/},  {\it \bibinfo{volume}{747}\/},
  \bibinfo{pages}{88}. \DOIprefix\doi{10.1088/0004-637X/747/2/88}.
  \href{http://arxiv.org/abs/1106.2556}{\tt arXiv:1106.2556}.
\bibitem[{{Narayana Bhat} et~al.(2016){Narayana Bhat}, {Meegan}, {von Kienlin},
  {Paciesas}, {Briggs}, {Burgess}, {Burns}, {Chaplin}, {Cleveland}, {Collazzi},
  {Connaughton}, {Diekmann}, {Fitzpatrick}, {Gibby}, {Giles}, {Goldstein},
  {Greiner}, {Jenke}, {Kippen}, {Kouveliotou}, {Mailyan}, {McBreen}, {Pelassa},
  {Preece}, {Roberts}, {Sparke}, {Stanbro}, {Veres}, {Wilson-Hodge}, {Xiong},
  {Younes}, {Yu} \& {Zhang}}]{Narayana2016}
\bibinfo{author}{{Narayana Bhat}, P.}, \bibinfo{author}{{Meegan}, C.~A.},
  \bibinfo{author}{{von Kienlin}, A.}, \bibinfo{author}{{Paciesas}, W.~S.},
  \bibinfo{author}{{Briggs}, M.~S.}, \bibinfo{author}{{Burgess}, J.~M.},
  \bibinfo{author}{{Burns}, E.}, \bibinfo{author}{{Chaplin}, V.},
  \bibinfo{author}{{Cleveland}, W.~H.}, \bibinfo{author}{{Collazzi}, A.~C.},
  \bibinfo{author}{{Connaughton}, V.}, \bibinfo{author}{{Diekmann}, A.~M.},
  \bibinfo{author}{{Fitzpatrick}, G.}, \bibinfo{author}{{Gibby}, M.~H.},
  \bibinfo{author}{{Giles}, M.~M.}, \bibinfo{author}{{Goldstein}, A.~M.},
  \bibinfo{author}{{Greiner}, J.}, \bibinfo{author}{{Jenke}, P.~A.},
  \bibinfo{author}{{Kippen}, R.~M.}, \bibinfo{author}{{Kouveliotou}, C.},
  \bibinfo{author}{{Mailyan}, B.}, \bibinfo{author}{{McBreen}, S.},
  \bibinfo{author}{{Pelassa}, V.}, \bibinfo{author}{{Preece}, R.~D.},
  \bibinfo{author}{{Roberts}, O.~J.}, \bibinfo{author}{{Sparke}, L.~S.},
  \bibinfo{author}{{Stanbro}, M.}, \bibinfo{author}{{Veres}, P.},
  \bibinfo{author}{{Wilson-Hodge}, C.~A.}, \bibinfo{author}{{Xiong}, S.},
  \bibinfo{author}{{Younes}, G.}, \bibinfo{author}{{Yu}, H.-F.}, \&
  \bibinfo{author}{{Zhang}, B.} (\bibinfo{year}{2016}).
\newblock \bibinfo{title}{{The Third Fermi GBM Gamma-Ray Burst Catalog: The
  First Six Years}}.
\newblock {\it \bibinfo{journal}{\apjs}\/},  {\it \bibinfo{volume}{223}\/},
  \bibinfo{pages}{28}. \DOIprefix\doi{10.3847/0067-0049/223/2/28}.
  \href{http://arxiv.org/abs/1603.07612}{\tt arXiv:1603.07612}.
\bibitem[{{Nissanke} et~al.(2010){Nissanke}, {Holz}, {Hughes}, {Dalal} \&
  {Sievers}}]{Nissanke2010}
\bibinfo{author}{{Nissanke}, S.}, \bibinfo{author}{{Holz}, D.~E.},
  \bibinfo{author}{{Hughes}, S.~A.}, \bibinfo{author}{{Dalal}, N.}, \&
  \bibinfo{author}{{Sievers}, J.~L.} (\bibinfo{year}{2010}).
\newblock \bibinfo{title}{{Exploring Short Gamma-ray Bursts as
  Gravitational-wave Standard Sirens}}.
\newblock {\it \bibinfo{journal}{\apj}\/},  {\it \bibinfo{volume}{725}\/},
  \bibinfo{pages}{496--514}. \DOIprefix\doi{10.1088/0004-637X/725/1/496}.
  \href{http://arxiv.org/abs/0904.1017}{\tt arXiv:0904.1017}.
\bibitem[{{Nousek} et~al.(2006){Nousek}, {Kouveliotou}, {Grupe}, {Page},
  {Granot}, {Ramirez-Ruiz}, {Patel}, {Burrows}, {Mangano}, {Barthelmy},
  {Beardmore}, {Campana}, {Capalbi}, {Chincarini}, {Cusumano}, {Falcone},
  {Gehrels}, {Giommi}, {Goad}, {Godet}, {Hurkett}, {Kennea}, {Moretti},
  {O'Brien}, {Osborne}, {Romano}, {Tagliaferri} \& {Wells}}]{Nousek2006}
\bibinfo{author}{{Nousek}, J.~A.}, \bibinfo{author}{{Kouveliotou}, C.},
  \bibinfo{author}{{Grupe}, D.}, \bibinfo{author}{{Page}, K.~L.},
  \bibinfo{author}{{Granot}, J.}, \bibinfo{author}{{Ramirez-Ruiz}, E.},
  \bibinfo{author}{{Patel}, S.~K.}, \bibinfo{author}{{Burrows}, D.~N.},
  \bibinfo{author}{{Mangano}, V.}, \bibinfo{author}{{Barthelmy}, S.},
  \bibinfo{author}{{Beardmore}, A.~P.}, \bibinfo{author}{{Campana}, S.},
  \bibinfo{author}{{Capalbi}, M.}, \bibinfo{author}{{Chincarini}, G.},
  \bibinfo{author}{{Cusumano}, G.}, \bibinfo{author}{{Falcone}, A.~D.},
  \bibinfo{author}{{Gehrels}, N.}, \bibinfo{author}{{Giommi}, P.},
  \bibinfo{author}{{Goad}, M.~R.}, \bibinfo{author}{{Godet}, O.},
  \bibinfo{author}{{Hurkett}, C.~P.}, \bibinfo{author}{{Kennea}, J.~A.},
  \bibinfo{author}{{Moretti}, A.}, \bibinfo{author}{{O'Brien}, P.~T.},
  \bibinfo{author}{{Osborne}, J.~P.}, \bibinfo{author}{{Romano}, P.},
  \bibinfo{author}{{Tagliaferri}, G.}, \& \bibinfo{author}{{Wells}, A.~A.}
  (\bibinfo{year}{2006}).
\newblock \bibinfo{title}{{Evidence for a Canonical Gamma-Ray Burst Afterglow
  Light Curve in the Swift XRT Data}}.
\newblock {\it \bibinfo{journal}{\apj}\/},  {\it \bibinfo{volume}{642}\/},
  \bibinfo{pages}{389--400}. \DOIprefix\doi{10.1086/500724}.
  \href{http://arxiv.org/abs/astro-ph/0508332}{\tt arXiv:astro-ph/0508332}.
\bibitem[{{Osborne}(2015)}]{osborne15}
\bibinfo{author}{{Osborne}, J.~P.} (\bibinfo{year}{2015}).
\newblock \bibinfo{title}{{Getting to know classical novae with Swift}}.
\newblock {\it \bibinfo{journal}{Journal of High Energy Astrophysics}\/},  {\it
  \bibinfo{volume}{7}\/}, \bibinfo{pages}{117--125}.
  \DOIprefix\doi{10.1016/j.jheap.2015.06.005}.
  \href{http://arxiv.org/abs/1507.02153}{\tt arXiv:1507.02153}.
\bibitem[{{Osborne} et~al.(2011){Osborne}, {Page}, {Beardmore}, {Bode}, {Goad},
  {O'Brien}, {Starrfield}, {Rauch}, {Ness}, {Krautter}, {Schwarz}, {Burrows},
  {Gehrels}, {Drake}, {Evans} \& {Eyres}}]{Osborne2011}
\bibinfo{author}{{Osborne}, J.~P.}, \bibinfo{author}{{Page}, K.~L.},
  \bibinfo{author}{{Beardmore}, A.~P.}, \bibinfo{author}{{Bode}, M.~F.},
  \bibinfo{author}{{Goad}, M.~R.}, \bibinfo{author}{{O'Brien}, T.~J.},
  \bibinfo{author}{{Starrfield}, S.}, \bibinfo{author}{{Rauch}, T.},
  \bibinfo{author}{{Ness}, J.-U.}, \bibinfo{author}{{Krautter}, J.},
  \bibinfo{author}{{Schwarz}, G.}, \bibinfo{author}{{Burrows}, D.~N.},
  \bibinfo{author}{{Gehrels}, N.}, \bibinfo{author}{{Drake}, J.~J.},
  \bibinfo{author}{{Evans}, A.}, \& \bibinfo{author}{{Eyres}, S.~P.~S.}
  (\bibinfo{year}{2011}).
\newblock \bibinfo{title}{{The Supersoft X-ray Phase of Nova RS Ophiuchi
  2006}}.
\newblock {\it \bibinfo{journal}{\apj}\/},  {\it \bibinfo{volume}{727}\/},
  \bibinfo{pages}{124}. \DOIprefix\doi{10.1088/0004-637X/727/2/124}.
  \href{http://arxiv.org/abs/1011.5327}{\tt arXiv:1011.5327}.
\bibitem[{{Osten} et~al.(2007){Osten}, {Drake}, {Tueller}, {Cummings}, {Perri},
  {Moretti} \& {Covino}}]{Osten2007}
\bibinfo{author}{{Osten}, R.~A.}, \bibinfo{author}{{Drake}, S.},
  \bibinfo{author}{{Tueller}, J.}, \bibinfo{author}{{Cummings}, J.},
  \bibinfo{author}{{Perri}, M.}, \bibinfo{author}{{Moretti}, A.}, \&
  \bibinfo{author}{{Covino}, S.} (\bibinfo{year}{2007}).
\newblock \bibinfo{title}{{Nonthermal Hard X-Ray Emission and Iron K{$\alpha$}
  Emission from a Superflare on II Pegasi}}.
\newblock {\it \bibinfo{journal}{\apj}\/},  {\it \bibinfo{volume}{654}\/},
  \bibinfo{pages}{1052--1067}. \DOIprefix\doi{10.1086/509252}.
  \href{http://arxiv.org/abs/astro-ph/0609205}{\tt arXiv:astro-ph/0609205}.
\bibitem[{{Osten} et~al.(2010){Osten}, {Godet}, {Drake}, {Tueller}, {Cummings},
  {Krimm}, {Pye}, {Pal'shin}, {Golenetskii}, {Reale}, {Oates}, {Page} \&
  {Melandri}}]{Osten2010}
\bibinfo{author}{{Osten}, R.~A.}, \bibinfo{author}{{Godet}, O.},
  \bibinfo{author}{{Drake}, S.}, \bibinfo{author}{{Tueller}, J.},
  \bibinfo{author}{{Cummings}, J.}, \bibinfo{author}{{Krimm}, H.},
  \bibinfo{author}{{Pye}, J.}, \bibinfo{author}{{Pal'shin}, V.},
  \bibinfo{author}{{Golenetskii}, S.}, \bibinfo{author}{{Reale}, F.},
  \bibinfo{author}{{Oates}, S.~R.}, \bibinfo{author}{{Page}, M.~J.}, \&
  \bibinfo{author}{{Melandri}, A.} (\bibinfo{year}{2010}).
\newblock \bibinfo{title}{{The Mouse That Roared: A Superflare from the dMe
  Flare Star EV Lac Detected by Swift and Konus-Wind}}.
\newblock {\it \bibinfo{journal}{\apj}\/},  {\it \bibinfo{volume}{721}\/},
  \bibinfo{pages}{785--801}. \DOIprefix\doi{10.1088/0004-637X/721/1/785}.
  \href{http://arxiv.org/abs/1007.5300}{\tt arXiv:1007.5300}.
\bibitem[{{Paczy{\'n}ski}(1998)}]{Paczynski1998}
\bibinfo{author}{{Paczy{\'n}ski}, B.} (\bibinfo{year}{1998}).
\newblock \bibinfo{title}{{Are Gamma-Ray Bursts in Star-Forming Regions?}}
\newblock {\it \bibinfo{journal}{\apjl}\/},  {\it \bibinfo{volume}{494}\/},
  \bibinfo{pages}{L45--L48}. \DOIprefix\doi{10.1086/311148}.
  \href{http://arxiv.org/abs/astro-ph/9710086}{\tt arXiv:astro-ph/9710086}.
\bibitem[{{Patil} et~al.(2014){Patil}, {Zaroubi}, {Chapman}, {Jeli{\'c}},
  {Harker}, {Abdalla}, {Asad}, {Bernardi}, {Brentjens}, {de Bruyn}, {Bus},
  {Ciardi}, {Daiboo}, {Fernandez}, {Ghosh}, {Jensen}, {Kazemi}, {Koopmans},
  {Labropoulos}, {Mevius}, {Martinez}, {Mellema}, {Offringa}, {Pandey},
  {Schaye}, {Thomas}, {Vedantham}, {Veligatla}, {Wijnholds} \&
  {Yatawatta}}]{Patil2014}
\bibinfo{author}{{Patil}, A.~H.}, \bibinfo{author}{{Zaroubi}, S.},
  \bibinfo{author}{{Chapman}, E.}, \bibinfo{author}{{Jeli{\'c}}, V.},
  \bibinfo{author}{{Harker}, G.}, \bibinfo{author}{{Abdalla}, F.~B.},
  \bibinfo{author}{{Asad}, K.~M.~B.}, \bibinfo{author}{{Bernardi}, G.},
  \bibinfo{author}{{Brentjens}, M.~A.}, \bibinfo{author}{{de Bruyn}, A.~G.},
  \bibinfo{author}{{Bus}, S.}, \bibinfo{author}{{Ciardi}, B.},
  \bibinfo{author}{{Daiboo}, S.}, \bibinfo{author}{{Fernandez}, E.~R.},
  \bibinfo{author}{{Ghosh}, A.}, \bibinfo{author}{{Jensen}, H.},
  \bibinfo{author}{{Kazemi}, S.}, \bibinfo{author}{{Koopmans}, L.~V.~E.},
  \bibinfo{author}{{Labropoulos}, P.}, \bibinfo{author}{{Mevius}, M.},
  \bibinfo{author}{{Martinez}, O.}, \bibinfo{author}{{Mellema}, G.},
  \bibinfo{author}{{Offringa}, A.~R.}, \bibinfo{author}{{Pandey}, V.~N.},
  \bibinfo{author}{{Schaye}, J.}, \bibinfo{author}{{Thomas}, R.~M.},
  \bibinfo{author}{{Vedantham}, H.~K.}, \bibinfo{author}{{Veligatla}, V.},
  \bibinfo{author}{{Wijnholds}, S.~J.}, \& \bibinfo{author}{{Yatawatta}, S.}
  (\bibinfo{year}{2014}).
\newblock \bibinfo{title}{{Constraining the epoch of reionization with the
  variance statistic: simulations of the LOFAR case}}.
\newblock {\it \bibinfo{journal}{\mnras}\/},  {\it \bibinfo{volume}{443}\/},
  \bibinfo{pages}{1113--1124}. \DOIprefix\doi{10.1093/mnras/stu1178}.
  \href{http://arxiv.org/abs/1401.4172}{\tt arXiv:1401.4172}.
\bibitem[{{Perley} et~al.(2016){Perley}, {Tanvir}, {Hjorth}, {Laskar},
  {Berger}, {Chary}, {de Ugarte Postigo}, {Fynbo}, {Kr{\"u}hler}, {Levan},
  {Micha{\l}owski} \& {Schulze}}]{Perley2016}
\bibinfo{author}{{Perley}, D.~A.}, \bibinfo{author}{{Tanvir}, N.~R.},
  \bibinfo{author}{{Hjorth}, J.}, \bibinfo{author}{{Laskar}, T.},
  \bibinfo{author}{{Berger}, E.}, \bibinfo{author}{{Chary}, R.},
  \bibinfo{author}{{de Ugarte Postigo}, A.}, \bibinfo{author}{{Fynbo},
  J.~P.~U.}, \bibinfo{author}{{Kr{\"u}hler}, T.}, \bibinfo{author}{{Levan},
  A.~J.}, \bibinfo{author}{{Micha{\l}owski}, M.~J.}, \&
  \bibinfo{author}{{Schulze}, S.} (\bibinfo{year}{2016}).
\newblock \bibinfo{title}{{The Swift GRB Host Galaxy Legacy Survey. II.
  Rest-frame Near-IR Luminosity Distribution and Evidence for a Near-solar
  Metallicity Threshold}}.
\newblock {\it \bibinfo{journal}{\apj}\/},  {\it \bibinfo{volume}{817}\/},
  \bibinfo{pages}{8}. \DOIprefix\doi{10.3847/0004-637X/817/1/8}.
  \href{http://arxiv.org/abs/1504.02479}{\tt arXiv:1504.02479}.
\bibitem[{{Pescalli} et~al.(2016){Pescalli}, {Ghirlanda}, {Salvaterra},
  {Ghisellini}, {Vergani}, {Nappo}, {Salafia}, {Melandri}, {Covino} \&
  {G{\"o}tz}}]{Pescalli2016}
\bibinfo{author}{{Pescalli}, A.}, \bibinfo{author}{{Ghirlanda}, G.},
  \bibinfo{author}{{Salvaterra}, R.}, \bibinfo{author}{{Ghisellini}, G.},
  \bibinfo{author}{{Vergani}, S.~D.}, \bibinfo{author}{{Nappo}, F.},
  \bibinfo{author}{{Salafia}, O.~S.}, \bibinfo{author}{{Melandri}, A.},
  \bibinfo{author}{{Covino}, S.}, \& \bibinfo{author}{{G{\"o}tz}, D.}
  (\bibinfo{year}{2016}).
\newblock \bibinfo{title}{{The rate and luminosity function of long gamma ray
  bursts}}.
\newblock {\it \bibinfo{journal}{\aap}\/},  {\it \bibinfo{volume}{587}\/},
  \bibinfo{pages}{A40}. \DOIprefix\doi{10.1051/0004-6361/201526760}.
  \href{http://arxiv.org/abs/1506.05463}{\tt arXiv:1506.05463}.
\bibitem[{{Petrosian} et~al.(2015){Petrosian}, {Kitanidis} \&
  {Kocevski}}]{Petrosian2015}
\bibinfo{author}{{Petrosian}, V.}, \bibinfo{author}{{Kitanidis}, E.}, \&
  \bibinfo{author}{{Kocevski}, D.} (\bibinfo{year}{2015}).
\newblock \bibinfo{title}{{Cosmological Evolution of Long Gamma-Ray Bursts and
  the Star Formation Rate}}.
\newblock {\it \bibinfo{journal}{\apj}\/},  {\it \bibinfo{volume}{806}\/},
  \bibinfo{pages}{44}. \DOIprefix\doi{10.1088/0004-637X/806/1/44}.
  \href{http://arxiv.org/abs/1504.01414}{\tt arXiv:1504.01414}.
\bibitem[{{Phinney}(2009)}]{Phinney2009}
\bibinfo{author}{{Phinney}, E.~S.} (\bibinfo{year}{2009}).
\newblock \bibinfo{title}{{Finding and Using Electromagnetic Counterparts of
  Gravitational Wave Sources}}.
\newblock In {\it \bibinfo{booktitle}{astro2010: The Astronomy and Astrophysics
  Decadal Survey}\/}.
\newblock volume \bibinfo{volume}{2010} of {\it \bibinfo{series}{ArXiv
  Astrophysics e-prints}\/}.
\newblock \href{http://arxiv.org/abs/0903.0098}{\tt arXiv:0903.0098}.
\bibitem[{{Pian} et~al.(2017){Pian}, {D'Avanzo}, {Benetti}, {Branchesi},
  {Brocato}, {Campana}, {Cappellaro}, {Covino}, {D'Elia}, {Fynbo}, {Getman},
  {Ghirlanda}, {Ghisellini}, {Grado}, {Greco}, {Hjorth}, {Kouveliotou},
  {Levan}, {Limatola}, {Malesani}, {Mazzali}, {Melandri}, {M{\o}ller},
  {Nicastro}, {Palazzi}, {Piranomonte}, {Rossi}, {Salafia}, {Selsing},
  {Stratta}, {Tanaka}, {Tanvir}, {Tomasella}, {Watson}, {Yang}, {Amati},
  {Antonelli}, {Ascenzi}, {Bernardini}, {Bo{\"e}r}, {Bufano}, {Bulgarelli},
  {Capaccioli}, {Casella}, {Castro-Tirado}, {Chassande-Mottin}, {Ciolfi},
  {Copperwheat}, {Dadina}, {De Cesare}, {di Paola}, {Fan}, {Gendre},
  {Giuffrida}, {Giunta}, {Hunt}, {Israel}, {Jin}, {Kasliwal}, {Klose}, {Lisi},
  {Longo}, {Maiorano}, {Mapelli}, {Masetti}, {Nava}, {Patricelli}, {Perley},
  {Pescalli}, {Piran}, {Possenti}, {Pulone}, {Razzano}, {Salvaterra},
  {Schipani}, {Spera}, {Stamerra}, {Stella}, {Tagliaferri}, {Testa}, {Troja},
  {Turatto}, {Vergani} \& {Vergani}}]{Pian2017}
\bibinfo{author}{{Pian}, E.}, \bibinfo{author}{{D'Avanzo}, P.},
  \bibinfo{author}{{Benetti}, S.}, \bibinfo{author}{{Branchesi}, M.},
  \bibinfo{author}{{Brocato}, E.}, \bibinfo{author}{{Campana}, S.},
  \bibinfo{author}{{Cappellaro}, E.}, \bibinfo{author}{{Covino}, S.},
  \bibinfo{author}{{D'Elia}, V.}, \bibinfo{author}{{Fynbo}, J.~P.~U.},
  \bibinfo{author}{{Getman}, F.}, \bibinfo{author}{{Ghirlanda}, G.},
  \bibinfo{author}{{Ghisellini}, G.}, \bibinfo{author}{{Grado}, A.},
  \bibinfo{author}{{Greco}, G.}, \bibinfo{author}{{Hjorth}, J.},
  \bibinfo{author}{{Kouveliotou}, C.}, \bibinfo{author}{{Levan}, A.},
  \bibinfo{author}{{Limatola}, L.}, \bibinfo{author}{{Malesani}, D.},
  \bibinfo{author}{{Mazzali}, P.~A.}, \bibinfo{author}{{Melandri}, A.},
  \bibinfo{author}{{M{\o}ller}, P.}, \bibinfo{author}{{Nicastro}, L.},
  \bibinfo{author}{{Palazzi}, E.}, \bibinfo{author}{{Piranomonte}, S.},
  \bibinfo{author}{{Rossi}, A.}, \bibinfo{author}{{Salafia}, O.~S.},
  \bibinfo{author}{{Selsing}, J.}, \bibinfo{author}{{Stratta}, G.},
  \bibinfo{author}{{Tanaka}, M.}, \bibinfo{author}{{Tanvir}, N.~R.},
  \bibinfo{author}{{Tomasella}, L.}, \bibinfo{author}{{Watson}, D.},
  \bibinfo{author}{{Yang}, S.}, \bibinfo{author}{{Amati}, L.},
  \bibinfo{author}{{Antonelli}, L.~A.}, \bibinfo{author}{{Ascenzi}, S.},
  \bibinfo{author}{{Bernardini}, M.~G.}, \bibinfo{author}{{Bo{\"e}r}, M.},
  \bibinfo{author}{{Bufano}, F.}, \bibinfo{author}{{Bulgarelli}, A.},
  \bibinfo{author}{{Capaccioli}, M.}, \bibinfo{author}{{Casella}, P.},
  \bibinfo{author}{{Castro-Tirado}, A.~J.},
  \bibinfo{author}{{Chassande-Mottin}, E.}, \bibinfo{author}{{Ciolfi}, R.},
  \bibinfo{author}{{Copperwheat}, C.~M.}, \bibinfo{author}{{Dadina}, M.},
  \bibinfo{author}{{De Cesare}, G.}, \bibinfo{author}{{di Paola}, A.},
  \bibinfo{author}{{Fan}, Y.~Z.}, \bibinfo{author}{{Gendre}, B.},
  \bibinfo{author}{{Giuffrida}, G.}, \bibinfo{author}{{Giunta}, A.},
  \bibinfo{author}{{Hunt}, L.~K.}, \bibinfo{author}{{Israel}, G.~L.},
  \bibinfo{author}{{Jin}, Z.-P.}, \bibinfo{author}{{Kasliwal}, M.~M.},
  \bibinfo{author}{{Klose}, S.}, \bibinfo{author}{{Lisi}, M.},
  \bibinfo{author}{{Longo}, F.}, \bibinfo{author}{{Maiorano}, E.},
  \bibinfo{author}{{Mapelli}, M.}, \bibinfo{author}{{Masetti}, N.},
  \bibinfo{author}{{Nava}, L.}, \bibinfo{author}{{Patricelli}, B.},
  \bibinfo{author}{{Perley}, D.}, \bibinfo{author}{{Pescalli}, A.},
  \bibinfo{author}{{Piran}, T.}, \bibinfo{author}{{Possenti}, A.},
  \bibinfo{author}{{Pulone}, L.}, \bibinfo{author}{{Razzano}, M.},
  \bibinfo{author}{{Salvaterra}, R.}, \bibinfo{author}{{Schipani}, P.},
  \bibinfo{author}{{Spera}, M.}, \bibinfo{author}{{Stamerra}, A.},
  \bibinfo{author}{{Stella}, L.}, \bibinfo{author}{{Tagliaferri}, G.},
  \bibinfo{author}{{Testa}, V.}, \bibinfo{author}{{Troja}, E.},
  \bibinfo{author}{{Turatto}, M.}, \bibinfo{author}{{Vergani}, S.~D.}, \&
  \bibinfo{author}{{Vergani}, D.} (\bibinfo{year}{2017}).
\newblock \bibinfo{title}{{Spectroscopic identification of r-process
  nucleosynthesis in a double neutron-star merger}}.
\newblock {\it \bibinfo{journal}{\nat}\/},  {\it \bibinfo{volume}{551}\/},
  \bibinfo{pages}{67--70}. \DOIprefix\doi{10.1038/nature24298}.
  \href{http://arxiv.org/abs/1710.05858}{\tt arXiv:1710.05858}.
\bibitem[{{Piro} et~al.(2017){Piro}, {Giacomazzo} \& {Perna}}]{Piro2017}
\bibinfo{author}{{Piro}, A.~L.}, \bibinfo{author}{{Giacomazzo}, B.}, \&
  \bibinfo{author}{{Perna}, R.} (\bibinfo{year}{2017}).
\newblock \bibinfo{title}{{The Fate of Neutron Star Binary Mergers}}.
\newblock {\it \bibinfo{journal}{\apjl}\/},  {\it \bibinfo{volume}{844}\/},
  \bibinfo{pages}{L19}. \DOIprefix\doi{10.3847/2041-8213/aa7f2f}.
  \href{http://arxiv.org/abs/1704.08697}{\tt arXiv:1704.08697}.
\bibitem[{{Planck Collaboration} et~al.(2016){Planck Collaboration}, {Adam},
  {Aghanim}, {Ashdown}, {Aumont}, {Baccigalupi}, {Ballardini}, {Banday},
  {Barreiro}, {Bartolo}, {Basak}, {Battye}, {Benabed}, {Bernard}, {Bersanelli},
  {Bielewicz}, {Bock}, {Bonaldi}, {Bonavera}, {Bond}, {Borrill}, {Bouchet},
  {Boulanger}, {Bucher}, {Burigana}, {Calabrese}, {Cardoso}, {Carron},
  {Chiang}, {Colombo}, {Combet}, {Comis}, {Couchot}, {Coulais}, {Crill},
  {Curto}, {Cuttaia}, {Davis}, {de Bernardis}, {de Rosa}, {de Zotti},
  {Delabrouille}, {Di Valentino}, {Dickinson}, {Diego}, {Dor{\'e}}, {Douspis},
  {Ducout}, {Dupac}, {Elsner}, {En{\ss}lin}, {Eriksen}, {Falgarone}, {Fantaye},
  {Finelli}, {Forastieri}, {Frailis}, {Fraisse}, {Franceschi}, {Frolov},
  {Galeotta}, {Galli}, {Ganga}, {G{\'e}nova-Santos}, {Gerbino}, {Ghosh},
  {Gonz{\'a}lez-Nuevo}, {G{\'o}rski}, {Gruppuso}, {Gudmundsson}, {Hansen},
  {Helou}, {Henrot-Versill{\'e}}, {Herranz}, {Hivon}, {Huang}, {Ili{\'c}},
  {Jaffe}, {Jones}, {Keih{\"a}nen}, {Keskitalo}, {Kisner}, {Knox},
  {Krachmalnicoff}, {Kunz}, {Kurki-Suonio}, {Lagache}, {L{\"a}hteenm{\"a}ki},
  {Lamarre}, {Langer}, {Lasenby}, {Lattanzi}, {Lawrence}, {Le Jeune},
  {Levrier}, {Lewis}, {Liguori}, {Lilje}, {L{\'o}pez-Caniego}
  et~al.}]{Planck2016}
\bibinfo{author}{{Planck Collaboration}}, \bibinfo{author}{{Adam}, R.},
  \bibinfo{author}{{Aghanim}, N.}, \bibinfo{author}{{Ashdown}, M.},
  \bibinfo{author}{{Aumont}, J.}, \bibinfo{author}{{Baccigalupi}, C.},
  \bibinfo{author}{{Ballardini}, M.}, \bibinfo{author}{{Banday}, A.~J.},
  \bibinfo{author}{{Barreiro}, R.~B.}, \bibinfo{author}{{Bartolo}, N.},
  \bibinfo{author}{{Basak}, S.}, \bibinfo{author}{{Battye}, R.},
  \bibinfo{author}{{Benabed}, K.}, \bibinfo{author}{{Bernard}, J.-P.},
  \bibinfo{author}{{Bersanelli}, M.}, \bibinfo{author}{{Bielewicz}, P.},
  \bibinfo{author}{{Bock}, J.~J.}, \bibinfo{author}{{Bonaldi}, A.},
  \bibinfo{author}{{Bonavera}, L.}, \bibinfo{author}{{Bond}, J.~R.},
  \bibinfo{author}{{Borrill}, J.}, \bibinfo{author}{{Bouchet}, F.~R.},
  \bibinfo{author}{{Boulanger}, F.}, \bibinfo{author}{{Bucher}, M.},
  \bibinfo{author}{{Burigana}, C.}, \bibinfo{author}{{Calabrese}, E.},
  \bibinfo{author}{{Cardoso}, J.-F.}, \bibinfo{author}{{Carron}, J.},
  \bibinfo{author}{{Chiang}, H.~C.}, \bibinfo{author}{{Colombo}, L.~P.~L.},
  \bibinfo{author}{{Combet}, C.}, \bibinfo{author}{{Comis}, B.},
  \bibinfo{author}{{Couchot}, F.}, \bibinfo{author}{{Coulais}, A.},
  \bibinfo{author}{{Crill}, B.~P.}, \bibinfo{author}{{Curto}, A.},
  \bibinfo{author}{{Cuttaia}, F.}, \bibinfo{author}{{Davis}, R.~J.},
  \bibinfo{author}{{de Bernardis}, P.}, \bibinfo{author}{{de Rosa}, A.},
  \bibinfo{author}{{de Zotti}, G.}, \bibinfo{author}{{Delabrouille}, J.},
  \bibinfo{author}{{Di Valentino}, E.}, \bibinfo{author}{{Dickinson}, C.},
  \bibinfo{author}{{Diego}, J.~M.}, \bibinfo{author}{{Dor{\'e}}, O.},
  \bibinfo{author}{{Douspis}, M.}, \bibinfo{author}{{Ducout}, A.},
  \bibinfo{author}{{Dupac}, X.}, \bibinfo{author}{{Elsner}, F.},
  \bibinfo{author}{{En{\ss}lin}, T.~A.}, \bibinfo{author}{{Eriksen}, H.~K.},
  \bibinfo{author}{{Falgarone}, E.}, \bibinfo{author}{{Fantaye}, Y.},
  \bibinfo{author}{{Finelli}, F.}, \bibinfo{author}{{Forastieri}, F.},
  \bibinfo{author}{{Frailis}, M.}, \bibinfo{author}{{Fraisse}, A.~A.},
  \bibinfo{author}{{Franceschi}, E.}, \bibinfo{author}{{Frolov}, A.},
  \bibinfo{author}{{Galeotta}, S.}, \bibinfo{author}{{Galli}, S.},
  \bibinfo{author}{{Ganga}, K.}, \bibinfo{author}{{G{\'e}nova-Santos}, R.~T.},
  \bibinfo{author}{{Gerbino}, M.}, \bibinfo{author}{{Ghosh}, T.},
  \bibinfo{author}{{Gonz{\'a}lez-Nuevo}, J.}, \bibinfo{author}{{G{\'o}rski},
  K.~M.}, \bibinfo{author}{{Gruppuso}, A.}, \bibinfo{author}{{Gudmundsson},
  J.~E.}, \bibinfo{author}{{Hansen}, F.~K.}, \bibinfo{author}{{Helou}, G.},
  \bibinfo{author}{{Henrot-Versill{\'e}}, S.}, \bibinfo{author}{{Herranz}, D.},
  \bibinfo{author}{{Hivon}, E.}, \bibinfo{author}{{Huang}, Z.},
  \bibinfo{author}{{Ili{\'c}}, S.}, \bibinfo{author}{{Jaffe}, A.~H.},
  \bibinfo{author}{{Jones}, W.~C.}, \bibinfo{author}{{Keih{\"a}nen}, E.},
  \bibinfo{author}{{Keskitalo}, R.}, \bibinfo{author}{{Kisner}, T.~S.},
  \bibinfo{author}{{Knox}, L.}, \bibinfo{author}{{Krachmalnicoff}, N.},
  \bibinfo{author}{{Kunz}, M.}, \bibinfo{author}{{Kurki-Suonio}, H.},
  \bibinfo{author}{{Lagache}, G.}, \bibinfo{author}{{L{\"a}hteenm{\"a}ki}, A.},
  \bibinfo{author}{{Lamarre}, J.-M.}, \bibinfo{author}{{Langer}, M.},
  \bibinfo{author}{{Lasenby}, A.}, \bibinfo{author}{{Lattanzi}, M.},
  \bibinfo{author}{{Lawrence}, C.~R.}, \bibinfo{author}{{Le Jeune}, M.},
  \bibinfo{author}{{Levrier}, F.}, \bibinfo{author}{{Lewis}, A.},
  \bibinfo{author}{{Liguori}, M.}, \bibinfo{author}{{Lilje}, P.~B.},
  \bibinfo{author}{{L{\'o}pez-Caniego}, M.} et~al. (\bibinfo{year}{2016}).
\newblock \bibinfo{title}{{Planck intermediate results. XLVII. Planck
  constraints on reionization history}}.
\newblock {\it \bibinfo{journal}{\aap}\/},  {\it \bibinfo{volume}{596}\/},
  \bibinfo{pages}{A108}. \DOIprefix\doi{10.1051/0004-6361/201628897}.
  \href{http://arxiv.org/abs/1605.03507}{\tt arXiv:1605.03507}.
\bibitem[{{Punturo} et~al.(2010){Punturo}, {Abernathy}, {Acernese}, {Allen},
  {Andersson}, {Arun}, {Barone}, {Barr}, {Barsuglia}, {Beker}, {Beveridge},
  {Birindelli}, {Bose}, {Bosi}, {Braccini}, {Bradaschia}, {Bulik}, {Calloni},
  {Cella}, {Chassande Mottin}, {Chelkowski}, {Chincarini}, {Clark}, {Coccia},
  {Colacino}, {Colas}, {Cumming}, {Cunningham}, {Cuoco}, {Danilishin},
  {Danzmann}, {De Luca}, {De Salvo}, {Dent}, {De Rosa}, {Di Fiore}, {Di
  Virgilio}, {Doets}, {Fafone}, {Falferi}, {Flaminio}, {Franc}, {Frasconi},
  {Freise}, {Fulda}, {Gair}, {Gemme}, {Gennai}, {Giazotto}, {Glampedakis},
  {Granata}, {Grote}, {Guidi}, {Hammond}, {Hannam}, {Harms}, {Heinert},
  {Hendry}, {Heng}, {Hennes}, {Hild}, {Hough}, {Husa}, {Huttner}, {Jones},
  {Khalili}, {Kokeyama}, {Kokkotas}, {Krishnan}, {Lorenzini}, {L{\"u}ck},
  {Majorana}, {Mandel}, {Mandic}, {Martin}, {Michel}, {Minenkov}, {Morgado},
  {Mosca}, {Mours}, {M{\"u}ller-Ebhardt}, {Murray}, {Nawrodt}, {Nelson},
  {Oshaughnessy}, {Ott}, {Palomba}, {Paoli}, {Parguez}, {Pasqualetti},
  {Passaquieti}, {Passuello}, {Pinard}, {Poggiani}, {Popolizio}, {Prato},
  {Puppo}, {Rabeling}, {Rapagnani} et~al.}]{Punturo2010}
\bibinfo{author}{{Punturo}, M.}, \bibinfo{author}{{Abernathy}, M.},
  \bibinfo{author}{{Acernese}, F.}, \bibinfo{author}{{Allen}, B.},
  \bibinfo{author}{{Andersson}, N.}, \bibinfo{author}{{Arun}, K.},
  \bibinfo{author}{{Barone}, F.}, \bibinfo{author}{{Barr}, B.},
  \bibinfo{author}{{Barsuglia}, M.}, \bibinfo{author}{{Beker}, M.},
  \bibinfo{author}{{Beveridge}, N.}, \bibinfo{author}{{Birindelli}, S.},
  \bibinfo{author}{{Bose}, S.}, \bibinfo{author}{{Bosi}, L.},
  \bibinfo{author}{{Braccini}, S.}, \bibinfo{author}{{Bradaschia}, C.},
  \bibinfo{author}{{Bulik}, T.}, \bibinfo{author}{{Calloni}, E.},
  \bibinfo{author}{{Cella}, G.}, \bibinfo{author}{{Chassande Mottin}, E.},
  \bibinfo{author}{{Chelkowski}, S.}, \bibinfo{author}{{Chincarini}, A.},
  \bibinfo{author}{{Clark}, J.}, \bibinfo{author}{{Coccia}, E.},
  \bibinfo{author}{{Colacino}, C.}, \bibinfo{author}{{Colas}, J.},
  \bibinfo{author}{{Cumming}, A.}, \bibinfo{author}{{Cunningham}, L.},
  \bibinfo{author}{{Cuoco}, E.}, \bibinfo{author}{{Danilishin}, S.},
  \bibinfo{author}{{Danzmann}, K.}, \bibinfo{author}{{De Luca}, G.},
  \bibinfo{author}{{De Salvo}, R.}, \bibinfo{author}{{Dent}, T.},
  \bibinfo{author}{{De Rosa}, R.}, \bibinfo{author}{{Di Fiore}, L.},
  \bibinfo{author}{{Di Virgilio}, A.}, \bibinfo{author}{{Doets}, M.},
  \bibinfo{author}{{Fafone}, V.}, \bibinfo{author}{{Falferi}, P.},
  \bibinfo{author}{{Flaminio}, R.}, \bibinfo{author}{{Franc}, J.},
  \bibinfo{author}{{Frasconi}, F.}, \bibinfo{author}{{Freise}, A.},
  \bibinfo{author}{{Fulda}, P.}, \bibinfo{author}{{Gair}, J.},
  \bibinfo{author}{{Gemme}, G.}, \bibinfo{author}{{Gennai}, A.},
  \bibinfo{author}{{Giazotto}, A.}, \bibinfo{author}{{Glampedakis}, K.},
  \bibinfo{author}{{Granata}, M.}, \bibinfo{author}{{Grote}, H.},
  \bibinfo{author}{{Guidi}, G.}, \bibinfo{author}{{Hammond}, G.},
  \bibinfo{author}{{Hannam}, M.}, \bibinfo{author}{{Harms}, J.},
  \bibinfo{author}{{Heinert}, D.}, \bibinfo{author}{{Hendry}, M.},
  \bibinfo{author}{{Heng}, I.}, \bibinfo{author}{{Hennes}, E.},
  \bibinfo{author}{{Hild}, S.}, \bibinfo{author}{{Hough}, J.},
  \bibinfo{author}{{Husa}, S.}, \bibinfo{author}{{Huttner}, S.},
  \bibinfo{author}{{Jones}, G.}, \bibinfo{author}{{Khalili}, F.},
  \bibinfo{author}{{Kokeyama}, K.}, \bibinfo{author}{{Kokkotas}, K.},
  \bibinfo{author}{{Krishnan}, B.}, \bibinfo{author}{{Lorenzini}, M.},
  \bibinfo{author}{{L{\"u}ck}, H.}, \bibinfo{author}{{Majorana}, E.},
  \bibinfo{author}{{Mandel}, I.}, \bibinfo{author}{{Mandic}, V.},
  \bibinfo{author}{{Martin}, I.}, \bibinfo{author}{{Michel}, C.},
  \bibinfo{author}{{Minenkov}, Y.}, \bibinfo{author}{{Morgado}, N.},
  \bibinfo{author}{{Mosca}, S.}, \bibinfo{author}{{Mours}, B.},
  \bibinfo{author}{{M{\"u}ller-Ebhardt}, H.}, \bibinfo{author}{{Murray}, P.},
  \bibinfo{author}{{Nawrodt}, R.}, \bibinfo{author}{{Nelson}, J.},
  \bibinfo{author}{{Oshaughnessy}, R.}, \bibinfo{author}{{Ott}, C.~D.},
  \bibinfo{author}{{Palomba}, C.}, \bibinfo{author}{{Paoli}, A.},
  \bibinfo{author}{{Parguez}, G.}, \bibinfo{author}{{Pasqualetti}, A.},
  \bibinfo{author}{{Passaquieti}, R.}, \bibinfo{author}{{Passuello}, D.},
  \bibinfo{author}{{Pinard}, L.}, \bibinfo{author}{{Poggiani}, R.},
  \bibinfo{author}{{Popolizio}, P.}, \bibinfo{author}{{Prato}, M.},
  \bibinfo{author}{{Puppo}, P.}, \bibinfo{author}{{Rabeling}, D.},
  \bibinfo{author}{{Rapagnani}, P.} et~al. (\bibinfo{year}{2010}).
\newblock \bibinfo{title}{{The Einstein Telescope: a third-generation
  gravitational wave observatory}}.
\newblock {\it \bibinfo{journal}{Classical and Quantum Gravity}\/},  {\it
  \bibinfo{volume}{27}\/}, \bibinfo{pages}{194002}.
  \DOIprefix\doi{10.1088/0264-9381/27/19/194002}.
\bibitem[{{Pye} \& {McHardy}(1983)}]{Pye1983}
\bibinfo{author}{{Pye}, J.~P.}, \& \bibinfo{author}{{McHardy}, I.~M.}
  (\bibinfo{year}{1983}).
\newblock \bibinfo{title}{{The Ariel V sky survey of fast-transient X-ray
  sources}}.
\newblock {\it \bibinfo{journal}{\mnras}\/},  {\it \bibinfo{volume}{205}\/},
  \bibinfo{pages}{875--888}. \DOIprefix\doi{10.1093/mnras/205.3.875}.
\bibitem[{{Racusin} et~al.(2009){Racusin}, {Liang}, {Burrows}, {Falcone},
  {Sakamoto}, {Zhang}, {Zhang}, {Evans} \& {Osborne}}]{Racusin2009}
\bibinfo{author}{{Racusin}, J.~L.}, \bibinfo{author}{{Liang}, E.~W.},
  \bibinfo{author}{{Burrows}, D.~N.}, \bibinfo{author}{{Falcone}, A.},
  \bibinfo{author}{{Sakamoto}, T.}, \bibinfo{author}{{Zhang}, B.~B.},
  \bibinfo{author}{{Zhang}, B.}, \bibinfo{author}{{Evans}, P.}, \&
  \bibinfo{author}{{Osborne}, J.} (\bibinfo{year}{2009}).
\newblock \bibinfo{title}{{Jet Breaks and Energetics of Swift Gamma-Ray Burst
  X-Ray Afterglows}}.
\newblock {\it \bibinfo{journal}{\apj}\/},  {\it \bibinfo{volume}{698}\/},
  \bibinfo{pages}{43--74}. \DOIprefix\doi{10.1088/0004-637X/698/1/43}.
  \href{http://arxiv.org/abs/0812.4780}{\tt arXiv:0812.4780}.
\bibitem[{{Rea} et~al.(2015){Rea}, {Gull{\'o}n}, {Pons}, {Perna}, {Dainotti},
  {Miralles} \& {Torres}}]{Rea2015}
\bibinfo{author}{{Rea}, N.}, \bibinfo{author}{{Gull{\'o}n}, M.},
  \bibinfo{author}{{Pons}, J.~A.}, \bibinfo{author}{{Perna}, R.},
  \bibinfo{author}{{Dainotti}, M.~G.}, \bibinfo{author}{{Miralles}, J.~A.}, \&
  \bibinfo{author}{{Torres}, D.~F.} (\bibinfo{year}{2015}).
\newblock \bibinfo{title}{{Constraining the GRB-Magnetar Model by Means of the
  Galactic Pulsar Population}}.
\newblock {\it \bibinfo{journal}{\apj}\/},  {\it \bibinfo{volume}{813}\/},
  \bibinfo{pages}{92}. \DOIprefix\doi{10.1088/0004-637X/813/2/92}.
  \href{http://arxiv.org/abs/1510.01430}{\tt arXiv:1510.01430}.
\bibitem[{{Reale}(2007)}]{Reale2007}
\bibinfo{author}{{Reale}, F.} (\bibinfo{year}{2007}).
\newblock \bibinfo{title}{{Diagnostics of stellar flares from X-ray
  observations: from the decay to the rise phase}}.
\newblock {\it \bibinfo{journal}{\aap}\/},  {\it \bibinfo{volume}{471}\/},
  \bibinfo{pages}{271--279}. \DOIprefix\doi{10.1051/0004-6361:20077223}.
  \href{http://arxiv.org/abs/0705.3254}{\tt arXiv:0705.3254}.
\bibitem[{{Rees}(1988)}]{Rees1988}
\bibinfo{author}{{Rees}, M.~J.} (\bibinfo{year}{1988}).
\newblock \bibinfo{title}{{Tidal disruption of stars by black holes of 10 to
  the 6th-10 to the 8th solar masses in nearby galaxies}}.
\newblock {\it \bibinfo{journal}{\nat}\/},  {\it \bibinfo{volume}{333}\/},
  \bibinfo{pages}{523--528}. \DOIprefix\doi{10.1038/333523a0}.
\bibitem[{{Rezzolla} et~al.(2011){Rezzolla}, {Giacomazzo}, {Baiotti}, {Granot},
  {Kouveliotou} \& {Aloy}}]{Rezzolla2011}
\bibinfo{author}{{Rezzolla}, L.}, \bibinfo{author}{{Giacomazzo}, B.},
  \bibinfo{author}{{Baiotti}, L.}, \bibinfo{author}{{Granot}, J.},
  \bibinfo{author}{{Kouveliotou}, C.}, \& \bibinfo{author}{{Aloy}, M.~A.}
  (\bibinfo{year}{2011}).
\newblock \bibinfo{title}{{The Missing Link: Merging Neutron Stars Naturally
  Produce Jet-like Structures and Can Power Short Gamma-ray Bursts}}.
\newblock {\it \bibinfo{journal}{\apjl}\/},  {\it \bibinfo{volume}{732}\/},
  \bibinfo{pages}{L6}. \DOIprefix\doi{10.1088/2041-8205/732/1/L6}.
  \href{http://arxiv.org/abs/1101.4298}{\tt arXiv:1101.4298}.
\bibitem[{{Rezzolla} \& {Kumar}(2015)}]{Rezzolla2015}
\bibinfo{author}{{Rezzolla}, L.}, \& \bibinfo{author}{{Kumar}, P.}
  (\bibinfo{year}{2015}).
\newblock \bibinfo{title}{{A Novel Paradigm for Short Gamma-Ray Bursts With
  Extended X-Ray Emission}}.
\newblock {\it \bibinfo{journal}{\apj}\/},  {\it \bibinfo{volume}{802}\/},
  \bibinfo{pages}{95}. \DOIprefix\doi{10.1088/0004-637X/802/2/95}.
  \href{http://arxiv.org/abs/1410.8560}{\tt arXiv:1410.8560}.
\bibitem[{{Rezzolla} \& {Takami}(2016)}]{Rezzolla2016}
\bibinfo{author}{{Rezzolla}, L.}, \& \bibinfo{author}{{Takami}, K.}
  (\bibinfo{year}{2016}).
\newblock \bibinfo{title}{{Gravitational-wave signal from binary neutron stars:
  A systematic analysis of the spectral properties}}.
\newblock {\it \bibinfo{journal}{\prd}\/},  {\it \bibinfo{volume}{93}\/},
  \bibinfo{pages}{124051}. \DOIprefix\doi{10.1103/PhysRevD.93.124051}.
  \href{http://arxiv.org/abs/1604.00246}{\tt arXiv:1604.00246}.
\bibitem[{{Robertson} \& {Ellis}(2012)}]{Robertson2012}
\bibinfo{author}{{Robertson}, B.~E.}, \& \bibinfo{author}{{Ellis}, R.~S.}
  (\bibinfo{year}{2012}).
\newblock \bibinfo{title}{{Connecting the Gamma Ray Burst Rate and the Cosmic
  Star Formation History: Implications for Reionization and Galaxy Evolution}}.
\newblock {\it \bibinfo{journal}{\apj}\/},  {\it \bibinfo{volume}{744}\/},
  \bibinfo{pages}{95}. \DOIprefix\doi{10.1088/0004-637X/744/2/95}.
  \href{http://arxiv.org/abs/1109.0990}{\tt arXiv:1109.0990}.
\bibitem[{{Robertson} et~al.(2013){Robertson}, {Furlanetto}, {Schneider},
  {Charlot}, {Ellis}, {Stark}, {McLure}, {Dunlop}, {Koekemoer}, {Schenker},
  {Ouchi}, {Ono}, {Curtis-Lake}, {Rogers}, {Bowler} \&
  {Cirasuolo}}]{Robertson2013}
\bibinfo{author}{{Robertson}, B.~E.}, \bibinfo{author}{{Furlanetto}, S.~R.},
  \bibinfo{author}{{Schneider}, E.}, \bibinfo{author}{{Charlot}, S.},
  \bibinfo{author}{{Ellis}, R.~S.}, \bibinfo{author}{{Stark}, D.~P.},
  \bibinfo{author}{{McLure}, R.~J.}, \bibinfo{author}{{Dunlop}, J.~S.},
  \bibinfo{author}{{Koekemoer}, A.}, \bibinfo{author}{{Schenker}, M.~A.},
  \bibinfo{author}{{Ouchi}, M.}, \bibinfo{author}{{Ono}, Y.},
  \bibinfo{author}{{Curtis-Lake}, E.}, \bibinfo{author}{{Rogers}, A.~B.},
  \bibinfo{author}{{Bowler}, R.~A.~A.}, \& \bibinfo{author}{{Cirasuolo}, M.}
  (\bibinfo{year}{2013}).
\newblock \bibinfo{title}{{New Constraints on Cosmic Reionization from the 2012
  Hubble Ultra Deep Field Campaign}}.
\newblock {\it \bibinfo{journal}{\apj}\/},  {\it \bibinfo{volume}{768}\/},
  \bibinfo{pages}{71}. \DOIprefix\doi{10.1088/0004-637X/768/1/71}.
  \href{http://arxiv.org/abs/1301.1228}{\tt arXiv:1301.1228}.
\bibitem[{{Romano} et~al.(2015){Romano}, {Bozzo}, {Mangano}, {Esposito},
  {Israel}, {Tiengo}, {Campana}, {Ducci}, {Ferrigno} \& {Kennea}}]{rom2015}
\bibinfo{author}{{Romano}, P.}, \bibinfo{author}{{Bozzo}, E.},
  \bibinfo{author}{{Mangano}, V.}, \bibinfo{author}{{Esposito}, P.},
  \bibinfo{author}{{Israel}, G.}, \bibinfo{author}{{Tiengo}, A.},
  \bibinfo{author}{{Campana}, S.}, \bibinfo{author}{{Ducci}, L.},
  \bibinfo{author}{{Ferrigno}, C.}, \& \bibinfo{author}{{Kennea}, J.~A.}
  (\bibinfo{year}{2015}).
\newblock \bibinfo{title}{{Giant outburst from the supergiant fast X-ray
  transient IGR J17544-2619: accretion from a transient disc?}}
\newblock {\it \bibinfo{journal}{\aap}\/},  {\it \bibinfo{volume}{576}\/},
  \bibinfo{pages}{L4}. \DOIprefix\doi{10.1051/0004-6361/201525749}.
  \href{http://arxiv.org/abs/1502.04717}{\tt arXiv:1502.04717}.
\bibitem[{{Rossi} et~al.(2012){Rossi}, {Klose}, {Ferrero}, {Greiner}, {Arnold},
  {Gonsalves}, {Hartmann}, {Updike}, {Kann}, {Kr{\"u}hler}, {Palazzi},
  {Savaglio}, {Schulze}, {Afonso}, {Amati}, {Castro-Tirado}, {Clemens},
  {Filgas}, {Gorosabel}, {Hunt}, {K{\"u}pc{\"u} Yolda{\c s}}, {Masetti},
  {Nardini}, {Nicuesa Guelbenzu}, {Olivares}, {Pian}, {Rau}, {Schady},
  {Schmidl}, {Yolda{\c s}} \& {de Ugarte Postigo}}]{Rossi2012}
\bibinfo{author}{{Rossi}, A.}, \bibinfo{author}{{Klose}, S.},
  \bibinfo{author}{{Ferrero}, P.}, \bibinfo{author}{{Greiner}, J.},
  \bibinfo{author}{{Arnold}, L.~A.}, \bibinfo{author}{{Gonsalves}, E.},
  \bibinfo{author}{{Hartmann}, D.~H.}, \bibinfo{author}{{Updike}, A.~C.},
  \bibinfo{author}{{Kann}, D.~A.}, \bibinfo{author}{{Kr{\"u}hler}, T.},
  \bibinfo{author}{{Palazzi}, E.}, \bibinfo{author}{{Savaglio}, S.},
  \bibinfo{author}{{Schulze}, S.}, \bibinfo{author}{{Afonso}, P.~M.~J.},
  \bibinfo{author}{{Amati}, L.}, \bibinfo{author}{{Castro-Tirado}, A.~J.},
  \bibinfo{author}{{Clemens}, C.}, \bibinfo{author}{{Filgas}, R.},
  \bibinfo{author}{{Gorosabel}, J.}, \bibinfo{author}{{Hunt}, L.~K.},
  \bibinfo{author}{{K{\"u}pc{\"u} Yolda{\c s}}, A.},
  \bibinfo{author}{{Masetti}, N.}, \bibinfo{author}{{Nardini}, M.},
  \bibinfo{author}{{Nicuesa Guelbenzu}, A.}, \bibinfo{author}{{Olivares},
  F.~E.}, \bibinfo{author}{{Pian}, E.}, \bibinfo{author}{{Rau}, A.},
  \bibinfo{author}{{Schady}, P.}, \bibinfo{author}{{Schmidl}, S.},
  \bibinfo{author}{{Yolda{\c s}}, A.}, \& \bibinfo{author}{{de Ugarte Postigo},
  A.} (\bibinfo{year}{2012}).
\newblock \bibinfo{title}{{A deep search for the host galaxies of gamma-ray
  bursts with no detected optical afterglow}}.
\newblock {\it \bibinfo{journal}{\aap}\/},  {\it \bibinfo{volume}{545}\/},
  \bibinfo{pages}{A77}. \DOIprefix\doi{10.1051/0004-6361/201117201}.
  \href{http://arxiv.org/abs/1202.1434}{\tt arXiv:1202.1434}.
\bibitem[{{Rossi} et~al.(2011){Rossi}, {Schulze}, {Klose}, {Kann}, {Rau},
  {Krimm}, {J{\'o}hannesson}, {Panaitescu}, {Yuan}, {Ferrero}, {Kr{\"u}hler},
  {Greiner}, {Schady}, {Pandey}, {Amati}, {Afonso}, {Akerlof}, {Arnold},
  {Clemens}, {Filgas}, {Hartmann}, {K{\"u}pc{\"u} Yolda{\c s}}, {McBreen},
  {McKay}, {Nicuesa Guelbenzu}, {Olivares}, {Paciesas}, {Rykoff}, {Szokoly},
  {Updike} \& {Yolda{\c s}}}]{Rossi2011}
\bibinfo{author}{{Rossi}, A.}, \bibinfo{author}{{Schulze}, S.},
  \bibinfo{author}{{Klose}, S.}, \bibinfo{author}{{Kann}, D.~A.},
  \bibinfo{author}{{Rau}, A.}, \bibinfo{author}{{Krimm}, H.~A.},
  \bibinfo{author}{{J{\'o}hannesson}, G.}, \bibinfo{author}{{Panaitescu}, A.},
  \bibinfo{author}{{Yuan}, F.}, \bibinfo{author}{{Ferrero}, P.},
  \bibinfo{author}{{Kr{\"u}hler}, T.}, \bibinfo{author}{{Greiner}, J.},
  \bibinfo{author}{{Schady}, P.}, \bibinfo{author}{{Pandey}, S.~B.},
  \bibinfo{author}{{Amati}, L.}, \bibinfo{author}{{Afonso}, P.~M.~J.},
  \bibinfo{author}{{Akerlof}, C.~W.}, \bibinfo{author}{{Arnold}, L.~A.},
  \bibinfo{author}{{Clemens}, C.}, \bibinfo{author}{{Filgas}, R.},
  \bibinfo{author}{{Hartmann}, D.~H.}, \bibinfo{author}{{K{\"u}pc{\"u} Yolda{\c
  s}}, A.}, \bibinfo{author}{{McBreen}, S.}, \bibinfo{author}{{McKay}, T.~A.},
  \bibinfo{author}{{Nicuesa Guelbenzu}, A.}, \bibinfo{author}{{Olivares},
  F.~E.}, \bibinfo{author}{{Paciesas}, B.}, \bibinfo{author}{{Rykoff}, E.~S.},
  \bibinfo{author}{{Szokoly}, G.}, \bibinfo{author}{{Updike}, A.~C.}, \&
  \bibinfo{author}{{Yolda{\c s}}, A.} (\bibinfo{year}{2011}).
\newblock \bibinfo{title}{{The Swift/Fermi GRB 080928 from 1 eV to 150 keV}}.
\newblock {\it \bibinfo{journal}{\aap}\/},  {\it \bibinfo{volume}{529}\/},
  \bibinfo{pages}{A142}. \DOIprefix\doi{10.1051/0004-6361/201015324}.
  \href{http://arxiv.org/abs/1007.0383}{\tt arXiv:1007.0383}.
\bibitem[{{Rowlinson} et~al.(2014){Rowlinson}, {Gompertz}, {Dainotti},
  {O'Brien}, {Wijers} \& {van der Horst}}]{Rowlinson2014}
\bibinfo{author}{{Rowlinson}, A.}, \bibinfo{author}{{Gompertz}, B.~P.},
  \bibinfo{author}{{Dainotti}, M.}, \bibinfo{author}{{O'Brien}, P.~T.},
  \bibinfo{author}{{Wijers}, R.~A.~M.~J.}, \& \bibinfo{author}{{van der Horst},
  A.~J.} (\bibinfo{year}{2014}).
\newblock \bibinfo{title}{{Constraining properties of GRB magnetar central
  engines using the observed plateau luminosity and duration correlation}}.
\newblock {\it \bibinfo{journal}{\mnras}\/},  {\it \bibinfo{volume}{443}\/},
  \bibinfo{pages}{1779--1787}. \DOIprefix\doi{10.1093/mnras/stu1277}.
  \href{http://arxiv.org/abs/1407.1053}{\tt arXiv:1407.1053}.
\bibitem[{{Ryde} et~al.(2010){Ryde}, {Axelsson}, {Zhang}, {McGlynn}, {Pe'er},
  {Lundman}, {Larsson}, {Battelino}, {Zhang}, {Bissaldi}, {Bregeon}, {Briggs},
  {Chiang}, {de Palma}, {Guiriec}, {Larsson}, {Longo}, {McBreen}, {Omodei},
  {Petrosian}, {Preece} \& {van der Horst}}]{ryde2010}
\bibinfo{author}{{Ryde}, F.}, \bibinfo{author}{{Axelsson}, M.},
  \bibinfo{author}{{Zhang}, B.~B.}, \bibinfo{author}{{McGlynn}, S.},
  \bibinfo{author}{{Pe'er}, A.}, \bibinfo{author}{{Lundman}, C.},
  \bibinfo{author}{{Larsson}, S.}, \bibinfo{author}{{Battelino}, M.},
  \bibinfo{author}{{Zhang}, B.}, \bibinfo{author}{{Bissaldi}, E.},
  \bibinfo{author}{{Bregeon}, J.}, \bibinfo{author}{{Briggs}, M.~S.},
  \bibinfo{author}{{Chiang}, J.}, \bibinfo{author}{{de Palma}, F.},
  \bibinfo{author}{{Guiriec}, S.}, \bibinfo{author}{{Larsson}, J.},
  \bibinfo{author}{{Longo}, F.}, \bibinfo{author}{{McBreen}, S.},
  \bibinfo{author}{{Omodei}, N.}, \bibinfo{author}{{Petrosian}, V.},
  \bibinfo{author}{{Preece}, R.}, \& \bibinfo{author}{{van der Horst}, A.~J.}
  (\bibinfo{year}{2010}).
\newblock \bibinfo{title}{{Identification and Properties of the Photospheric
  Emission in GRB090902B}}.
\newblock {\it \bibinfo{journal}{\apjl}\/},  {\it \bibinfo{volume}{709}\/},
  \bibinfo{pages}{L172--L177}. \DOIprefix\doi{10.1088/2041-8205/709/2/L172}.
  \href{http://arxiv.org/abs/0911.2025}{\tt arXiv:0911.2025}.
\bibitem[{{Sakamoto} et~al.(2005){Sakamoto}, {Lamb}, {Kawai}, {Yoshida},
  {Graziani}, {Fenimore}, {Donaghy}, {Matsuoka}, {Suzuki}, {Ricker}, {Atteia},
  {Shirasaki}, {Tamagawa}, {Torii}, {Galassi}, {Doty}, {Vanderspek}, {Crew},
  {Villasenor}, {Butler}, {Prigozhin}, {Jernigan}, {Barraud}, {Boer},
  {Dezalay}, {Olive}, {Hurley}, {Levine}, {Monnelly}, {Martel}, {Morgan},
  {Woosley}, {Cline}, {Braga}, {Manchanda}, {Pizzichini}, {Takagishi} \&
  {Yamauchi}}]{Sakamoto2005}
\bibinfo{author}{{Sakamoto}, T.}, \bibinfo{author}{{Lamb}, D.~Q.},
  \bibinfo{author}{{Kawai}, N.}, \bibinfo{author}{{Yoshida}, A.},
  \bibinfo{author}{{Graziani}, C.}, \bibinfo{author}{{Fenimore}, E.~E.},
  \bibinfo{author}{{Donaghy}, T.~Q.}, \bibinfo{author}{{Matsuoka}, M.},
  \bibinfo{author}{{Suzuki}, M.}, \bibinfo{author}{{Ricker}, G.},
  \bibinfo{author}{{Atteia}, J.-L.}, \bibinfo{author}{{Shirasaki}, Y.},
  \bibinfo{author}{{Tamagawa}, T.}, \bibinfo{author}{{Torii}, K.},
  \bibinfo{author}{{Galassi}, M.}, \bibinfo{author}{{Doty}, J.},
  \bibinfo{author}{{Vanderspek}, R.}, \bibinfo{author}{{Crew}, G.~B.},
  \bibinfo{author}{{Villasenor}, J.}, \bibinfo{author}{{Butler}, N.},
  \bibinfo{author}{{Prigozhin}, G.}, \bibinfo{author}{{Jernigan}, J.~G.},
  \bibinfo{author}{{Barraud}, C.}, \bibinfo{author}{{Boer}, M.},
  \bibinfo{author}{{Dezalay}, J.-P.}, \bibinfo{author}{{Olive}, J.-F.},
  \bibinfo{author}{{Hurley}, K.}, \bibinfo{author}{{Levine}, A.},
  \bibinfo{author}{{Monnelly}, G.}, \bibinfo{author}{{Martel}, F.},
  \bibinfo{author}{{Morgan}, E.}, \bibinfo{author}{{Woosley}, S.~E.},
  \bibinfo{author}{{Cline}, T.}, \bibinfo{author}{{Braga}, J.},
  \bibinfo{author}{{Manchanda}, R.}, \bibinfo{author}{{Pizzichini}, G.},
  \bibinfo{author}{{Takagishi}, K.}, \& \bibinfo{author}{{Yamauchi}, M.}
  (\bibinfo{year}{2005}).
\newblock \bibinfo{title}{{Global Characteristics of X-Ray Flashes and
  X-Ray-Rich Gamma-Ray Bursts Observed by HETE-2}}.
\newblock {\it \bibinfo{journal}{\apj}\/},  {\it \bibinfo{volume}{629}\/},
  \bibinfo{pages}{311--327}. \DOIprefix\doi{10.1086/431235}.
\bibitem[{{Salvaterra} et~al.(2012){Salvaterra}, {Campana}, {Vergani},
  {Covino}, {D'Avanzo}, {Fugazza}, {Ghirlanda}, {Ghisellini}, {Melandri},
  {Nava}, {Sbarufatti}, {Flores}, {Piranomonte} \&
  {Tagliaferri}}]{Salvaterra2012}
\bibinfo{author}{{Salvaterra}, R.}, \bibinfo{author}{{Campana}, S.},
  \bibinfo{author}{{Vergani}, S.~D.}, \bibinfo{author}{{Covino}, S.},
  \bibinfo{author}{{D'Avanzo}, P.}, \bibinfo{author}{{Fugazza}, D.},
  \bibinfo{author}{{Ghirlanda}, G.}, \bibinfo{author}{{Ghisellini}, G.},
  \bibinfo{author}{{Melandri}, A.}, \bibinfo{author}{{Nava}, L.},
  \bibinfo{author}{{Sbarufatti}, B.}, \bibinfo{author}{{Flores}, H.},
  \bibinfo{author}{{Piranomonte}, S.}, \& \bibinfo{author}{{Tagliaferri}, G.}
  (\bibinfo{year}{2012}).
\newblock \bibinfo{title}{{A Complete Sample of Bright Swift Long Gamma-Ray
  Bursts. I. Sample Presentation, Luminosity Function and Evolution}}.
\newblock {\it \bibinfo{journal}{\apj}\/},  {\it \bibinfo{volume}{749}\/},
  \bibinfo{pages}{68}. \DOIprefix\doi{10.1088/0004-637X/749/1/68}.
  \href{http://arxiv.org/abs/1112.1700}{\tt arXiv:1112.1700}.
\bibitem[{{Salvaterra} et~al.(2013){Salvaterra}, {Maio}, {Ciardi} \&
  {Campisi}}]{Salvaterra2013}
\bibinfo{author}{{Salvaterra}, R.}, \bibinfo{author}{{Maio}, U.},
  \bibinfo{author}{{Ciardi}, B.}, \& \bibinfo{author}{{Campisi}, M.~A.}
  (\bibinfo{year}{2013}).
\newblock \bibinfo{title}{{Simulating high-z gamma-ray burst host galaxies}}.
\newblock {\it \bibinfo{journal}{\mnras}\/},  {\it \bibinfo{volume}{429}\/},
  \bibinfo{pages}{2718--2726}. \DOIprefix\doi{10.1093/mnras/sts541}.
  \href{http://arxiv.org/abs/1212.0856}{\tt arXiv:1212.0856}.
\bibitem[{{Santander}(2016)}]{Santander2016}
\bibinfo{author}{{Santander}, M.} (\bibinfo{year}{2016}).
\newblock \bibinfo{title}{{The Dawn of Multi-Messenger Astronomy}}.
\newblock {\it \bibinfo{journal}{ArXiv e-prints}\/}, .
  \href{http://arxiv.org/abs/1606.09335}{\tt arXiv:1606.09335}.
\bibitem[{{Sari} \& {Piran}(1999)}]{Sari1999}
\bibinfo{author}{{Sari}, R.}, \& \bibinfo{author}{{Piran}, T.}
  (\bibinfo{year}{1999}).
\newblock \bibinfo{title}{{Predictions for the Very Early Afterglow and the
  Optical Flash}}.
\newblock {\it \bibinfo{journal}{\apj}\/},  {\it \bibinfo{volume}{520}\/},
  \bibinfo{pages}{641--649}. \DOIprefix\doi{10.1086/307508}.
  \href{http://arxiv.org/abs/astro-ph/9901338}{\tt arXiv:astro-ph/9901338}.
\bibitem[{{Sathyaprakash} et~al.(2012){Sathyaprakash}, {Abernathy}, {Acernese},
  {Ajith}, {Allen}, {Amaro-Seoane}, {Andersson}, {Aoudia}, {Arun}, {Astone} \&
  et~al.}]{Sathyaprakash2012}
\bibinfo{author}{{Sathyaprakash}, B.}, \bibinfo{author}{{Abernathy}, M.},
  \bibinfo{author}{{Acernese}, F.}, \bibinfo{author}{{Ajith}, P.},
  \bibinfo{author}{{Allen}, B.}, \bibinfo{author}{{Amaro-Seoane}, P.},
  \bibinfo{author}{{Andersson}, N.}, \bibinfo{author}{{Aoudia}, S.},
  \bibinfo{author}{{Arun}, K.}, \bibinfo{author}{{Astone}, P.}, \&
  \bibinfo{author}{et~al.} (\bibinfo{year}{2012}).
\newblock \bibinfo{title}{{Scientific objectives of Einstein Telescope}}.
\newblock {\it \bibinfo{journal}{Classical and Quantum Gravity}\/},  {\it
  \bibinfo{volume}{29}\/}, \bibinfo{pages}{124013}.
  \DOIprefix\doi{10.1088/0264-9381/29/12/124013}.
  \href{http://arxiv.org/abs/1206.0331}{\tt arXiv:1206.0331}.
\bibitem[{{Savaglio} et~al.(2009){Savaglio}, {Glazebrook} \& {Le
  Borgne}}]{Savaglio2009}
\bibinfo{author}{{Savaglio}, S.}, \bibinfo{author}{{Glazebrook}, K.}, \&
  \bibinfo{author}{{Le Borgne}, D.} (\bibinfo{year}{2009}).
\newblock \bibinfo{title}{{The Galaxy Population Hosting Gamma-Ray Bursts}}.
\newblock {\it \bibinfo{journal}{\apj}\/},  {\it \bibinfo{volume}{691}\/},
  \bibinfo{pages}{182--211}. \DOIprefix\doi{10.1088/0004-637X/691/1/182}.
  \href{http://arxiv.org/abs/0803.2718}{\tt arXiv:0803.2718}.
\bibitem[{{Schenker} et~al.(2014){Schenker}, {Ellis}, {Konidaris} \&
  {Stark}}]{Schenker2014}
\bibinfo{author}{{Schenker}, M.~A.}, \bibinfo{author}{{Ellis}, R.~S.},
  \bibinfo{author}{{Konidaris}, N.~P.}, \& \bibinfo{author}{{Stark}, D.~P.}
  (\bibinfo{year}{2014}).
\newblock \bibinfo{title}{{Line-emitting Galaxies beyond a Redshift of 7: An
  Improved Method for Estimating the Evolving Neutrality of the Intergalactic
  Medium}}.
\newblock {\it \bibinfo{journal}{\apj}\/},  {\it \bibinfo{volume}{795}\/},
  \bibinfo{pages}{20}. \DOIprefix\doi{10.1088/0004-637X/795/1/20}.
  \href{http://arxiv.org/abs/1404.4632}{\tt arXiv:1404.4632}.
\bibitem[{{Schulze} et~al.(2011){Schulze}, {Klose}, {Bj{\"o}rnsson},
  {Jakobsson}, {Kann}, {Rossi}, {Kr{\"u}hler}, {Greiner} \&
  {Ferrero}}]{Schulze2011}
\bibinfo{author}{{Schulze}, S.}, \bibinfo{author}{{Klose}, S.},
  \bibinfo{author}{{Bj{\"o}rnsson}, G.}, \bibinfo{author}{{Jakobsson}, P.},
  \bibinfo{author}{{Kann}, D.~A.}, \bibinfo{author}{{Rossi}, A.},
  \bibinfo{author}{{Kr{\"u}hler}, T.}, \bibinfo{author}{{Greiner}, J.}, \&
  \bibinfo{author}{{Ferrero}, P.} (\bibinfo{year}{2011}).
\newblock \bibinfo{title}{{The circumburst density profile around GRB
  progenitors: a statistical study}}.
\newblock {\it \bibinfo{journal}{\aap}\/},  {\it \bibinfo{volume}{526}\/},
  \bibinfo{pages}{A23}. \DOIprefix\doi{10.1051/0004-6361/201015581}.
  \href{http://arxiv.org/abs/1010.4057}{\tt arXiv:1010.4057}.
\bibitem[{{Schulze} et~al.(2016){Schulze}, {Kr{\"u}hler}, {Leloudas},
  {Gorosabel}, {Mehner}, {Buchner}, {Kim}, {Ibar}, {Amor{\'{\i}}n},
  {Herrero-Illana}, {Anderson}, {Bauer}, {Christensen}, {de Pasquale}, {de
  Ugarte Postigo}, {Gallazzi}, {Hjorth}, {Morrell}, {Malesani}, {Sparre},
  {Stalder}, {Stark}, {Th{\"o}ne} \& {Wheeler}}]{Schulze2016}
\bibinfo{author}{{Schulze}, S.}, \bibinfo{author}{{Kr{\"u}hler}, T.},
  \bibinfo{author}{{Leloudas}, G.}, \bibinfo{author}{{Gorosabel}, J.},
  \bibinfo{author}{{Mehner}, A.}, \bibinfo{author}{{Buchner}, J.},
  \bibinfo{author}{{Kim}, S.}, \bibinfo{author}{{Ibar}, E.},
  \bibinfo{author}{{Amor{\'{\i}}n}, R.}, \bibinfo{author}{{Herrero-Illana},
  R.}, \bibinfo{author}{{Anderson}, J.~P.}, \bibinfo{author}{{Bauer}, F.~E.},
  \bibinfo{author}{{Christensen}, L.}, \bibinfo{author}{{de Pasquale}, M.},
  \bibinfo{author}{{de Ugarte Postigo}, A.}, \bibinfo{author}{{Gallazzi}, A.},
  \bibinfo{author}{{Hjorth}, J.}, \bibinfo{author}{{Morrell}, N.},
  \bibinfo{author}{{Malesani}, D.}, \bibinfo{author}{{Sparre}, M.},
  \bibinfo{author}{{Stalder}, B.}, \bibinfo{author}{{Stark}, A.~A.},
  \bibinfo{author}{{Th{\"o}ne}, C.~C.}, \& \bibinfo{author}{{Wheeler}, J.~C.}
  (\bibinfo{year}{2016}).
\newblock \bibinfo{title}{{Cosmic evolution and metal aversion in
  super-luminous supernova host galaxies}}.
\newblock {\it \bibinfo{journal}{ArXiv e-prints}\/}, .
  \href{http://arxiv.org/abs/1612.05978}{\tt arXiv:1612.05978}.
\bibitem[{{Schutz}(1986)}]{Schutz1986}
\bibinfo{author}{{Schutz}, B.~F.} (\bibinfo{year}{1986}).
\newblock \bibinfo{title}{{Determining the Hubble constant from gravitational
  wave observations}}.
\newblock {\it \bibinfo{journal}{\nat}\/},  {\it \bibinfo{volume}{323}\/},
  \bibinfo{pages}{310}. \DOIprefix\doi{10.1038/323310a0}.
\bibitem[{{Schwarz} et~al.(2011){Schwarz}, {Ness}, {Osborne}, {Page}, {Evans},
  {Beardmore}, {Walter}, {Helton}, {Woodward}, {Bode}, {Starrfield} \&
  {Drake}}]{Schwarz2011}
\bibinfo{author}{{Schwarz}, G.~J.}, \bibinfo{author}{{Ness}, J.-U.},
  \bibinfo{author}{{Osborne}, J.~P.}, \bibinfo{author}{{Page}, K.~L.},
  \bibinfo{author}{{Evans}, P.~A.}, \bibinfo{author}{{Beardmore}, A.~P.},
  \bibinfo{author}{{Walter}, F.~M.}, \bibinfo{author}{{Helton}, L.~A.},
  \bibinfo{author}{{Woodward}, C.~E.}, \bibinfo{author}{{Bode}, M.},
  \bibinfo{author}{{Starrfield}, S.}, \& \bibinfo{author}{{Drake}, J.~J.}
  (\bibinfo{year}{2011}).
\newblock \bibinfo{title}{{Swift X-Ray Observations of Classical Novae. II. The
  Super Soft Source Sample}}.
\newblock {\it \bibinfo{journal}{\apjs}\/},  {\it \bibinfo{volume}{197}\/},
  \bibinfo{pages}{31}. \DOIprefix\doi{10.1088/0067-0049/197/2/31}.
  \href{http://arxiv.org/abs/1110.6224}{\tt arXiv:1110.6224}.
\bibitem[{{Siegel} \& {Ciolfi}(2016{\natexlab{a}})}]{Siegel2016a}
\bibinfo{author}{{Siegel}, D.~M.}, \& \bibinfo{author}{{Ciolfi}, R.}
  (\bibinfo{year}{2016}{\natexlab{a}}).
\newblock \bibinfo{title}{{Electromagnetic Emission from Long-lived Binary
  Neutron Star Merger Remnants. I. Formulation of the Problem}}.
\newblock {\it \bibinfo{journal}{\apj}\/},  {\it \bibinfo{volume}{819}\/},
  \bibinfo{pages}{14}. \DOIprefix\doi{10.3847/0004-637X/819/1/14}.
  \href{http://arxiv.org/abs/1508.07911}{\tt arXiv:1508.07911}.
\bibitem[{{Siegel} \& {Ciolfi}(2016{\natexlab{b}})}]{Siegel2016b}
\bibinfo{author}{{Siegel}, D.~M.}, \& \bibinfo{author}{{Ciolfi}, R.}
  (\bibinfo{year}{2016}{\natexlab{b}}).
\newblock \bibinfo{title}{{Electromagnetic Emission from Long-lived Binary
  Neutron Star Merger Remnants. II. Lightcurves and Spectra}}.
\newblock {\it \bibinfo{journal}{\apj}\/},  {\it \bibinfo{volume}{819}\/},
  \bibinfo{pages}{15}. \DOIprefix\doi{10.3847/0004-637X/819/1/15}.
  \href{http://arxiv.org/abs/1508.07939}{\tt arXiv:1508.07939}.
\bibitem[{{Singer} et~al.(2016){Singer}, {Chen}, {Holz}, {Farr}, {Price},
  {Raymond}, {Cenko}, {Gehrels}, {Cannizzo}, {Kasliwal}, {Nissanke},
  {Coughlin}, {Farr}, {Urban}, {Vitale}, {Veitch}, {Graff}, {Berry},
  {Mohapatra} \& {Mandel}}]{Singer2016}
\bibinfo{author}{{Singer}, L.~P.}, \bibinfo{author}{{Chen}, H.-Y.},
  \bibinfo{author}{{Holz}, D.~E.}, \bibinfo{author}{{Farr}, W.~M.},
  \bibinfo{author}{{Price}, L.~R.}, \bibinfo{author}{{Raymond}, V.},
  \bibinfo{author}{{Cenko}, S.~B.}, \bibinfo{author}{{Gehrels}, N.},
  \bibinfo{author}{{Cannizzo}, J.}, \bibinfo{author}{{Kasliwal}, M.~M.},
  \bibinfo{author}{{Nissanke}, S.}, \bibinfo{author}{{Coughlin}, M.},
  \bibinfo{author}{{Farr}, B.}, \bibinfo{author}{{Urban}, A.~L.},
  \bibinfo{author}{{Vitale}, S.}, \bibinfo{author}{{Veitch}, J.},
  \bibinfo{author}{{Graff}, P.}, \bibinfo{author}{{Berry}, C.~P.~L.},
  \bibinfo{author}{{Mohapatra}, S.}, \& \bibinfo{author}{{Mandel}, I.}
  (\bibinfo{year}{2016}).
\newblock \bibinfo{title}{{Going the Distance: Mapping Host Galaxies of LIGO
  and Virgo Sources in Three Dimensions Using Local Cosmography and Targeted
  Follow-up}}.
\newblock {\it \bibinfo{journal}{\apjl}\/},  {\it \bibinfo{volume}{829}\/},
  \bibinfo{pages}{L15}. \DOIprefix\doi{10.3847/2041-8205/829/1/L15}.
  \href{http://arxiv.org/abs/1603.07333}{\tt arXiv:1603.07333}.
\bibitem[{{Smartt} et~al.(2017){Smartt}, {Chen}, {Jerkstrand}, {Coughlin},
  {Kankare}, {Sim}, {Fraser}, {Inserra}, {Maguire}, {Chambers}, {Huber},
  {Kr{\"u}hler}, {Leloudas}, {Magee}, {Shingles}, {Smith}, {Young}, {Tonry},
  {Kotak}, {Gal-Yam}, {Lyman}, {Homan}, {Agliozzo}, {Anderson}, {Angus},
  {Ashall}, {Barbarino}, {Bauer}, {Berton}, {Botticella}, {Bulla}, {Bulger},
  {Cannizzaro}, {Cano}, {Cartier}, {Cikota}, {Clark}, {De Cia}, {Della Valle},
  {Denneau}, {Dennefeld}, {Dessart}, {Dimitriadis}, {Elias-Rosa}, {Firth},
  {Flewelling}, {Fl{\"o}rs}, {Franckowiak}, {Frohmaier}, {Galbany},
  {Gonz{\'a}lez-Gait{\'a}n}, {Greiner}, {Gromadzki}, {Guelbenzu},
  {Guti{\'e}rrez}, {Hamanowicz}, {Hanlon}, {Harmanen}, {Heintz}, {Heinze},
  {Hernandez}, {Hodgkin}, {Hook}, {Izzo}, {James}, {Jonker}, {Kerzendorf},
  {Klose}, {Kostrzewa-Rutkowska}, {Kowalski}, {Kromer}, {Kuncarayakti},
  {Lawrence}, {Lowe}, {Magnier}, {Manulis}, {Martin-Carrillo}, {Mattila},
  {McBrien}, {M{\"u}ller}, {Nordin}, {O'Neill}, {Onori}, {Palmerio},
  {Pastorello}, {Patat}, {Pignata}, {Podsiadlowski}, {Pumo}, {Prentice}, {Rau},
  {Razza}, {Rest}, {Reynolds}, {Roy}, {Ruiter}, {Rybicki}, {Salmon}, {Schady}
  et~al.}]{Smartt2017}
\bibinfo{author}{{Smartt}, S.~J.}, \bibinfo{author}{{Chen}, T.-W.},
  \bibinfo{author}{{Jerkstrand}, A.}, \bibinfo{author}{{Coughlin}, M.},
  \bibinfo{author}{{Kankare}, E.}, \bibinfo{author}{{Sim}, S.~A.},
  \bibinfo{author}{{Fraser}, M.}, \bibinfo{author}{{Inserra}, C.},
  \bibinfo{author}{{Maguire}, K.}, \bibinfo{author}{{Chambers}, K.~C.},
  \bibinfo{author}{{Huber}, M.~E.}, \bibinfo{author}{{Kr{\"u}hler}, T.},
  \bibinfo{author}{{Leloudas}, G.}, \bibinfo{author}{{Magee}, M.},
  \bibinfo{author}{{Shingles}, L.~J.}, \bibinfo{author}{{Smith}, K.~W.},
  \bibinfo{author}{{Young}, D.~R.}, \bibinfo{author}{{Tonry}, J.},
  \bibinfo{author}{{Kotak}, R.}, \bibinfo{author}{{Gal-Yam}, A.},
  \bibinfo{author}{{Lyman}, J.~D.}, \bibinfo{author}{{Homan}, D.~S.},
  \bibinfo{author}{{Agliozzo}, C.}, \bibinfo{author}{{Anderson}, J.~P.},
  \bibinfo{author}{{Angus}, C.~R.}, \bibinfo{author}{{Ashall}, C.},
  \bibinfo{author}{{Barbarino}, C.}, \bibinfo{author}{{Bauer}, F.~E.},
  \bibinfo{author}{{Berton}, M.}, \bibinfo{author}{{Botticella}, M.~T.},
  \bibinfo{author}{{Bulla}, M.}, \bibinfo{author}{{Bulger}, J.},
  \bibinfo{author}{{Cannizzaro}, G.}, \bibinfo{author}{{Cano}, Z.},
  \bibinfo{author}{{Cartier}, R.}, \bibinfo{author}{{Cikota}, A.},
  \bibinfo{author}{{Clark}, P.}, \bibinfo{author}{{De Cia}, A.},
  \bibinfo{author}{{Della Valle}, M.}, \bibinfo{author}{{Denneau}, L.},
  \bibinfo{author}{{Dennefeld}, M.}, \bibinfo{author}{{Dessart}, L.},
  \bibinfo{author}{{Dimitriadis}, G.}, \bibinfo{author}{{Elias-Rosa}, N.},
  \bibinfo{author}{{Firth}, R.~E.}, \bibinfo{author}{{Flewelling}, H.},
  \bibinfo{author}{{Fl{\"o}rs}, A.}, \bibinfo{author}{{Franckowiak}, A.},
  \bibinfo{author}{{Frohmaier}, C.}, \bibinfo{author}{{Galbany}, L.},
  \bibinfo{author}{{Gonz{\'a}lez-Gait{\'a}n}, S.}, \bibinfo{author}{{Greiner},
  J.}, \bibinfo{author}{{Gromadzki}, M.}, \bibinfo{author}{{Guelbenzu}, A.~N.},
  \bibinfo{author}{{Guti{\'e}rrez}, C.~P.}, \bibinfo{author}{{Hamanowicz}, A.},
  \bibinfo{author}{{Hanlon}, L.}, \bibinfo{author}{{Harmanen}, J.},
  \bibinfo{author}{{Heintz}, K.~E.}, \bibinfo{author}{{Heinze}, A.},
  \bibinfo{author}{{Hernandez}, M.-S.}, \bibinfo{author}{{Hodgkin}, S.~T.},
  \bibinfo{author}{{Hook}, I.~M.}, \bibinfo{author}{{Izzo}, L.},
  \bibinfo{author}{{James}, P.~A.}, \bibinfo{author}{{Jonker}, P.~G.},
  \bibinfo{author}{{Kerzendorf}, W.~E.}, \bibinfo{author}{{Klose}, S.},
  \bibinfo{author}{{Kostrzewa-Rutkowska}, Z.}, \bibinfo{author}{{Kowalski},
  M.}, \bibinfo{author}{{Kromer}, M.}, \bibinfo{author}{{Kuncarayakti}, H.},
  \bibinfo{author}{{Lawrence}, A.}, \bibinfo{author}{{Lowe}, T.~B.},
  \bibinfo{author}{{Magnier}, E.~A.}, \bibinfo{author}{{Manulis}, I.},
  \bibinfo{author}{{Martin-Carrillo}, A.}, \bibinfo{author}{{Mattila}, S.},
  \bibinfo{author}{{McBrien}, O.}, \bibinfo{author}{{M{\"u}ller}, A.},
  \bibinfo{author}{{Nordin}, J.}, \bibinfo{author}{{O'Neill}, D.},
  \bibinfo{author}{{Onori}, F.}, \bibinfo{author}{{Palmerio}, J.~T.},
  \bibinfo{author}{{Pastorello}, A.}, \bibinfo{author}{{Patat}, F.},
  \bibinfo{author}{{Pignata}, G.}, \bibinfo{author}{{Podsiadlowski}, P.},
  \bibinfo{author}{{Pumo}, M.~L.}, \bibinfo{author}{{Prentice}, S.~J.},
  \bibinfo{author}{{Rau}, A.}, \bibinfo{author}{{Razza}, A.},
  \bibinfo{author}{{Rest}, A.}, \bibinfo{author}{{Reynolds}, T.},
  \bibinfo{author}{{Roy}, R.}, \bibinfo{author}{{Ruiter}, A.~J.},
  \bibinfo{author}{{Rybicki}, K.~A.}, \bibinfo{author}{{Salmon}, L.},
  \bibinfo{author}{{Schady}, P.} et~al. (\bibinfo{year}{2017}).
\newblock \bibinfo{title}{{A kilonova as the electromagnetic counterpart to a
  gravitational-wave source}}.
\newblock {\it \bibinfo{journal}{\nat}\/},  {\it \bibinfo{volume}{551}\/},
  \bibinfo{pages}{75--79}. \DOIprefix\doi{10.1038/nature24303}.
  \href{http://arxiv.org/abs/1710.05841}{\tt arXiv:1710.05841}.
\bibitem[{{Soderberg} et~al.(2008){Soderberg}, {Berger}, {Page}, {Schady},
  {Parrent}, {Pooley}, {Wang}, {Ofek}, {Cucchiara}, {Rau}, {Waxman}, {Simon},
  {Bock}, {Milne}, {Page}, {Barentine}, {Barthelmy}, {Beardmore}, {Bietenholz},
  {Brown}, {Burrows}, {Burrows}, {Byrngelson}, {Cenko}, {Chandra}, {Cummings},
  {Fox}, {Gal-Yam}, {Gehrels}, {Immler}, {Kasliwal}, {Kong}, {Krimm},
  {Kulkarni}, {Maccarone}, {M{\'e}sz{\'a}ros}, {Nakar}, {O'Brien}, {Overzier},
  {de Pasquale}, {Racusin}, {Rea} \& {York}}]{Soderberg2008}
\bibinfo{author}{{Soderberg}, A.~M.}, \bibinfo{author}{{Berger}, E.},
  \bibinfo{author}{{Page}, K.~L.}, \bibinfo{author}{{Schady}, P.},
  \bibinfo{author}{{Parrent}, J.}, \bibinfo{author}{{Pooley}, D.},
  \bibinfo{author}{{Wang}, X.-Y.}, \bibinfo{author}{{Ofek}, E.~O.},
  \bibinfo{author}{{Cucchiara}, A.}, \bibinfo{author}{{Rau}, A.},
  \bibinfo{author}{{Waxman}, E.}, \bibinfo{author}{{Simon}, J.~D.},
  \bibinfo{author}{{Bock}, D.~C.-J.}, \bibinfo{author}{{Milne}, P.~A.},
  \bibinfo{author}{{Page}, M.~J.}, \bibinfo{author}{{Barentine}, J.~C.},
  \bibinfo{author}{{Barthelmy}, S.~D.}, \bibinfo{author}{{Beardmore}, A.~P.},
  \bibinfo{author}{{Bietenholz}, M.~F.}, \bibinfo{author}{{Brown}, P.},
  \bibinfo{author}{{Burrows}, A.}, \bibinfo{author}{{Burrows}, D.~N.},
  \bibinfo{author}{{Byrngelson}, G.}, \bibinfo{author}{{Cenko}, S.~B.},
  \bibinfo{author}{{Chandra}, P.}, \bibinfo{author}{{Cummings}, J.~R.},
  \bibinfo{author}{{Fox}, D.~B.}, \bibinfo{author}{{Gal-Yam}, A.},
  \bibinfo{author}{{Gehrels}, N.}, \bibinfo{author}{{Immler}, S.},
  \bibinfo{author}{{Kasliwal}, M.}, \bibinfo{author}{{Kong}, A.~K.~H.},
  \bibinfo{author}{{Krimm}, H.~A.}, \bibinfo{author}{{Kulkarni}, S.~R.},
  \bibinfo{author}{{Maccarone}, T.~J.}, \bibinfo{author}{{M{\'e}sz{\'a}ros},
  P.}, \bibinfo{author}{{Nakar}, E.}, \bibinfo{author}{{O'Brien}, P.~T.},
  \bibinfo{author}{{Overzier}, R.~A.}, \bibinfo{author}{{de Pasquale}, M.},
  \bibinfo{author}{{Racusin}, J.}, \bibinfo{author}{{Rea}, N.}, \&
  \bibinfo{author}{{York}, D.~G.} (\bibinfo{year}{2008}).
\newblock \bibinfo{title}{{An extremely luminous X-ray outburst at the birth of
  a supernova}}.
\newblock {\it \bibinfo{journal}{\nat}\/},  {\it \bibinfo{volume}{453}\/},
  \bibinfo{pages}{469--474}. \DOIprefix\doi{10.1038/nature06997}.
  \href{http://arxiv.org/abs/0802.1712}{\tt arXiv:0802.1712}.
\bibitem[{{Soderberg} et~al.(2010){Soderberg}, {Chakraborti}, {Pignata},
  {Chevalier}, {Chandra}, {Ray}, {Wieringa}, {Copete}, {Chaplin},
  {Connaughton}, {Barthelmy}, {Bietenholz}, {Chugai}, {Stritzinger}, {Hamuy},
  {Fransson}, {Fox}, {Levesque}, {Grindlay}, {Challis}, {Foley}, {Kirshner},
  {Milne} \& {Torres}}]{Soderberg2010}
\bibinfo{author}{{Soderberg}, A.~M.}, \bibinfo{author}{{Chakraborti}, S.},
  \bibinfo{author}{{Pignata}, G.}, \bibinfo{author}{{Chevalier}, R.~A.},
  \bibinfo{author}{{Chandra}, P.}, \bibinfo{author}{{Ray}, A.},
  \bibinfo{author}{{Wieringa}, M.~H.}, \bibinfo{author}{{Copete}, A.},
  \bibinfo{author}{{Chaplin}, V.}, \bibinfo{author}{{Connaughton}, V.},
  \bibinfo{author}{{Barthelmy}, S.~D.}, \bibinfo{author}{{Bietenholz}, M.~F.},
  \bibinfo{author}{{Chugai}, N.}, \bibinfo{author}{{Stritzinger}, M.~D.},
  \bibinfo{author}{{Hamuy}, M.}, \bibinfo{author}{{Fransson}, C.},
  \bibinfo{author}{{Fox}, O.}, \bibinfo{author}{{Levesque}, E.~M.},
  \bibinfo{author}{{Grindlay}, J.~E.}, \bibinfo{author}{{Challis}, P.},
  \bibinfo{author}{{Foley}, R.~J.}, \bibinfo{author}{{Kirshner}, R.~P.},
  \bibinfo{author}{{Milne}, P.~A.}, \& \bibinfo{author}{{Torres}, M.~A.~P.}
  (\bibinfo{year}{2010}).
\newblock \bibinfo{title}{{A relativistic type Ibc supernova without a detected
  {$\gamma$}-ray burst}}.
\newblock {\it \bibinfo{journal}{\nat}\/},  {\it \bibinfo{volume}{463}\/},
  \bibinfo{pages}{513--515}. \DOIprefix\doi{10.1038/nature08714}.
  \href{http://arxiv.org/abs/0908.2817}{\tt arXiv:0908.2817}.
\bibitem[{{Sokoloski} et~al.(2006){Sokoloski}, {Kenyon}, {Espey}, {Keyes},
  {McCandliss}, {Kong}, {Aufdenberg}, {Filippenko}, {Li}, {Brocksopp},
  {Kaiser}, {Charles}, {Rupen} \& {Stone}}]{Sokoloski2006}
\bibinfo{author}{{Sokoloski}, J.~L.}, \bibinfo{author}{{Kenyon}, S.~J.},
  \bibinfo{author}{{Espey}, B.~R.}, \bibinfo{author}{{Keyes}, C.~D.},
  \bibinfo{author}{{McCandliss}, S.~R.}, \bibinfo{author}{{Kong}, A.~K.~H.},
  \bibinfo{author}{{Aufdenberg}, J.~P.}, \bibinfo{author}{{Filippenko}, A.~V.},
  \bibinfo{author}{{Li}, W.}, \bibinfo{author}{{Brocksopp}, C.},
  \bibinfo{author}{{Kaiser}, C.~R.}, \bibinfo{author}{{Charles}, P.~A.},
  \bibinfo{author}{{Rupen}, M.~P.}, \& \bibinfo{author}{{Stone}, R.~P.~S.}
  (\bibinfo{year}{2006}).
\newblock \bibinfo{title}{{A ``Combination Nova'' Outburst in Z Andromedae:
  Nuclear Shell Burning Triggered by a Disk Instability}}.
\newblock {\it \bibinfo{journal}{\apj}\/},  {\it \bibinfo{volume}{636}\/},
  \bibinfo{pages}{1002--1019}. \DOIprefix\doi{10.1086/498206}.
  \href{http://arxiv.org/abs/astro-ph/0509638}{\tt arXiv:astro-ph/0509638}.
\bibitem[{{Sparre} et~al.(2014){Sparre}, {Hartoog}, {Kr{\"u}hler}, {Fynbo},
  {Watson}, {Wiersema}, {D'Elia}, {Zafar}, {Afonso}, {Covino}, {de Ugarte
  Postigo}, {Flores}, {Goldoni}, {Greiner}, {Hjorth}, {Jakobsson}, {Kaper},
  {Klose}, {Levan}, {Malesani}, {Milvang-Jensen}, {Nardini}, {Piranomonte},
  {Sollerman}, {S{\'a}nchez-Ram{\'{\i}}rez}, {Schulze}, {Tanvir}, {Vergani} \&
  {Wijers}}]{Sparre2014}
\bibinfo{author}{{Sparre}, M.}, \bibinfo{author}{{Hartoog}, O.~E.},
  \bibinfo{author}{{Kr{\"u}hler}, T.}, \bibinfo{author}{{Fynbo}, J.~P.~U.},
  \bibinfo{author}{{Watson}, D.~J.}, \bibinfo{author}{{Wiersema}, K.},
  \bibinfo{author}{{D'Elia}, V.}, \bibinfo{author}{{Zafar}, T.},
  \bibinfo{author}{{Afonso}, P.~M.~J.}, \bibinfo{author}{{Covino}, S.},
  \bibinfo{author}{{de Ugarte Postigo}, A.}, \bibinfo{author}{{Flores}, H.},
  \bibinfo{author}{{Goldoni}, P.}, \bibinfo{author}{{Greiner}, J.},
  \bibinfo{author}{{Hjorth}, J.}, \bibinfo{author}{{Jakobsson}, P.},
  \bibinfo{author}{{Kaper}, L.}, \bibinfo{author}{{Klose}, S.},
  \bibinfo{author}{{Levan}, A.~J.}, \bibinfo{author}{{Malesani}, D.},
  \bibinfo{author}{{Milvang-Jensen}, B.}, \bibinfo{author}{{Nardini}, M.},
  \bibinfo{author}{{Piranomonte}, S.}, \bibinfo{author}{{Sollerman}, J.},
  \bibinfo{author}{{S{\'a}nchez-Ram{\'{\i}}rez}, R.},
  \bibinfo{author}{{Schulze}, S.}, \bibinfo{author}{{Tanvir}, N.~R.},
  \bibinfo{author}{{Vergani}, S.~D.}, \& \bibinfo{author}{{Wijers},
  R.~A.~M.~J.} (\bibinfo{year}{2014}).
\newblock \bibinfo{title}{{The Metallicity and Dust Content of a Redshift 5
  Gamma-Ray Burst Host Galaxy}}.
\newblock {\it \bibinfo{journal}{\apj}\/},  {\it \bibinfo{volume}{785}\/},
  \bibinfo{pages}{150}. \DOIprefix\doi{10.1088/0004-637X/785/2/150}.
  \href{http://arxiv.org/abs/1309.2940}{\tt arXiv:1309.2940}.
\bibitem[{{Starling} et~al.(2011){Starling}, {Wiersema}, {Levan}, {Sakamoto},
  {Bersier}, {Goldoni}, {Oates}, {Rowlinson}, {Campana}, {Sollerman}, {Tanvir},
  {Malesani}, {Fynbo}, {Covino}, {D'Avanzo}, {O'Brien}, {Page}, {Osborne},
  {Vergani}, {Barthelmy}, {Burrows}, {Cano}, {Curran}, {de Pasquale}, {D'Elia},
  {Evans}, {Flores}, {Fruchter}, {Garnavich}, {Gehrels}, {Gorosabel}, {Hjorth},
  {Holland}, {van der Horst}, {Hurkett}, {Jakobsson}, {Kamble}, {Kouveliotou},
  {Kuin}, {Kaper}, {Mazzali}, {Nugent}, {Pian}, {Stamatikos}, {Th{\"o}ne} \&
  {Woosley}}]{Starling2011}
\bibinfo{author}{{Starling}, R.~L.~C.}, \bibinfo{author}{{Wiersema}, K.},
  \bibinfo{author}{{Levan}, A.~J.}, \bibinfo{author}{{Sakamoto}, T.},
  \bibinfo{author}{{Bersier}, D.}, \bibinfo{author}{{Goldoni}, P.},
  \bibinfo{author}{{Oates}, S.~R.}, \bibinfo{author}{{Rowlinson}, A.},
  \bibinfo{author}{{Campana}, S.}, \bibinfo{author}{{Sollerman}, J.},
  \bibinfo{author}{{Tanvir}, N.~R.}, \bibinfo{author}{{Malesani}, D.},
  \bibinfo{author}{{Fynbo}, J.~P.~U.}, \bibinfo{author}{{Covino}, S.},
  \bibinfo{author}{{D'Avanzo}, P.}, \bibinfo{author}{{O'Brien}, P.~T.},
  \bibinfo{author}{{Page}, K.~L.}, \bibinfo{author}{{Osborne}, J.~P.},
  \bibinfo{author}{{Vergani}, S.~D.}, \bibinfo{author}{{Barthelmy}, S.},
  \bibinfo{author}{{Burrows}, D.~N.}, \bibinfo{author}{{Cano}, Z.},
  \bibinfo{author}{{Curran}, P.~A.}, \bibinfo{author}{{de Pasquale}, M.},
  \bibinfo{author}{{D'Elia}, V.}, \bibinfo{author}{{Evans}, P.~A.},
  \bibinfo{author}{{Flores}, H.}, \bibinfo{author}{{Fruchter}, A.~S.},
  \bibinfo{author}{{Garnavich}, P.}, \bibinfo{author}{{Gehrels}, N.},
  \bibinfo{author}{{Gorosabel}, J.}, \bibinfo{author}{{Hjorth}, J.},
  \bibinfo{author}{{Holland}, S.~T.}, \bibinfo{author}{{van der Horst}, A.~J.},
  \bibinfo{author}{{Hurkett}, C.~P.}, \bibinfo{author}{{Jakobsson}, P.},
  \bibinfo{author}{{Kamble}, A.~P.}, \bibinfo{author}{{Kouveliotou}, C.},
  \bibinfo{author}{{Kuin}, N.~P.~M.}, \bibinfo{author}{{Kaper}, L.},
  \bibinfo{author}{{Mazzali}, P.~A.}, \bibinfo{author}{{Nugent}, P.~E.},
  \bibinfo{author}{{Pian}, E.}, \bibinfo{author}{{Stamatikos}, M.},
  \bibinfo{author}{{Th{\"o}ne}, C.~C.}, \& \bibinfo{author}{{Woosley}, S.~E.}
  (\bibinfo{year}{2011}).
\newblock \bibinfo{title}{{Discovery of the nearby long, soft GRB 100316D with
  an associated supernova}}.
\newblock {\it \bibinfo{journal}{\mnras}\/},  {\it \bibinfo{volume}{411}\/},
  \bibinfo{pages}{2792--2803}.
  \DOIprefix\doi{10.1111/j.1365-2966.2010.17879.x}.
  \href{http://arxiv.org/abs/1004.2919}{\tt arXiv:1004.2919}.
\bibitem[{{Starrfield} et~al.(2009){Starrfield}, {Iliadis}, {Hix}, {Timmes} \&
  {Sparks}}]{Starrfield2009}
\bibinfo{author}{{Starrfield}, S.}, \bibinfo{author}{{Iliadis}, C.},
  \bibinfo{author}{{Hix}, W.~R.}, \bibinfo{author}{{Timmes}, F.~X.}, \&
  \bibinfo{author}{{Sparks}, W.~M.} (\bibinfo{year}{2009}).
\newblock \bibinfo{title}{{The Effects of the pep Nuclear Reaction and Other
  Improvements in the Nuclear Reaction Rate Library on Simulations of the
  Classical Nova Outburst}}.
\newblock {\it \bibinfo{journal}{\apj}\/},  {\it \bibinfo{volume}{692}\/},
  \bibinfo{pages}{1532--1542}. \DOIprefix\doi{10.1088/0004-637X/692/2/1532}.
  \href{http://arxiv.org/abs/0811.0197}{\tt arXiv:0811.0197}.
\bibitem[{{Stratta} et~al.(2017){Stratta}, {Ciolfi}, {Amati}, {Ghirlanda},
  {Tanvir}, {Bozzo}, {Gotz}, {O'Brien}, {Frontera}, {Osborne}, {Rezzolla},
  {Rossi}, {Maiorano}, {Vinciguerra}, {Guidorzi}, {Drago}, {Nicastro},
  {Palazzi}, {Branchesi}, {Boer}, {Brocato}, {Bulgarelli}, {Covino}, {D'Elia},
  {Dainotti}, {De Pasquale}, {Gendre}, {Jonker}, {Longo}, {Mereghetti},
  {Mignani}, {Mundell}, {Piranomonte}, {Razzano}, {Sz{\'e}csi}, {van Putten},
  {Zhang}, {Hudec}, {Vergani}, {Malesani}, {D'Avanzo}, {Colafrancesco},
  {Stamerra}, {Caruana}, {Starling}, {Willingale}, {Salvaterra}, {Maio},
  {Greiner}, {Rosati}, {Labanti}, {Fuschino}, {Riccardo}, {Grado}, {Colpi},
  {Rodic}, {Patricelli} \& {Bernardini}}]{Stratta2017}
\bibinfo{author}{{Stratta}, G.}, \bibinfo{author}{{Ciolfi}, R.},
  \bibinfo{author}{{Amati}, L.}, \bibinfo{author}{{Ghirlanda}, G.},
  \bibinfo{author}{{Tanvir}, N.}, \bibinfo{author}{{Bozzo}, E.},
  \bibinfo{author}{{Gotz}, D.}, \bibinfo{author}{{O'Brien}, P.},
  \bibinfo{author}{{Frontera}, F.}, \bibinfo{author}{{Osborne}, J.~P.},
  \bibinfo{author}{{Rezzolla}, L.}, \bibinfo{author}{{Rossi}, A.},
  \bibinfo{author}{{Maiorano}, E.}, \bibinfo{author}{{Vinciguerra}, S.},
  \bibinfo{author}{{Guidorzi}, C.}, \bibinfo{author}{{Drago}, A.},
  \bibinfo{author}{{Nicastro}, L.}, \bibinfo{author}{{Palazzi}, E.},
  \bibinfo{author}{{Branchesi}, M.}, \bibinfo{author}{{Boer}, M.},
  \bibinfo{author}{{Brocato}, E.}, \bibinfo{author}{{Bulgarelli}, A.},
  \bibinfo{author}{{Covino}, S.}, \bibinfo{author}{{D'Elia}, V.},
  \bibinfo{author}{{Dainotti}, M.~G.}, \bibinfo{author}{{De Pasquale}, M.},
  \bibinfo{author}{{Gendre}, B.}, \bibinfo{author}{{Jonker}, P.},
  \bibinfo{author}{{Longo}, F.}, \bibinfo{author}{{Mereghetti}, S.},
  \bibinfo{author}{{Mignani}, R.}, \bibinfo{author}{{Mundell}, C.~G.},
  \bibinfo{author}{{Piranomonte}, S.}, \bibinfo{author}{{Razzano}, M.},
  \bibinfo{author}{{Sz{\'e}csi}, D.}, \bibinfo{author}{{van Putten}, M.},
  \bibinfo{author}{{Zhang}, B.}, \bibinfo{author}{{Hudec}, R.},
  \bibinfo{author}{{Vergani}, S.}, \bibinfo{author}{{Malesani}, D.},
  \bibinfo{author}{{D'Avanzo}, P.}, \bibinfo{author}{{Colafrancesco}, S.},
  \bibinfo{author}{{Stamerra}, A.}, \bibinfo{author}{{Caruana}, J.},
  \bibinfo{author}{{Starling}, R.}, \bibinfo{author}{{Willingale}, R.},
  \bibinfo{author}{{Salvaterra}, R.}, \bibinfo{author}{{Maio}, U.},
  \bibinfo{author}{{Greiner}, J.}, \bibinfo{author}{{Rosati}, P.},
  \bibinfo{author}{{Labanti}, C.}, \bibinfo{author}{{Fuschino}, F.},
  \bibinfo{author}{{Riccardo}, C.}, \bibinfo{author}{{Grado}, A.},
  \bibinfo{author}{{Colpi}, M.}, \bibinfo{author}{{Rodic}, T.},
  \bibinfo{author}{{Patricelli}, B.}, \& \bibinfo{author}{{Bernardini}, M.}
  (\bibinfo{year}{2017}).
\newblock \bibinfo{title}{{THESEUS: a key space mission for Multi-Messenger
  Astrophysics}}.
\newblock {\it \bibinfo{journal}{ArXiv e-prints}\/}, .
  \href{http://arxiv.org/abs/1712.08153}{\tt arXiv:1712.08153}.
\bibitem[{{Stratta} et~al.(2013){Stratta}, {Gendre}, {Atteia}, {Bo{\"e}r},
  {Coward}, {De Pasquale}, {Howell}, {Klotz}, {Oates} \& {Piro}}]{Stratta2013}
\bibinfo{author}{{Stratta}, G.}, \bibinfo{author}{{Gendre}, B.},
  \bibinfo{author}{{Atteia}, J.~L.}, \bibinfo{author}{{Bo{\"e}r}, M.},
  \bibinfo{author}{{Coward}, D.~M.}, \bibinfo{author}{{De Pasquale}, M.},
  \bibinfo{author}{{Howell}, E.}, \bibinfo{author}{{Klotz}, A.},
  \bibinfo{author}{{Oates}, S.}, \& \bibinfo{author}{{Piro}, L.}
  (\bibinfo{year}{2013}).
\newblock \bibinfo{title}{{The Ultra-long GRB 111209A. II. Prompt to Afterglow
  and Afterglow Properties}}.
\newblock {\it \bibinfo{journal}{\apj}\/},  {\it \bibinfo{volume}{779}\/},
  \bibinfo{pages}{66}. \DOIprefix\doi{10.1088/0004-637X/779/1/66}.
  \href{http://arxiv.org/abs/1306.1699}{\tt arXiv:1306.1699}.
\bibitem[{{Strohmayer} \& {Watts}(2006)}]{Strohmayer2006}
\bibinfo{author}{{Strohmayer}, T.~E.}, \& \bibinfo{author}{{Watts}, A.~L.}
  (\bibinfo{year}{2006}).
\newblock \bibinfo{title}{{The 2004 Hyperflare from SGR 1806-20: Further
  Evidence for Global Torsional Vibrations}}.
\newblock {\it \bibinfo{journal}{\apj}\/},  {\it \bibinfo{volume}{653}\/},
  \bibinfo{pages}{593--601}. \DOIprefix\doi{10.1086/508703}.
  \href{http://arxiv.org/abs/astro-ph/0608463}{\tt arXiv:astro-ph/0608463}.
\bibitem[{{Sullivan} et~al.(2011){Sullivan}, {Guy}, {Conley}, {Regnault},
  {Astier}, {Balland}, {Basa}, {Carlberg}, {Fouchez}, {Hardin}, {Hook},
  {Howell}, {Pain}, {Palanque-Delabrouille}, {Perrett}, {Pritchet}, {Rich},
  {Ruhlmann-Kleider}, {Balam}, {Baumont}, {Ellis}, {Fabbro}, {Fakhouri},
  {Fourmanoit}, {Gonz{\'a}lez-Gait{\'a}n}, {Graham}, {Hudson}, {Hsiao},
  {Kronborg}, {Lidman}, {Mourao}, {Neill}, {Perlmutter}, {Ripoche}, {Suzuki} \&
  {Walker}}]{Sullivan2011}
\bibinfo{author}{{Sullivan}, M.}, \bibinfo{author}{{Guy}, J.},
  \bibinfo{author}{{Conley}, A.}, \bibinfo{author}{{Regnault}, N.},
  \bibinfo{author}{{Astier}, P.}, \bibinfo{author}{{Balland}, C.},
  \bibinfo{author}{{Basa}, S.}, \bibinfo{author}{{Carlberg}, R.~G.},
  \bibinfo{author}{{Fouchez}, D.}, \bibinfo{author}{{Hardin}, D.},
  \bibinfo{author}{{Hook}, I.~M.}, \bibinfo{author}{{Howell}, D.~A.},
  \bibinfo{author}{{Pain}, R.}, \bibinfo{author}{{Palanque-Delabrouille}, N.},
  \bibinfo{author}{{Perrett}, K.~M.}, \bibinfo{author}{{Pritchet}, C.~J.},
  \bibinfo{author}{{Rich}, J.}, \bibinfo{author}{{Ruhlmann-Kleider}, V.},
  \bibinfo{author}{{Balam}, D.}, \bibinfo{author}{{Baumont}, S.},
  \bibinfo{author}{{Ellis}, R.~S.}, \bibinfo{author}{{Fabbro}, S.},
  \bibinfo{author}{{Fakhouri}, H.~K.}, \bibinfo{author}{{Fourmanoit}, N.},
  \bibinfo{author}{{Gonz{\'a}lez-Gait{\'a}n}, S.}, \bibinfo{author}{{Graham},
  M.~L.}, \bibinfo{author}{{Hudson}, M.~J.}, \bibinfo{author}{{Hsiao}, E.},
  \bibinfo{author}{{Kronborg}, T.}, \bibinfo{author}{{Lidman}, C.},
  \bibinfo{author}{{Mourao}, A.~M.}, \bibinfo{author}{{Neill}, J.~D.},
  \bibinfo{author}{{Perlmutter}, S.}, \bibinfo{author}{{Ripoche}, P.},
  \bibinfo{author}{{Suzuki}, N.}, \& \bibinfo{author}{{Walker}, E.~S.}
  (\bibinfo{year}{2011}).
\newblock \bibinfo{title}{{SNLS3: Constraints on Dark Energy Combining the
  Supernova Legacy Survey Three-year Data with Other Probes}}.
\newblock {\it \bibinfo{journal}{\apj}\/},  {\it \bibinfo{volume}{737}\/},
  \bibinfo{pages}{102}. \DOIprefix\doi{10.1088/0004-637X/737/2/102}.
  \href{http://arxiv.org/abs/1104.1444}{\tt arXiv:1104.1444}.
\bibitem[{{Sz{\'e}csi}(2017)}]{Szecsi2017}
\bibinfo{author}{{Sz{\'e}csi}, D.} (\bibinfo{year}{2017}).
\newblock \bibinfo{title}{{How may short-duration GRBs form? A review of
  progenitor theories.}}
\newblock {\it \bibinfo{journal}{Contributions of the Astronomical Observatory
  Skalnate Pleso}\/},  {\it \bibinfo{volume}{47}\/}, \bibinfo{pages}{108--115}.
\bibitem[{{Sz{\'e}csi} et~al.(2015){Sz{\'e}csi}, {Langer}, {Yoon}, {Sanyal},
  {de Mink}, {Evans} \& {Dermine}}]{Szecsi2015}
\bibinfo{author}{{Sz{\'e}csi}, D.}, \bibinfo{author}{{Langer}, N.},
  \bibinfo{author}{{Yoon}, S.-C.}, \bibinfo{author}{{Sanyal}, D.},
  \bibinfo{author}{{de Mink}, S.}, \bibinfo{author}{{Evans}, C.~J.}, \&
  \bibinfo{author}{{Dermine}, T.} (\bibinfo{year}{2015}).
\newblock \bibinfo{title}{{Low-metallicity massive single stars with rotation.
  Evolutionary models applicable to I Zwicky 18}}.
\newblock {\it \bibinfo{journal}{\aap}\/},  {\it \bibinfo{volume}{581}\/},
  \bibinfo{pages}{A15}. \DOIprefix\doi{10.1051/0004-6361/201526617}.
  \href{http://arxiv.org/abs/1506.09132}{\tt arXiv:1506.09132}.
\bibitem[{{Takami} et~al.(2014){Takami}, {Rezzolla} \& {Baiotti}}]{Takami2014}
\bibinfo{author}{{Takami}, K.}, \bibinfo{author}{{Rezzolla}, L.}, \&
  \bibinfo{author}{{Baiotti}, L.} (\bibinfo{year}{2014}).
\newblock \bibinfo{title}{{Constraining the Equation of State of Neutron Stars
  from Binary Mergers}}.
\newblock {\it \bibinfo{journal}{Physical Review Letters}\/},  {\it
  \bibinfo{volume}{113}\/}, \bibinfo{pages}{091104}.
  \DOIprefix\doi{10.1103/PhysRevLett.113.091104}.
  \href{http://arxiv.org/abs/1403.5672}{\tt arXiv:1403.5672}.
\bibitem[{{Tanvir} et~al.(2012){Tanvir}, {Levan}, {Fruchter}, {Fynbo},
  {Hjorth}, {Wiersema}, {Bremer}, {Rhoads}, {Jakobsson}, {O'Brien}, {Stanway},
  {Bersier}, {Natarajan}, {Greiner}, {Watson}, {Castro-Tirado}, {Wijers},
  {Starling}, {Misra}, {Graham} \& {Kouveliotou}}]{Tanvir2012}
\bibinfo{author}{{Tanvir}, N.~R.}, \bibinfo{author}{{Levan}, A.~J.},
  \bibinfo{author}{{Fruchter}, A.~S.}, \bibinfo{author}{{Fynbo}, J.~P.~U.},
  \bibinfo{author}{{Hjorth}, J.}, \bibinfo{author}{{Wiersema}, K.},
  \bibinfo{author}{{Bremer}, M.~N.}, \bibinfo{author}{{Rhoads}, J.},
  \bibinfo{author}{{Jakobsson}, P.}, \bibinfo{author}{{O'Brien}, P.~T.},
  \bibinfo{author}{{Stanway}, E.~R.}, \bibinfo{author}{{Bersier}, D.},
  \bibinfo{author}{{Natarajan}, P.}, \bibinfo{author}{{Greiner}, J.},
  \bibinfo{author}{{Watson}, D.}, \bibinfo{author}{{Castro-Tirado}, A.~J.},
  \bibinfo{author}{{Wijers}, R.~A.~M.~J.}, \bibinfo{author}{{Starling},
  R.~L.~C.}, \bibinfo{author}{{Misra}, K.}, \bibinfo{author}{{Graham}, J.~F.},
  \& \bibinfo{author}{{Kouveliotou}, C.} (\bibinfo{year}{2012}).
\newblock \bibinfo{title}{{Star Formation in the Early Universe: Beyond the Tip
  of the Iceberg}}.
\newblock {\it \bibinfo{journal}{\apj}\/},  {\it \bibinfo{volume}{754}\/},
  \bibinfo{pages}{46}. \DOIprefix\doi{10.1088/0004-637X/754/1/46}.
  \href{http://arxiv.org/abs/1201.6074}{\tt arXiv:1201.6074}.
\bibitem[{{Tanvir} et~al.(2013){Tanvir}, {Levan}, {Fruchter}, {Hjorth},
  {Hounsell}, {Wiersema} \& {Tunnicliffe}}]{Tanvir2013}
\bibinfo{author}{{Tanvir}, N.~R.}, \bibinfo{author}{{Levan}, A.~J.},
  \bibinfo{author}{{Fruchter}, A.~S.}, \bibinfo{author}{{Hjorth}, J.},
  \bibinfo{author}{{Hounsell}, R.~A.}, \bibinfo{author}{{Wiersema}, K.}, \&
  \bibinfo{author}{{Tunnicliffe}, R.~L.} (\bibinfo{year}{2013}).
\newblock \bibinfo{title}{{A `kilonova' associated with the short-duration
  {$\gamma$}-ray burst GRB 130603B}}.
\newblock {\it \bibinfo{journal}{\nat}\/},  {\it \bibinfo{volume}{500}\/},
  \bibinfo{pages}{547--549}. \DOIprefix\doi{10.1038/nature12505}.
  \href{http://arxiv.org/abs/1306.4971}{\tt arXiv:1306.4971}.
\bibitem[{{Tanvir} et~al.(2017){Tanvir}, {Levan}, {Gonz{\'a}lez-Fern{\'a}ndez},
  {Korobkin}, {Mandel}, {Rosswog}, {Hjorth}, {D'Avanzo}, {Fruchter}, {Fryer},
  {Kangas}, {Milvang-Jensen}, {Rosetti}, {Steeghs}, {Wollaeger}, {Cano},
  {Copperwheat}, {Covino}, {D'Elia}, {de Ugarte Postigo}, {Evans}, {Even},
  {Fairhurst}, {Figuera Jaimes}, {Fontes}, {Fujii}, {Fynbo}, {Gompertz},
  {Greiner}, {Hodosan}, {Irwin}, {Jakobsson}, {J{\o}rgensen}, {Kann}, {Lyman},
  {Malesani}, {McMahon}, {Melandri}, {O'Brien}, {Osborne}, {Palazzi}, {Perley},
  {Pian}, {Piranomonte}, {Rabus}, {Rol}, {Rowlinson}, {Schulze}, {Sutton},
  {Th{\"o}ne}, {Ulaczyk}, {Watson}, {Wiersema} \& {Wijers}}]{Tanvir2017}
\bibinfo{author}{{Tanvir}, N.~R.}, \bibinfo{author}{{Levan}, A.~J.},
  \bibinfo{author}{{Gonz{\'a}lez-Fern{\'a}ndez}, C.},
  \bibinfo{author}{{Korobkin}, O.}, \bibinfo{author}{{Mandel}, I.},
  \bibinfo{author}{{Rosswog}, S.}, \bibinfo{author}{{Hjorth}, J.},
  \bibinfo{author}{{D'Avanzo}, P.}, \bibinfo{author}{{Fruchter}, A.~S.},
  \bibinfo{author}{{Fryer}, C.~L.}, \bibinfo{author}{{Kangas}, T.},
  \bibinfo{author}{{Milvang-Jensen}, B.}, \bibinfo{author}{{Rosetti}, S.},
  \bibinfo{author}{{Steeghs}, D.}, \bibinfo{author}{{Wollaeger}, R.~T.},
  \bibinfo{author}{{Cano}, Z.}, \bibinfo{author}{{Copperwheat}, C.~M.},
  \bibinfo{author}{{Covino}, S.}, \bibinfo{author}{{D'Elia}, V.},
  \bibinfo{author}{{de Ugarte Postigo}, A.}, \bibinfo{author}{{Evans}, P.~A.},
  \bibinfo{author}{{Even}, W.~P.}, \bibinfo{author}{{Fairhurst}, S.},
  \bibinfo{author}{{Figuera Jaimes}, R.}, \bibinfo{author}{{Fontes}, C.~J.},
  \bibinfo{author}{{Fujii}, Y.~I.}, \bibinfo{author}{{Fynbo}, J.~P.~U.},
  \bibinfo{author}{{Gompertz}, B.~P.}, \bibinfo{author}{{Greiner}, J.},
  \bibinfo{author}{{Hodosan}, G.}, \bibinfo{author}{{Irwin}, M.~J.},
  \bibinfo{author}{{Jakobsson}, P.}, \bibinfo{author}{{J{\o}rgensen}, U.~G.},
  \bibinfo{author}{{Kann}, D.~A.}, \bibinfo{author}{{Lyman}, J.~D.},
  \bibinfo{author}{{Malesani}, D.}, \bibinfo{author}{{McMahon}, R.~G.},
  \bibinfo{author}{{Melandri}, A.}, \bibinfo{author}{{O'Brien}, P.~T.},
  \bibinfo{author}{{Osborne}, J.~P.}, \bibinfo{author}{{Palazzi}, E.},
  \bibinfo{author}{{Perley}, D.~A.}, \bibinfo{author}{{Pian}, E.},
  \bibinfo{author}{{Piranomonte}, S.}, \bibinfo{author}{{Rabus}, M.},
  \bibinfo{author}{{Rol}, E.}, \bibinfo{author}{{Rowlinson}, A.},
  \bibinfo{author}{{Schulze}, S.}, \bibinfo{author}{{Sutton}, P.},
  \bibinfo{author}{{Th{\"o}ne}, C.~C.}, \bibinfo{author}{{Ulaczyk}, K.},
  \bibinfo{author}{{Watson}, D.}, \bibinfo{author}{{Wiersema}, K.}, \&
  \bibinfo{author}{{Wijers}, R.~A.~M.~J.} (\bibinfo{year}{2017}).
\newblock \bibinfo{title}{{The Emergence of a Lanthanide-rich Kilonova
  Following the Merger of Two Neutron Stars}}.
\newblock {\it \bibinfo{journal}{\apjl}\/},  {\it \bibinfo{volume}{848}\/},
  \bibinfo{pages}{L27}. \DOIprefix\doi{10.3847/2041-8213/aa90b6}.
  \href{http://arxiv.org/abs/1710.05455}{\tt arXiv:1710.05455}.
\bibitem[{{Thompson} \& {Duncan}(1995)}]{Thompson1995}
\bibinfo{author}{{Thompson}, C.}, \& \bibinfo{author}{{Duncan}, R.~C.}
  (\bibinfo{year}{1995}).
\newblock \bibinfo{title}{{The soft gamma repeaters as very strongly magnetized
  neutron stars - I. Radiative mechanism for outbursts}}.
\newblock {\it \bibinfo{journal}{\mnras}\/},  {\it \bibinfo{volume}{275}\/},
  \bibinfo{pages}{255--300}. \DOIprefix\doi{10.1093/mnras/275.2.255}.
\bibitem[{{Tramacere} et~al.(2007){Tramacere}, {Giommi}, {Massaro}, {Perri},
  {Nesci}, {Colafrancesco}, {Tagliaferri}, {Chincarini}, {Falcone}, {Burrows},
  {Roming}, {McMath Chester} \& {Gehrels}}]{Tramacere2007}
\bibinfo{author}{{Tramacere}, A.}, \bibinfo{author}{{Giommi}, P.},
  \bibinfo{author}{{Massaro}, E.}, \bibinfo{author}{{Perri}, M.},
  \bibinfo{author}{{Nesci}, R.}, \bibinfo{author}{{Colafrancesco}, S.},
  \bibinfo{author}{{Tagliaferri}, G.}, \bibinfo{author}{{Chincarini}, G.},
  \bibinfo{author}{{Falcone}, A.}, \bibinfo{author}{{Burrows}, D.~N.},
  \bibinfo{author}{{Roming}, P.}, \bibinfo{author}{{McMath Chester}, M.}, \&
  \bibinfo{author}{{Gehrels}, N.} (\bibinfo{year}{2007}).
\newblock \bibinfo{title}{{SWIFT observations of TeV BL Lacertae objects}}.
\newblock {\it \bibinfo{journal}{\aap}\/},  {\it \bibinfo{volume}{467}\/},
  \bibinfo{pages}{501--508}. \DOIprefix\doi{10.1051/0004-6361:20066226}.
  \href{http://arxiv.org/abs/astro-ph/0611276}{\tt arXiv:astro-ph/0611276}.
\bibitem[{{Troja} et~al.(2017){Troja}, {Lipunov}, {Mundell}, {Butler},
  {Watson}, {Kobayashi}, {Cenko}, {Marshall}, {Ricci}, {Fruchter}, {Wieringa},
  {Gorbovskoy}, {Kornilov}, {Kutyrev}, {Lee}, {Toy}, {Tyurina}, {Budnev},
  {Buckley}, {Gonzalez}, {Gress}, {Horesh}, {Panasyuk}, {Prochaska},
  {Ramirez-Ruiz}, {Rebolo Lopez}, {Richer}, {Roman-Zuniga}, {Serra-Ricart},
  {Yurkov} \& {Gehrels}}]{Troja2017}
\bibinfo{author}{{Troja}, E.}, \bibinfo{author}{{Lipunov}, V.~M.},
  \bibinfo{author}{{Mundell}, C.~G.}, \bibinfo{author}{{Butler}, N.~R.},
  \bibinfo{author}{{Watson}, A.~M.}, \bibinfo{author}{{Kobayashi}, S.},
  \bibinfo{author}{{Cenko}, S.~B.}, \bibinfo{author}{{Marshall}, F.~E.},
  \bibinfo{author}{{Ricci}, R.}, \bibinfo{author}{{Fruchter}, A.},
  \bibinfo{author}{{Wieringa}, M.~H.}, \bibinfo{author}{{Gorbovskoy}, E.~S.},
  \bibinfo{author}{{Kornilov}, V.}, \bibinfo{author}{{Kutyrev}, A.},
  \bibinfo{author}{{Lee}, W.~H.}, \bibinfo{author}{{Toy}, V.},
  \bibinfo{author}{{Tyurina}, N.~V.}, \bibinfo{author}{{Budnev}, N.~M.},
  \bibinfo{author}{{Buckley}, D.~A.~H.}, \bibinfo{author}{{Gonzalez}, J.},
  \bibinfo{author}{{Gress}, O.}, \bibinfo{author}{{Horesh}, A.},
  \bibinfo{author}{{Panasyuk}, M.~I.}, \bibinfo{author}{{Prochaska}, J.~X.},
  \bibinfo{author}{{Ramirez-Ruiz}, E.}, \bibinfo{author}{{Rebolo Lopez}, R.},
  \bibinfo{author}{{Richer}, M.~G.}, \bibinfo{author}{{Roman-Zuniga}, C.},
  \bibinfo{author}{{Serra-Ricart}, M.}, \bibinfo{author}{{Yurkov}, V.}, \&
  \bibinfo{author}{{Gehrels}, N.} (\bibinfo{year}{2017}).
\newblock \bibinfo{title}{{Significant and variable linear polarization during
  the prompt optical flash of GRB 160625B.}}
\newblock {\it \bibinfo{journal}{\nat}\/},  {\it \bibinfo{volume}{547}\/},
  \bibinfo{pages}{425--427}.
\bibitem[{{Troja} et~al.(2016){Troja}, {Sakamoto}, {Cenko}, {Lien}, {Gehrels},
  {Castro-Tirado}, {Ricci}, {Capone}, {Toy}, {Kutyrev}, {Kawai}, {Cucchiara},
  {Fruchter}, {Gorosabel}, {Jeong}, {Levan}, {Perley}, {Sanchez-Ramirez},
  {Tanvir} \& {Veilleux}}]{Troja2016}
\bibinfo{author}{{Troja}, E.}, \bibinfo{author}{{Sakamoto}, T.},
  \bibinfo{author}{{Cenko}, S.~B.}, \bibinfo{author}{{Lien}, A.},
  \bibinfo{author}{{Gehrels}, N.}, \bibinfo{author}{{Castro-Tirado}, A.~J.},
  \bibinfo{author}{{Ricci}, R.}, \bibinfo{author}{{Capone}, J.},
  \bibinfo{author}{{Toy}, V.}, \bibinfo{author}{{Kutyrev}, A.},
  \bibinfo{author}{{Kawai}, N.}, \bibinfo{author}{{Cucchiara}, A.},
  \bibinfo{author}{{Fruchter}, A.}, \bibinfo{author}{{Gorosabel}, J.},
  \bibinfo{author}{{Jeong}, S.}, \bibinfo{author}{{Levan}, A.},
  \bibinfo{author}{{Perley}, D.}, \bibinfo{author}{{Sanchez-Ramirez}, R.},
  \bibinfo{author}{{Tanvir}, N.}, \& \bibinfo{author}{{Veilleux}, S.}
  (\bibinfo{year}{2016}).
\newblock \bibinfo{title}{{An Achromatic Break in the Afterglow of the Short
  GRB 140903A: Evidence for a Narrow Jet}}.
\newblock {\it \bibinfo{journal}{\apj}\/},  {\it \bibinfo{volume}{827}\/},
  \bibinfo{pages}{102}. \DOIprefix\doi{10.3847/0004-637X/827/2/102}.
  \href{http://arxiv.org/abs/1605.03573}{\tt arXiv:1605.03573}.
\bibitem[{{Ueda} et~al.(2005){Ueda}, {Ishisaki}, {Takahashi}, {Makishima} \&
  {Ohashi}}]{Ueda2005}
\bibinfo{author}{{Ueda}, Y.}, \bibinfo{author}{{Ishisaki}, Y.},
  \bibinfo{author}{{Takahashi}, T.}, \bibinfo{author}{{Makishima}, K.}, \&
  \bibinfo{author}{{Ohashi}, T.} (\bibinfo{year}{2005}).
\newblock \bibinfo{title}{{The ASCA Medium Sensitivity Survey (The GIS Catalog
  Project): Source Catalog II.}}
\newblock {\it \bibinfo{journal}{\apjs}\/},  {\it \bibinfo{volume}{161}\/},
  \bibinfo{pages}{185--223}. \DOIprefix\doi{10.1086/468187}.
\bibitem[{{van Eerten} et~al.(2010){van Eerten}, {Zhang} \&
  {MacFadyen}}]{vanEerten2010}
\bibinfo{author}{{van Eerten}, H.}, \bibinfo{author}{{Zhang}, W.}, \&
  \bibinfo{author}{{MacFadyen}, A.} (\bibinfo{year}{2010}).
\newblock \bibinfo{title}{{Off-axis Gamma-ray Burst Afterglow Modeling Based on
  a Two-dimensional Axisymmetric Hydrodynamics Simulation}}.
\newblock {\it \bibinfo{journal}{\apj}\/},  {\it \bibinfo{volume}{722}\/},
  \bibinfo{pages}{235--247}. \DOIprefix\doi{10.1088/0004-637X/722/1/235}.
  \href{http://arxiv.org/abs/1006.5125}{\tt arXiv:1006.5125}.
\bibitem[{{van Putten} et~al.(2014){van Putten}, {Guidorzi} \&
  {Frontera}}]{vanPutten2014}
\bibinfo{author}{{van Putten}, M.~H.~P.~M.}, \bibinfo{author}{{Guidorzi}, C.},
  \& \bibinfo{author}{{Frontera}, F.} (\bibinfo{year}{2014}).
\newblock \bibinfo{title}{{Broadband Turbulent Spectra in Gamma-Ray Burst Light
  Curves}}.
\newblock {\it \bibinfo{journal}{\apj}\/},  {\it \bibinfo{volume}{786}\/},
  \bibinfo{pages}{146}. \DOIprefix\doi{10.1088/0004-637X/786/2/146}.
  \href{http://arxiv.org/abs/1411.6940}{\tt arXiv:1411.6940}.
\bibitem[{{Vanzella} et~al.(2017{\natexlab{a}}){Vanzella}, {Calura},
  {Meneghetti}, {Mercurio}, {Castellano}, {Caminha}, {Balestra}, {Rosati},
  {Tozzi}, {De Barros}, {Grazian}, {D'Ercole}, {Ciotti}, {Caputi}, {Grillo},
  {Merlin}, {Pentericci}, {Fontana}, {Cristiani} \& {Coe}}]{Vanzella2017b}
\bibinfo{author}{{Vanzella}, E.}, \bibinfo{author}{{Calura}, F.},
  \bibinfo{author}{{Meneghetti}, M.}, \bibinfo{author}{{Mercurio}, A.},
  \bibinfo{author}{{Castellano}, M.}, \bibinfo{author}{{Caminha}, G.~B.},
  \bibinfo{author}{{Balestra}, I.}, \bibinfo{author}{{Rosati}, P.},
  \bibinfo{author}{{Tozzi}, P.}, \bibinfo{author}{{De Barros}, S.},
  \bibinfo{author}{{Grazian}, A.}, \bibinfo{author}{{D'Ercole}, A.},
  \bibinfo{author}{{Ciotti}, L.}, \bibinfo{author}{{Caputi}, K.},
  \bibinfo{author}{{Grillo}, C.}, \bibinfo{author}{{Merlin}, E.},
  \bibinfo{author}{{Pentericci}, L.}, \bibinfo{author}{{Fontana}, A.},
  \bibinfo{author}{{Cristiani}, S.}, \& \bibinfo{author}{{Coe}, D.}
  (\bibinfo{year}{2017}{\natexlab{a}}).
\newblock \bibinfo{title}{{Paving the way for the JWST: witnessing globular
  cluster formation at z 3}}.
\newblock {\it \bibinfo{journal}{\mnras}\/},  {\it \bibinfo{volume}{467}\/},
  \bibinfo{pages}{4304--4321}. \DOIprefix\doi{10.1093/mnras/stx351}.
  \href{http://arxiv.org/abs/1612.01526}{\tt arXiv:1612.01526}.
\bibitem[{{Vanzella} et~al.(2017{\natexlab{b}}){Vanzella}, {Castellano},
  {Meneghetti}, {Mercurio}, {Caminha}, {Cupani}, {Calura}, {Christensen},
  {Merlin}, {Rosati}, {Gronke}, {Dijkstra}, {Mignoli}, {Gilli}, {De Barros},
  {Caputi}, {Grillo}, {Balestra}, {Cristiani}, {Nonino}, {Giallongo},
  {Grazian}, {Pentericci}, {Fontana}, {Comastri}, {Vignali}, {Zamorani},
  {Brusa}, {Bergamini} \& {Tozzi}}]{Vanzella2017}
\bibinfo{author}{{Vanzella}, E.}, \bibinfo{author}{{Castellano}, M.},
  \bibinfo{author}{{Meneghetti}, M.}, \bibinfo{author}{{Mercurio}, A.},
  \bibinfo{author}{{Caminha}, G.~B.}, \bibinfo{author}{{Cupani}, G.},
  \bibinfo{author}{{Calura}, F.}, \bibinfo{author}{{Christensen}, L.},
  \bibinfo{author}{{Merlin}, E.}, \bibinfo{author}{{Rosati}, P.},
  \bibinfo{author}{{Gronke}, M.}, \bibinfo{author}{{Dijkstra}, M.},
  \bibinfo{author}{{Mignoli}, M.}, \bibinfo{author}{{Gilli}, R.},
  \bibinfo{author}{{De Barros}, S.}, \bibinfo{author}{{Caputi}, K.},
  \bibinfo{author}{{Grillo}, C.}, \bibinfo{author}{{Balestra}, I.},
  \bibinfo{author}{{Cristiani}, S.}, \bibinfo{author}{{Nonino}, M.},
  \bibinfo{author}{{Giallongo}, E.}, \bibinfo{author}{{Grazian}, A.},
  \bibinfo{author}{{Pentericci}, L.}, \bibinfo{author}{{Fontana}, A.},
  \bibinfo{author}{{Comastri}, A.}, \bibinfo{author}{{Vignali}, C.},
  \bibinfo{author}{{Zamorani}, G.}, \bibinfo{author}{{Brusa}, M.},
  \bibinfo{author}{{Bergamini}, P.}, \& \bibinfo{author}{{Tozzi}, P.}
  (\bibinfo{year}{2017}{\natexlab{b}}).
\newblock \bibinfo{title}{{Magnifying the Early Episodes of Star Formation:
  Super Star Clusters at Cosmological Distances}}.
\newblock {\it \bibinfo{journal}{\apj}\/},  {\it \bibinfo{volume}{842}\/},
  \bibinfo{pages}{47}. \DOIprefix\doi{10.3847/1538-4357/aa74ae}.
  \href{http://arxiv.org/abs/1703.02044}{\tt arXiv:1703.02044}.
\bibitem[{{Vergani} et~al.(2017){Vergani}, {Palmerio}, {Salvaterra}, {Japelj},
  {Mannucci}, {Perley}, {D'Avanzo}, {Kr{\"u}hler}, {Puech}, {Boissier},
  {Campana}, {Covino}, {Hunt}, {Petitjean} \& {Tagliaferri}}]{Vergani2017}
\bibinfo{author}{{Vergani}, S.~D.}, \bibinfo{author}{{Palmerio}, J.},
  \bibinfo{author}{{Salvaterra}, R.}, \bibinfo{author}{{Japelj}, J.},
  \bibinfo{author}{{Mannucci}, F.}, \bibinfo{author}{{Perley}, D.~A.},
  \bibinfo{author}{{D'Avanzo}, P.}, \bibinfo{author}{{Kr{\"u}hler}, T.},
  \bibinfo{author}{{Puech}, M.}, \bibinfo{author}{{Boissier}, S.},
  \bibinfo{author}{{Campana}, S.}, \bibinfo{author}{{Covino}, S.},
  \bibinfo{author}{{Hunt}, L.~K.}, \bibinfo{author}{{Petitjean}, P.}, \&
  \bibinfo{author}{{Tagliaferri}, G.} (\bibinfo{year}{2017}).
\newblock \bibinfo{title}{{The chemical enrichment of long gamma-ray bursts
  nurseries up to z = 2}}.
\newblock {\it \bibinfo{journal}{\aap}\/},  {\it \bibinfo{volume}{599}\/},
  \bibinfo{pages}{A120}. \DOIprefix\doi{10.1051/0004-6361/201629759}.
  \href{http://arxiv.org/abs/1701.02312}{\tt arXiv:1701.02312}.
\bibitem[{{Wanderman} \& {Piran}(2015)}]{Wanderman2015}
\bibinfo{author}{{Wanderman}, D.}, \& \bibinfo{author}{{Piran}, T.}
  (\bibinfo{year}{2015}).
\newblock \bibinfo{title}{{The rate, luminosity function and time delay of
  non-Collapsar short GRBs}}.
\newblock {\it \bibinfo{journal}{\mnras}\/},  {\it \bibinfo{volume}{448}\/},
  \bibinfo{pages}{3026--3037}. \DOIprefix\doi{10.1093/mnras/stv123}.
  \href{http://arxiv.org/abs/1405.5878}{\tt arXiv:1405.5878}.
\bibitem[{{Waxman} \& {Bahcall}(1997)}]{Waxman1997}
\bibinfo{author}{{Waxman}, E.}, \& \bibinfo{author}{{Bahcall}, J.}
  (\bibinfo{year}{1997}).
\newblock \bibinfo{title}{{High Energy Neutrinos from Cosmological Gamma-Ray
  Burst Fireballs}}.
\newblock {\it \bibinfo{journal}{Physical Review Letters}\/},  {\it
  \bibinfo{volume}{78}\/}, \bibinfo{pages}{2292--2295}.
  \DOIprefix\doi{10.1103/PhysRevLett.78.2292}.
  \href{http://arxiv.org/abs/astro-ph/9701231}{\tt arXiv:astro-ph/9701231}.
\bibitem[{{Willingale} \& {M{\'e}sz{\'a}ros}(2017)}]{Willingale2017}
\bibinfo{author}{{Willingale}, R.}, \& \bibinfo{author}{{M{\'e}sz{\'a}ros}, P.}
  (\bibinfo{year}{2017}).
\newblock \bibinfo{title}{{Gamma-Ray Bursts and Fast Transients.
  Multi-wavelength Observations and Multi-messenger Signals}}.
\newblock {\it \bibinfo{journal}{\ssr}\/},  {\it \bibinfo{volume}{207}\/},
  \bibinfo{pages}{63--86}. \DOIprefix\doi{10.1007/s11214-017-0366-4}.
\bibitem[{{Woosley}(1993)}]{Woosley1993}
\bibinfo{author}{{Woosley}, S.~E.} (\bibinfo{year}{1993}).
\newblock \bibinfo{title}{{Gamma-ray bursts from stellar mass accretion disks
  around black holes}}.
\newblock {\it \bibinfo{journal}{\apj}\/},  {\it \bibinfo{volume}{405}\/},
  \bibinfo{pages}{273--277}. \DOIprefix\doi{10.1086/172359}.
\bibitem[{{Yonetoku} et~al.(2014){Yonetoku}, {Mihara}, {Sawano}, {Ikeda},
  {Harayama}, {Takata}, {Yoshida}, {Seta}, {Toyanago}, {Kagawa}, {Kawai},
  {Kawai}, {Sakamoto}, {Serino}, {Kurosawa}, {Gunji}, {Tanimori}, {Murakami},
  {Yatsu}, {Yamaoka}, {Yoshida}, {Kawabata}, {Matsumoto}, {Tsumura},
  {Matsuura}, {Shirahata}, {Okita}, {Yanagisawa}, {Yoshida} \&
  {Motohara}}]{yonetoku14}
\bibinfo{author}{{Yonetoku}, D.}, \bibinfo{author}{{Mihara}, T.},
  \bibinfo{author}{{Sawano}, T.}, \bibinfo{author}{{Ikeda}, H.},
  \bibinfo{author}{{Harayama}, A.}, \bibinfo{author}{{Takata}, S.},
  \bibinfo{author}{{Yoshida}, K.}, \bibinfo{author}{{Seta}, H.},
  \bibinfo{author}{{Toyanago}, A.}, \bibinfo{author}{{Kagawa}, Y.},
  \bibinfo{author}{{Kawai}, K.}, \bibinfo{author}{{Kawai}, N.},
  \bibinfo{author}{{Sakamoto}, T.}, \bibinfo{author}{{Serino}, M.},
  \bibinfo{author}{{Kurosawa}, S.}, \bibinfo{author}{{Gunji}, S.},
  \bibinfo{author}{{Tanimori}, T.}, \bibinfo{author}{{Murakami}, T.},
  \bibinfo{author}{{Yatsu}, Y.}, \bibinfo{author}{{Yamaoka}, K.},
  \bibinfo{author}{{Yoshida}, A.}, \bibinfo{author}{{Kawabata}, K.},
  \bibinfo{author}{{Matsumoto}, T.}, \bibinfo{author}{{Tsumura}, K.},
  \bibinfo{author}{{Matsuura}, S.}, \bibinfo{author}{{Shirahata}, M.},
  \bibinfo{author}{{Okita}, H.}, \bibinfo{author}{{Yanagisawa}, K.},
  \bibinfo{author}{{Yoshida}, M.}, \& \bibinfo{author}{{Motohara}, K.}
  (\bibinfo{year}{2014}).
\newblock \bibinfo{title}{{High-z gamma-ray bursts for unraveling the dark ages
  mission HiZ-GUNDAM}}.
\newblock In {\it \bibinfo{booktitle}{Space Telescopes and Instrumentation
  2014: Ultraviolet to Gamma Ray}\/} (p. \bibinfo{pages}{91442S}).
\newblock volume \bibinfo{volume}{9144} of {\it \bibinfo{series}{\procspie}\/}.
\newblock \DOIprefix\doi{10.1117/12.2055041}.
  \href{http://arxiv.org/abs/1406.4202}{\tt arXiv:1406.4202}.
\bibitem[{{Yonetoku} et~al.(2004){Yonetoku}, {Murakami}, {Nakamura},
  {Yamazaki}, {Inoue} \& {Ioka}}]{Yonetoku2004}
\bibinfo{author}{{Yonetoku}, D.}, \bibinfo{author}{{Murakami}, T.},
  \bibinfo{author}{{Nakamura}, T.}, \bibinfo{author}{{Yamazaki}, R.},
  \bibinfo{author}{{Inoue}, A.~K.}, \& \bibinfo{author}{{Ioka}, K.}
  (\bibinfo{year}{2004}).
\newblock \bibinfo{title}{{Gamma-Ray Burst Formation Rate Inferred from the
  Spectral Peak Energy-Peak Luminosity Relation}}.
\newblock {\it \bibinfo{journal}{\apj}\/},  {\it \bibinfo{volume}{609}\/},
  \bibinfo{pages}{935--951}. \DOIprefix\doi{10.1086/421285}.
  \href{http://arxiv.org/abs/astro-ph/0309217}{\tt arXiv:astro-ph/0309217}.
\bibitem[{{Yoon} et~al.(2012){Yoon}, {Dierks} \& {Langer}}]{Yoon2012}
\bibinfo{author}{{Yoon}, S.-C.}, \bibinfo{author}{{Dierks}, A.}, \&
  \bibinfo{author}{{Langer}, N.} (\bibinfo{year}{2012}).
\newblock \bibinfo{title}{{Evolution of massive Population III stars with
  rotation and magnetic fields}}.
\newblock {\it \bibinfo{journal}{\aap}\/},  {\it \bibinfo{volume}{542}\/},
  \bibinfo{pages}{A113}. \DOIprefix\doi{10.1051/0004-6361/201117769}.
  \href{http://arxiv.org/abs/1201.2364}{\tt arXiv:1201.2364}.
\bibitem[{{Yoon} et~al.(2006){Yoon}, {Langer} \& {Norman}}]{Yoon2006}
\bibinfo{author}{{Yoon}, S.-C.}, \bibinfo{author}{{Langer}, N.}, \&
  \bibinfo{author}{{Norman}, C.} (\bibinfo{year}{2006}).
\newblock \bibinfo{title}{{Single star progenitors of long gamma-ray bursts. I.
  Model grids and redshift dependent GRB rate}}.
\newblock {\it \bibinfo{journal}{\aap}\/},  {\it \bibinfo{volume}{460}\/},
  \bibinfo{pages}{199--208}. \DOIprefix\doi{10.1051/0004-6361:20065912}.
  \href{http://arxiv.org/abs/astro-ph/0606637}{\tt arXiv:astro-ph/0606637}.
\bibitem[{{Zhang}(2013)}]{Zhang2013}
\bibinfo{author}{{Zhang}, B.} (\bibinfo{year}{2013}).
\newblock \bibinfo{title}{{Early X-Ray and Optical Afterglow of Gravitational
  Wave Bursts from Mergers of Binary Neutron Stars}}.
\newblock {\it \bibinfo{journal}{\apjl}\/},  {\it \bibinfo{volume}{763}\/},
  \bibinfo{pages}{L22}. \DOIprefix\doi{10.1088/2041-8205/763/1/L22}.
  \href{http://arxiv.org/abs/1212.0773}{\tt arXiv:1212.0773}.
\bibitem[{{Zhang} \& {M{\'e}sz{\'a}ros}(2001)}]{Zhang2001}
\bibinfo{author}{{Zhang}, B.}, \& \bibinfo{author}{{M{\'e}sz{\'a}ros}, P.}
  (\bibinfo{year}{2001}).
\newblock \bibinfo{title}{{Gamma-Ray Burst Afterglow with Continuous Energy
  Injection: Signature of a Highly Magnetized Millisecond Pulsar}}.
\newblock {\it \bibinfo{journal}{\apjl}\/},  {\it \bibinfo{volume}{552}\/},
  \bibinfo{pages}{L35--L38}. \DOIprefix\doi{10.1086/320255}.
  \href{http://arxiv.org/abs/astro-ph/0011133}{\tt arXiv:astro-ph/0011133}.

\end{thebibliography}

\end{document}